\documentclass[journal=iecred,manuscript=article,layout=twocolumn]{achemso}

\usepackage[version=3]{mhchem} 
\usepackage[svgnames]{pstricks}
\usepackage{pst-solides3d}
\usepackage{subcaption}
\usepackage{inputenc}
\usepackage{soul}
\usepackage{subcaption}
\DeclareUnicodeCharacter{00F8}{\o}
\DeclareUnicodeCharacter{00D8}{\O}
\DeclareUnicodeCharacter{2212}{-}
\DeclareUnicodeCharacter{2009}{ }
\DeclareUnicodeCharacter{03B3}{$\gamma$}
\DeclareUnicodeCharacter{221E}{$\infty$}
\DeclareUnicodeCharacter{1D31}{$^\mathrm{E}$}
\DeclareUnicodeCharacter{1AB2}{$^\infty$}
\SectionNumbersOn

\author{Pierre J. Walker}
\affiliation[Caltech]
{Division of Chemistry and Chemical Engineering, California Institute of Technology, Pasadena, California, USA}
\alsoaffiliation[ICL]{Department of Chemical Engineering, Imperial College London, SW7 2AZ, United Kingdom}
\email{pjwalker@caltech.edu}
\author{Simon Mueller}
\affiliation[Hamburg]
{Institute of Thermal Separation Processes, Hamburg University of Technology, Eißendorfer Straße 38, 21073 Hamburg, Germany}
\email{simon.mueller@tuhh.de}
\author{Irina Smirnova}
\affiliation[Hamburg]
{Institute of Thermal Separation Processes, Hamburg University of Technology, Eißendorfer Straße 38, 21073 Hamburg, Germany}
\email{irina.smirnova@tuhh.de}

\title{Confidence Interval and Uncertainty Propagation Analysis of SAFT-type Equations of State}

\keywords{Molecular Modelling, SAFT, Equation of State, Parameter Degeneracy}

\begin{document}

\begin{tocentry}
\centering
\includegraphics{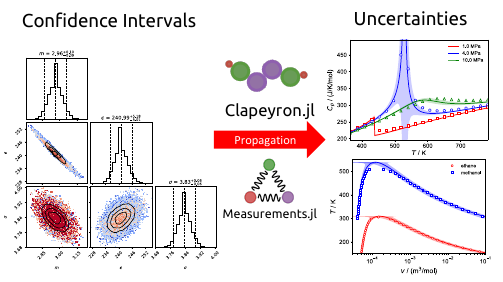}

\end{tocentry}

\begin{abstract}
  Thermodynamic models and, in particular, SAFT-type equations are vital in characterizing complex systems. This paper presents a framework for sampling parameter distributions in PC-SAFT and SAFT-VR Mie equations of state to understand parameter confidence intervals and correlations. We identify conserved quantities contributing to significant correlations. Comparing the equations of state, we find that additional parameters introduced in the SAFT-VR Mie equation increase relative uncertainties (1\%-2\% to 3\%-4\%) and introduce more correlations. When incorporating association through additional parameters, relative uncertainties increase, but correlations slightly decrease. We investigate how uncertainties propagate to derived properties and observe small uncertainties for that data with which the parameters were regressed, especially for saturated-liquid volumes. However, extrapolating to saturated-vapour volumes yields larger uncertainties due to the larger isothermal compressibility. Near the critical point, uncertainties in saturated volumes diverge due to increased sensitivity of the isothermal compressibility to parameter uncertainties. This effect significantly impacts bulk properties, particularly isobaric heat capacity, where uncertainties near the critical point become extremely large, even when these uncertainties are small. We emphasize that even small uncertainties near the critical point lead to divergences in predicted properties.
\end{abstract}

\section{Introduction}
Over the past three decades, the SAFT\cite{chapman_saft_1989, chapman_new_1990} (Statistical Associating Fluid Theory) equation of state has emerged as a powerful tool in the field of molecular simulations and thermodynamics\citep{kontogeorgisThermodynamicModelsIndustrial2010}. Since then it has undergone significant advancements, offering valuable insights into the behavior of complex fluids and materials leading to a whole family of equations, such as the Perturbed-Chain SAFT (PC-SAFT) equation\citep{grossPerturbedChainSAFTEquation2001,grossApplicationPerturbedChainSAFT2002} and SAFT with a variable range Mie potential (SAFT-VR Mie)\citep{lafitte_accurate_2013,dufal_saft_2015}. 

While the correlations of the pure component properties can be outstanding, the predictive application for components without enough thermodynamic data available to regress its pure component parameters can be challenging. Prior works have shown that these parameters can be correlated to molecular properties such as molecular weight\cite{albers_reducing_2012, albers_reducing_2012-1}. Some also propose approaches to predict the pure-component properties based on molecular simulation\cite{ferrando_prediction_2012}, from critical properties/acentric factor \cite{cismondi_rescaling_2005, privat_application_2019, anoune_new_2021}, information from quantum-chemistry calculations\cite{lucia_new_2009, leonhard_making_2007, leonhard_making_2007-1, singh_making_2007, van_nhu_quantum_2008, kaminski_sepp_2020, kaminski_quantum-mechanics-based_2019, mahmoudabadi_predictive_2021,walker_ab_2022} or machine learning\cite{matsukawa_estimation_2021, chungGroupContributionMachine2022, felton_ml-saft_2023, habicht_predicting_2023}.

However, even if the thermodynamic data are sufficient for a meaningful regression of the parameters, several parameter sets might lead to almost the same model performance with respect to the properties evaluated. This phenomenon, known as parameter degeneracy, is a problem for the consistent development of thermodynamic models that are compatible with each other. Even for equations of state with only two parameters like the Soave--Redlich--Kwong equation of state, a certain degeneracy might be observed\cite{swaminathan_evaluation_2006}. In the case of SAFT-based equations of state, just for water, several publications exist in which this has been examined in detail to try to select the best set of parameters\cite{clark_developing_2006, forte_multi-criteria_2018, graham_multi-objective_2022, klajmon_assessing_2023}. If one takes the usual average deviations of equations of state on pure-component densities and vapor pressures (0-2\%) and the value of the objective function in the case of water from \citet{clark_developing_2006}, a very large range of values for $\epsilon_{HB}$ and $\epsilon$ would lead to a similar model for the density and vapor pressure of water. Parameter degeneracy can also be observed in alkanes \cite{swaminathan_evaluation_2006, swaminathan_demonstrating_2007, de_villiers_evaluation_2013}, alkanols\cite{de_villiers_evaluation_2013} and ammonia\cite{mac_dowell_transferable_2011}. \citet{ramirez-velez_parametrization_2022} even showed that, for a large variety of associating components, the density and vapor pressure can be regressed without the need of an association term. Nevertheless, these degeneracies become more pronounced with flavors of SAFT which have a greater number of parameters\cite{dufal_prediction_2014, ramrattan_corresponding-states_2015, dufal_developing_2015, dufal_saft_2015, cripwell_saft-vr-mie_2018}. The problem of parameter degeneracy carries over on to the prediction of mixture thermodynamics as mixing rules are applied\cite{swaminathan_thermodynamic_2005,grenner_modeling_2007, swaminathan_demonstrating_2007,walkerAdvancedPredictiveMixing2022}. This leads to the need for different binary-interaction parameters ($k_{ij}$) for each combination of pure-component parameters. Of the prediction methods developed for $k_{ij}$ based on QSPR\cite{von_solms_capabilities_2006, stavrou_estimation_2016} or other binary thermodynamic data\cite{schacht_application_2010}, these would only be consistently applicable using the pure-component parameters used for their correlation.

In the literature, several strategies have been suggested to try to mitigate the problem of parameter degeneracy. When several local minima are present, instead of local solvers, solvers that search a larger parameter space have been used\cite{behzadi_application_2005, swaminathan_demonstrating_2007, klajmon_investigating_2020}. To explore a larger area of the parameter space and obtain a compromise between several properties, multiobjective optimisation\cite{swaminathan_demonstrating_2007, fuenzalida_improved_2016, rehner_multiobjective_2020, graham_multi-objective_2022} has also been applied. Others have tried exploring the impact of adding more pure-component thermodynamic data such as vapourisation enthalpies, heat capacities, speeds of sound and surface tensions to improve the quality of the fit\cite{kontogeorgisThermodynamicModelsIndustrial2010,oliveira_new_2016, ramirez-velez_parameterization_2020, rehner_multiobjective_2020}. Another approach used also includes adding mixture data to correlate the pure-component parameters and discriminate between model parameter sets leading to models that predict pure-component and mixture phase equilibria with high accuracy\cite{grandjean_application_2014, hutacharoen_predicting_2017, cripwell_saft-vr-mie_2018}. For more information of what has been tried in the past to combat parameter degeneracy, we invite the reader to examine the work of \citet{cripwell_saft-vr_2018}.

However, despite all these attempts to remedy the issue of parameter degeneracy, very few authors have considered the impact of these degeneracies on the validity of their model parameters. More specifically, rigorous confidence-interval analysis of the model parameters within SAFT equations of state has rarely been performed in literature. This is quite surprising considering how common a procedure this is in other fields such as chemical kinetics\citep{wangCombustionKineticModel2015,scaliaEvaluatingScalableUncertainty2020,heidCharacterizingUncertaintyMachine2023}. Some of the few examples in the literature include the work by \citet{kaminski_trade-off_2016} where the covariance matrix between SAFT parameters was determined \textit{a priori}, but they did not consider how to obtain confidence intervals from this. More recently, \citet{creton_assessment_2023} performed a holistic analysis of the model sensitivity on the parameters, which did provide a correlation matrix between parameters. However, as of yet, no one has devised a rigorous strategy to evaluate the confidence intervals of the SAFT parameters and, even more, no one has examined how the uncertainties in the parameters might then impact the predictions made using the models. It is these two objectives that we aim to satisfy within this work. We focus on both the PC-SAFT and SAFT-VR Mie equations of state, primarily due to the former's popularity and the fact that the latter requires two additional pure-component parameters. 

The remainder of this article is structured as follows. In section \ref{sect:methods}, we provide a high-level overview of the SAFT equations considered and their parameters, followed by a description of the methods used to perform confidence-interval and uncertainty-propagation analysis. In section \ref{sect:r&d}, we provide representative examples of the results obtained for non-associating and associating species when performing confidence-interval and uncertainty-propagation analysis. Finally, the results are summarised and potential future work is provided in section \ref{sect:conc}.


\begin{figure*}[h!]
  \centering
      \includegraphics[width=0.8\textwidth]{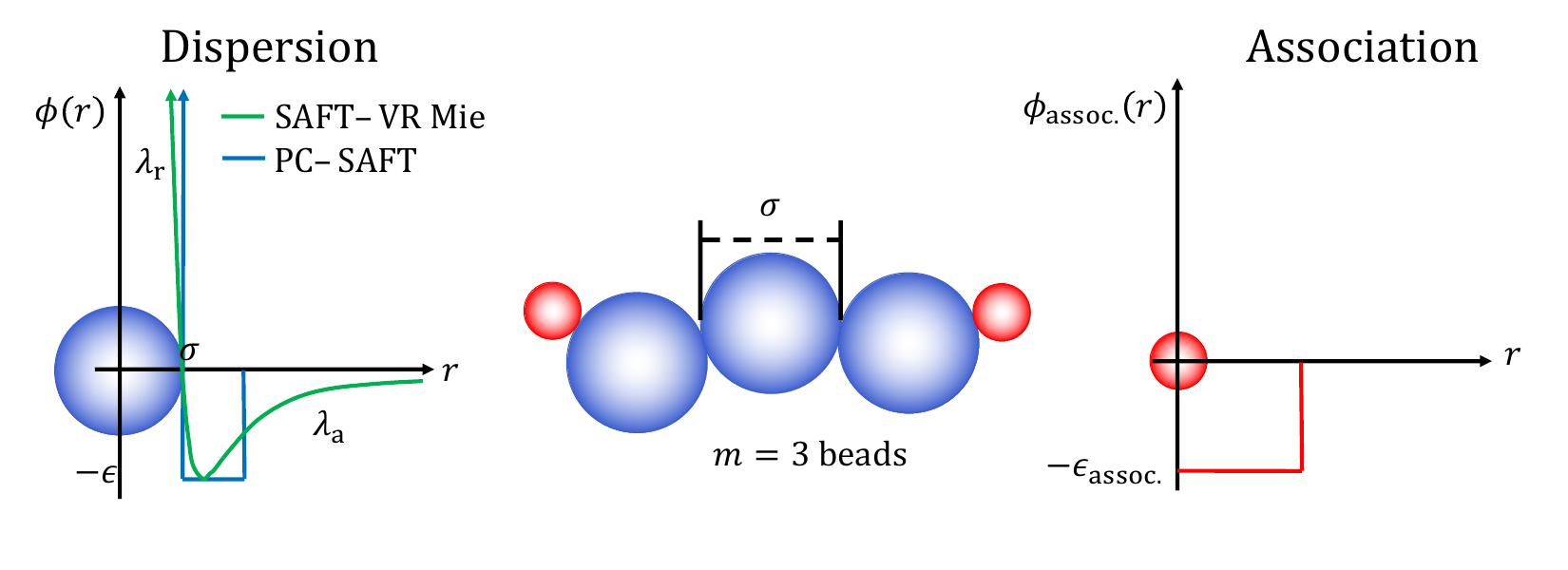}
      \caption{Visual illustration of the representation of molecules within the PC-SAFT and SAFT-VR Mie equations of state. Blue spheres indicate hard-spheres and red spheres indicate association sites.}
      \label{fig:saft_params}
  \end{figure*}


\section{Methods}
\label{sect:methods}
In this section, we will describe the model parameters used to characterise species within both the PC-SAFT and SAFT-VR Mie equation of state. Subsequently, we will describe the methodologies used to determine the confidence intervals. Finally, we will explain how uncertainties in the model parameters are propagated to the predicted properties. Implementations of equations of state and uncertainty propagation have been provided within the Clapeyron.jl\citep{walkerClapeyronJlExtensible2022} package.

\subsection{Model Parameters}

For the purposes of this work, we will not rigorously derive the PC-SAFT and SAFT-VR Mie equations of state (interested readers are directed to read the original works\citep{grossPerturbedChainSAFTEquation2001,grossApplicationPerturbedChainSAFT2002,lafitte_accurate_2013,dufal_saft_2015}). Instead, we provide a high-level overview of how species are represented within these approaches. 

Within both the PC-SAFT and SAFT-VR Mie approaches species are modelled as chains of hard-spheres of length $m$ where each sphere has a diameter $\sigma$. We note that, although when rigorously derived, both approaches would require that $m$ be an integer, in practise, this is left as freely adjustable (with the exception of species which should only have a single segment, such as the noble gases\citep{dufal_development_2013,walker_ab_2022}). These chains can interact through dispersive interactions characterised by an energetic parameter, $\epsilon$. Whilst the PC-SAFT equation of state assumes that the pair-wise potential takes the form of a square-well, in the SAFT-VR Mie equation of state a Mie potential is used. This introduces two additional parameters, $\lambda_\mathrm{a}$ and $\lambda_\mathrm{r}$, which are the attractive and repulsive exponents of the Mie potential between segments. Typically, to reduce the number of parameters to fit and its relationship to van der Waals dispersive interactions, the attractive exponent is often set \textit{a priori} to six.

Furthermore, both equations of state allow for the incorporation of association interactions between sites on species. To characterise this interaction, an additional energetic parameter, $\epsilon_\mathrm{assoc.}$, is introduced. However, the PC-SAFT and SAFT-VR Mie equations slightly differ when characterising the length scale of the associative interactions. In PC-SAFT, a dimensionless bonding volume (normalised by the segment diameter of the species), $\kappa$ is used, whilst, in SAFT-VR Mie, the bonding volume, $K$, carries real units. If one were to define $K=\kappa\sigma^3$ within the PC-SAFT equation of state, then we could treat this parameter as equivalent to the one used in the SAFT-VR Mie equation of state. Nevertheless, to remain consistent with the nomenclature used in both equations of state, we will use their respective parameter for the bonding volume.  A visual illustration of how species are represented in these equations is provided in figure \ref{fig:saft_params}.

At this stage, we highlight that, intuitively, some degeneracies between parameters can already be derived. For example, in the case of the chain length, $m$, and segment diameter, $\sigma$, we can imagine that the set of parameters which conserve a certain volume and/or area of the species would lead to a roughly equivalent representation. In addition, we can imagine that species such as water, which certainly experience both dispersive and associative interaction, will require fitting both $\epsilon$ and $\epsilon_\mathrm{assoc.}$. However, it is more-complicated to argue \textit{how strong} either of these interactions should be (for example, if we reduce $\epsilon_\mathrm{assoc.}$ slightly, we could compensate by increasing $\epsilon$). As such, we can already see that parameter degeneracies are almost inherent to these two equations of state.

\subsection{Confidence Interval Analysis}
At this stage, both the PC-SAFT and SAFT-VR Mie equation of state can be treated generally as a model where, for given set of inputs, $\mathbf{x}$, and parameters, $\boldsymbol{\theta}$, we can provide predictions of outputs, $\mathbf{y}$:
\begin{equation}
    \mathbf{y}=\mathbf{f}(\mathbf{x};\boldsymbol{\theta})\,,
\end{equation}
where $\mathbf{f}$ can be seen as our model (including both the equation of state and algorithms needed to solve for the outputs). However, in most cases, we must perform parameter estimation to estimate the values of $\boldsymbol{\theta}$. In this case, our objective is to maximise the likelihood of a set of parameters given the various inputs and outputs:
\begin{equation}
\boldsymbol{\theta}^*=\arg\max_{\boldsymbol{\theta}}p(\boldsymbol{\theta}|\mathbf{x},\mathbf{y})\,,
\end{equation}
where $\boldsymbol{\theta}^*$ is typically referred to as the maximum likelihood estimate. Unfortunately, it is almost impossible to derive a functional form for $p(\boldsymbol{\theta}|\mathbf{x},\mathbf{y})$. However, using Bayes' rule, we can re-arrange the above expression to give a more-workable form:
\begin{equation}
\label{eq:mle_max}
\boldsymbol{\theta}^*=\arg\max_{\boldsymbol{\theta}}\frac{p_{\boldsymbol{\theta}}(\boldsymbol{\theta})}{Z}p_\mathbf{y}(\mathbf{y}|\mathbf{x},\boldsymbol{\theta})\,,
\end{equation}
where $Z$ is a normalisation constant. In the case of $p_{\boldsymbol{\theta}}(\boldsymbol{\theta})$, which is the probability of a set of parameters $\boldsymbol{\theta}$ occurring, as we have no prior knowledge for this distribution, we simply set this quantity to unity. We shall treat $p_\mathbf{y}(\mathbf{y}|\mathbf{x},\boldsymbol{\theta})$ as a normally distributed multivariate function of dimension $K$:
\begin{equation}
    p_\mathbf{y}(\mathbf{y}|\mathbf{x},\boldsymbol{\theta})=\frac{\exp(-\frac{1}{2}(\mathbf{y}-\boldsymbol{\mu}_\mathbf{y})^T\mathbf{C}_\mathbf{y}^{-1}(\mathbf{y}-\boldsymbol{\mu}_\mathbf{y})}{\sqrt{(2\pi)^K|\det(\mathbf{C}_\mathbf{y})|}}\,,
\end{equation}
where $\mathbf{C}_\mathbf{y}$ is the covariance matrix corresponding to our experimental data. If our model is accurate enough, we would expect that the expected value for our output, $\boldsymbol{\mu}_\mathbf{y}$, to be equivalent to our model predictions ($\boldsymbol{\mu}_\mathbf{y}\approx \mathbf{f}(\mathbf{x}|\boldsymbol{\theta})$). This allows us to write:
\begin{equation}
    p_\mathbf{y}(\mathbf{y}|\mathbf{x},\boldsymbol{\theta})\approx\frac{\exp(-\frac{1}{2}(\mathbf{y}-\mathbf{f}(\mathbf{x};\boldsymbol{\theta}))^T\mathbf{C}_\mathbf{y}^{-1}(\mathbf{y}-\mathbf{f}(\mathbf{x};\boldsymbol{\theta}))}{\sqrt{(2\pi)^K|\det(\mathbf{C}_\mathbf{y})|}}\,.
\end{equation}
From here, we can convert equation \ref{eq:mle_max} to a minimisation by taking the negative logarithm:
\begin{equation}
\boldsymbol{\theta}^*=\arg\min_{\boldsymbol{\theta}}A+(\mathbf{y}-\mathbf{f}(\mathbf{x};\boldsymbol{\theta}))^T\mathbf{C}_\mathbf{y}^{-1}(\mathbf{y}-\mathbf{f}(\mathbf{x};\boldsymbol{\theta}))\,,
\end{equation}
where $A$ is some constant which does not depend on $\boldsymbol{\theta}$ and, as such, we will ignore for the remainder of this section. If all our experiments are independent, we would expect the covariance matrix, $\mathbf{C}_\mathbf{y}$ to be diagonal. As such, we could re-write the above equation as a more-familiar least-squares optimisation:
\begin{equation}
\label{eq:mle_simp}
    \boldsymbol{\theta}^*=\arg\min_{\boldsymbol{\theta}} \sum_i \frac{1}{\sigma_{i}^2}(y_i-f_i(x_i;\boldsymbol{\theta}))^2\,,
\end{equation}
where $\sigma_{i}$ is the standard deviation of output $y_i$ and the index $i$ denotes a specific experiment. Often, in regressing SAFT parameters, the prefactor $\frac{1}{\sigma_{i}^2}$ is ignored and replaced with a weighting factor. 

Equation \ref{eq:mle_simp} can be solved using a range of optimisation algorithms. Due to the non-convex nature of this objective function for equations of state\citep{dufal_development_2013,dufal_saft_2015}, in many cases, a combination of derivative-free and derivative-based algorithms is used to obtain the global minimum (or maximum likelihood estimate). 

This is the stage at which most authors stop and simply report the numerical values for their optimised parameters. As of yet, no one has taken the time to fully analyse the confidence intervals of the parameters they have just obtained. One of the objectives of this study is to perform this analysis for the PC-SAFT and SAFT-VR Mie equations of state. To do this, let us first define a new variable, $\chi^2$:
\begin{equation}
\label{eq:chi_squared}
    \chi^2(\boldsymbol{\theta}) = (\mathbf{y}-\mathbf{f}(\mathbf{x};\boldsymbol{\theta}))^T\mathbf{C}_\mathbf{y}^{-1}(\mathbf{y}-\mathbf{f}(\mathbf{x};\boldsymbol{\theta}))\,.
\end{equation}
In the case of linear models, the $\chi^2$ term will follow a $\chi^2$-distribution, allowing us to determine an upper bound for $\chi^2$ where the following will be true:
\begin{equation}
    p(\chi^2(\boldsymbol{\theta})\leq \chi^2_\mathrm{max}) = \alpha\,,
\end{equation}
where $\alpha$ is our degree of confidence (typically set to 95\%). This will result in hyperellipsoid confidence intervals, which will allow us to provide bounds for all our parameters, giving us uncertainties. This approach does not hold for non-linear models. In these cases, as is done in tools such as scipy\citep{2020SciPy-NMeth}, a linear approximation of $\chi^2$ is used where we Taylor expand around $\boldsymbol{\theta}^*$ (note that we ignore the first-order term as, given we are at the minimum, it will be equal to zero):
\begin{align}
    \chi^2 &\approx (\mathbf{y}-\mathbf{f}(\mathbf{x};\boldsymbol{\theta}^*))^T\mathbf{C}_\mathbf{y}^{-1}(\mathbf{y}-\mathbf{f}(\mathbf{x};\boldsymbol{\theta}^*))\\\nonumber
    &+2(\boldsymbol{\theta}-\boldsymbol{\theta}^*)^T\mathbf{S}^T_{\boldsymbol{\theta}}\mathbf{C}_\mathbf{y}^{-1}\mathbf{S}_{\boldsymbol{\theta}}(\boldsymbol{\theta}-\boldsymbol{\theta}^*)\,,
\end{align}
where $\mathbf{S}_{\boldsymbol{\theta}}$ are the sensitivities of our model to changes in $\boldsymbol{\theta}$:
\begin{figure}[h!]
  \centering
      \includegraphics[width=0.45\textwidth]{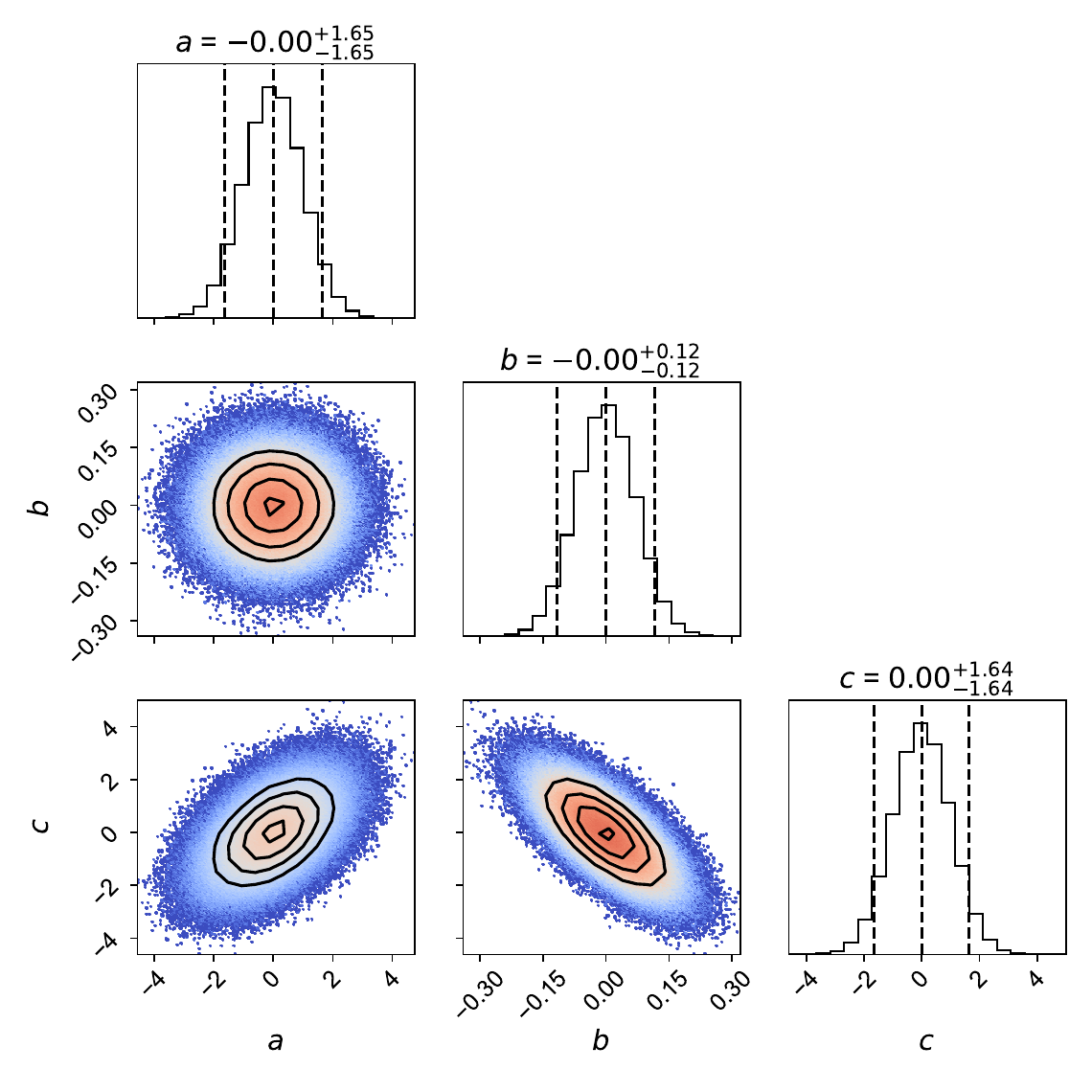}
      \caption{Example corner plot for normally distributed variables $a$, $b$, $c$. All have an expected value of zero. The covariance matrix was set up such that: $\mathrm{cov}[aa]=\mathrm{cov}[cc]=1$, $\mathrm{cov}[bb]=0.005$, $\mathrm{cov}[ac]=0.5$, $\mathrm{cov}[ab]=0$ and $\mathrm{cov}[bc]=-0.05$. Diagonal plots correspond to the distribution of the parameters, where the dashed lines correspond to the 95\% confidence interval (which is given in the titles of the figures). The off-diagonal plots correspond to the distribution between two parameters where the solid lines correspond to the confidence interval contours. The last contour corresponds to the 95\% confidence interval between the two parameters.}
      \label{fig:example_ci}
  \end{figure}
\begin{equation}
    \mathbf{S}_{\boldsymbol{\theta}} = \nabla_{\boldsymbol{\theta}}\mathbf{f}\,.
\end{equation}
Typically, a covariance matrix for the parameters $\boldsymbol{\theta}$ can be defined as:
\begin{equation}
    \mathrm{C}^{-1}_{\boldsymbol{\theta}}=2\mathbf{S}^T_{\boldsymbol{\theta}}\mathbf{C}_\mathbf{y}^{-1}\mathbf{S}_{\boldsymbol{\theta}}\,.
\end{equation}
From here, we can obtain the uncertainties for $\boldsymbol{\theta}$ assuming a $\chi^2$ distribution. This approach has been shown to be appropriate for a range of non-linear systems, assuming the objective function is convex\citep{vugrin_confidence_2007}. If the parameters are normally distributed, the confidence intervals should look something like those shown in figure \ref{fig:example_ci}. We can see that, in this case, parameters with narrow distributions have redder cores within their confidence intervals (such as the case of $b$). If two parameters are not correlated (such as $a$ and $b$), the distribution between these parameters is spherical. Conversely, if two parameters are correlated, they will have ellispoidal distributions where the slope of the ellipsoid is determined by the covariance between the two parameters. Ideally, if there are no degeneracies between our parameters, we would have perfectly spherical distributions with red cores.

However, in the case of equations of state, as mentioned in the introduction, there are many examples in the literature which highlight that the objective function is highly non-convex, often containing multiple minima\citep{dufal_prediction_2014}. Furthermore, due to the fact that iterative solvers are often needed to obtain the model outputs, it is essentially impossible to evaluate the parameter sensitivities without encountering large numerical errors. As such, an alternative strategy is required to determine the confidence intervals of our parameters. 
Within this work, we leverage Monte Carlo Markov-Chain (MCMC) simulations to sample the distribution $p(\boldsymbol{\theta}|\mathbf{x},\mathbf{y})$. 

As a first step, we will optimise a set of parameters to obtain the maximum likelihood estimate ($\boldsymbol{\theta}^*$) using equation \ref{eq:mle_simp}. Consistent with most other parameterisation of SAFT equations, we will use experimental saturation pressures and saturated-liquid densities (our outputs, $\mathbf{y}_i$) over a range of temperature (our inputs, $\mathbf{x}_i$), from either the triple point or 30\% of the critical temperature, to 90\% of the critical temperature. To examine the `natural' uncertainties of these parameters, we will use high-accuracy equations of state\citep{kunz_gerg-2008_2012} to generate pseudo-experimental data. For both saturation pressures and saturated-liquid densities, the relative uncertainties of these properties have been set to 0.1\%\citep{kunz_gerg-2008_2012} (or $\sigma_i\approx 0.001y_i$), allowing us to write our objective function as (factoring our constant pre-factors):
\begin{equation}
    \label{eq:mle_simp_rel}
    \boldsymbol{\theta}^*=\arg\min_{\boldsymbol{\theta}} \sum_i \left(\frac{y_i-f_i(x_i;\boldsymbol{\theta})}{y_i}\right)^2\,.
\end{equation}
We solve equation \ref{eq:mle_simp_rel} using the Evolutionary Centers algorithm\citep{mejia-de-diosNewEvolutionaryOptimization2019} provided by the Metaheuristics.jl package\citep{mejia-de-diosMetaheuristicsJuliaPackage2022}. 

Subsequently, we use the Metropolis-Hastings algorithm to perform the MCMC simulations. Using 1000 samples run in parallel, all initialised at a value of $\boldsymbol{\theta}_0=\boldsymbol{\theta}^*$, we perform the following steps at each iteration, $i$:
\begin{enumerate}
    \item Propose a $\boldsymbol{\theta}'$ sampled from a normal distribution $\mathcal{N}(\boldsymbol{\theta}_i,\boldsymbol{\sigma}_\theta)$. 
    \item Evaluate $\chi^2(\boldsymbol{\theta}')$ using equation \ref{eq:chi_squared}. Assuming that $\chi^2$ is normally distributed, evaluate the ratio:
    \begin{equation}
        \alpha = \frac{\exp(-\chi^2(\boldsymbol{\theta}'))}{\exp(-\chi^2(\boldsymbol{\theta}_i))}\,.
    \end{equation}
    \item If $\alpha>1$, we accept $\boldsymbol{\theta}'$ and set $\boldsymbol{\theta}_{i+1}=\boldsymbol{\theta}'$. Otherwise, we accept $\boldsymbol{\theta}'$ with probability $\alpha$ by sampling from a uniform distribution.
\end{enumerate}
The above steps are repeated 1000 times. To allow for sufficient equilibration, we ignore the first 100 steps of all samples. Furthermore, to ensure that we sample a sufficiently large number of values for $\boldsymbol{\theta}$, we set $\boldsymbol{\sigma}_\theta=c\boldsymbol{\theta}$ where we find that setting $c=0.04$ allows for an acceptance rate of approximately 10\%--20\%. After a sufficient number of steps, we will have obtained a distribution for $\boldsymbol{\theta}$ sampled from $p(\boldsymbol{\theta}|\mathbf{x},\mathbf{y})$, from which we can directly evaluate the 95\% confidence interval. We will also be able to visualise the cross-correlation between parameters to determine whether we recover the typical hyperellipsoidal confidence intervals.
\subsection{Uncertainty Propagation}
To date, no prior studies incorporate an examination of how uncertainties in the model parameters of SAFT equations influence the accuracy of the predictions. One of the reasons for this is the fact that the propagation of uncertainties through these equations would require significant effort to perform an examination analytically. According the to linear error propagation theory\citep{taylorIntroductionErrorAnalysis1996}, for a given function $g(\mathbf{x})$, if our inputs carry some uncertainty $\Delta\mathbf{x}$, then the uncertainty in our output, $\Delta g$, can be written as:
\begin{equation}
    \Delta g = |\nabla g \odot \Delta\mathbf{x}|\,.
\end{equation}

\begin{figure*}[h!]
  \centering
    \begin{subfigure}[b]{0.49\textwidth}
      \includegraphics[width=1\textwidth]{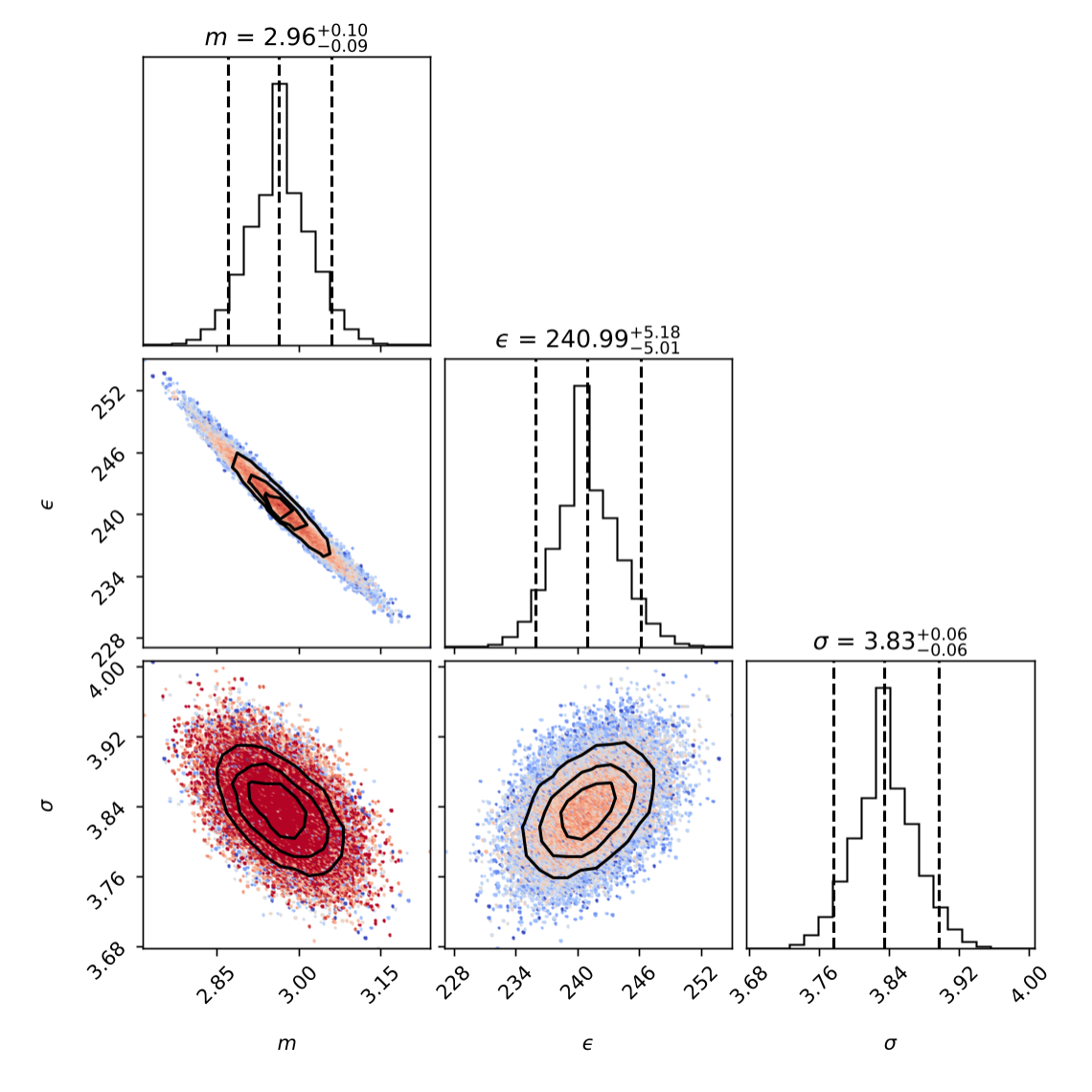}
      \caption{PC-SAFT}
      \label{fig:hexane_pc}
    \end{subfigure}
    \begin{subfigure}[b]{0.49\textwidth}
      \includegraphics[width=1\textwidth]{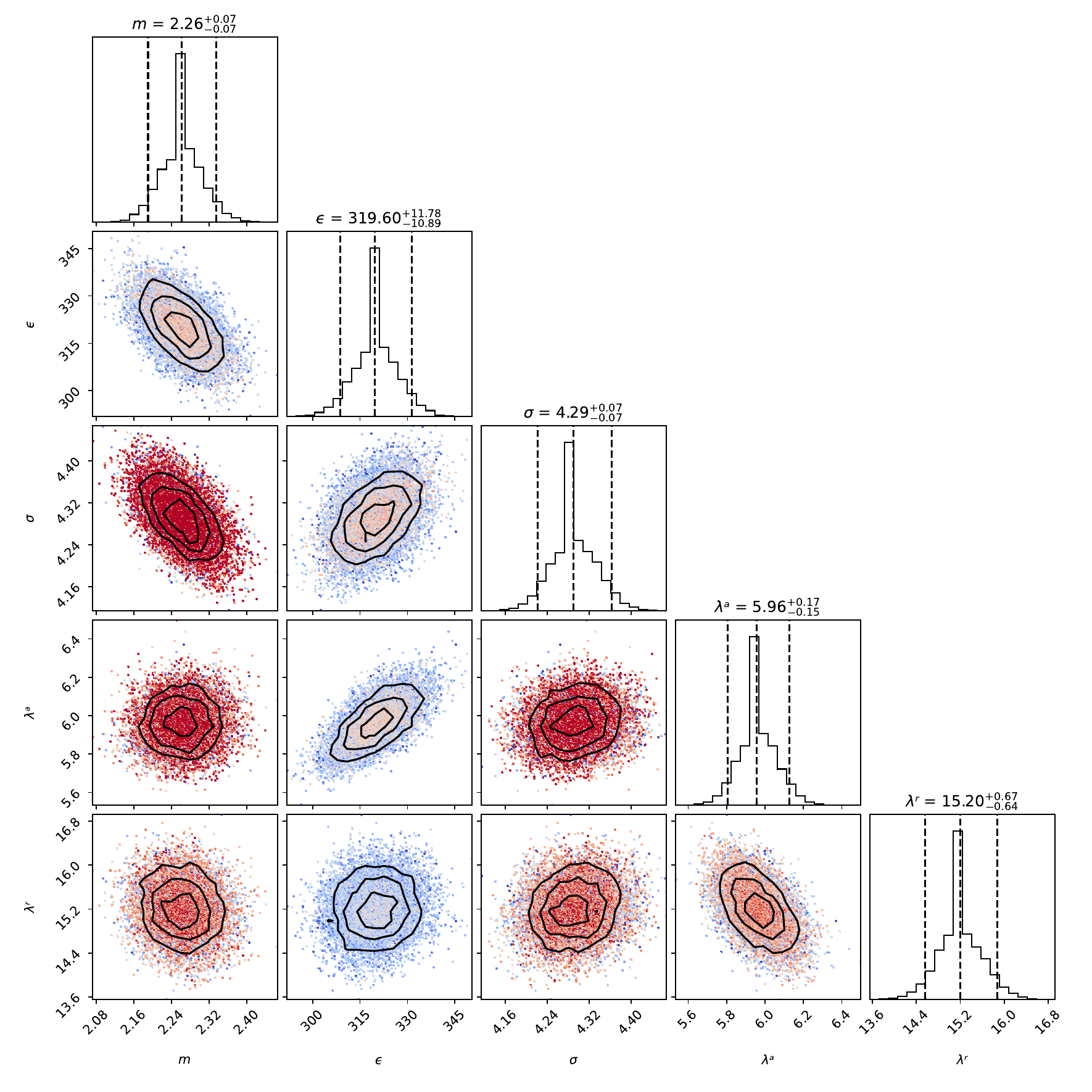}
      \caption{SAFT-VR Mie}
      \label{fig:hexane_vr_mie}
    \end{subfigure}
    \caption{Confidence intervals obtained for the pure-component parameters of $n$-hexane in PC-SAFT and SAFT-VR Mie. Colors and styles are identical to figure \ref{fig:example_ci}.}
    \label{fig:ci_results}
  \end{figure*}
  
As mentioned previously, evaluating the derivative of the outputs from equations of state with respect to model parameters is very challenging due to the complexities of these equations and the iterative solvers involved. However, Measurements.jl\citep{giordano_uncertainty_2016} is a package that allows for the automated propagation of uncertainties and is compatible with Clapeyron.jl\citep{walkerClapeyronJlExtensible2022}. This compatibility arises as a result of the multiple-dispatch paradigm in the Julia coding language.

Unfortunately, the introduction of uncertainties to the model parameters greatly increases the computational cost of evaluating properties. As such, this feature is left as a conditional extension of the base Clapeyron.jl package in version 0.5.0.

\section{Results and Discussions}
\label{sect:r&d}
Within this work, we have performed both confidence-interval and uncertainty-propagation analysis of a range of species in both the PC-SAFT and SAFT-VR Mie equations of state. We considered species from the alkane homologous series (methane to $n$-decane), carbon dioxide and argon so that we could examine the case in which we only have dispersive interactions. We also considered both water and methanol to examine the impact of introducing two additional parameters to account for associative interactions. However, including all of these results within this article would be very repetitive given many of the figures generated for these species are qualitatively similar. As such, most of the results will be provided in the supplementary information. Within this article, we present the results for $n$-hexane and methanol, as these proved to be sufficiently representative of the results obtained in all other species we considered.  
\subsection{Confidence Interval Analysis}
\label{sect:ci}
\paragraph{Non-associating species}
The confidence intervals obtained for $n$-hexane in both the PC-SAFT and SAFT-VR Mie equations of state are shown in figure \ref{fig:ci_results}. Let us first consider the case with the PC-SAFT equation of state in which we have the fewest parameters (figure \ref{fig:hexane_pc}). In the case of the individual parameters, we note that the optimised parameters are almost identical to the ones obtained by \citet{grossPerturbedChainSAFTEquation2001} (which is the case for most of species). We can see that the distributions of these parameters are close to symmetric, with $\epsilon$ being the only exception as it is slightly skewed towards larger values. This most likely as a result of the fact that, given the theoretical critical temperature given by equations of state scales with $\epsilon$, if this parameter is underestimated, there is a chance that the predicted critical point will fall below the highest temperature in our experimental data. This behaviour is observed in all other figures shown here (for both equations of state and even when association is introduced). Nevertheless, the skeweness of this parameter is quite minor and could safely still be treated as symmetric. We also note that, in terms of relative uncertainties, most of these parameters have a value of about 2\%--3\% (which is consistent for all species considered). This error is larger than the uncertainties of the experimental data used, highlighting that they arise primarily from inherent degeneracies within the PC-SAFT equation of state. 

These degeneracies can be better observed by examining the off-diagonal plots in figure \ref{fig:hexane_pc}. At the outset, we highlight that these distributions are all ellipsoidal, indicating that it is likely that a linearised approximation of the $\chi^2(\boldsymbol{\theta})$ function may be appropriate. Unsurprisingly, there is indeed a negative correlation between the segment length, $m$, and bead size $\sigma$ where, as mentioned previously, to converge the species volume ($\mathcal{V}\sim m\sigma^3$), it should be possible to vary both parameters in opposite directions. However, what is quite surprising is that, whilst we previously discussed that a red core within these distributions would be a good thing, in this case, the entire distribution is red, highlighting that the entire distribution is almost evenly sampled with a similar level of intensity, with the exceptions of the outskirts of the distribution. This would mean that parameters within this space could be treated as almost equivalent in terms of representing $n$-hexane. As such, this figure highlights that the segment length and bead size are \textit{highly} degenerate (consistent with other studies\citep{dufal_prediction_2014}). This is an observation that has been made for all species, regardless of equation of state or whether or not association is included.

In addition, a slightly surprising correlation is the one between the segment length, $m$ and the potential depth, $\epsilon$, where the two parameters have a strongly negative correlation. One potential explanation for where this correlation arises from is due to the fact that $\epsilon$ is the depth of the pair-wise potential well between two segments, not the whole molecule. As such, another conserved quantity within the PC-SAFT equation would be a sort of `total interaction energy':
\begin{equation}
    \mathcal{E}\sim m\epsilon\,.
\end{equation}
This conserved quantity would lead to the negative correlation between the two parameters. Furthermore, both quantities, $\mathcal{V}$ and $\mathcal{E}$ appear within the PC-SAFT equation state (within the dispersion term). Interestingly, this correlation is not as degenerate as the one between $m$ and $\sigma$. This is perhaps due to the fact that $\epsilon$ is more-strongly constrained by the experimental data used, due to its relationship with the critical temperature. This highlights the importance of using experimental data close to the critical temperature. Finally, the last correlation to consider is between $\sigma$ and $\epsilon$, where we observe a positive correlation. This is most likely due to the fact that, with the two preserved quantities we've defined previously, we can define a new one as:
\begin{equation}
    \mathcal{P} = \frac{\mathcal{E}}{\mathcal{V}} \sim \frac{\epsilon}{\sigma^3}\,,
\end{equation}
which would lead to the positive correlation observed. Interestingly, the above analysis seems to hint that perhaps we are optimising the wrong parameters. A less degenerate set of parameters might be obtained by fitting the preserved quantities ($\mathcal{V}$ and $\mathcal{E}$) and a third parameter. It would be interesting to see if fitting these quantities instead would remove some of the correlations observed.

Subsequently, we consider the case of parameterising $n$-hexane within the SAFT-VR Mie equation of state, where we introduce two new parameters, $\lambda_\mathrm{a}$ and $\lambda_\mathrm{r}$. These results are shown within figure \ref{fig:hexane_vr_mie}. Initially, most of the observations made for $m$, $\epsilon$, and $\sigma$ remain true here. Examining the individual parameters, most of the distributions remain symmetrical, with the relative uncertainties now ranging from 2\%--4\%, which is to be expected with a larger number of parameters. Interestingly, both $\lambda_\mathrm{a}$ and $\lambda_\mathrm{r}$ are slightly skewed to larger values. In the case of $\lambda_\mathrm{r}$, this can be rationalised given that it must always obey the relationship:
\begin{equation}
\label{eq:lambda}
    \lambda_\mathrm{r}>\lambda_\mathrm{a}\,.
\end{equation}
\begin{figure}[h!]
  \centering
      \includegraphics[width=0.5\textwidth]{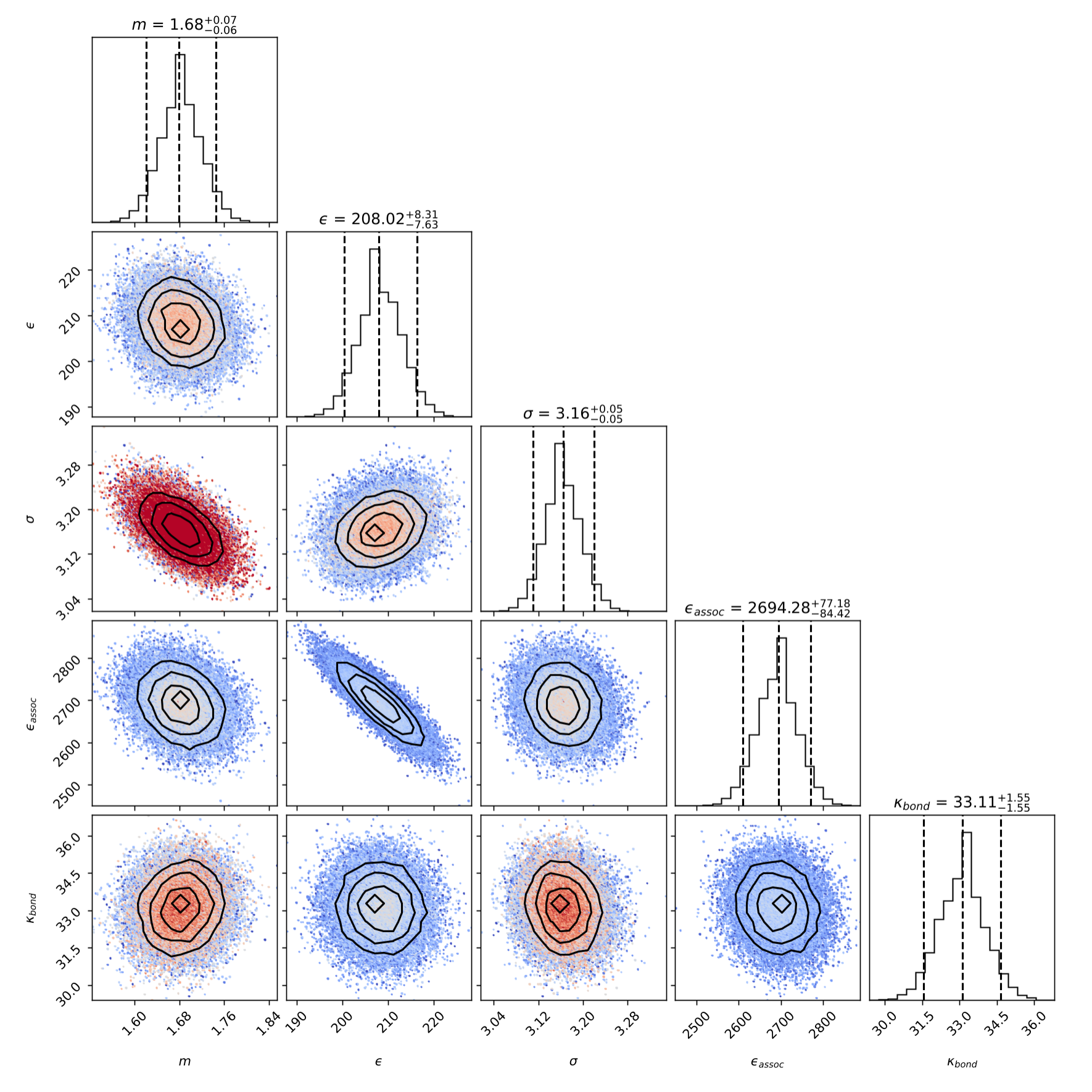}
      \caption{Confidence intervals obtained for the pure component parameters of methanol in PC-SAFT. Colors and styles are identical to figure \ref{fig:example_ci}.}
      \label{fig:methanol_pc}
  \end{figure}
As such, it is naturally biased to larger values. On the other hand, $\lambda_\mathrm{a}$ being skewed towards larger values is more-difficult to rationalise. One potential reason could be that, as we are representing dispersive attractions with this exponent, these must be short-range interactions which will generally favour larger exponents (if we have $\lambda_\mathrm{a}$ closer to four, this would be closer to dipolar interactions, which are not present within this species). Interestingly, the correlation between $\lambda_\mathrm{a}$ and $\lambda_\mathrm{r}$ is actually negative. Whilst the skeweness of the individual distributions doesn't directly relate to the correlation between parameters, this does hint towards another conserved quantity we have not observed yet. Indeed, examining the SAFT-VR Mie equation itself, a common quantity is the van-der Waals integral, given by:
\begin{align}
\label{eq:vdw}
    -\frac{\psi}{4\pi} &= \int_\sigma^\infty\phi(r)r^2dr \\ 
    &= \epsilon\sigma^3 \mathcal{C}\frac{(\lambda_\mathrm{r}-\lambda_\mathrm{a})}{(\lambda_\mathrm{r}-3)(\lambda_\mathrm{a}-3)}\,,\nonumber
\end{align}
where:
\begin{equation}
    \mathcal{C} = \left(\frac{\lambda_{\mathrm{r}}}{\lambda_{r}-\lambda_{\mathrm{a}}}\right)\left(
    \frac{\lambda_{r}}{\lambda_{\mathrm{a}}}\right)^{\frac{\lambda_{\mathrm{a}}}{\lambda_{\mathrm{r}}-\lambda_{\mathrm{a}}}}\,.
\end{equation}
What may not be too apparent in the above expressions is that treating $\psi$ as a conserved quantity results in a negative correlation between $\lambda_\mathrm{a}$ and $\lambda_\mathrm{r}$, thus explaining the observations made in relation to figure \ref{fig:hexane_vr_mie}.

Most of the correlations between $\lambda_\mathrm{a}$ or $\lambda_\mathrm{r}$ and the other parameters are quite weak, highlighting that, for the most part, these can be optimised independently. An exception to this is the strong positive correlation between $\lambda_\mathrm{a}$ and $\epsilon$. Thankfully, this too can be explained using equation \ref{eq:vdw} where this also results in an asymptotic positive relationship between $\lambda_\mathrm{a}$ and $\epsilon$. However, it also results in an asymptotic positive correlation between $\lambda_\mathrm{r}$ and $\epsilon$. This does not manifest itself within figure \ref{fig:hexane_vr_mie} because of equation \ref{eq:lambda} where we are already in the region where this positive correlation has effectively becomes negligible. 

Overall, in increasing the number of parameters grom from the PC-SAFT to the SAFT-VR Mie equation of state, we have not significantly increase the relative uncertainties of parameters but have introduced new correlations between parameters, worsening the degeneracy of the parameter space.
\paragraph{Associating species}
Now considering an associating species, methanol, in conjunction with the PC-SAFT equation of state, we obtain the distributions shown in figure \ref{fig:methanol_pc}. Once again, in the main, the correlations observed for $n$-hexane carry over here, particularly for $m$, $\epsilon$ and $\sigma$. In terms of the individual parameters, the distributions remain mostly symmetric, with the noteworth exceptions of $\epsilon$ and $\epsilon_\mathrm{assoc.}$ which are skewed towards larger values. The reason for this is likely similar to that of $n$-hexane where these two parameters scale with the critical temperature of the species. Unfortunately, the relative uncertainties of the parameters has increased to approximately 2\%--5\%.

  \begin{figure*}[h!]
  \centering
      \includegraphics[width=0.9\textwidth]{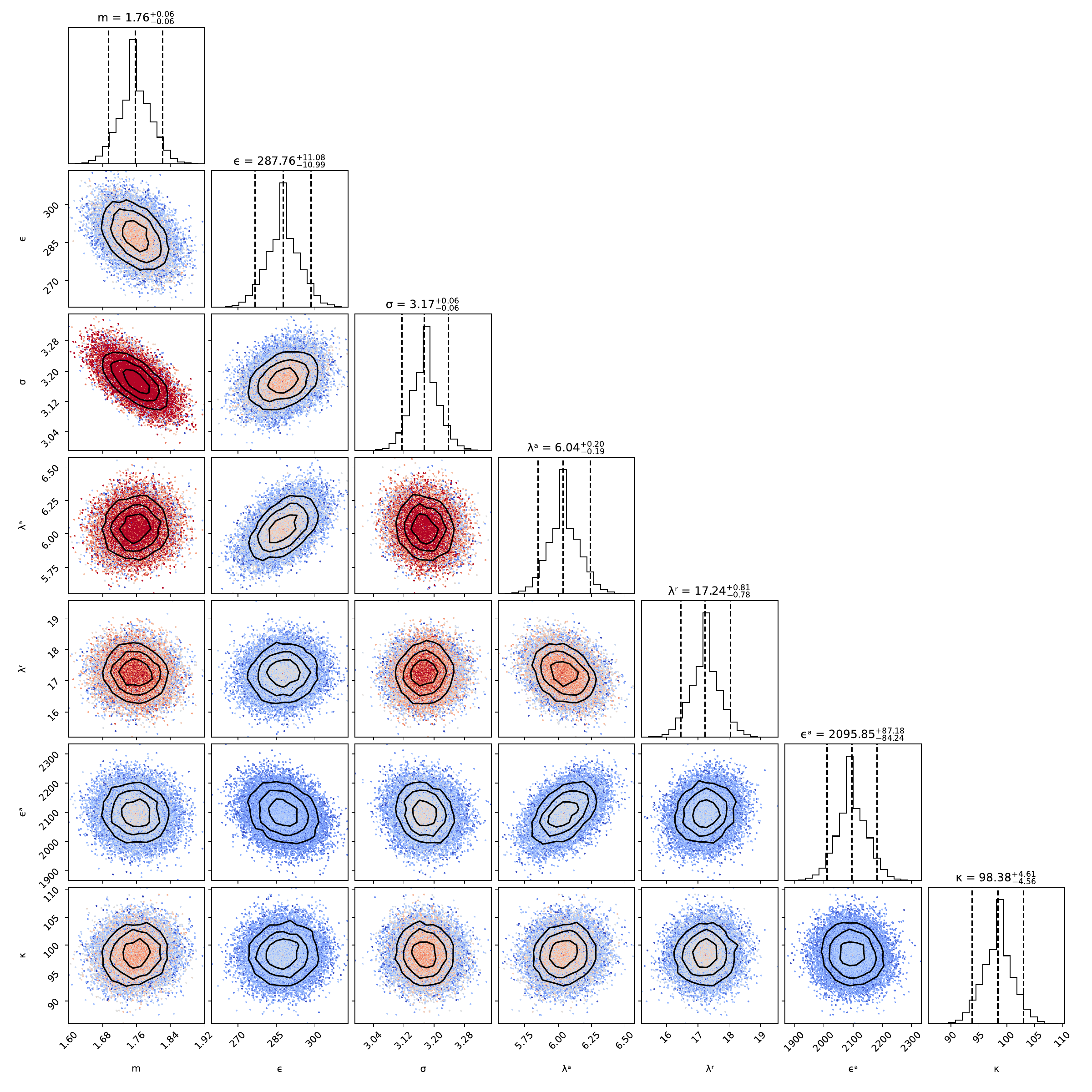}
      \caption{Confidence intervals obtained for the pure component parameters of methanol in SAFT-VR Mie. Colors and styles are identical to figure \ref{fig:example_ci}.}
      \label{fig:methanol_vr}
  \end{figure*}

Interestingly, the correlation between $\epsilon$ and $m$, while still present, is less significant than it was in the case of $n$-hexane. One potential explanation for this might be the emergence of the strong negative correlation between $\epsilon$ and $\epsilon_\mathrm{assoc.}$. We had anticipated these correlations as we can compensate for a weaker dispersive interaction with a stronger associative interaction, and \textit{vice versa}. This has also been observed in prior works\citep{dufal_saft_2015}. This strong correlation means that the correlation between $\epsilon$ and $m$ has now been weakened as $\epsilon_\mathrm{assoc.}$ can compensate for any changes in $\epsilon$. We can see this has resulted in a slight negative correlation between $m$ and $\epsilon_\mathrm{assoc.}$ for a similar reason. 

Furthermore, the dimensionless bonding volume, $\kappa_\mathrm{bond}$, does not appear to be strongly correlated within any of the other variables. This is partly to be expected as its role within the association term is to account for the length scale / steric effects of the associative interactions, which should be independent of the association energy, $\epsilon_\mathrm{assoc.}$.  

\begin{figure*}[h!]
  \centering
    \begin{subfigure}[b]{0.49\textwidth}
      \includegraphics[width=1\textwidth]{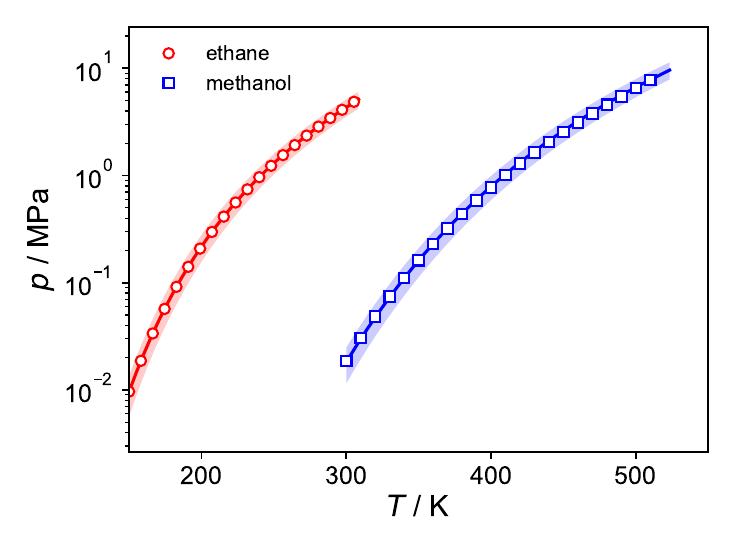}
      \caption{Saturation Curve (PC-SAFT)}
      \label{fig:pc_psat_error}
    \end{subfigure}
    \begin{subfigure}[b]{0.49\textwidth}
      \includegraphics[width=1\textwidth]{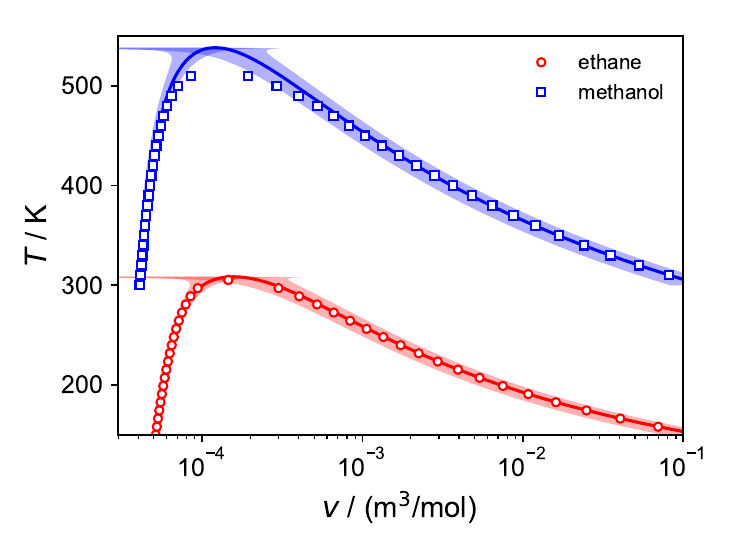}
      \caption{Saturation Envelope (PC-SAFT)}
      \label{fig:pc_rhosat_error}
    \end{subfigure}
    \begin{subfigure}[b]{0.49\textwidth}
      \includegraphics[width=1\textwidth]{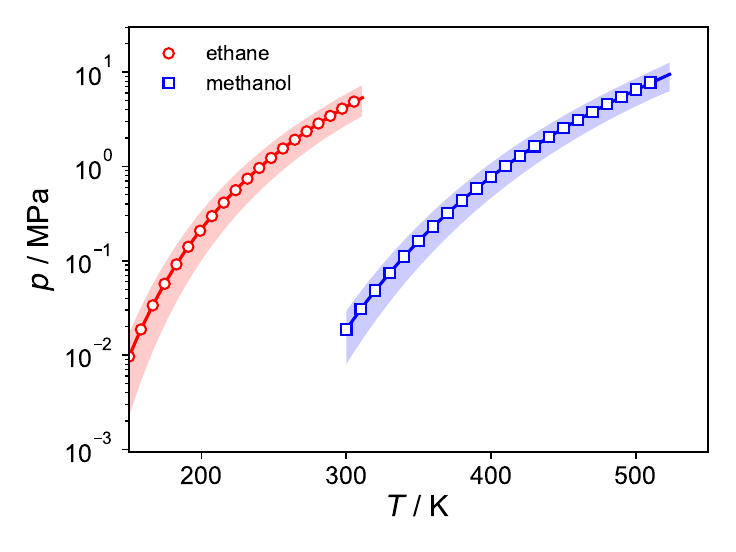}
      \caption{Saturation Curve (SAFT-VR Mie)}
      \label{fig:vr_psat_error}
    \end{subfigure}
    \begin{subfigure}[b]{0.49\textwidth}
      \includegraphics[width=1\textwidth]{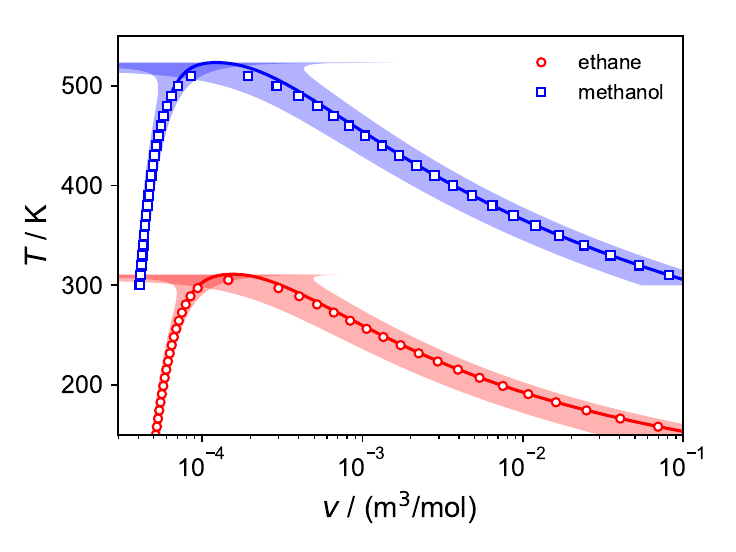}
      \caption{Saturation Envelope (SAFT-VR Mie)}
      \label{fig:vr_rho_sat_error}
    \end{subfigure}
    \caption{Predicted values for the saturated volumes (\ref{fig:pc_rhosat_error} and \ref{fig:vr_rho_sat_error}) and saturation pressure (\ref{fig:pc_psat_error} and \ref{fig:vr_psat_error}) for ethane (red) and methanol (blue) using PC-SAFT (\ref{fig:pc_rhosat_error} and \ref{fig:pc_psat_error}) and SAFT-VR Mie (\ref{fig:vr_rho_sat_error} to \ref{fig:vr_psat_error}). Shaded regions correspond to the uncertainty interval for the predicted properties using the parameters and confidence intervals obtained in section \ref{sect:ci}.}
    \label{fig:ep_results_sat}
  \end{figure*}

If we now consider methanol in conjunction with the SAFT-VR Mie equation of state, where we now have seven parameters total, we obtain the distributions shown in figure \ref{fig:methanol_vr}. Whilst this figure is quite intimidating, much of the behaviour we observed previously carries over. Once again, most of the parameters have symmetric distributions with some expected exceptions. Unsurprisingly, having the most parameters of any model considered here, the relative uncertainties are larger than they were in the case of $n$-hexane, being in the range of 3\%--6\%. Interestingly, as it was in the case of methanol with the PC-SAFT equation of state, due to the increased number of parameters, the correlations between most parameters has weakened. A clear example is the correlation between $\lambda_\mathrm{a}$ and $\lambda_\mathrm{r}$ which is still present, but not as significant as it was in figure \ref{fig:hexane_pc}. We can also see that the correlation between $\lambda_\mathrm{a}$ and $\epsilon$ has partially transferred to the correlation between $\lambda_\mathrm{a}$ and $\epsilon_\mathrm{assoc.}$. 

Overall, in introducing association interactions, we slightly reduce the correlation between parameters at the cost of increasing the relative uncertainty of the parameters.

\subsection{Uncertainty Propagation}
\paragraph{Equilibirum properties}
\begin{figure*}[h!]
  \centering
    \begin{subfigure}[b]{0.32\textwidth}
      \includegraphics[width=1\textwidth]{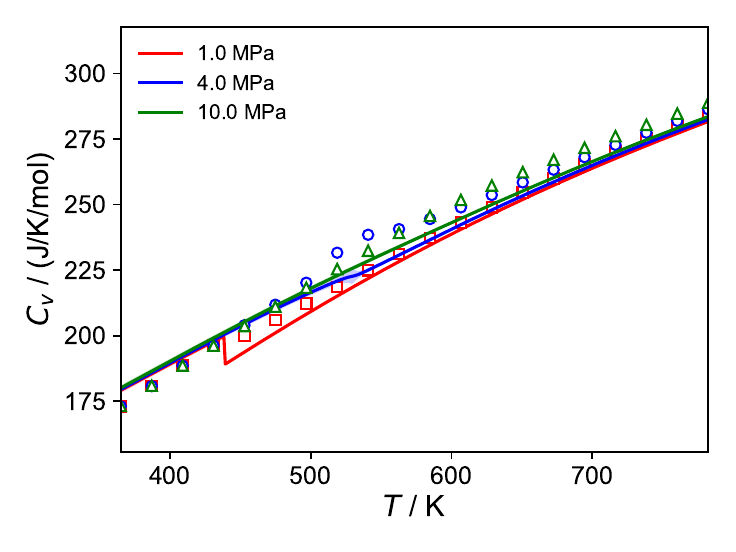}
      \caption{\centering Isochoric Heat Capacity (PC-SAFT)}
      \label{fig:pc_cv_error}
    \end{subfigure}
    \begin{subfigure}[b]{0.32\textwidth}
      \includegraphics[width=1\textwidth]{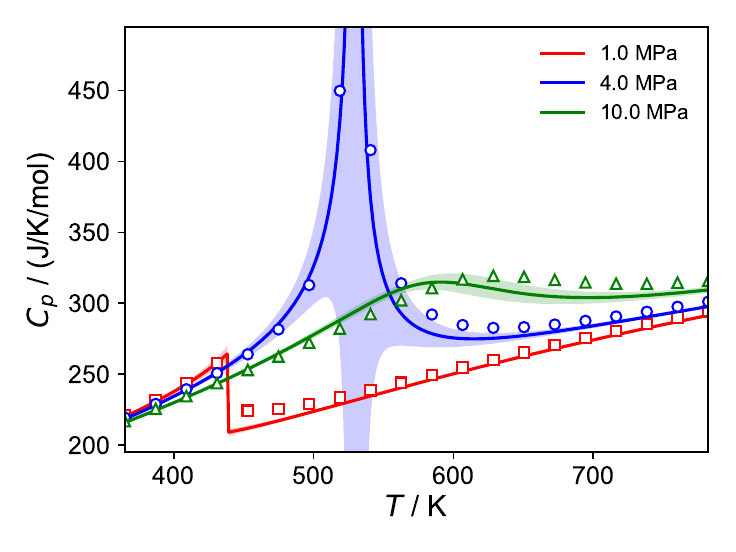}
      \caption{\centering Isobaric Heat Capacity (PC-SAFT)}
      \label{fig:pc_cp_error}
    \end{subfigure}
    \begin{subfigure}[b]{0.32\textwidth}
      \includegraphics[width=1\textwidth]{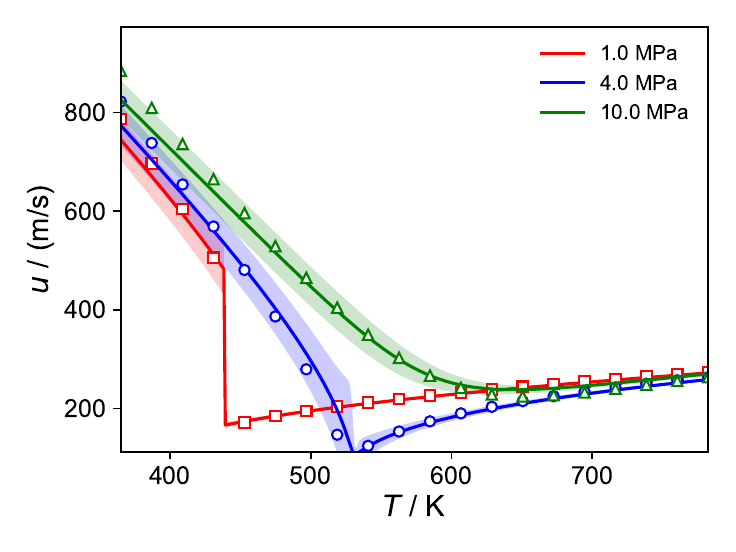}
      \caption{\centering Speed of Sound \quad\quad (PC-SAFT)}
      \label{fig:pc_sos_error}
    \end{subfigure}
    \begin{subfigure}[b]{0.32\textwidth}
      \includegraphics[width=1\textwidth]{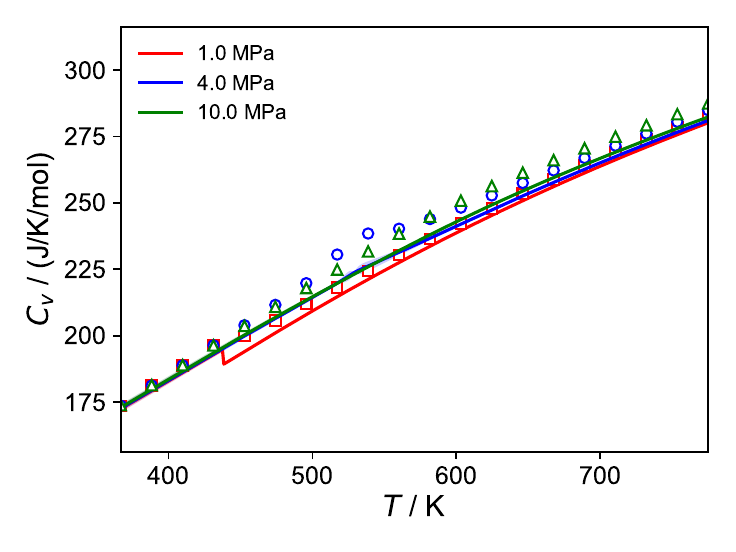}
      \caption{\centering Isochoric Heat Capacity (SAFT-VR Mie)}
      \label{fig:vr_cv_error}
    \end{subfigure}
    \begin{subfigure}[b]{0.32\textwidth}
      \includegraphics[width=1\textwidth]{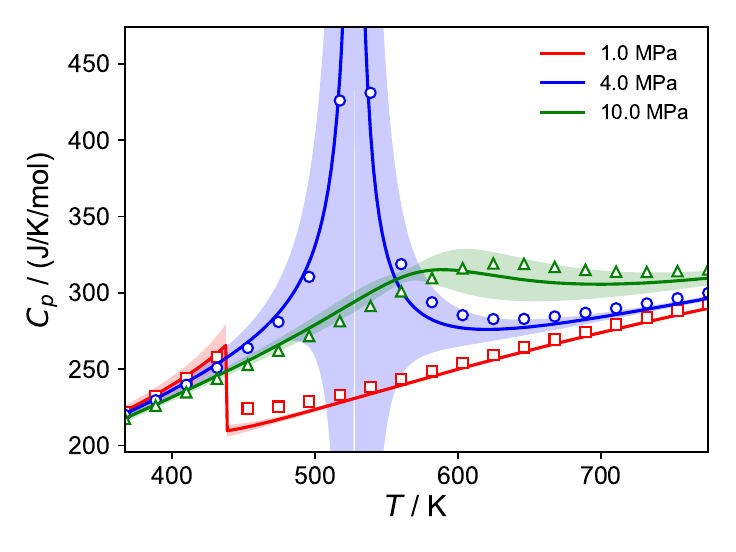}
      \caption{\centering Isobaric Heat Capacity (SAFT-VR Mie)}
      \label{fig:vr_cp_error}
    \end{subfigure}
    \begin{subfigure}[b]{0.32\textwidth}
      \includegraphics[width=1\textwidth]{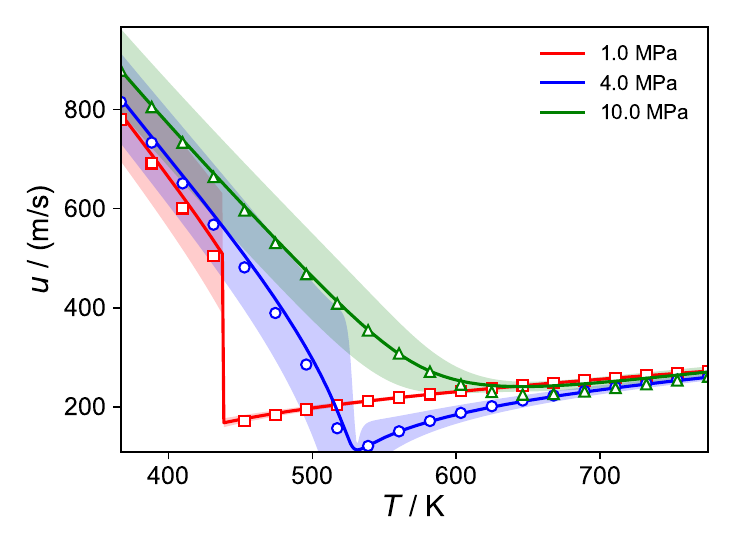}
      \caption{\centering Speed of Sound \quad\quad (SAFT-VR Mie)}
      \label{fig:vr_sos_error}
    \end{subfigure}
    \caption{Predicted values for the isochoric heat capacity (\ref{fig:pc_cv_error} and \ref{fig:vr_cv_error}), isobaric heat capacity (\ref{fig:pc_cp_error} and \ref{fig:vr_cp_error}) and speed of sound (\ref{fig:pc_sos_error} and \ref{fig:vr_sos_error}) for $n$-hexane using PC-SAFT (\ref{fig:pc_cv_error} to \ref{fig:pc_sos_error}) and SAFT-VR Mie (\ref{fig:vr_cv_error} to \ref{fig:vr_sos_error}). Shaded regions correspond to the uncertainty interval for the predicted properties using the parameters and confidence intervals obtained in section \ref{sect:ci}.}
    \label{fig:ep_results}
  \end{figure*}
Having now obtained the uncertainties for each parameter in the previous section, we are able to obtain the uncertainties of our outputs. Initially, we will consider the uncertainty of the properties we used to regress these parameters (the saturation pressure and saturated-liquid density), as shown in figure \ref{fig:ep_results_sat} in the case of ethane and methanol with the PC-SAFT and SAFT-VR Mie equations of state. 

If we first consider the saturation pressure obtained from the PC-SAFT equation (figure \ref{fig:pc_psat_error}), we can see that, on a logarithmic scale, the uncertainties in our saturation pressure appear relatively small. However, it is important to note that these uncertainties are small in \textit{logarithmic} space. On a linear scale, the relative uncertainties at lower saturation pressures will appear much larger. The uncertainties become slightly larger between ethane and methanol, although this is to be expected as we observed how the introduction of association does increase the relative uncertainty of our parameters. This is worsened when switching to the SAFT-VR Mie equation of state (figure \ref{fig:vr_psat_error}) where, with the additional parameter and greater uncertainty in these parameters, it is unsurprising that the uncertainties in our outputs also become larger. Nevertheless, we had previously observed only about a 1\% difference in relative uncertainties between the PC-SAFT and SAFT-VR Mie parameters. The difference in uncertainties in our saturation pressures between the two equations of state is larger than this. This is as a result of the fact that, with more parameters, the uncertainties accumulate much faster than they do if we had fewer parameters. This highlights the advantage of setting some of these parameter \textit{a priori}.

Next, we examine the saturated volumes, remembering that we only used the saturated-liquid volumes (densities) to regress the model parameters. In this case, as shown in figures \ref{fig:pc_rhosat_error} and \ref{fig:vr_rho_sat_error}, the uncertainties in the saturated-liquid volumes away from the critical temperature are very small, regardless of whether the PC-SAFT or SAFT-VR Mie equation of state is used or, whether or not association is included. However, examining the saturated-vapour volumes, despite the fact that the model predictions are very accurate, the uncertainties are much larger. This is quite surprising as one might anticipate that the saturated-vapour volumes are `easier' to estimate. One potential explanation for this is the fact that the isothermal compressibility is much smaller in the liquid phase, meaning that, even accounting for uncertainties, only a small range of liquid volumes will satisfy the conditions for phase equilibrium. Once again, the introduction of additional parameters also worsens the uncertainties of the predicted vapour volumes.

Another interesting behaviour to note is that, in all cases, as we approach the critical temperature, the uncertainties in the saturated volumes begin to diverge, becoming exceptionally large. This arises from the fact that, as we approach the critical temperature, fluctuations become extremely large, with the isothermal compressibility diverging to infinity. As such, slight changes in our model parameters will lead to substantial uncertainties in the predicted properties. This divergence in uncertainties of the saturated volumes is present regardless of how small the parameter uncertainties are. We can anticipate that this behaviour will appear in other properties as we approach the critical point.
\paragraph{Bulk properties}
We now extrapolate to a completely different set of properties: bulk properties. Within figure \ref{fig:ep_results}, we show results for the case of $n$-hexane where we consider the isochoric heat capacity, isobaric heat capacity and speed of sound given by both the PC-SAFT and SAFT-VR Mie equations of state. Note that we also provide such figures for the Joule--Thomson coefficient in the supplementary information. For the purposes of this section, we will not consider associating species as much of the behaviour observed here carries over to these species. 

Initially, we consider the isochoric heat capacity where we can see that the uncertainties are incredibly small and can hardly be observed within figures \ref{fig:pc_cv_error} and \ref{fig:vr_cv_error}. The only region in which they become substantial is near the critical point (around 530\,K and 4\,MPa). While this is clearly related to the proximity to the critical point, neither equation of state explicitly accounts for the divergence of the isochoric heat capacity at the critical point, without modifications\citep{llovell_second-order_2006,walkerNewPredictiveGroupContribution2020}. As such, we wouldn't expect any changes in the uncertainties near the critical point for the isochoric heat capacity. We can then deduce that these uncertainties arise from the iterative solvers used to solve for the volume at a given pressure and temperature. Given the large fluctuations in derivatives with respect to volume near the critical point, it is naturally harder to solve for the volume in this region. Nevertheless, it is reassuring to see that these uncertainties remain quite small. This highlights the need for robust numerical solvers when performing these calculations as, if too many iterations are needed, we can anticipate that uncertainties will accumulate very quickly.

Next, we examine the isobaric heat capacity, as shown in figures \ref{fig:pc_cp_error} and \ref{fig:vr_cp_error}. At this stage, we remind readers of the relationship between the isochoric ($C_V$) and isobaric ($C_p$) heat capacity:
\begin{equation}
    C_p-C_V = \frac{TV\alpha_V^2}{\beta_T}\,,
\end{equation}
where $\alpha_V$ is the isobaric expansivity and $\beta_T$ is the isothermal compressibility. If we now examine conditions far from the critical point, we again see that the uncertainties remain quite small. This isn't too surprising as, at these conditions, the difference between the two heat capacities remains relatively constant and, as a result, the uncertainties remain of similar magnitudes. The exception to this is when, in the liquid phase, as we approach the saturation temperature, the uncertainties become larger, especially when using the SAFT-VR Mie equation of state. This is to be expected as the isothermal compressibility will become smaller as we approach the saturation point, leading to a greater sensitivity to uncertainties in the parameters.

However, as we approach the critical point, due to this dependence on the isothermal compressibility, the isobaric heat capacity becomes very sensitive to the uncertainties in the parameters. This sensitivity leads to a rapid divergence in the isobaric heat capacity, as observed when $p=1\,\mathrm{MPa}$. Even as we move further from the critical point at higher pressures, these uncertainties remain large in the regions close to the maximum in the isobaric heat capacity (which would correspond to the locality of the Widom line). This situation becomes rather interesting when we consider the speed of sound ($u$), which is inversely related to the isothermal compressibility as:
\begin{equation}
    u^2 = \frac{1}{\beta_T}\frac{V}{m}\frac{C_p}{C_V}\,,
\end{equation}
where $m$ is the mass. As shown in figures \ref{fig:pc_sos_error} and \ref{fig:vr_sos_error}, this inverse relationship leads to very large uncertainties within the liquid phase. However, rather interestingly, as we approach the critical point, the divergences in the uncertainty of both the isobaric heat capacity and isothermal compressibility appear to cancel out, leading to a reduction in the certain near the critical point. We can even see that, around 530\,K and 4\,MPa, the uncertainties become very small compared to the rest of the temperature range. 

\section{Conclusion}
\label{sect:conc}
Within this article, we have provided a framework in which one can sample the distribution of parameters within the PC-SAFT and SAFT-VR Mie equations of state so as to better understand the confidence intervals of these parameters and the correlations between them. In the case where these correlations are significant, we have defined potential conserved quantities which lead to these correlations (such as the particle volume, total interaction energy and the van der Waals integral). In comparing the PC-SAFT and SAFT-VR Mie equations of state, we have highlighted that the addition of more parameters leads to larger relative uncertainties in the parameters (from 1\%--2\% to 3\%--4\%), and introduces more correlations between parameters. However, when introducing more parameters through association, an increase in the relative uncertainties is still observed, although, correlations between parameters are slightly reduced. A potential avenue for future work would be to consider how these confidence intervals vary when the type of data used to train the parameters is changed as this has been shown to change the shape of the objective function\citep{de_villiers_evaluation_2013}.

With the uncertainties of the model parameters obtained, we were then able to examine how these uncertainties propagate to properties we might be interested in obtaining from these equations of state. For the properties used to regress the model parameters, we found that their uncertainties remained reasonably small, particularly for the saturated-liquid volumes. However, when extrapolating to the saturated-vapour volumes, we found that the uncertainties here were much larger, primarily due to the large isothermal compressibility in the vapour phase. Furthermore, as we approached the critical point, the uncertainties in the saturated volumes diverged, primarily due to the increased sensitivity of the isothermal compressibility to the uncertainty in the model parameters. This effect also had a significant impact on bulk properties, particularly the isobaric heat capacity where, near the critical point, the uncertainties diverge to extremely large values, as a result of this sensitivity of the isothermal compressibility. This wasn't as significant of an issue in the case of the isochoric heat capacity, which does not depend on the isothermal compressibility, or the speed of sound where the divergences actually cancel out.

In general, we have shown that, due to the nature of fluctuations near the critical point, the presence of any uncertainty in our model parameters will lead to a divergence in the uncertainties of the predicted properties. As such, an interesting point to raise at this stage is that, a great deal of work has been done to improve the modelling of equations of state near the critical point\citep{llovell_second-order_2006,vanwestenAccurateFirstorderPerturbation2021,vanwestenAccurateThermodynamicsSimple2021}. Having now established how the inherent divergences near the critical point lead to equally large divergences in the uncertainties, regardless of how small the uncertainties in the model parameters may be, it does raise the question as to how significant these improvements can be, considering the inherent limitation in the parameters? The authors believe that these efforts are still worthwhile and needed, as long as this limitation is acknowledged. 
\section{Acknowledgements} 
P.J.W. would like to thank Andrés Riedemann for the development of ForwardDiffOverMeasurements.jl, which ensured the compability between Clapeyron.jl and Measurements.jl, and Dr. Andrew J. Haslam for useful feedback in writing this manuscript.

\section{Supplementary Information}
Within \textbf{si.pdf}, we provide all the corner and uncertainty propagation plots generated for all species and equations of state considered within this work.

All of codes used to generate these figures will be made available on the GitHub repository for Clapeyron.jl (version 0.5.0): \url{https://github.com/ClapeyronThermo/Clapeyron.jl}.

\bibliography{zotero_references,references_pierre}

\providecommand{\latin}[1]{#1}
\makeatletter
\providecommand{\doi}
  {\begingroup\let\do\@makeother\dospecials
  \catcode`\{=1 \catcode`\}=2 \doi@aux}
\providecommand{\doi@aux}[1]{\endgroup\texttt{#1}}
\makeatother
\providecommand*\mcitethebibliography{\thebibliography}
\csname @ifundefined\endcsname{endmcitethebibliography}
  {\let\endmcitethebibliography\endthebibliography}{}
\begin{mcitethebibliography}{73}
\providecommand*\natexlab[1]{#1}
\providecommand*\mciteSetBstSublistMode[1]{}
\providecommand*\mciteSetBstMaxWidthForm[2]{}
\providecommand*\mciteBstWouldAddEndPuncttrue
  {\def\EndOfBibitem{\unskip.}}
\providecommand*\mciteBstWouldAddEndPunctfalse
  {\let\EndOfBibitem\relax}
\providecommand*\mciteSetBstMidEndSepPunct[3]{}
\providecommand*\mciteSetBstSublistLabelBeginEnd[3]{}
\providecommand*\EndOfBibitem{}
\mciteSetBstSublistMode{f}
\mciteSetBstMaxWidthForm{subitem}{(\alph{mcitesubitemcount})}
\mciteSetBstSublistLabelBeginEnd
  {\mcitemaxwidthsubitemform\space}
  {\relax}
  {\relax}

\bibitem[Chapman \latin{et~al.}(1989)Chapman, Gubbins, Jackson, and
  Radosz]{chapman_saft_1989}
Chapman,~W.; Gubbins,~K.; Jackson,~G.; Radosz,~M. {SAFT}: {Equation}-of-state
  solution model for associating fluids. \emph{Fluid Phase Equilib.}
  \textbf{1989}, \emph{52}, 31--38\relax
\mciteBstWouldAddEndPuncttrue
\mciteSetBstMidEndSepPunct{\mcitedefaultmidpunct}
{\mcitedefaultendpunct}{\mcitedefaultseppunct}\relax
\EndOfBibitem
\bibitem[Chapman \latin{et~al.}(1990)Chapman, Gubbins, Jackson, and
  Radosz]{chapman_new_1990}
Chapman,~W.~G.; Gubbins,~K.~E.; Jackson,~G.; Radosz,~M. New reference equation
  of state for associating liquids. \emph{Ind. Eng. Chem. Res.} \textbf{1990},
  \emph{29}, 1709--1721\relax
\mciteBstWouldAddEndPuncttrue
\mciteSetBstMidEndSepPunct{\mcitedefaultmidpunct}
{\mcitedefaultendpunct}{\mcitedefaultseppunct}\relax
\EndOfBibitem
\bibitem[Kontogeorgis and Folas(2010)Kontogeorgis, and
  Folas]{kontogeorgisThermodynamicModelsIndustrial2010}
Kontogeorgis,~G.~M.; Folas,~G.~K. \emph{Thermodynamic {{Models}} for
  {{Industrial Applications}}: {{From Classical}} and {{Advanced Mixing Rules}}
  to {{Association Theories}}}, 1st ed.; {Wiley}: {West Sussex}, 2010\relax
\mciteBstWouldAddEndPuncttrue
\mciteSetBstMidEndSepPunct{\mcitedefaultmidpunct}
{\mcitedefaultendpunct}{\mcitedefaultseppunct}\relax
\EndOfBibitem
\bibitem[Gross and Sadowski(2001)Gross, and
  Sadowski]{grossPerturbedChainSAFTEquation2001}
Gross,~J.; Sadowski,~G. Perturbed-{{Chain SAFT}}: {{An Equation}} of {{State
  Based}} on a {{Perturbation Theory}} for {{Chain Molecules}}. \emph{Ind. Eng.
  Chem. Res.} \textbf{2001}, \emph{40}, 1244--1260\relax
\mciteBstWouldAddEndPuncttrue
\mciteSetBstMidEndSepPunct{\mcitedefaultmidpunct}
{\mcitedefaultendpunct}{\mcitedefaultseppunct}\relax
\EndOfBibitem
\bibitem[Gross and Sadowski(2002)Gross, and
  Sadowski]{grossApplicationPerturbedChainSAFT2002}
Gross,~J.; Sadowski,~G. Application of the {{Perturbed-Chain SAFT Equation}} of
  {{State}} to {{Associating Systems}}. \emph{Ind. Eng. Chem. Res.}
  \textbf{2002}, \emph{41}, 5510--5515\relax
\mciteBstWouldAddEndPuncttrue
\mciteSetBstMidEndSepPunct{\mcitedefaultmidpunct}
{\mcitedefaultendpunct}{\mcitedefaultseppunct}\relax
\EndOfBibitem
\bibitem[Lafitte \latin{et~al.}(2013)Lafitte, Apostolakou, Avendaño, Galindo,
  Adjiman, Müller, and Jackson]{lafitte_accurate_2013}
Lafitte,~T.; Apostolakou,~A.; Avendaño,~C.; Galindo,~A.; Adjiman,~C.~S.;
  Müller,~E.~A.; Jackson,~G. Accurate statistical associating fluid theory for
  chain molecules formed from {Mie} segments. \emph{J. Chem. Phys.}
  \textbf{2013}, \emph{139}, 154504\relax
\mciteBstWouldAddEndPuncttrue
\mciteSetBstMidEndSepPunct{\mcitedefaultmidpunct}
{\mcitedefaultendpunct}{\mcitedefaultseppunct}\relax
\EndOfBibitem
\bibitem[Dufal \latin{et~al.}(2015)Dufal, Lafitte, Haslam, Galindo, Clark,
  Vega, and Jackson]{dufal_saft_2015}
Dufal,~S.; Lafitte,~T.; Haslam,~A.~J.; Galindo,~A.; Clark,~G.~N.; Vega,~C.;
  Jackson,~G. The {A} in {SAFT}: developing the contribution of association to
  the {Helmholtz} free energy within a {Wertheim} {TPT1} treatment of generic
  {Mie} fluids. \emph{Mol. Phys.} \textbf{2015}, \emph{113}, 948--984,
  Publisher: Taylor \& Francis \_eprint:
  https://doi.org/10.1080/00268976.2015.1029027\relax
\mciteBstWouldAddEndPuncttrue
\mciteSetBstMidEndSepPunct{\mcitedefaultmidpunct}
{\mcitedefaultendpunct}{\mcitedefaultseppunct}\relax
\EndOfBibitem
\bibitem[Albers and Sadowski(2012)Albers, and Sadowski]{albers_reducing_2012}
Albers,~K.; Sadowski,~G. Reducing the amount of {PCP}-{SAFT} fitting
  parameters. 1. {Non}-polar and dipolar components. \emph{Fluid Phase
  Equilib.} \textbf{2012}, \emph{326}, 21--30\relax
\mciteBstWouldAddEndPuncttrue
\mciteSetBstMidEndSepPunct{\mcitedefaultmidpunct}
{\mcitedefaultendpunct}{\mcitedefaultseppunct}\relax
\EndOfBibitem
\bibitem[Albers \latin{et~al.}(2012)Albers, Heilig, and
  Sadowski]{albers_reducing_2012-1}
Albers,~K.; Heilig,~M.; Sadowski,~G. Reducing the amount of {PCP}–{SAFT}
  fitting parameters. 2. {Associating} components. \emph{Fluid Phase Equilib.}
  \textbf{2012}, \emph{326}, 31--44\relax
\mciteBstWouldAddEndPuncttrue
\mciteSetBstMidEndSepPunct{\mcitedefaultmidpunct}
{\mcitedefaultendpunct}{\mcitedefaultseppunct}\relax
\EndOfBibitem
\bibitem[Ferrando \latin{et~al.}(2012)Ferrando, de~Hemptinne, Mougin, and
  Passarello]{ferrando_prediction_2012}
Ferrando,~N.; de~Hemptinne,~J.-C.; Mougin,~P.; Passarello,~J.-P. Prediction of
  the {PC}-{SAFT} {Associating} {Parameters} by {Molecular} {Simulation}.
  \emph{J. Phys. Chem. B} \textbf{2012}, \emph{116}, 367--377, Publisher:
  American Chemical Society\relax
\mciteBstWouldAddEndPuncttrue
\mciteSetBstMidEndSepPunct{\mcitedefaultmidpunct}
{\mcitedefaultendpunct}{\mcitedefaultseppunct}\relax
\EndOfBibitem
\bibitem[Cismondi \latin{et~al.}(2005)Cismondi, Brignole, and
  Mollerup]{cismondi_rescaling_2005}
Cismondi,~M.; Brignole,~E.~A.; Mollerup,~J. Rescaling of three-parameter
  equations of state: {PC}-{SAFT} and {SPHCT}. \emph{Fluid Phase Equilib.}
  \textbf{2005}, \emph{234}, 108--121\relax
\mciteBstWouldAddEndPuncttrue
\mciteSetBstMidEndSepPunct{\mcitedefaultmidpunct}
{\mcitedefaultendpunct}{\mcitedefaultseppunct}\relax
\EndOfBibitem
\bibitem[Privat \latin{et~al.}(2019)Privat, Moine, Sirjean, Gani, and
  Jaubert]{privat_application_2019}
Privat,~R.; Moine,~E.; Sirjean,~B.; Gani,~R.; Jaubert,~J.-N. Application of the
  {Corresponding}-{State} {Law} to the {Parametrization} of {Statistical}
  {Associating} {Fluid} {Theory} ({SAFT})-{Type} {Models}: {Generation} and
  {Use} of “{Generalized} {Charts}”. \emph{Ind. Eng. Chem. Res.}
  \textbf{2019}, \emph{58}, 9127--9139, Publisher: American Chemical
  Society\relax
\mciteBstWouldAddEndPuncttrue
\mciteSetBstMidEndSepPunct{\mcitedefaultmidpunct}
{\mcitedefaultendpunct}{\mcitedefaultseppunct}\relax
\EndOfBibitem
\bibitem[Anoune \latin{et~al.}(2021)Anoune, Mimoune, Madani, and
  Merzougui]{anoune_new_2021}
Anoune,~I.; Mimoune,~Z.; Madani,~H.; Merzougui,~A. New modified {PC}-{SAFT}
  pure component parameters for accurate {VLE} and critical phenomena
  description. \emph{Fluid Phase Equilib.} \textbf{2021}, \emph{532},
  112916\relax
\mciteBstWouldAddEndPuncttrue
\mciteSetBstMidEndSepPunct{\mcitedefaultmidpunct}
{\mcitedefaultendpunct}{\mcitedefaultseppunct}\relax
\EndOfBibitem
\bibitem[Lucia \latin{et~al.}(2009)Lucia, Octavio, and Visco]{lucia_new_2009}
Lucia,~A.; Octavio,~L.~M.; Visco,~D.~P. A new algorithm for estimating
  association parameters in molecular-based equations of state by quantum
  chemistry. \emph{Comput. Chem. Eng.} \textbf{2009}, \emph{33}, 531--533\relax
\mciteBstWouldAddEndPuncttrue
\mciteSetBstMidEndSepPunct{\mcitedefaultmidpunct}
{\mcitedefaultendpunct}{\mcitedefaultseppunct}\relax
\EndOfBibitem
\bibitem[Leonhard \latin{et~al.}(2007)Leonhard, Van~Nhu, and
  Lucas]{leonhard_making_2007}
Leonhard,~K.; Van~Nhu,~N.; Lucas,~K. Making {Equation} of {State} {Models}
  {Predictive}−{Part} 3:  {Improved} {Treatment} of {Multipolar}
  {Interactions} in a {PC}-{SAFT} {Based} {Equation} of {State}. \emph{J. Phys.
  Chem. C} \textbf{2007}, \emph{111}, 15533--15543, Publisher: American
  Chemical Society\relax
\mciteBstWouldAddEndPuncttrue
\mciteSetBstMidEndSepPunct{\mcitedefaultmidpunct}
{\mcitedefaultendpunct}{\mcitedefaultseppunct}\relax
\EndOfBibitem
\bibitem[Leonhard \latin{et~al.}(2007)Leonhard, Van~Nhu, and
  Lucas]{leonhard_making_2007-1}
Leonhard,~K.; Van~Nhu,~N.; Lucas,~K. Making equation of state models
  predictive: {Part} 2: {An} improved {PCP}-{SAFT} equation of state.
  \emph{Fluid Phase Equilib.} \textbf{2007}, \emph{258}, 41--50\relax
\mciteBstWouldAddEndPuncttrue
\mciteSetBstMidEndSepPunct{\mcitedefaultmidpunct}
{\mcitedefaultendpunct}{\mcitedefaultseppunct}\relax
\EndOfBibitem
\bibitem[Singh \latin{et~al.}(2007)Singh, Leonhard, and
  Lucas]{singh_making_2007}
Singh,~M.; Leonhard,~K.; Lucas,~K. Making equation of state models predictive:
  {Part} 1: {Quantum} chemical computation of molecular properties. \emph{Fluid
  Phase Equilib.} \textbf{2007}, \emph{258}, 16--28\relax
\mciteBstWouldAddEndPuncttrue
\mciteSetBstMidEndSepPunct{\mcitedefaultmidpunct}
{\mcitedefaultendpunct}{\mcitedefaultseppunct}\relax
\EndOfBibitem
\bibitem[Van~Nhu \latin{et~al.}(2008)Van~Nhu, Singh, and
  Leonhard]{van_nhu_quantum_2008}
Van~Nhu,~N.; Singh,~M.; Leonhard,~K. Quantum {Mechanically} {Based}
  {Estimation} of {Perturbed}-{Chain} {Polar} {Statistical} {Associating}
  {Fluid} {Theory} {Parameters} for {Analyzing} {Their} {Physical}
  {Significance} and {Predicting} {Properties}. \emph{J. Phys. Chem. B}
  \textbf{2008}, \emph{112}, 5693--5701\relax
\mciteBstWouldAddEndPuncttrue
\mciteSetBstMidEndSepPunct{\mcitedefaultmidpunct}
{\mcitedefaultendpunct}{\mcitedefaultseppunct}\relax
\EndOfBibitem
\bibitem[Kaminski and Leonhard(2020)Kaminski, and Leonhard]{kaminski_sepp_2020}
Kaminski,~S.; Leonhard,~K. {SEPP}: {Segment}-{Based} {Equation} of {State}
  {Parameter} {Prediction}. \emph{J. Chem. Eng. Data} \textbf{2020}, Publisher:
  American Chemical Society\relax
\mciteBstWouldAddEndPuncttrue
\mciteSetBstMidEndSepPunct{\mcitedefaultmidpunct}
{\mcitedefaultendpunct}{\mcitedefaultseppunct}\relax
\EndOfBibitem
\bibitem[Kaminski(2019)]{kaminski_quantum-mechanics-based_2019}
Kaminski,~S. Quantum-mechanics-based prediction of {SAFT} parameters for
  non-associating and associating molecules containing carbon, hydrogen,
  oxygen, and nitrogen. Ph.D.\ thesis, Verlag Mainz, 2019\relax
\mciteBstWouldAddEndPuncttrue
\mciteSetBstMidEndSepPunct{\mcitedefaultmidpunct}
{\mcitedefaultendpunct}{\mcitedefaultseppunct}\relax
\EndOfBibitem
\bibitem[Mahmoudabadi and Pazuki(2021)Mahmoudabadi, and
  Pazuki]{mahmoudabadi_predictive_2021}
Mahmoudabadi,~S.~Z.; Pazuki,~G. A predictive {PC}-{SAFT} {EOS} based on {COSMO}
  for pharmaceutical compounds. \emph{Sci. Rep.} \textbf{2021}, \emph{11},
  6405\relax
\mciteBstWouldAddEndPuncttrue
\mciteSetBstMidEndSepPunct{\mcitedefaultmidpunct}
{\mcitedefaultendpunct}{\mcitedefaultseppunct}\relax
\EndOfBibitem
\bibitem[Walker \latin{et~al.}(2022)Walker, Zhao, Haslam, and
  Jackson]{walker_ab_2022}
Walker,~P.~J.; Zhao,~T.; Haslam,~A.~J.; Jackson,~G. Ab initio development of
  generalized {Lennard}-{Jones} ({Mie}) force fields for predictions of
  thermodynamic properties in advanced molecular-based {SAFT} equations of
  state. \emph{J. Chem. Phys.} \textbf{2022}, \emph{156}, 154106\relax
\mciteBstWouldAddEndPuncttrue
\mciteSetBstMidEndSepPunct{\mcitedefaultmidpunct}
{\mcitedefaultendpunct}{\mcitedefaultseppunct}\relax
\EndOfBibitem
\bibitem[Matsukawa \latin{et~al.}(2021)Matsukawa, Kitahara, and
  Otake]{matsukawa_estimation_2021}
Matsukawa,~H.; Kitahara,~M.; Otake,~K. Estimation of pure component parameters
  of {PC}-{SAFT} {EoS} by an artificial neural network based on a group
  contribution method. \emph{Fluid Phase Equilib.} \textbf{2021}, \emph{548},
  113179\relax
\mciteBstWouldAddEndPuncttrue
\mciteSetBstMidEndSepPunct{\mcitedefaultmidpunct}
{\mcitedefaultendpunct}{\mcitedefaultseppunct}\relax
\EndOfBibitem
\bibitem[Chung \latin{et~al.}(2022)Chung, Vermeire, Wu, Walker, Abraham, and
  Green]{chungGroupContributionMachine2022}
Chung,~Y.; Vermeire,~F.~H.; Wu,~H.; Walker,~P.; Abraham,~M.~H.; Green,~W.~H.
  Group {{Contribution}} and {{Machine Learning Approaches}} to {{Predict
  Abraham Solute Parameters}}, {{Solvation Free Energy}}, and {{Solvation
  Enthalpy}}. \emph{J. Chem. Inf. Model.} \textbf{2022}, \emph{62},
  433--446\relax
\mciteBstWouldAddEndPuncttrue
\mciteSetBstMidEndSepPunct{\mcitedefaultmidpunct}
{\mcitedefaultendpunct}{\mcitedefaultseppunct}\relax
\EndOfBibitem
\bibitem[Felton \latin{et~al.}(2023)Felton, Rasßpe-Lange, Rittig, Leonhard,
  Mitsos, Meyer-Kirschner, Knösche, and Lapkin]{felton_ml-saft_2023}
Felton,~K.; Rasßpe-Lange,~L.; Rittig,~J.; Leonhard,~K.; Mitsos,~A.;
  Meyer-Kirschner,~J.; Knösche,~C.; Lapkin,~A. {ML}-{SAFT}: {A} machine
  learning framework for {PCP}-{SAFT} parameter prediction. 2023;
  \url{https://chemrxiv.org/engage/chemrxiv/article-details/6456371107c3f029374e6608}\relax
\mciteBstWouldAddEndPuncttrue
\mciteSetBstMidEndSepPunct{\mcitedefaultmidpunct}
{\mcitedefaultendpunct}{\mcitedefaultseppunct}\relax
\EndOfBibitem
\bibitem[Habicht \latin{et~al.}(2023)Habicht, Brandenbusch, and
  Sadowski]{habicht_predicting_2023}
Habicht,~J.; Brandenbusch,~C.; Sadowski,~G. Predicting {PC}-{SAFT}
  pure-component parameters by machine learning using a molecular fingerprint
  as key input. \emph{Fluid Phase Equilib.} \textbf{2023}, \emph{565},
  113657\relax
\mciteBstWouldAddEndPuncttrue
\mciteSetBstMidEndSepPunct{\mcitedefaultmidpunct}
{\mcitedefaultendpunct}{\mcitedefaultseppunct}\relax
\EndOfBibitem
\bibitem[Swaminathan \latin{et~al.}(2006)Swaminathan, Visco, and
  Lucia]{swaminathan_evaluation_2006}
Swaminathan,~S.; Visco,~D.~P.; Lucia,~A. Evaluation of the {Pure} {Component}
  {Parameterization} {Methodology} on {Mixture} {Property} {Predictions} for
  {Thermodynamic} {Equations} of {State} {Using} {Terrain} {Methodology}. 2006;
  \url{https://doi.org/10.5281/zenodo.8061667}\relax
\mciteBstWouldAddEndPuncttrue
\mciteSetBstMidEndSepPunct{\mcitedefaultmidpunct}
{\mcitedefaultendpunct}{\mcitedefaultseppunct}\relax
\EndOfBibitem
\bibitem[Clark \latin{et~al.}(2006)Clark, Haslam, Galindo, and
  Jackson]{clark_developing_2006}
Clark,~G. N.~I.; Haslam,~A.~J.; Galindo,~A.; Jackson,~G. Developing optimal
  {Wertheim}-like models of water for use in {Statistical} {Associating}
  {Fluid} {Theory} ({SAFT}) and related approaches. \emph{Mol. Phys.}
  \textbf{2006}, \emph{104}, 3561--3581, Publisher: Taylor \& Francis \_eprint:
  https://doi.org/10.1080/00268970601081475\relax
\mciteBstWouldAddEndPuncttrue
\mciteSetBstMidEndSepPunct{\mcitedefaultmidpunct}
{\mcitedefaultendpunct}{\mcitedefaultseppunct}\relax
\EndOfBibitem
\bibitem[Forte \latin{et~al.}(2018)Forte, Burger, Langenbach, Hasse, and
  Bortz]{forte_multi-criteria_2018}
Forte,~E.; Burger,~J.; Langenbach,~K.; Hasse,~H.; Bortz,~M. Multi-criteria
  optimization for parameterization of {SAFT}-type equations of state for
  water. \emph{AIChE J.} \textbf{2018}, \emph{64}, 226--237, \_eprint:
  https://onlinelibrary.wiley.com/doi/pdf/10.1002/aic.15857\relax
\mciteBstWouldAddEndPuncttrue
\mciteSetBstMidEndSepPunct{\mcitedefaultmidpunct}
{\mcitedefaultendpunct}{\mcitedefaultseppunct}\relax
\EndOfBibitem
\bibitem[Graham \latin{et~al.}(2022)Graham, Forte, Burger, Galindo, Jackson,
  and Adjiman]{graham_multi-objective_2022}
Graham,~E.~J.; Forte,~E.; Burger,~J.; Galindo,~A.; Jackson,~G.; Adjiman,~C.~S.
  Multi-objective optimization of equation of state molecular parameters:
  {SAFT}-{VR} {Mie} models for water. \emph{Comput. Chem. Eng.} \textbf{2022},
  \emph{167}, 108015\relax
\mciteBstWouldAddEndPuncttrue
\mciteSetBstMidEndSepPunct{\mcitedefaultmidpunct}
{\mcitedefaultendpunct}{\mcitedefaultseppunct}\relax
\EndOfBibitem
\bibitem[Klajmon and Nezbeda(2023)Klajmon, and Nezbeda]{klajmon_assessing_2023}
Klajmon,~M.; Nezbeda,~I. Assessing the quality of {SAFT} equations for the
  vapor-liquid equilibrium of pure water. \emph{J. Mol. Liq.} \textbf{2023},
  \emph{376}, 121414\relax
\mciteBstWouldAddEndPuncttrue
\mciteSetBstMidEndSepPunct{\mcitedefaultmidpunct}
{\mcitedefaultendpunct}{\mcitedefaultseppunct}\relax
\EndOfBibitem
\bibitem[Swaminathan and Visco(2007)Swaminathan, and
  Visco]{swaminathan_demonstrating_2007}
Swaminathan,~S.; Visco,~D.~P. Demonstrating the {Effect} of {Multiple} {Sets}
  of {Parameters} on {Mixture} {Property} {Predictions} for a {SAFT}-based
  {EOS} using {Terrain} {Methodology}. 2007;
  \url{https://doi.org/10.5281/zenodo.8061655}\relax
\mciteBstWouldAddEndPuncttrue
\mciteSetBstMidEndSepPunct{\mcitedefaultmidpunct}
{\mcitedefaultendpunct}{\mcitedefaultseppunct}\relax
\EndOfBibitem
\bibitem[de~Villiers \latin{et~al.}(2013)de~Villiers, Schwarz, Burger, and
  Kontogeorgis]{de_villiers_evaluation_2013}
de~Villiers,~A.~J.; Schwarz,~C.~E.; Burger,~A.~J.; Kontogeorgis,~G.~M.
  Evaluation of the {PC}-{SAFT}, {SAFT} and {CPA} equations of state in
  predicting derivative properties of selected non-polar and hydrogen-bonding
  compounds. \emph{Fluid Phase Equilib.} \textbf{2013}, \emph{338}, 1--15,
  Publisher: Elsevier B.V.\relax
\mciteBstWouldAddEndPunctfalse
\mciteSetBstMidEndSepPunct{\mcitedefaultmidpunct}
{}{\mcitedefaultseppunct}\relax
\EndOfBibitem
\bibitem[Mac~Dowell \latin{et~al.}(2011)Mac~Dowell, Pereira, Llovell, Blas,
  Adjiman, Jackson, and Galindo]{mac_dowell_transferable_2011}
Mac~Dowell,~N.; Pereira,~F.~E.; Llovell,~F.; Blas,~F.~J.; Adjiman,~C.~S.;
  Jackson,~G.; Galindo,~A. Transferable {SAFT}-{VR} {Models} for the
  {Calculation} of the {Fluid} {Phase} {Equilibria} in {Reactive} {Mixtures} of
  {Carbon} {Dioxide}, {Water}, and n-{Alkylamines} in the {Context} of {Carbon}
  {Capture}. \emph{J. Phys. Chem. B} \textbf{2011}, \emph{115}, 8155--8168,
  Publisher: American Chemical Society\relax
\mciteBstWouldAddEndPuncttrue
\mciteSetBstMidEndSepPunct{\mcitedefaultmidpunct}
{\mcitedefaultendpunct}{\mcitedefaultseppunct}\relax
\EndOfBibitem
\bibitem[Ramírez-Vélez(2022)]{ramirez-velez_parametrization_2022}
Ramírez-Vélez,~N. Parametrization of equations of state : definition of a
  methodology applicable to {SAFT} models for pure species (with and without
  association term) and evaluation of their influence on model performances.
  phdthesis, Université de Lorraine, 2022\relax
\mciteBstWouldAddEndPuncttrue
\mciteSetBstMidEndSepPunct{\mcitedefaultmidpunct}
{\mcitedefaultendpunct}{\mcitedefaultseppunct}\relax
\EndOfBibitem
\bibitem[Dufal \latin{et~al.}(2014)Dufal, Papaioannou, Sadeqzadeh, Pogiatzis,
  Chremos, Adjiman, Jackson, and Galindo]{dufal_prediction_2014}
Dufal,~S.; Papaioannou,~V.; Sadeqzadeh,~M.; Pogiatzis,~T.; Chremos,~A.;
  Adjiman,~C.~S.; Jackson,~G.; Galindo,~A. Prediction of {Thermodynamic}
  {Properties} and {Phase} {Behavior} of {Fluids} and {Mixtures} with the
  {SAFT}-γ {Mie} {Group}-{Contribution} {Equation} of {State}. \emph{J. Chem.
  Eng. Data} \textbf{2014}, \emph{59}, 3272--3288, Publisher: American Chemical
  Society\relax
\mciteBstWouldAddEndPuncttrue
\mciteSetBstMidEndSepPunct{\mcitedefaultmidpunct}
{\mcitedefaultendpunct}{\mcitedefaultseppunct}\relax
\EndOfBibitem
\bibitem[Ramrattan \latin{et~al.}(2015)Ramrattan, Avendaño, Müller, and
  Galindo]{ramrattan_corresponding-states_2015}
Ramrattan,~N.; Avendaño,~C.; Müller,~E.; Galindo,~A. A corresponding-states
  framework for the description of the {Mie} family of intermolecular
  potentials. \emph{Mol. Phys.} \textbf{2015}, \emph{113}, 932--947, Publisher:
  Taylor \& Francis \_eprint:
  https://doi.org/10.1080/00268976.2015.1025112\relax
\mciteBstWouldAddEndPuncttrue
\mciteSetBstMidEndSepPunct{\mcitedefaultmidpunct}
{\mcitedefaultendpunct}{\mcitedefaultseppunct}\relax
\EndOfBibitem
\bibitem[Dufal \latin{et~al.}(2015)Dufal, Lafitte, Galindo, Jackson, and
  Haslam]{dufal_developing_2015}
Dufal,~S.; Lafitte,~T.; Galindo,~A.; Jackson,~G.; Haslam,~A.~J. Developing
  intermolecular-potential models for use with the {SAFT}-{VR} {Mie} equation
  of state. \emph{AIChE J.} \textbf{2015}, \emph{61}, 2891--2912\relax
\mciteBstWouldAddEndPuncttrue
\mciteSetBstMidEndSepPunct{\mcitedefaultmidpunct}
{\mcitedefaultendpunct}{\mcitedefaultseppunct}\relax
\EndOfBibitem
\bibitem[Cripwell \latin{et~al.}(2018)Cripwell, Schwarz, and
  Burger]{cripwell_saft-vr-mie_2018}
Cripwell,~J.~T.; Schwarz,~C.~E.; Burger,~A.~J. {SAFT}-{VR}-{Mie} with an
  incorporated polar term for accurate holistic prediction of the thermodynamic
  properties of polar components. \emph{Fluid Phase Equilib.} \textbf{2018},
  \emph{455}, 24--42\relax
\mciteBstWouldAddEndPuncttrue
\mciteSetBstMidEndSepPunct{\mcitedefaultmidpunct}
{\mcitedefaultendpunct}{\mcitedefaultseppunct}\relax
\EndOfBibitem
\bibitem[Swaminathan and Visco(2005)Swaminathan, and
  Visco]{swaminathan_thermodynamic_2005}
Swaminathan,~S.; Visco,~D.~P. Thermodynamic {Modeling} of {Refrigerants}
  {Using} the {Statistical} {Associating} {Fluid} {Theory} with {Variable}
  {Range}. 2. {Applications} to {Binary} {Mixtures}. \emph{Ind. Eng. Chem.
  Res.} \textbf{2005}, \emph{44}, 4806--4814, Publisher: American Chemical
  Society\relax
\mciteBstWouldAddEndPuncttrue
\mciteSetBstMidEndSepPunct{\mcitedefaultmidpunct}
{\mcitedefaultendpunct}{\mcitedefaultseppunct}\relax
\EndOfBibitem
\bibitem[Grenner \latin{et~al.}(2007)Grenner, Kontogeorgis, von Solms, and
  Michelsen]{grenner_modeling_2007}
Grenner,~A.; Kontogeorgis,~G.~M.; von Solms,~N.; Michelsen,~M.~L. Modeling
  phase equilibria of alkanols with the simplified {PC}-{SAFT} equation of
  state and generalized pure compound parameters. \emph{Fluid Phase Equilib.}
  \textbf{2007}, \emph{258}, 83--94\relax
\mciteBstWouldAddEndPuncttrue
\mciteSetBstMidEndSepPunct{\mcitedefaultmidpunct}
{\mcitedefaultendpunct}{\mcitedefaultseppunct}\relax
\EndOfBibitem
\bibitem[Walker(2022)]{walkerAdvancedPredictiveMixing2022}
Walker,~P.~J. Toward {{Advanced}}, {{Predictive Mixing Rules}} in {{SAFT
  Equations}} of {{State}}. \emph{Industrial \& Engineering Chemistry Research}
  \textbf{2022}, \emph{61}, 18165--18175\relax
\mciteBstWouldAddEndPuncttrue
\mciteSetBstMidEndSepPunct{\mcitedefaultmidpunct}
{\mcitedefaultendpunct}{\mcitedefaultseppunct}\relax
\EndOfBibitem
\bibitem[von Solms \latin{et~al.}(2006)von Solms, Kouskoumvekaki, Michelsen,
  and Kontogeorgis]{von_solms_capabilities_2006}
von Solms,~N.; Kouskoumvekaki,~I.~A.; Michelsen,~M.~L.; Kontogeorgis,~G.~M.
  Capabilities, limitations and challenges of a simplified {PC}-{SAFT} equation
  of state. \emph{Fluid Phase Equilib.} \textbf{2006}, \emph{241},
  344--353\relax
\mciteBstWouldAddEndPuncttrue
\mciteSetBstMidEndSepPunct{\mcitedefaultmidpunct}
{\mcitedefaultendpunct}{\mcitedefaultseppunct}\relax
\EndOfBibitem
\bibitem[Stavrou \latin{et~al.}(2016)Stavrou, Bardow, and
  Gross]{stavrou_estimation_2016}
Stavrou,~M.; Bardow,~A.; Gross,~J. Estimation of the binary interaction
  parameter kij of the {PC}-{SAFT} {Equation} of {State} based on pure
  component parameters using a {QSPR} method. \emph{Fluid Phase Equilib.}
  \textbf{2016}, \emph{416}, 138--149\relax
\mciteBstWouldAddEndPuncttrue
\mciteSetBstMidEndSepPunct{\mcitedefaultmidpunct}
{\mcitedefaultendpunct}{\mcitedefaultseppunct}\relax
\EndOfBibitem
\bibitem[Schacht \latin{et~al.}(2010)Schacht, Zubeir, de~Loos, and
  Gross]{schacht_application_2010}
Schacht,~C.~S.; Zubeir,~L.; de~Loos,~T.~W.; Gross,~J. Application of {Infinite}
  {Dilution} {Activity} {Coefficients} for {Determining} {Binary} {Equation} of
  {State} {Parameters}. \emph{Ind. Eng. Chem. Res.} \textbf{2010}, \emph{49},
  7646--7653, Publisher: American Chemical Society\relax
\mciteBstWouldAddEndPuncttrue
\mciteSetBstMidEndSepPunct{\mcitedefaultmidpunct}
{\mcitedefaultendpunct}{\mcitedefaultseppunct}\relax
\EndOfBibitem
\bibitem[Behzadi \latin{et~al.}(2005)Behzadi, Ghotbi, and
  Galindo]{behzadi_application_2005}
Behzadi,~B.; Ghotbi,~C.; Galindo,~A. Application of the simplex simulated
  annealing technique to nonlinear parameter optimization for the {SAFT}-{VR}
  equation of state. \emph{Chem. Eng. Sci.} \textbf{2005}, \emph{60},
  6607--6621\relax
\mciteBstWouldAddEndPuncttrue
\mciteSetBstMidEndSepPunct{\mcitedefaultmidpunct}
{\mcitedefaultendpunct}{\mcitedefaultseppunct}\relax
\EndOfBibitem
\bibitem[Klajmon(2020)]{klajmon_investigating_2020}
Klajmon,~M. Investigating {Various} {Parametrization} {Strategies} for
  {Pharmaceuticals} within the {PC}-{SAFT} {Equation} of {State}. \emph{J.
  Chem. Eng. Data} \textbf{2020}, \emph{65}, 5753--5767, Publisher: American
  Chemical Society\relax
\mciteBstWouldAddEndPuncttrue
\mciteSetBstMidEndSepPunct{\mcitedefaultmidpunct}
{\mcitedefaultendpunct}{\mcitedefaultseppunct}\relax
\EndOfBibitem
\bibitem[Fuenzalida \latin{et~al.}(2016)Fuenzalida, Cuevas-Valenzuela, and
  Pérez-Correa]{fuenzalida_improved_2016}
Fuenzalida,~M.; Cuevas-Valenzuela,~J.; Pérez-Correa,~J.~R. Improved estimation
  of {PC}-{SAFT} equation of state parameters using a multi-objective
  variable-weight cost function. \emph{Fluid Phase Equilib.} \textbf{2016},
  \emph{427}, 308--319\relax
\mciteBstWouldAddEndPuncttrue
\mciteSetBstMidEndSepPunct{\mcitedefaultmidpunct}
{\mcitedefaultendpunct}{\mcitedefaultseppunct}\relax
\EndOfBibitem
\bibitem[Rehner and Gross(2020)Rehner, and Gross]{rehner_multiobjective_2020}
Rehner,~P.; Gross,~J. Multiobjective {Optimization} of {PCP}-{SAFT}
  {Parameters} for {Water} and {Alcohols} {Using} {Surface} {Tension} {Data}.
  \emph{J. Chem. Eng. Data} \textbf{2020}, \emph{65}, 5698--5707, Publisher:
  American Chemical Society\relax
\mciteBstWouldAddEndPuncttrue
\mciteSetBstMidEndSepPunct{\mcitedefaultmidpunct}
{\mcitedefaultendpunct}{\mcitedefaultseppunct}\relax
\EndOfBibitem
\bibitem[Oliveira \latin{et~al.}(2016)Oliveira, Llovell, Coutinho, and
  Vega]{oliveira_new_2016}
Oliveira,~M.~B.; Llovell,~F.; Coutinho,~J. A.~P.; Vega,~L.~F. New {Procedure}
  for {Enhancing} the {Transferability} of {Statistical} {Associating} {Fluid}
  {Theory} ({SAFT}) {Molecular} {Parameters}: {The} {Role} of {Derivative}
  {Properties}. \emph{Ind. Eng. Chem. Res.} \textbf{2016}, \emph{55},
  10011--10024, Publisher: American Chemical Society\relax
\mciteBstWouldAddEndPuncttrue
\mciteSetBstMidEndSepPunct{\mcitedefaultmidpunct}
{\mcitedefaultendpunct}{\mcitedefaultseppunct}\relax
\EndOfBibitem
\bibitem[Ramírez-Vélez \latin{et~al.}(2020)Ramírez-Vélez, Piña-Martinez,
  Jaubert, and Privat]{ramirez-velez_parameterization_2020}
Ramírez-Vélez,~N.; Piña-Martinez,~A.; Jaubert,~J.-N.; Privat,~R.
  Parameterization of {SAFT} {Models}: {Analysis} of {Different} {Parameter}
  {Estimation} {Strategies} and {Application} to the {Development} of a
  {Comprehensive} {Database} of {PC}-{SAFT} {Molecular} {Parameters}. \emph{J.
  Chem. Eng. Data} \textbf{2020}, \emph{65}, 5920--5932, Publisher: American
  Chemical Society\relax
\mciteBstWouldAddEndPuncttrue
\mciteSetBstMidEndSepPunct{\mcitedefaultmidpunct}
{\mcitedefaultendpunct}{\mcitedefaultseppunct}\relax
\EndOfBibitem
\bibitem[Grandjean \latin{et~al.}(2014)Grandjean, de~Hemptinne, and
  Lugo]{grandjean_application_2014}
Grandjean,~L.; de~Hemptinne,~J.-C.; Lugo,~R. Application of {GC}-{PPC}-{SAFT}
  {EoS} to ammonia and its mixtures. \emph{Fluid Phase Equilib.} \textbf{2014},
  \emph{367}, 159--172\relax
\mciteBstWouldAddEndPuncttrue
\mciteSetBstMidEndSepPunct{\mcitedefaultmidpunct}
{\mcitedefaultendpunct}{\mcitedefaultseppunct}\relax
\EndOfBibitem
\bibitem[Hutacharoen \latin{et~al.}(2017)Hutacharoen, Dufal, Papaioannou,
  Shanker, Adjiman, Jackson, and Galindo]{hutacharoen_predicting_2017}
Hutacharoen,~P.; Dufal,~S.; Papaioannou,~V.; Shanker,~R.~M.; Adjiman,~C.~S.;
  Jackson,~G.; Galindo,~A. Predicting the {Solvation} of {Organic} {Compounds}
  in {Aqueous} {Environments}: {From} {Alkanes} and {Alcohols} to
  {Pharmaceuticals}. \emph{Ind. Eng. Chem. Res.} \textbf{2017}, \emph{56},
  10856--10876, Publisher: American Chemical Society\relax
\mciteBstWouldAddEndPuncttrue
\mciteSetBstMidEndSepPunct{\mcitedefaultmidpunct}
{\mcitedefaultendpunct}{\mcitedefaultseppunct}\relax
\EndOfBibitem
\bibitem[Cripwell \latin{et~al.}(2018)Cripwell, Smith, Schwarz, and
  Burger]{cripwell_saft-vr_2018}
Cripwell,~J.~T.; Smith,~S. A.~M.; Schwarz,~C.~E.; Burger,~A.~J. {SAFT}-{VR}
  {Mie}: {Application} to {Phase} {Equilibria} of {Alcohols} in {Mixtures} with
  n-{Alkanes} and {Water}. \emph{Ind. Eng. Chem. Res.} \textbf{2018},
  \emph{57}, 9693--9706, Publisher: American Chemical Society\relax
\mciteBstWouldAddEndPuncttrue
\mciteSetBstMidEndSepPunct{\mcitedefaultmidpunct}
{\mcitedefaultendpunct}{\mcitedefaultseppunct}\relax
\EndOfBibitem
\bibitem[Wang and Sheen(2015)Wang, and Sheen]{wangCombustionKineticModel2015}
Wang,~H.; Sheen,~D.~A. Combustion Kinetic Model Uncertainty Quantification,
  Propagation and Minimization. \emph{Progress in Energy and Combustion
  Science} \textbf{2015}, \emph{47}, 1--31\relax
\mciteBstWouldAddEndPuncttrue
\mciteSetBstMidEndSepPunct{\mcitedefaultmidpunct}
{\mcitedefaultendpunct}{\mcitedefaultseppunct}\relax
\EndOfBibitem
\bibitem[Scalia \latin{et~al.}(2020)Scalia, Grambow, Pernici, Li, and
  Green]{scaliaEvaluatingScalableUncertainty2020}
Scalia,~G.; Grambow,~C.~A.; Pernici,~B.; Li,~Y.-P.; Green,~W.~H. Evaluating
  {{Scalable Uncertainty Estimation Methods}} for {{Deep Learning-Based
  Molecular Property Prediction}}. \emph{Journal of Chemical Information and
  Modeling} \textbf{2020}, \emph{60}, 2697--2717\relax
\mciteBstWouldAddEndPuncttrue
\mciteSetBstMidEndSepPunct{\mcitedefaultmidpunct}
{\mcitedefaultendpunct}{\mcitedefaultseppunct}\relax
\EndOfBibitem
\bibitem[Heid \latin{et~al.}(2023)Heid, McGill, Vermeire, and
  Green]{heidCharacterizingUncertaintyMachine2023}
Heid,~E.; McGill,~C.~J.; Vermeire,~F.~H.; Green,~W.~H. Characterizing
  {{Uncertainty}} in {{Machine Learning}} for {{Chemistry}}. 2023\relax
\mciteBstWouldAddEndPuncttrue
\mciteSetBstMidEndSepPunct{\mcitedefaultmidpunct}
{\mcitedefaultendpunct}{\mcitedefaultseppunct}\relax
\EndOfBibitem
\bibitem[Kaminski \latin{et~al.}(2016)Kaminski, Bardow, and
  Leonhard]{kaminski_trade-off_2016}
Kaminski,~S.; Bardow,~A.; Leonhard,~K. The trade-off between experimental
  effort and accuracy for determination of {PCP}-{SAFT} parameters. \emph{Fluid
  Phase Equilib.} \textbf{2016}, \emph{428}, 182--189\relax
\mciteBstWouldAddEndPuncttrue
\mciteSetBstMidEndSepPunct{\mcitedefaultmidpunct}
{\mcitedefaultendpunct}{\mcitedefaultseppunct}\relax
\EndOfBibitem
\bibitem[Creton \latin{et~al.}(2023)Creton, Agoudjil, and
  de~Hemptinne]{creton_assessment_2023}
Creton,~B.; Agoudjil,~C.; de~Hemptinne,~J.-C. Assessment of two {PC}-{SAFT}
  parameterization strategies for pure compounds: {Model} accuracy and
  sensitivity analysis. \emph{Fluid Phase Equilib.} \textbf{2023}, \emph{565},
  113666\relax
\mciteBstWouldAddEndPuncttrue
\mciteSetBstMidEndSepPunct{\mcitedefaultmidpunct}
{\mcitedefaultendpunct}{\mcitedefaultseppunct}\relax
\EndOfBibitem
\bibitem[Walker \latin{et~al.}(2022)Walker, Yew, and
  Riedemann]{walkerClapeyronJlExtensible2022}
Walker,~P.~J.; Yew,~H.-W.; Riedemann,~A. Clapeyron.Jl: {{An Extensible}},
  {{Open-Source Fluid Thermodynamics Toolkit}}. \emph{Ind. Eng. Chem. Res.}
  \textbf{2022}, \emph{61}, 7130--7153\relax
\mciteBstWouldAddEndPuncttrue
\mciteSetBstMidEndSepPunct{\mcitedefaultmidpunct}
{\mcitedefaultendpunct}{\mcitedefaultseppunct}\relax
\EndOfBibitem
\bibitem[Dufal(2013)]{dufal_development_2013}
Dufal,~S. Development and application of advanced thermodynamic molecular
  description for complex reservoir fluids containing carbon dioxide and
  brines. {PhD} thesis, Imperial College London, London, GB, 2013; Accepted:
  2014-02-24T12:56:39Z Publisher: Imperial College London\relax
\mciteBstWouldAddEndPuncttrue
\mciteSetBstMidEndSepPunct{\mcitedefaultmidpunct}
{\mcitedefaultendpunct}{\mcitedefaultseppunct}\relax
\EndOfBibitem
\bibitem[Virtanen \latin{et~al.}(2020)Virtanen, Gommers, Oliphant, Haberland,
  Reddy, Cournapeau, Burovski, Peterson, Weckesser, Bright, {van der Walt},
  Brett, Wilson, Millman, Mayorov, Nelson, Jones, Kern, Larson, Carey, Polat,
  Feng, Moore, {VanderPlas}, Laxalde, Perktold, Cimrman, Henriksen, Quintero,
  Harris, Archibald, Ribeiro, Pedregosa, {van Mulbregt}, and {SciPy 1.0
  Contributors}]{2020SciPy-NMeth}
Virtanen,~P. \latin{et~al.}  {{SciPy} 1.0: Fundamental Algorithms for
  Scientific Computing in Python}. \emph{Nature Methods} \textbf{2020},
  \emph{17}, 261--272\relax
\mciteBstWouldAddEndPuncttrue
\mciteSetBstMidEndSepPunct{\mcitedefaultmidpunct}
{\mcitedefaultendpunct}{\mcitedefaultseppunct}\relax
\EndOfBibitem
\bibitem[Vugrin \latin{et~al.}(2007)Vugrin, Swiler, Roberts, Stucky-Mack, and
  Sullivan]{vugrin_confidence_2007}
Vugrin,~K.~W.; Swiler,~L.~P.; Roberts,~R.~M.; Stucky-Mack,~N.~J.;
  Sullivan,~S.~P. Confidence region estimation techniques for nonlinear
  regression in groundwater flow: {Three} case studies. \emph{Water Resources
  Research} \textbf{2007}, \emph{43}\relax
\mciteBstWouldAddEndPuncttrue
\mciteSetBstMidEndSepPunct{\mcitedefaultmidpunct}
{\mcitedefaultendpunct}{\mcitedefaultseppunct}\relax
\EndOfBibitem
\bibitem[Kunz and Wagner(2012)Kunz, and Wagner]{kunz_gerg-2008_2012}
Kunz,~O.; Wagner,~W. The {GERG}-2008 wide-range equation of state for natural
  gases and other mixtures: {An} expansion of {GERG}-2004. \emph{J. Chem. Eng.
  Data} \textbf{2012}, \emph{57}, 3032--3091\relax
\mciteBstWouldAddEndPuncttrue
\mciteSetBstMidEndSepPunct{\mcitedefaultmidpunct}
{\mcitedefaultendpunct}{\mcitedefaultseppunct}\relax
\EndOfBibitem
\bibitem[{Mej{\'i}a-de-Dios} and {Mezura-Montes}(2019){Mej{\'i}a-de-Dios}, and
  {Mezura-Montes}]{mejia-de-diosNewEvolutionaryOptimization2019}
{Mej{\'i}a-de-Dios},~J.-A.; {Mezura-Montes},~E. In \emph{Decision {{Science}}
  in {{Action}}: {{Theory}} and {{Applications}} of {{Modern Decision Analytic
  Optimisation}}}; Deep,~K., Jain,~M., Salhi,~S., Eds.; Asset {{Analytics}};
  {Springer}: {Singapore}, 2019; pp 65--74\relax
\mciteBstWouldAddEndPuncttrue
\mciteSetBstMidEndSepPunct{\mcitedefaultmidpunct}
{\mcitedefaultendpunct}{\mcitedefaultseppunct}\relax
\EndOfBibitem
\bibitem[{Mej{\'i}a-de-Dios} and {Mezura-Montes}(2022){Mej{\'i}a-de-Dios}, and
  {Mezura-Montes}]{mejia-de-diosMetaheuristicsJuliaPackage2022}
{Mej{\'i}a-de-Dios},~J.-A.; {Mezura-Montes},~E. Metaheuristics: {{A Julia
  Package}} for {{Single-}} and {{Multi-Objective Optimization}}. \emph{Journal
  of Open Source Software} \textbf{2022}, \emph{7}, 4723\relax
\mciteBstWouldAddEndPuncttrue
\mciteSetBstMidEndSepPunct{\mcitedefaultmidpunct}
{\mcitedefaultendpunct}{\mcitedefaultseppunct}\relax
\EndOfBibitem
\bibitem[Taylor(1996)]{taylorIntroductionErrorAnalysis1996}
Taylor,~J.~R. \emph{An {{Introduction}} to {{Error Analysis}}: {{The Study}} of
  {{Uncertainties}} in {{Physical Measurements}}}, 2nd ed.; {University Science
  Books}: {Sausalito, Calif}, 1996\relax
\mciteBstWouldAddEndPuncttrue
\mciteSetBstMidEndSepPunct{\mcitedefaultmidpunct}
{\mcitedefaultendpunct}{\mcitedefaultseppunct}\relax
\EndOfBibitem
\bibitem[Giordano(2016)]{giordano_uncertainty_2016}
Giordano,~M. Uncertainty propagation with functionally correlated quantities.
  2016; \url{http://arxiv.org/abs/1610.08716}, arXiv:1610.08716 [physics]\relax
\mciteBstWouldAddEndPuncttrue
\mciteSetBstMidEndSepPunct{\mcitedefaultmidpunct}
{\mcitedefaultendpunct}{\mcitedefaultseppunct}\relax
\EndOfBibitem
\bibitem[Llovell \latin{et~al.}(2006)Llovell, Peters, and
  Vega]{llovell_second-order_2006}
Llovell,~F.; Peters,~C.~J.; Vega,~L.~F. Second-order thermodynamic derivative
  properties of selected mixtures by the soft-{SAFT} equation of state.
  \emph{Fluid Phase Equilib.} \textbf{2006}, \emph{248}, 115--122\relax
\mciteBstWouldAddEndPuncttrue
\mciteSetBstMidEndSepPunct{\mcitedefaultmidpunct}
{\mcitedefaultendpunct}{\mcitedefaultseppunct}\relax
\EndOfBibitem
\bibitem[Walker and Haslam(2020)Walker, and
  Haslam]{walkerNewPredictiveGroupContribution2020}
Walker,~P.~J.; Haslam,~A.~J. A {{New Predictive Group-Contribution
  Ideal-Heat-Capacity Model}} and {{Its Influence}} on {{Second-Derivative
  Properties Calculated Using}} a {{Free-Energy Equation}} of {{State}}.
  \emph{J. Chem. Eng. Data} \textbf{2020}, \emph{65}, 5809--5829\relax
\mciteBstWouldAddEndPuncttrue
\mciteSetBstMidEndSepPunct{\mcitedefaultmidpunct}
{\mcitedefaultendpunct}{\mcitedefaultseppunct}\relax
\EndOfBibitem
\bibitem[{van Westen} and Gross(2021){van Westen}, and
  Gross]{vanwestenAccurateFirstorderPerturbation2021}
{van Westen},~T.; Gross,~J. Accurate First-Order Perturbation Theory for
  Fluids: {\emph{Uf}} -Theory. \emph{J. Chem. Phys.} \textbf{2021}, \emph{154},
  041102\relax
\mciteBstWouldAddEndPuncttrue
\mciteSetBstMidEndSepPunct{\mcitedefaultmidpunct}
{\mcitedefaultendpunct}{\mcitedefaultseppunct}\relax
\EndOfBibitem
\bibitem[{van Westen} and Gross(2021){van Westen}, and
  Gross]{vanwestenAccurateThermodynamicsSimple2021}
{van Westen},~T.; Gross,~J. Accurate Thermodynamics of Simple Fluids and Chain
  Fluids Based on First-Order Perturbation Theory and Second Virial
  Coefficients: {\emph{Uv}} -Theory. \emph{J. Chem. Phys.} \textbf{2021},
  \emph{155}, 244501\relax
\mciteBstWouldAddEndPuncttrue
\mciteSetBstMidEndSepPunct{\mcitedefaultmidpunct}
{\mcitedefaultendpunct}{\mcitedefaultseppunct}\relax
\EndOfBibitem
\end{mcitethebibliography}

\end{document}


\maketitle
\newpage
\section{PC-SAFT}
\subsection{Methane}

\begin{figure}[H]
  \centering
      \includegraphics[width=0.5\textwidth]{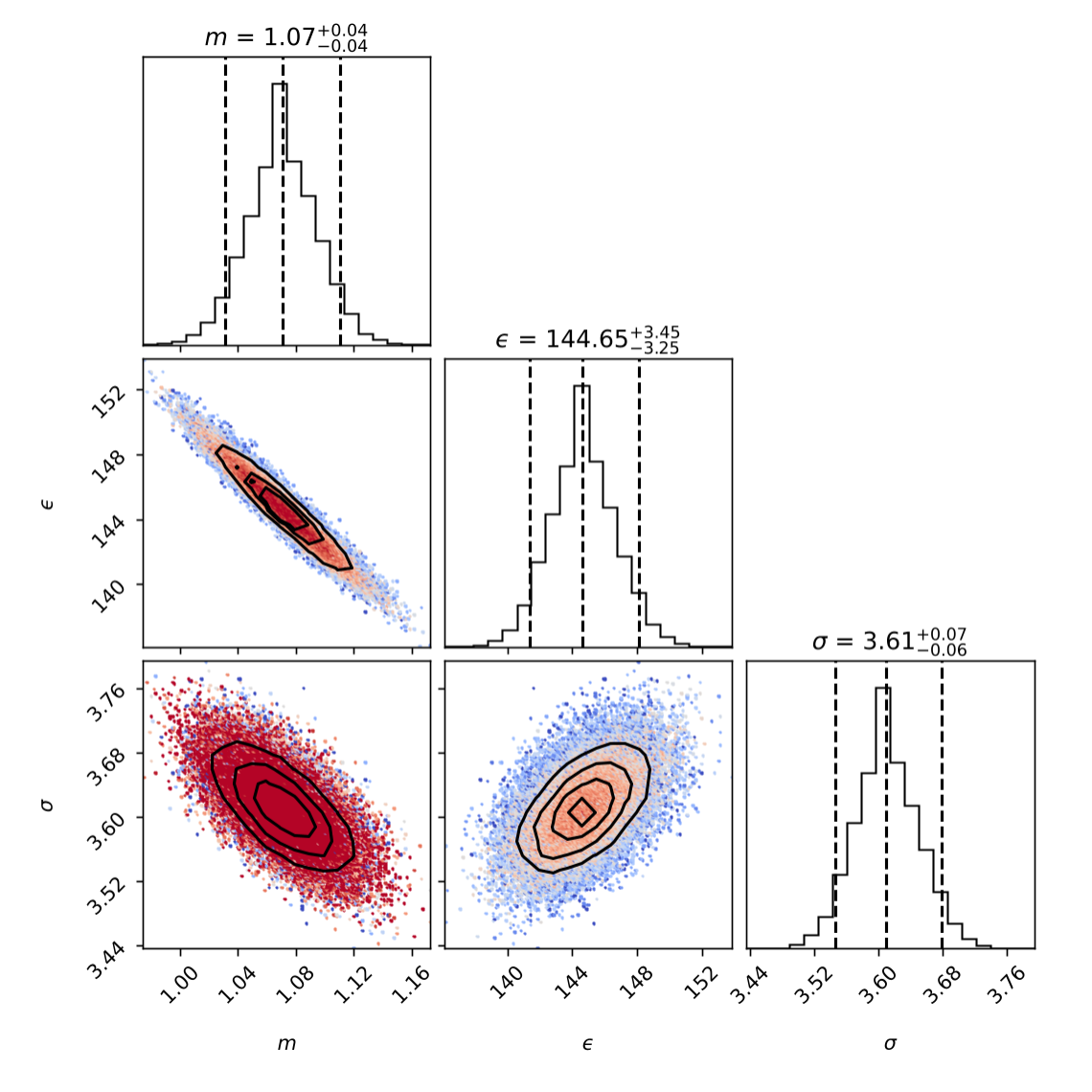}
      \caption{Confidence intervals obtained for the pure component parameters of methane in PC-SAFT. Colors and styles are identical to figure 2.}
  \end{figure}

\begin{figure}[H]
  \centering
    \begin{subfigure}[b]{0.49\textwidth}
      \includegraphics[width=1\textwidth]{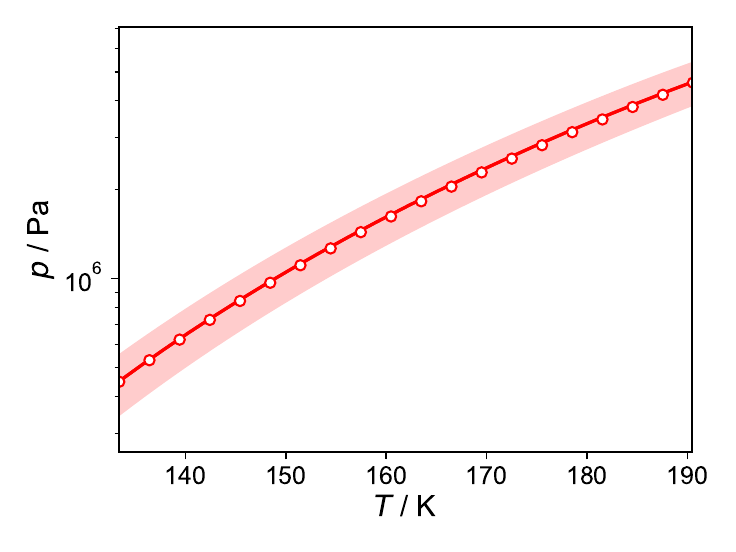}
      \caption{Saturation Curve}
    \end{subfigure}
    \begin{subfigure}[b]{0.49\textwidth}
      \includegraphics[width=1\textwidth]{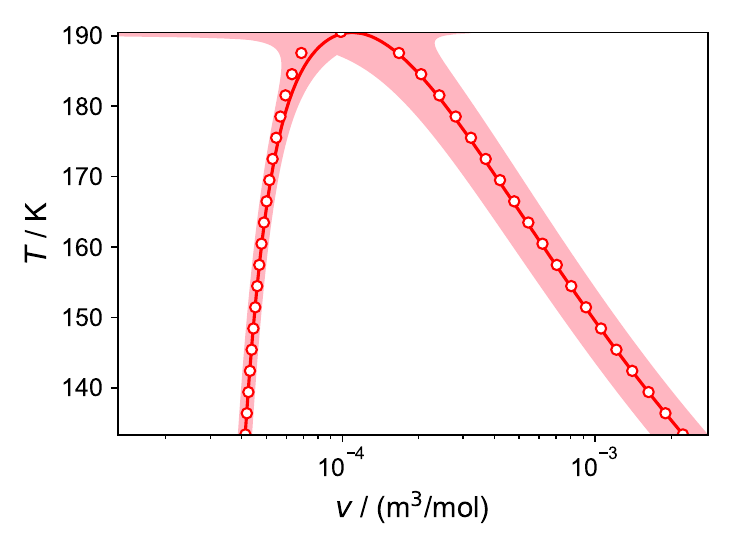}
      \caption{Saturation Envelope}

    \end{subfigure}
    \caption{Predicted values for the saturated volumes and saturation pressure for methane using PC-SAFT. Shaded regions correspond to the uncertainty interval for the predicted properties using the parameters and confidence intervals.}
  \end{figure}

\begin{figure}[H]
  \centering
    \begin{subfigure}[b]{0.49\textwidth}
      \includegraphics[width=1\textwidth]{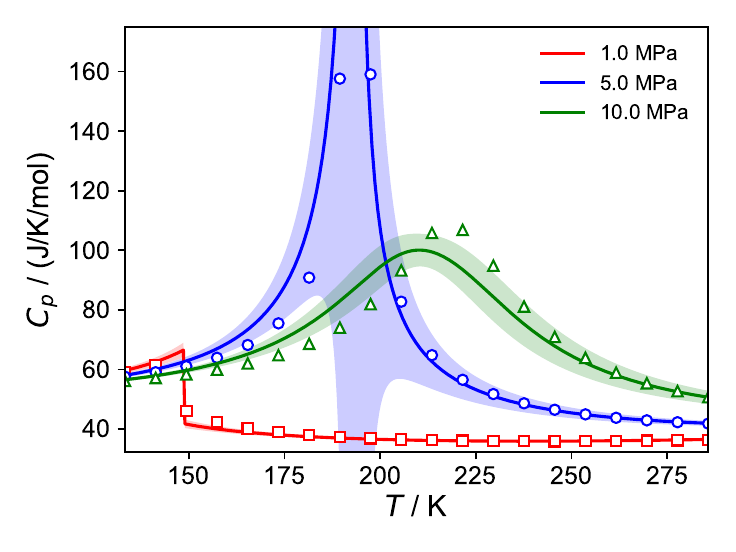}
      \caption{Isobaric heat capacity}
    \end{subfigure}
    \begin{subfigure}[b]{0.49\textwidth}
      \includegraphics[width=1\textwidth]{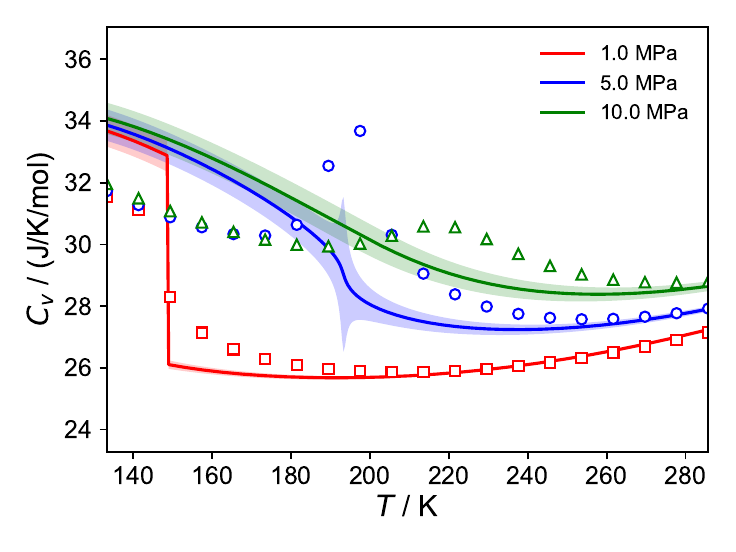}
      \caption{Isochoric heat capacity}
    \end{subfigure}
    \begin{subfigure}[b]{0.49\textwidth}
      \includegraphics[width=1\textwidth]{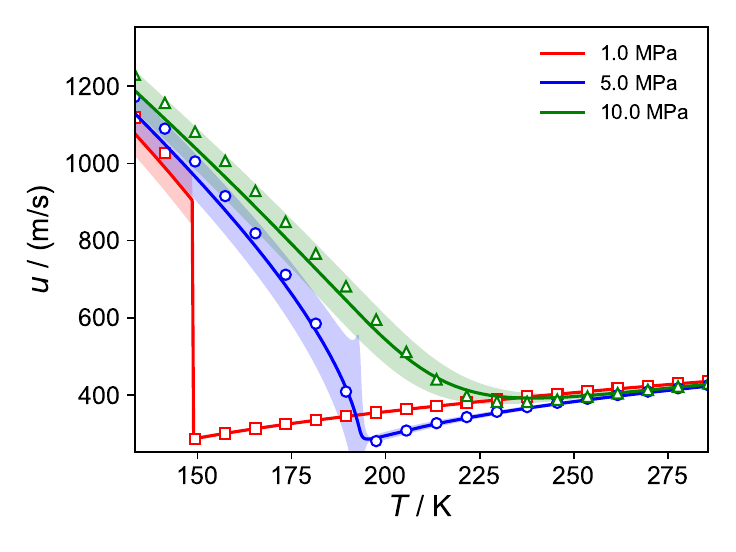}
      \caption{Speed of sound}
    \end{subfigure}
    \begin{subfigure}[b]{0.49\textwidth}
      \includegraphics[width=1\textwidth]{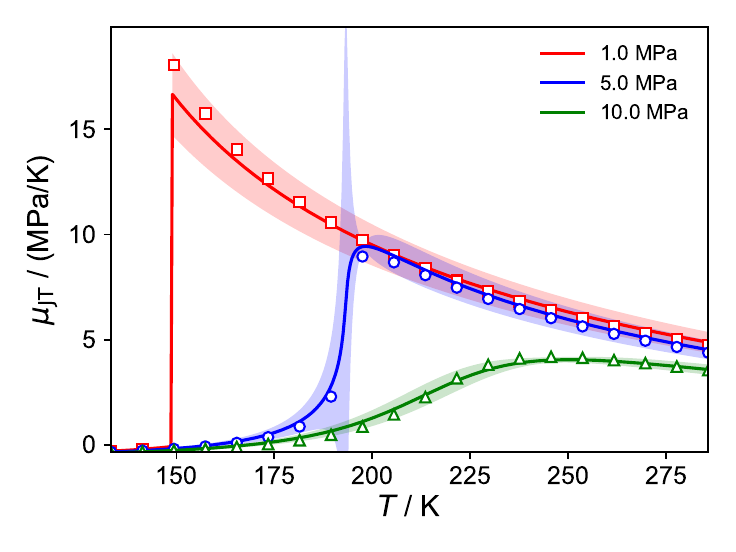}
      \caption{Joule--Thomson coefficient}
    \end{subfigure}
    \caption{Predicted values for the isobaric heat capacity, isochoric heat capacity, speed of sound and Joule--Thomson coefficient for methane using PC-SAFT at different pressures. Shaded regions correspond to the uncertainty interval for the predicted properties using the parameters and confidence intervals.}
  \end{figure}

\newpage
\subsection{Ethane}
\begin{figure}[H]
  \centering
      \includegraphics[width=0.5\textwidth]{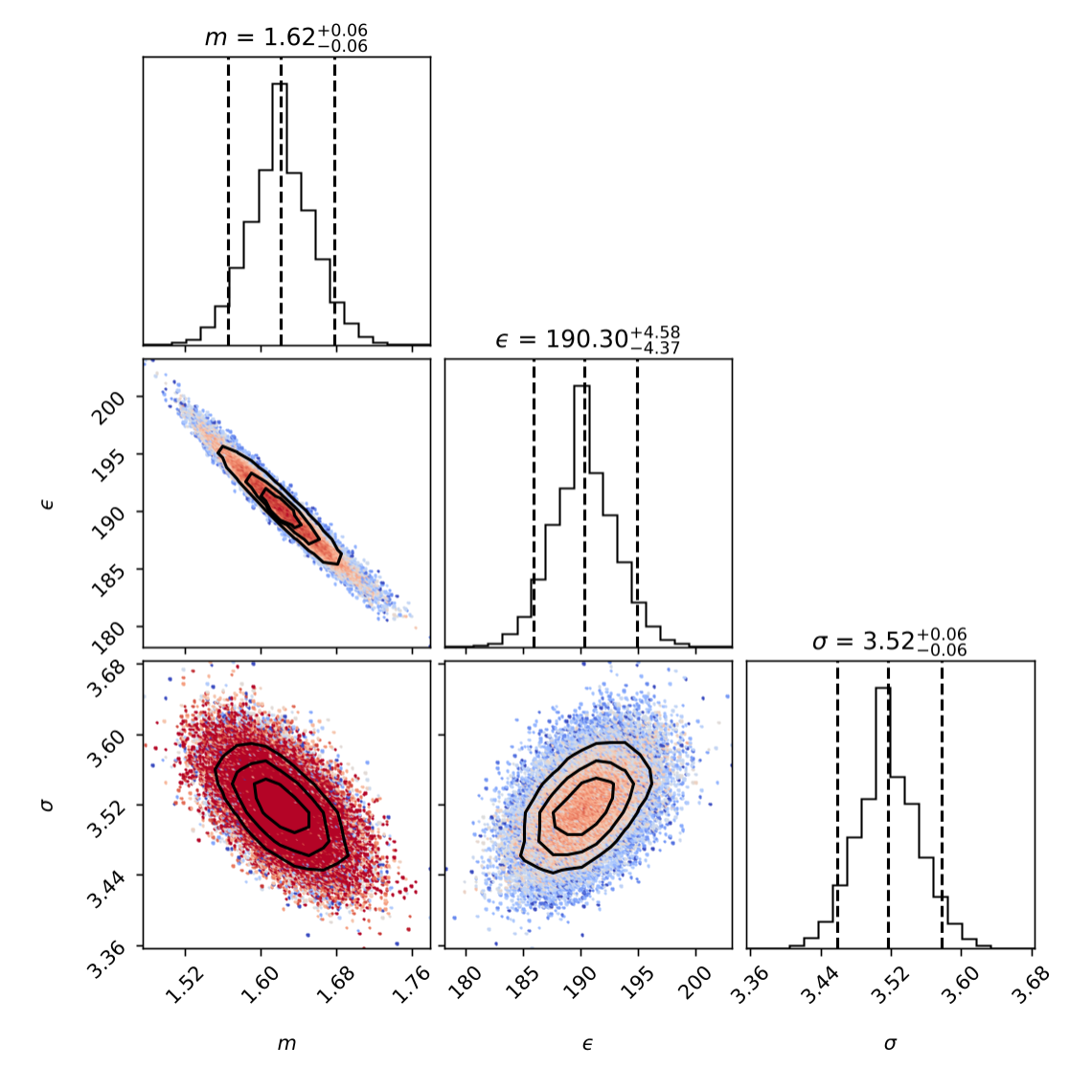}
      \caption{Confidence intervals obtained for the pure component parameters of ethane in PC-SAFT. Colors and styles are identical to figure 2.}
  \end{figure}

\begin{figure}[H]
  \centering
    \begin{subfigure}[b]{0.49\textwidth}
      \includegraphics[width=1\textwidth]{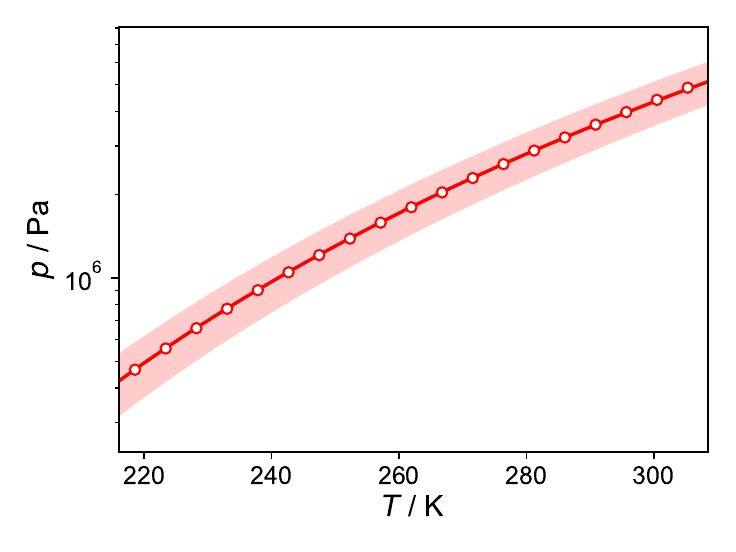}
      \caption{Saturation Curve}
    \end{subfigure}
    \begin{subfigure}[b]{0.49\textwidth}
      \includegraphics[width=1\textwidth]{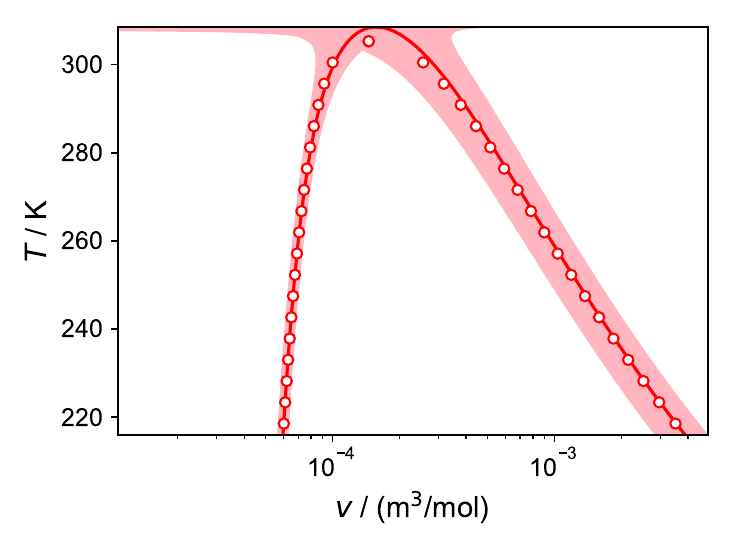}
      \caption{Saturation Envelope}

    \end{subfigure}
    \caption{Predicted values for the saturated volumes and saturation pressure for ethane using PC-SAFT. Shaded regions correspond to the uncertainty interval for the predicted properties using the parameters and confidence intervals.}
  \end{figure}

\begin{figure}[H]
  \centering
    \begin{subfigure}[b]{0.49\textwidth}
      \includegraphics[width=1\textwidth]{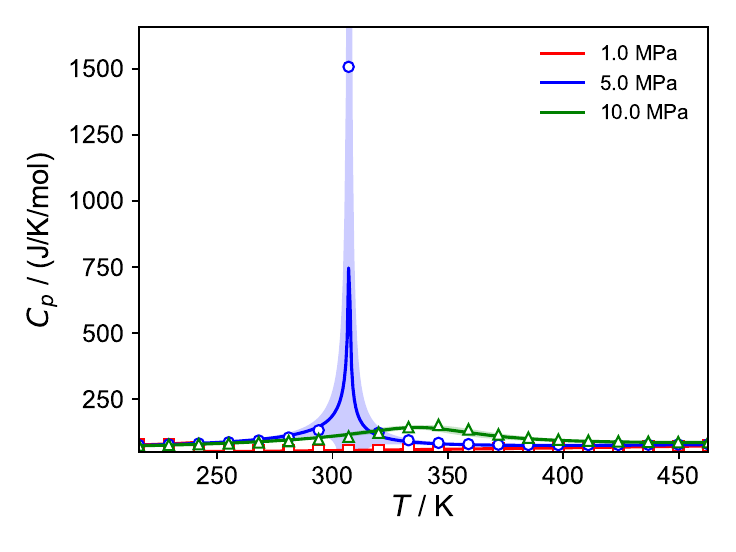}
      \caption{Isobaric heat capacity}
    \end{subfigure}
    \begin{subfigure}[b]{0.49\textwidth}
      \includegraphics[width=1\textwidth]{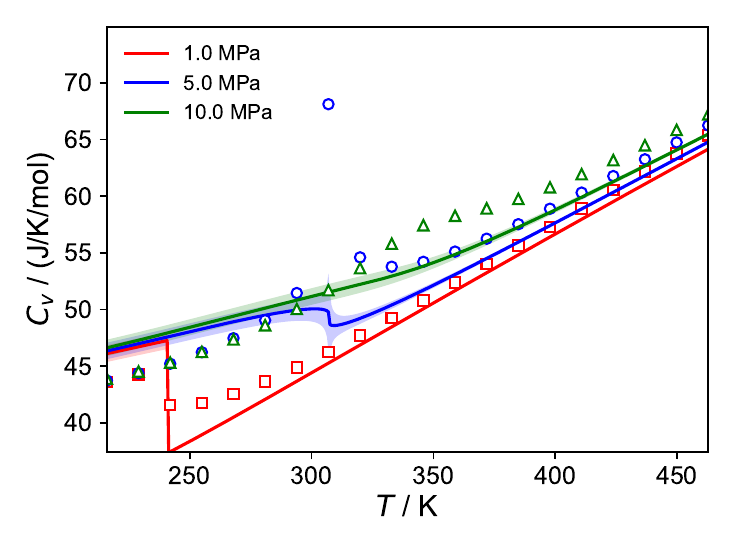}
      \caption{Isochoric heat capacity}
    \end{subfigure}
    \begin{subfigure}[b]{0.49\textwidth}
      \includegraphics[width=1\textwidth]{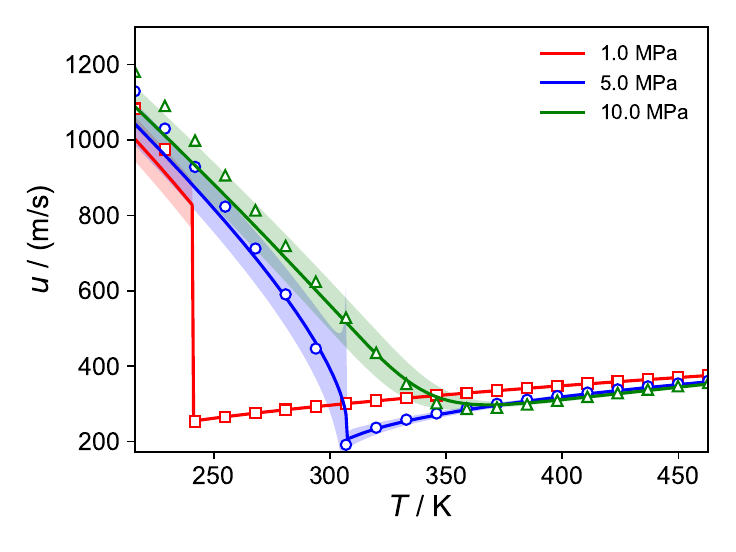}
      \caption{Speed of sound}
    \end{subfigure}
    \begin{subfigure}[b]{0.49\textwidth}
      \includegraphics[width=1\textwidth]{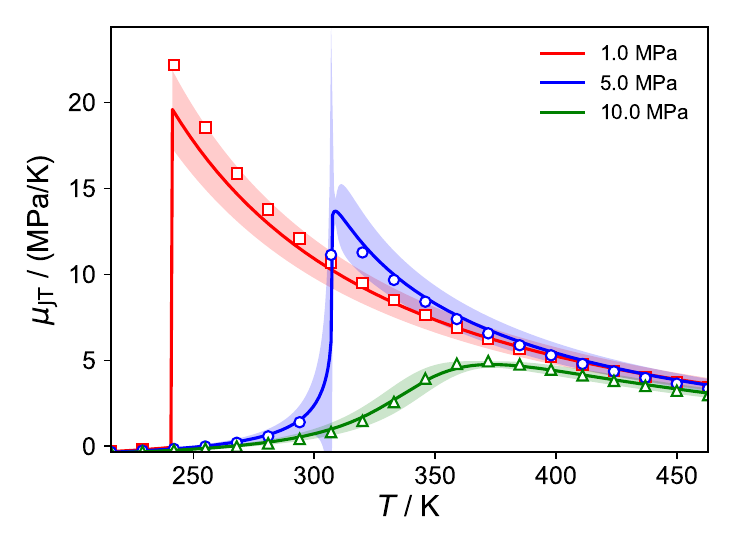}
      \caption{Joule--Thomson coefficient}
    \end{subfigure}
    \caption{Predicted values for the isobaric heat capacity, isochoric heat capacity, speed of sound and Joule--Thomson coefficient for ethane using PC-SAFT at different pressures. Shaded regions correspond to the uncertainty interval for the predicted properties using the parameters and confidence intervals.}
  \end{figure}
\newpage
\subsection{Propane}
\begin{figure}[H]
  \centering
      \includegraphics[width=0.5\textwidth]{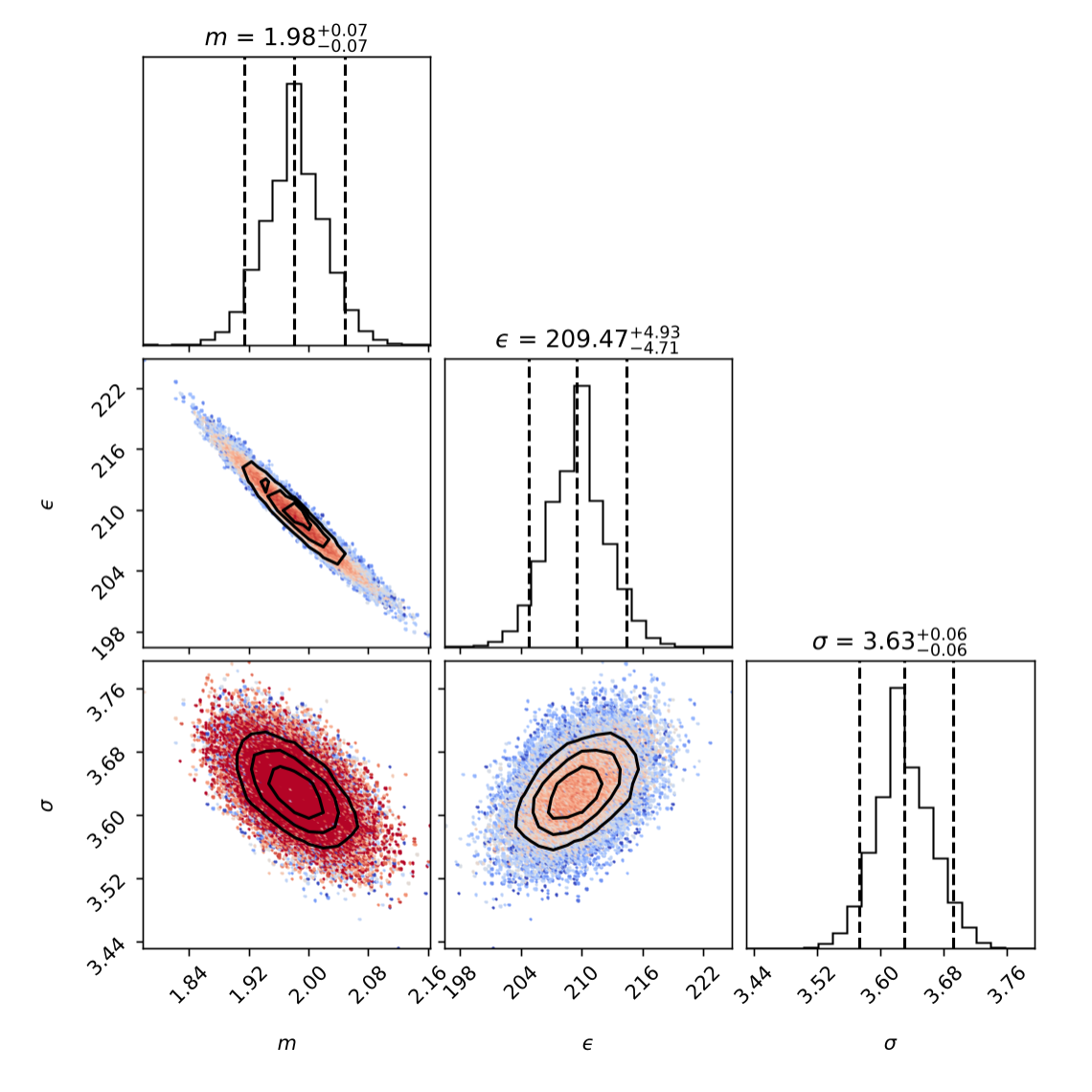}
      \caption{Confidence intervals obtained for the pure component parameters of propane in PC-SAFT. Colors and styles are identical to figure 2.}
  \end{figure}

\begin{figure}[H]
  \centering
    \begin{subfigure}[b]{0.49\textwidth}
      \includegraphics[width=1\textwidth]{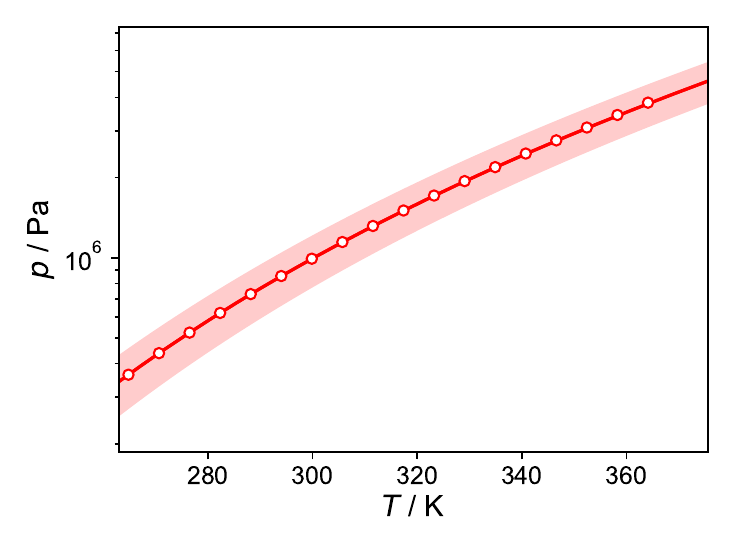}
      \caption{Saturation Curve}
    \end{subfigure}
    \begin{subfigure}[b]{0.49\textwidth}
      \includegraphics[width=1\textwidth]{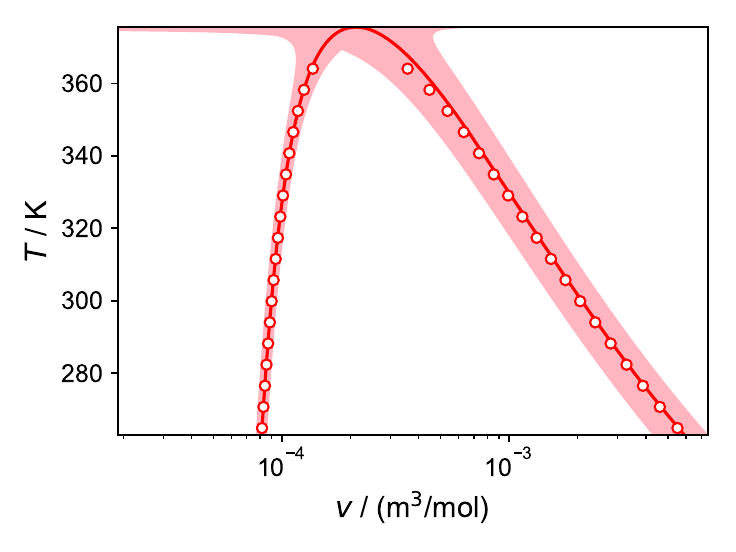}
      \caption{Saturation Envelope}

    \end{subfigure}
    \caption{Predicted values for the saturated volumes and saturation pressure for propane using PC-SAFT. Shaded regions correspond to the uncertainty interval for the predicted properties using the parameters and confidence intervals.}
  \end{figure}

\begin{figure}[H]
  \centering
    \begin{subfigure}[b]{0.49\textwidth}
      \includegraphics[width=1\textwidth]{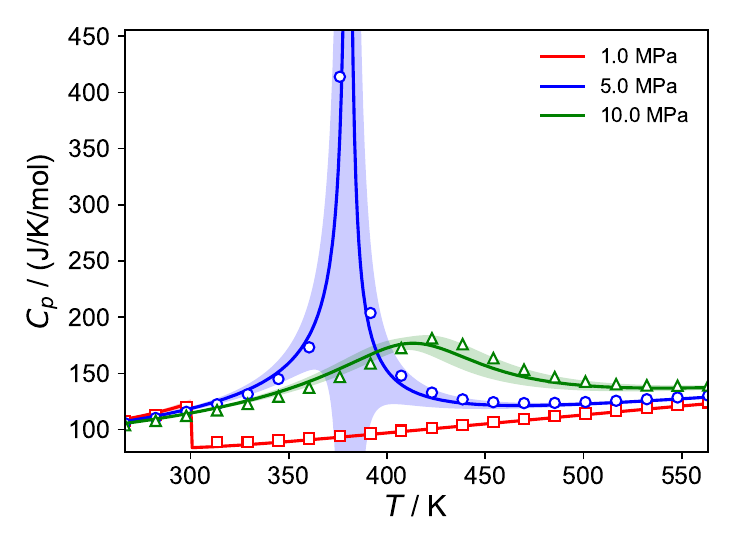}
      \caption{Isobaric heat capacity}
    \end{subfigure}
    \begin{subfigure}[b]{0.49\textwidth}
      \includegraphics[width=1\textwidth]{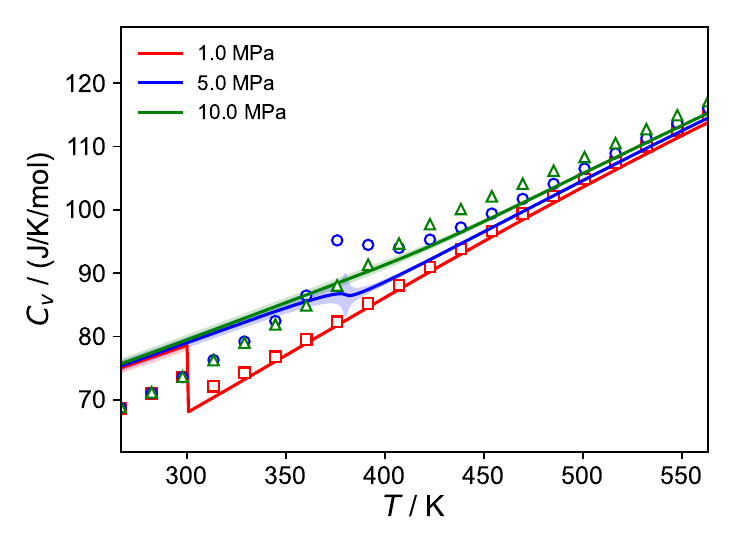}
      \caption{Isochoric heat capacity}
    \end{subfigure}
    \begin{subfigure}[b]{0.49\textwidth}
      \includegraphics[width=1\textwidth]{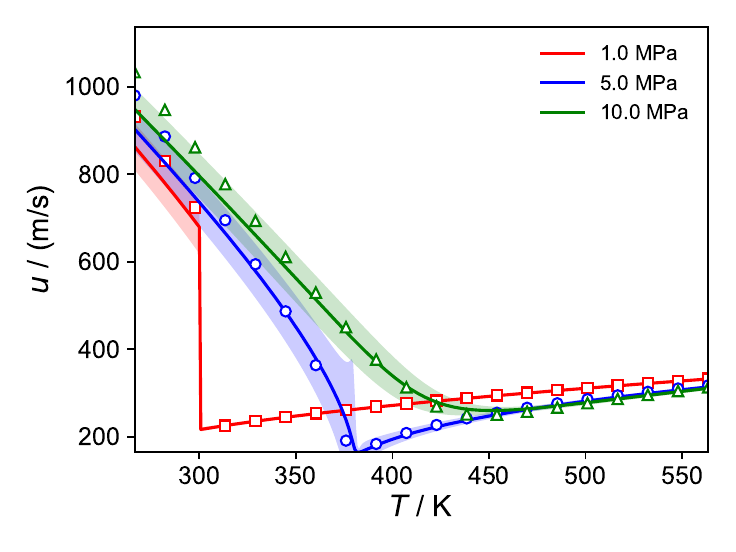}
      \caption{Speed of sound}
    \end{subfigure}
    \begin{subfigure}[b]{0.49\textwidth}
      \includegraphics[width=1\textwidth]{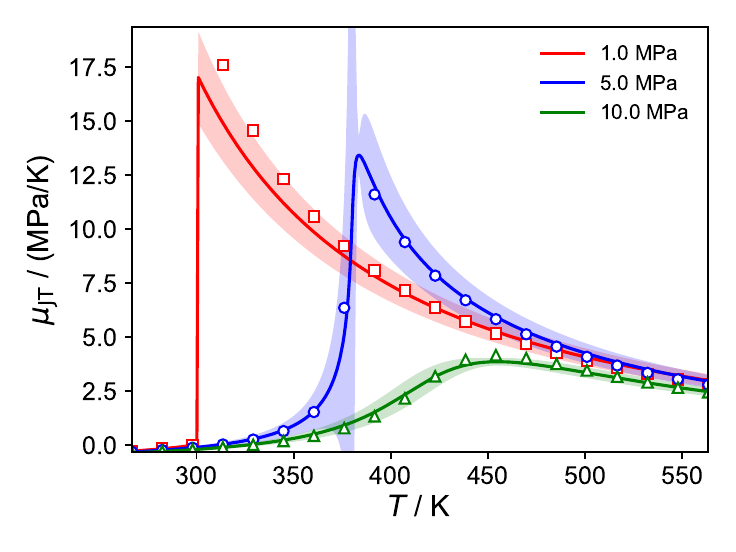}
      \caption{Joule--Thomson coefficient}
    \end{subfigure}
    \caption{Predicted values for the isobaric heat capacity, isochoric heat capacity, speed of sound and Joule--Thomson coefficient for propane using PC-SAFT at different pressures. Shaded regions correspond to the uncertainty interval for the predicted properties using the parameters and confidence intervals.}
  \end{figure}
\newpage

\subsection{$n$-butane}
\begin{figure}[H]
  \centering
      \includegraphics[width=0.5\textwidth]{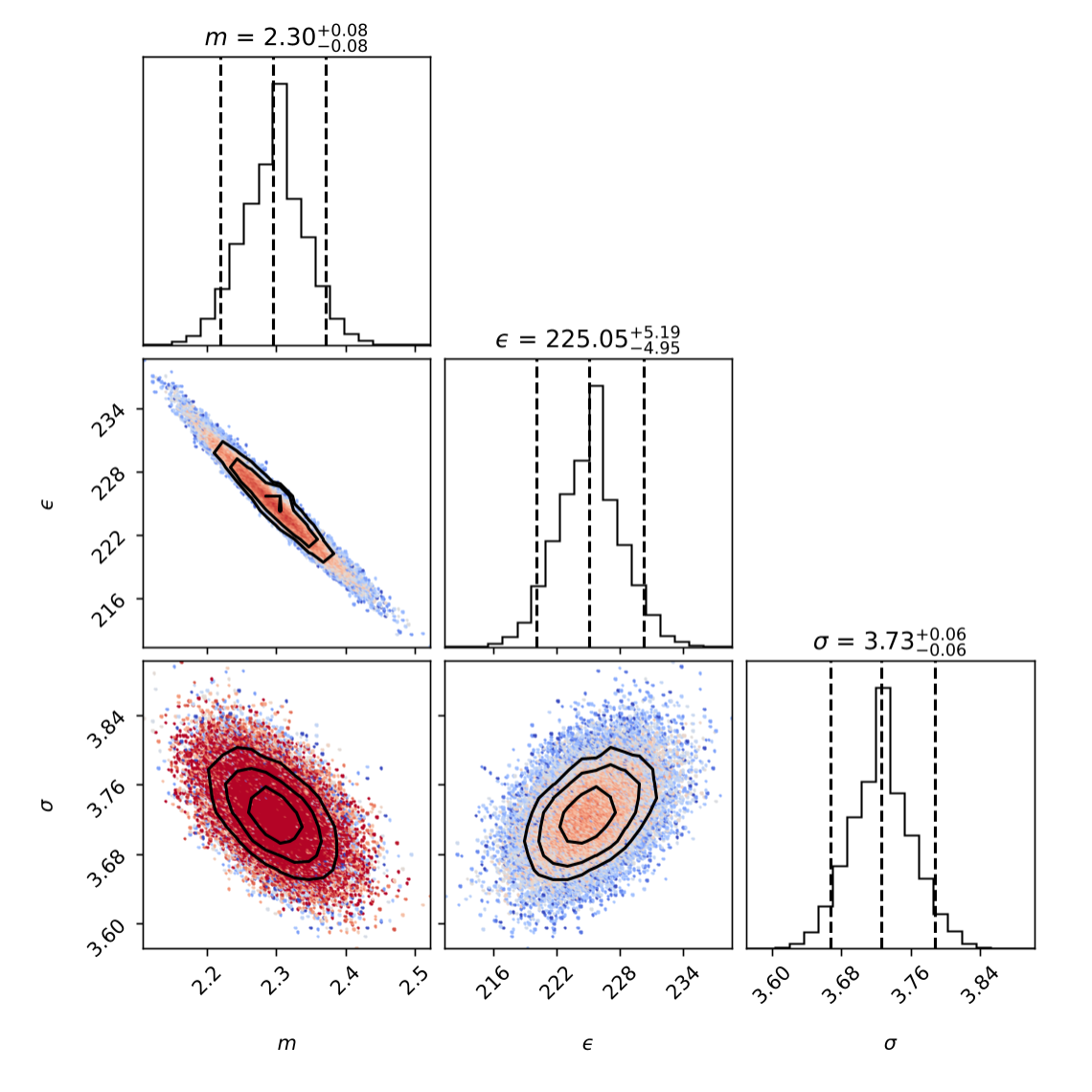}
      \caption{Confidence intervals obtained for the pure component parameters of $n$-butane in PC-SAFT. Colors and styles are identical to figure 2.}
  \end{figure}

\begin{figure}[H]
  \centering
    \begin{subfigure}[b]{0.49\textwidth}
      \includegraphics[width=1\textwidth]{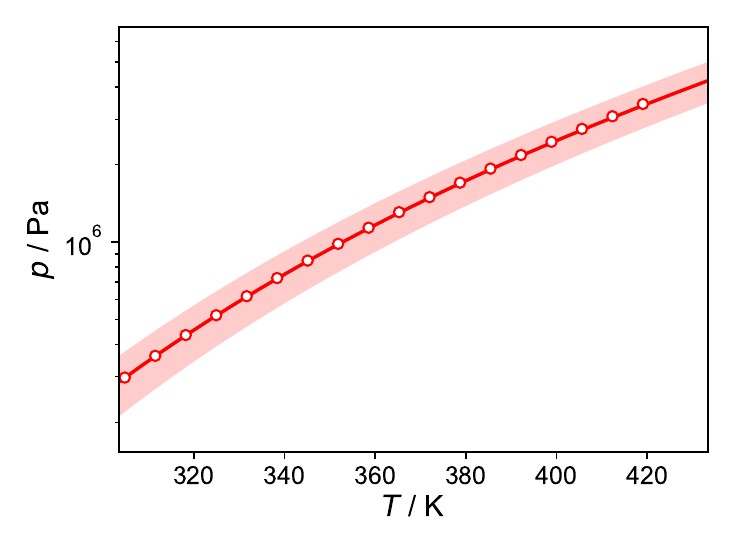}
      \caption{Saturation Curve}
    \end{subfigure}
    \begin{subfigure}[b]{0.49\textwidth}
      \includegraphics[width=1\textwidth]{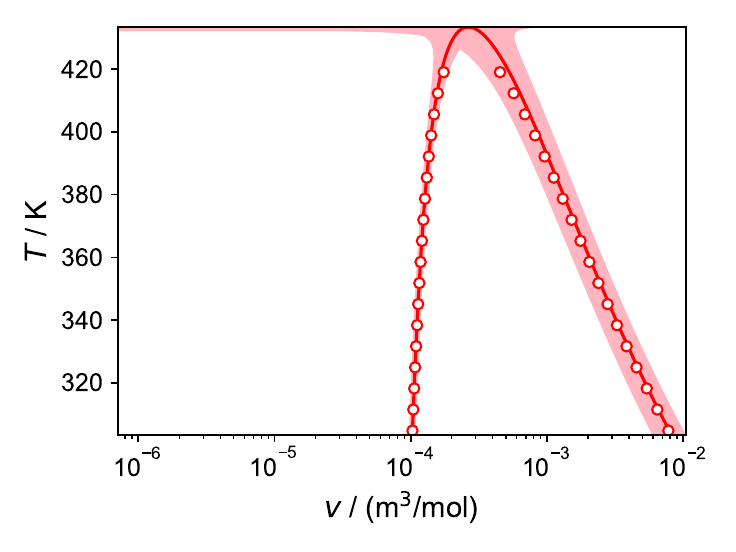}
      \caption{Saturation Envelope}

    \end{subfigure}
    \caption{Predicted values for the saturated volumes and saturation pressure for $n$-butane using PC-SAFT. Shaded regions correspond to the uncertainty interval for the predicted properties using the parameters and confidence intervals.}
  \end{figure}

\begin{figure}[H]
  \centering
    \begin{subfigure}[b]{0.49\textwidth}
      \includegraphics[width=1\textwidth]{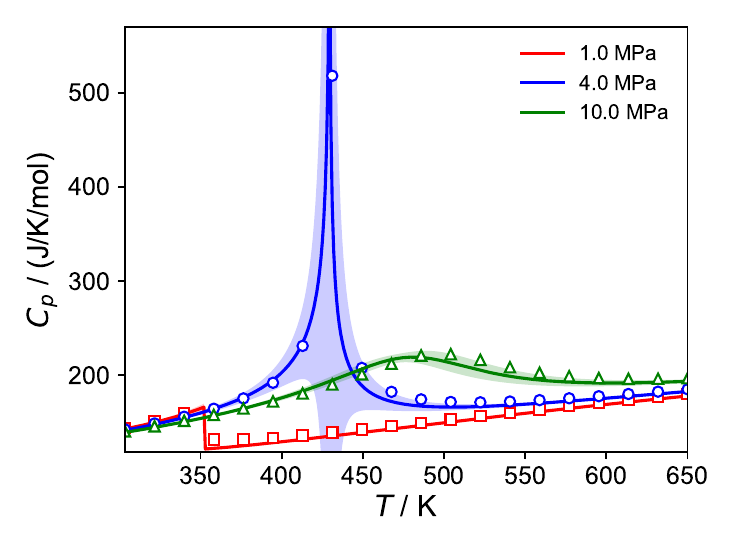}
      \caption{Isobaric heat capacity}
    \end{subfigure}
    \begin{subfigure}[b]{0.49\textwidth}
      \includegraphics[width=1\textwidth]{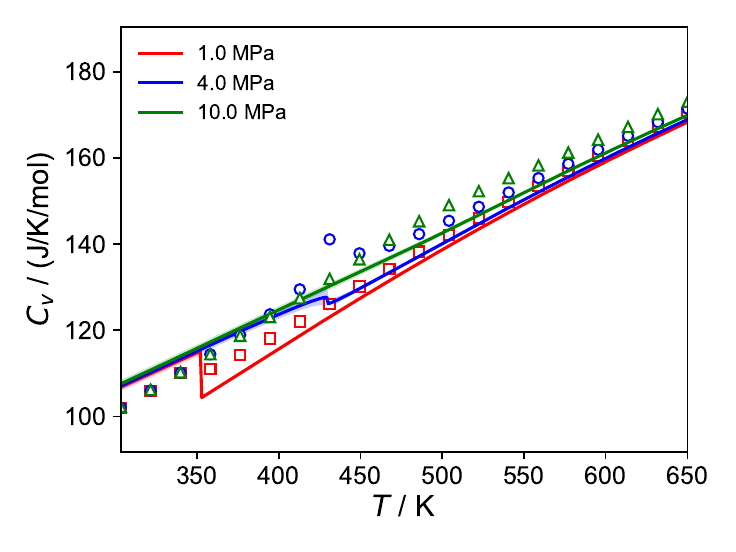}
      \caption{Isochoric heat capacity}
    \end{subfigure}
    \begin{subfigure}[b]{0.49\textwidth}
      \includegraphics[width=1\textwidth]{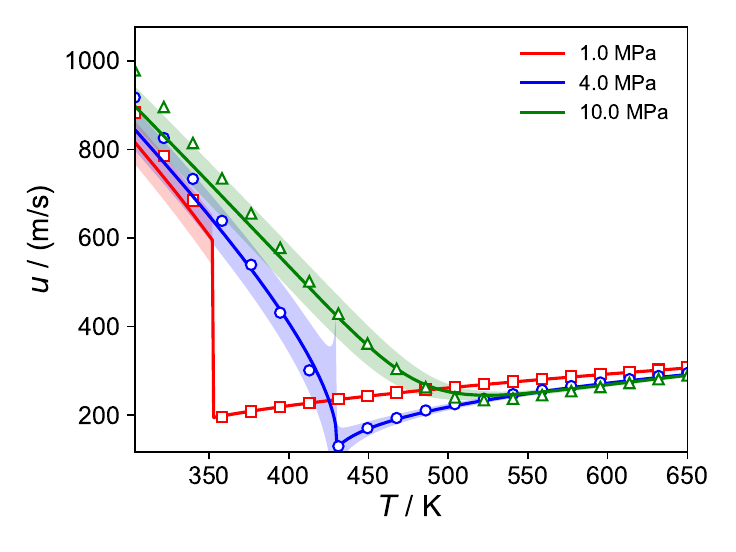}
      \caption{Speed of sound}
    \end{subfigure}
    \begin{subfigure}[b]{0.49\textwidth}
      \includegraphics[width=1\textwidth]{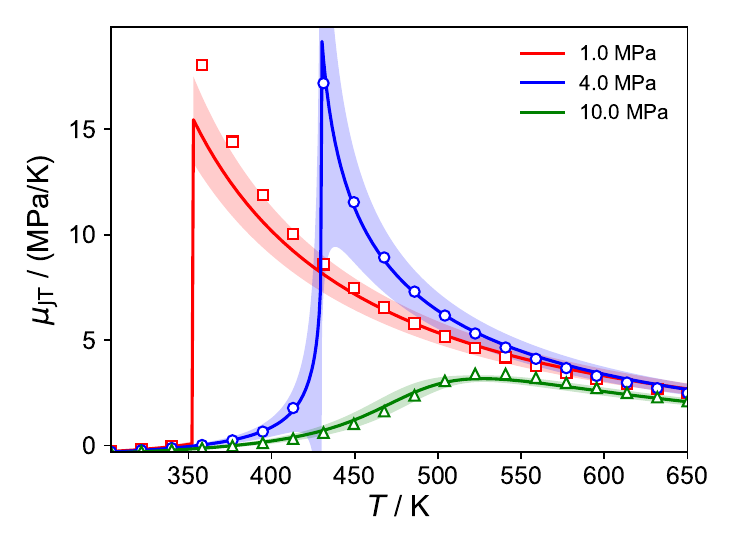}
      \caption{Joule--Thomson coefficient}
    \end{subfigure}
    \caption{Predicted values for the isobaric heat capacity, isochoric heat capacity, speed of sound and Joule--Thomson coefficient for $n$-butane using PC-SAFT at different pressures. Shaded regions correspond to the uncertainty interval for the predicted properties using the parameters and confidence intervals.}
  \end{figure}
\newpage

\subsection{$n$-pentane}
\begin{figure}[H]
  \centering
      \includegraphics[width=0.5\textwidth]{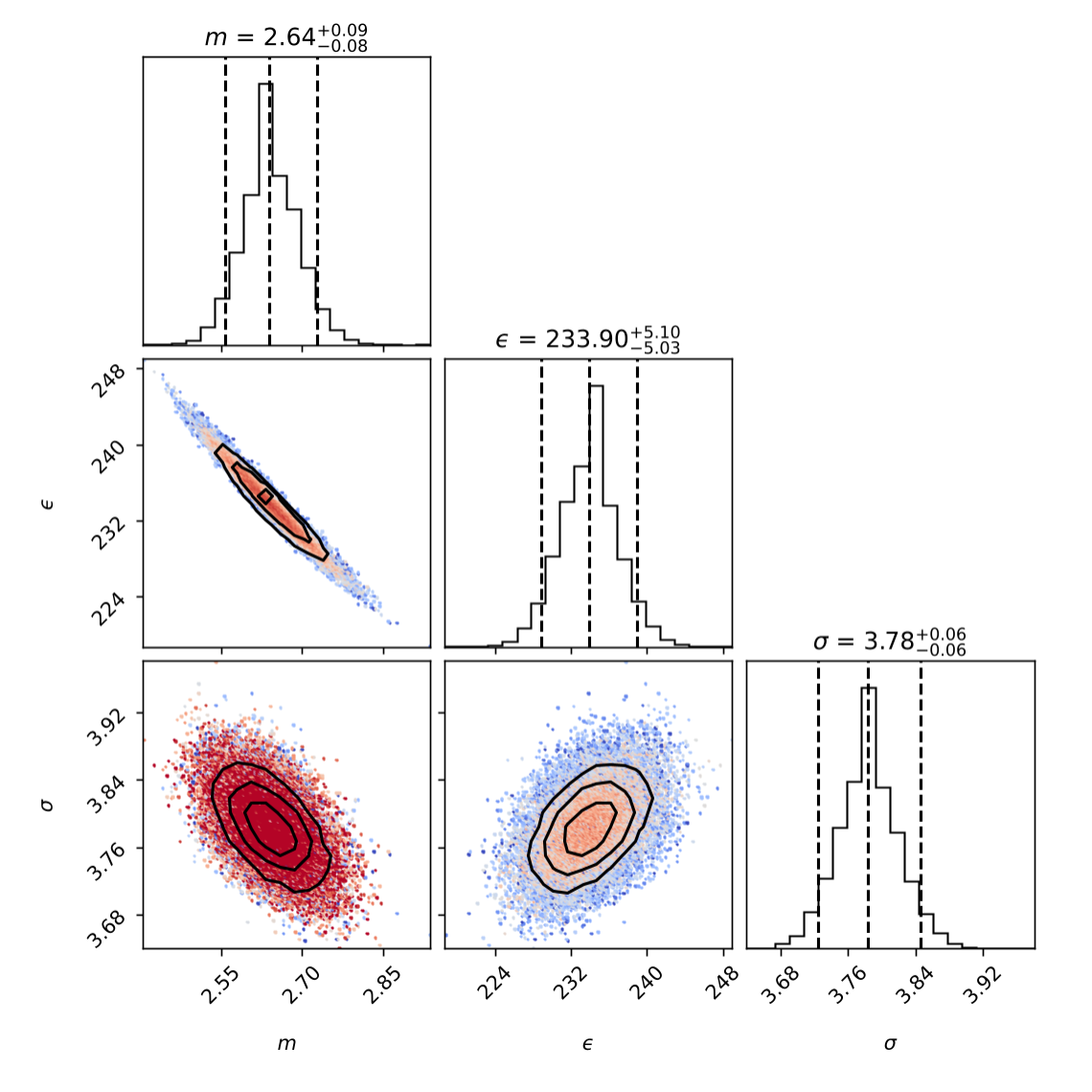}
      \caption{Confidence intervals obtained for the pure component parameters of $n$-pentane in PC-SAFT. Colors and styles are identical to figure 2.}
  \end{figure}

\begin{figure}[H]
  \centering
    \begin{subfigure}[b]{0.49\textwidth}
      \includegraphics[width=1\textwidth]{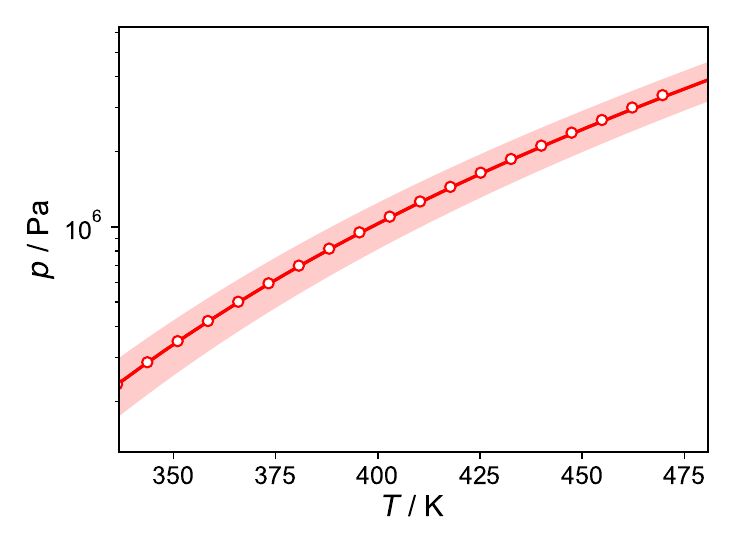}
      \caption{Saturation Curve}
    \end{subfigure}
    \begin{subfigure}[b]{0.49\textwidth}
      \includegraphics[width=1\textwidth]{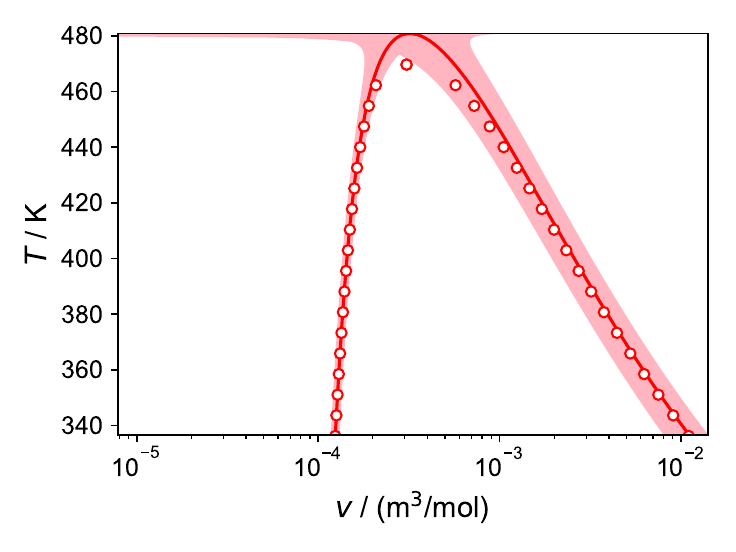}
      \caption{Saturation Envelope}

    \end{subfigure}
    \caption{Predicted values for the saturated volumes and saturation pressure for $n$-pentane using PC-SAFT. Shaded regions correspond to the uncertainty interval for the predicted properties using the parameters and confidence intervals.}
  \end{figure}

\begin{figure}[H]
  \centering
    \begin{subfigure}[b]{0.49\textwidth}
      \includegraphics[width=1\textwidth]{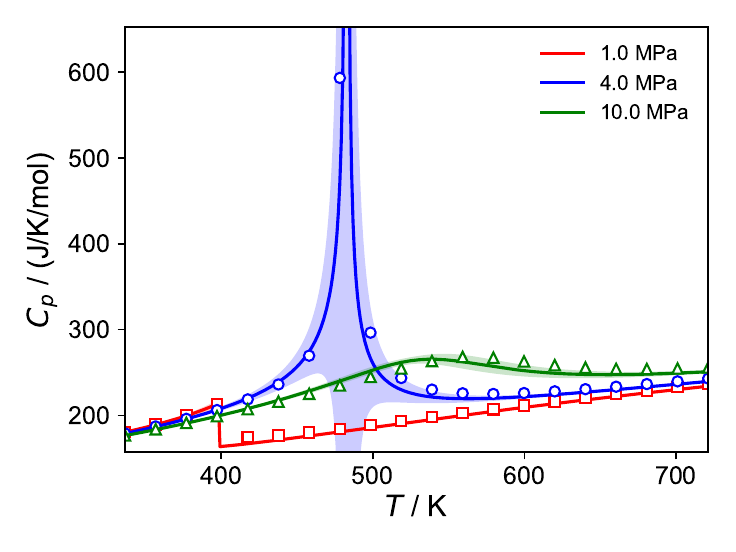}
      \caption{Isobaric heat capacity}
    \end{subfigure}
    \begin{subfigure}[b]{0.49\textwidth}
      \includegraphics[width=1\textwidth]{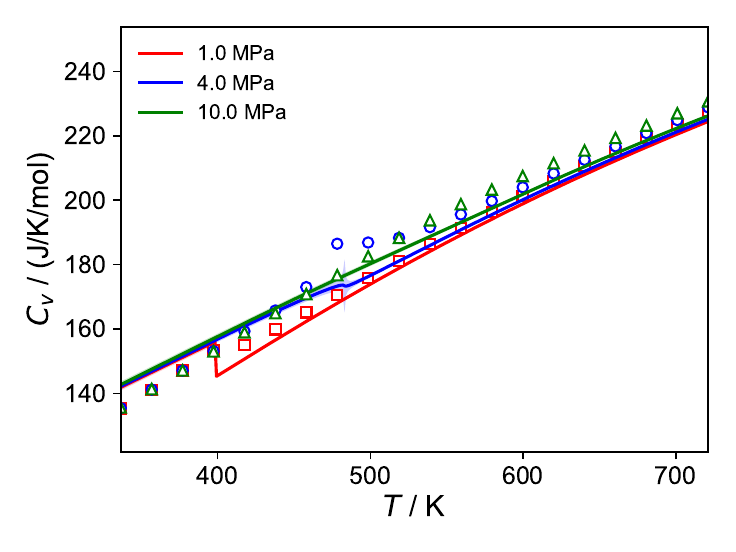}
      \caption{Isochoric heat capacity}
    \end{subfigure}
    \begin{subfigure}[b]{0.49\textwidth}
      \includegraphics[width=1\textwidth]{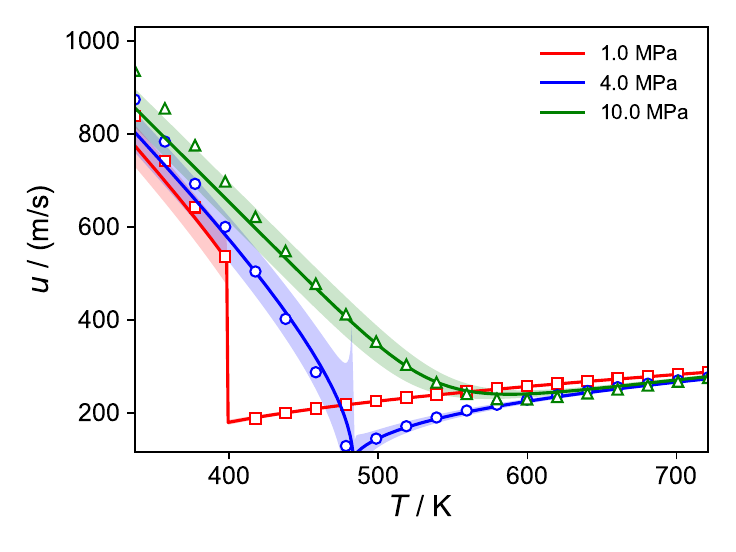}
      \caption{Speed of sound}
    \end{subfigure}
    \begin{subfigure}[b]{0.49\textwidth}
      \includegraphics[width=1\textwidth]{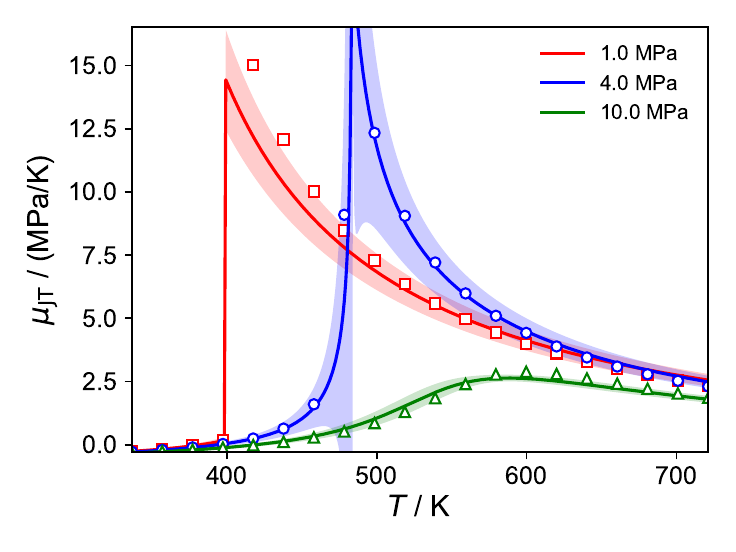}
      \caption{Joule--Thomson coefficient}
    \end{subfigure}
    \caption{Predicted values for the isobaric heat capacity, isochoric heat capacity, speed of sound and Joule--Thomson coefficient for $n$-pentane using PC-SAFT at different pressures. Shaded regions correspond to the uncertainty interval for the predicted properties using the parameters and confidence intervals.}
  \end{figure}
\newpage

\subsection{$n$-hexane}
\begin{figure}[H]
  \centering
      \includegraphics[width=0.5\textwidth]{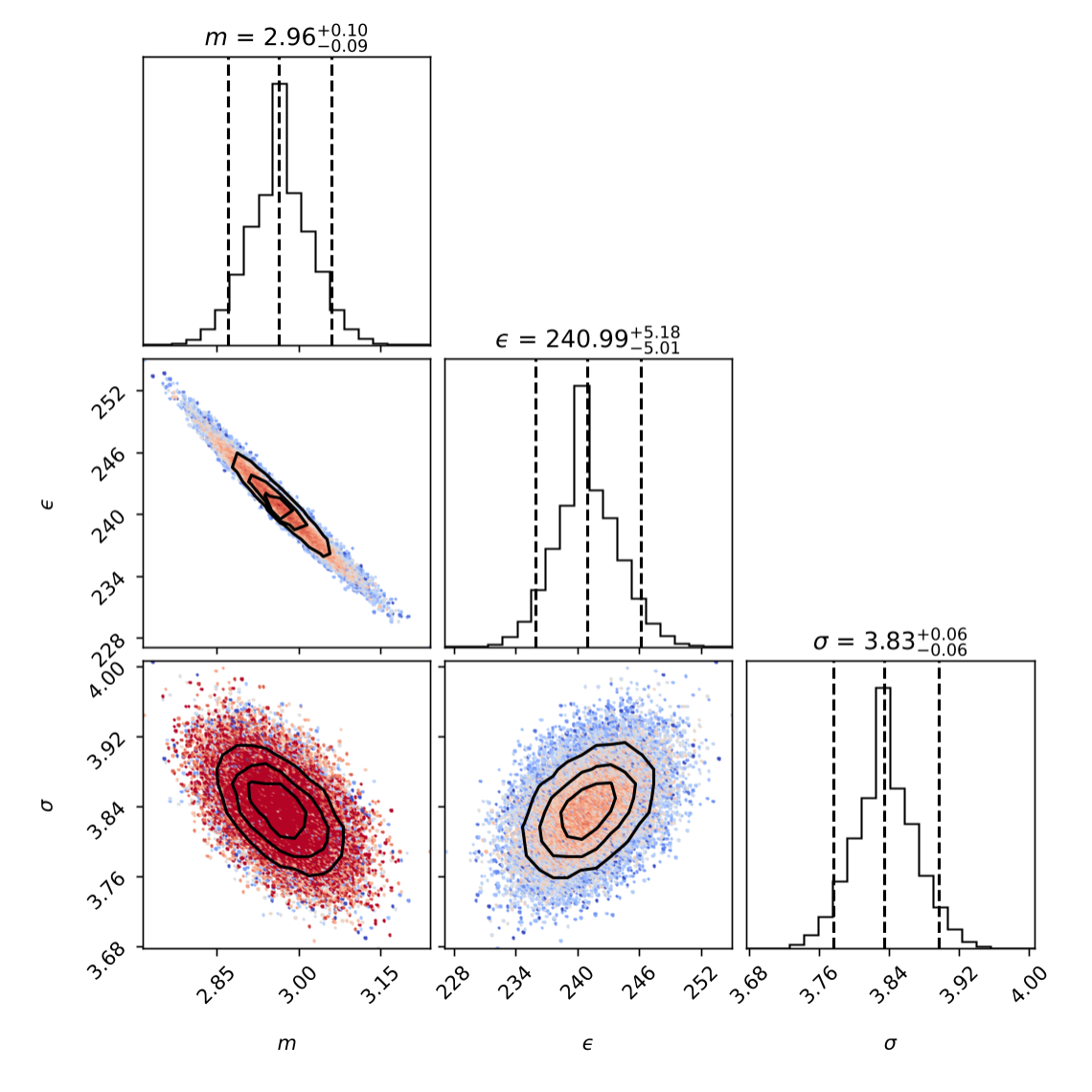}
      \caption{Confidence intervals obtained for the pure component parameters of $n$-hexane in PC-SAFT. Colors and styles are identical to figure 2.}
  \end{figure}

\begin{figure}[H]
  \centering
    \begin{subfigure}[b]{0.49\textwidth}
      \includegraphics[width=1\textwidth]{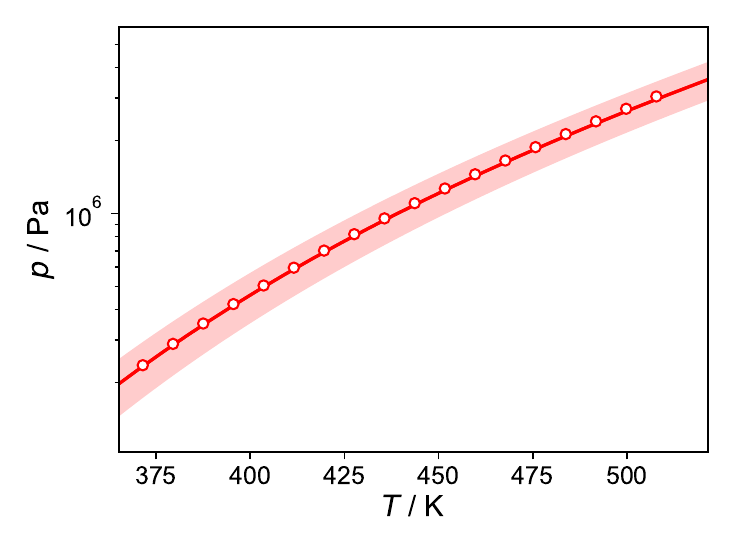}
      \caption{Saturation Curve}
    \end{subfigure}
    \begin{subfigure}[b]{0.49\textwidth}
      \includegraphics[width=1\textwidth]{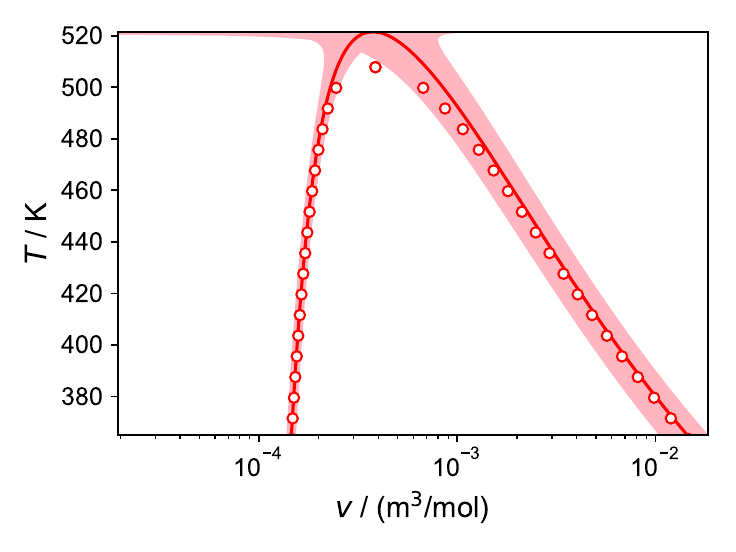}
      \caption{Saturation Envelope}

    \end{subfigure}
    \caption{Predicted values for the saturated volumes and saturation pressure for $n$-hexane using PC-SAFT. Shaded regions correspond to the uncertainty interval for the predicted properties using the parameters and confidence intervals.}
  \end{figure}

\begin{figure}[H]
  \centering
    \begin{subfigure}[b]{0.49\textwidth}
      \includegraphics[width=1\textwidth]{Figures/PCSAFT/Error/hexane_cp_error.pdf}
      \caption{Isobaric heat capacity}
    \end{subfigure}
    \begin{subfigure}[b]{0.49\textwidth}
      \includegraphics[width=1\textwidth]{Figures/PCSAFT/Error/hexane_cv_error.pdf}
      \caption{Isochoric heat capacity}
    \end{subfigure}
    \begin{subfigure}[b]{0.49\textwidth}
      \includegraphics[width=1\textwidth]{Figures/PCSAFT/Error/hexane_sos_error.pdf}
      \caption{Speed of sound}
    \end{subfigure}
    \begin{subfigure}[b]{0.49\textwidth}
      \includegraphics[width=1\textwidth]{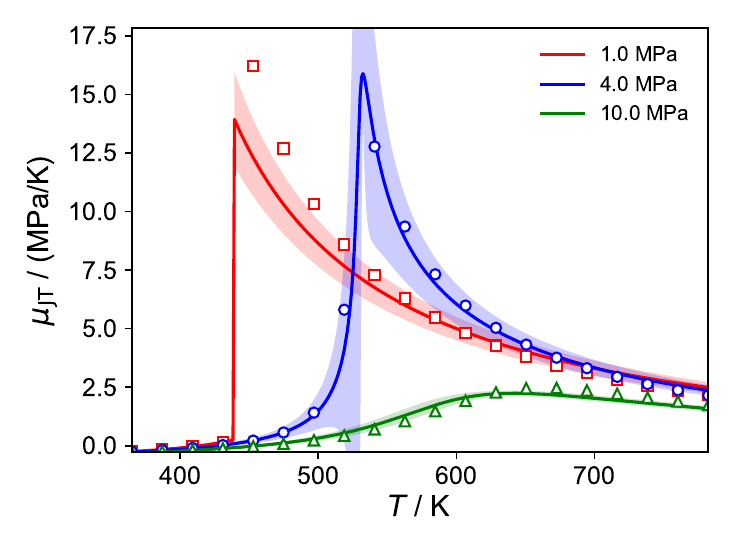}
      \caption{Joule--Thomson coefficient}
    \end{subfigure}
    \caption{Predicted values for the isobaric heat capacity, isochoric heat capacity, speed of sound and Joule--Thomson coefficient for $n$-hexane using PC-SAFT at different pressures. Shaded regions correspond to the uncertainty interval for the predicted properties using the parameters and confidence intervals.}
  \end{figure}
\newpage

\subsection{$n$-heptane}
\begin{figure}[H]
  \centering
      \includegraphics[width=0.5\textwidth]{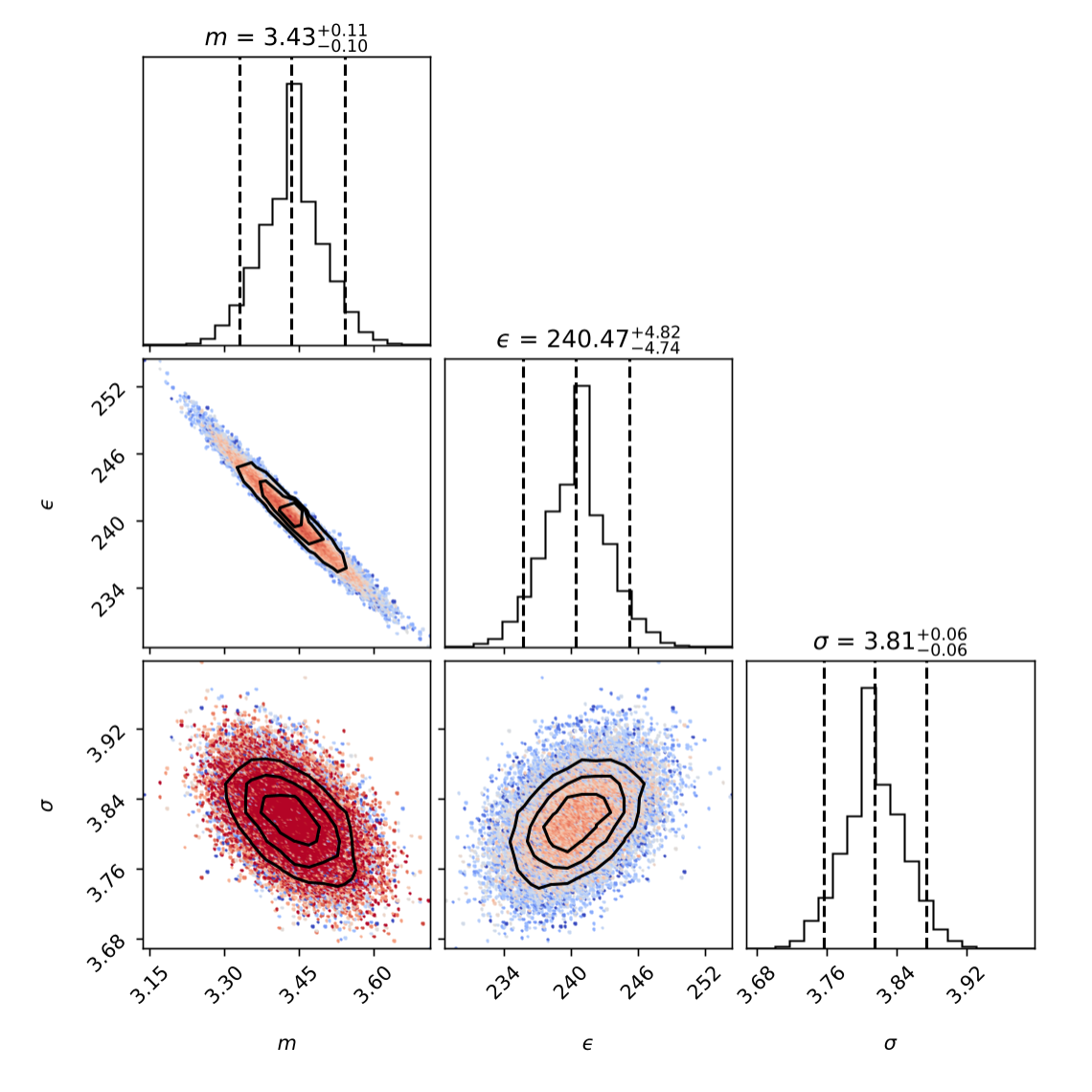}
      \caption{Confidence intervals obtained for the pure component parameters of $n$-heptane in PC-SAFT. Colors and styles are identical to figure 2.}
  \end{figure}

\begin{figure}[H]
  \centering
    \begin{subfigure}[b]{0.49\textwidth}
      \includegraphics[width=1\textwidth]{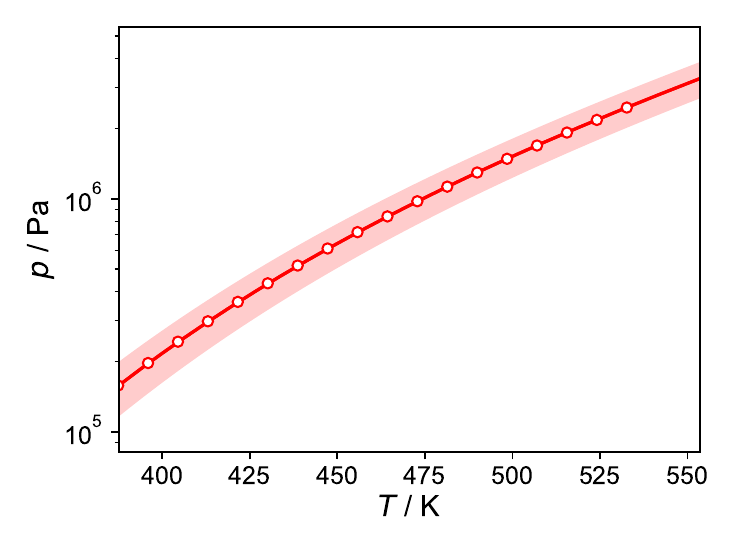}
      \caption{Saturation Curve}
    \end{subfigure}
    \begin{subfigure}[b]{0.49\textwidth}
      \includegraphics[width=1\textwidth]{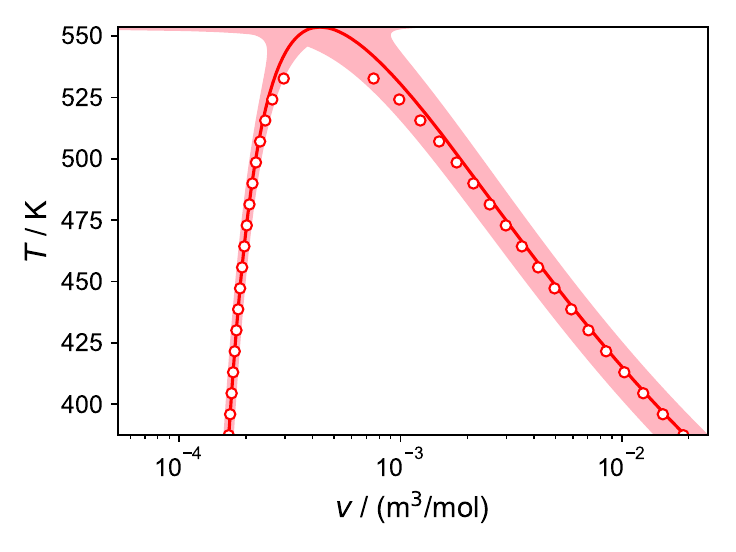}
      \caption{Saturation Envelope}

    \end{subfigure}
    \caption{Predicted values for the saturated volumes and saturation pressure for $n$-heptane using PC-SAFT. Shaded regions correspond to the uncertainty interval for the predicted properties using the parameters and confidence intervals.}
  \end{figure}

\begin{figure}[H]
  \centering
    \begin{subfigure}[b]{0.49\textwidth}
      \includegraphics[width=1\textwidth]{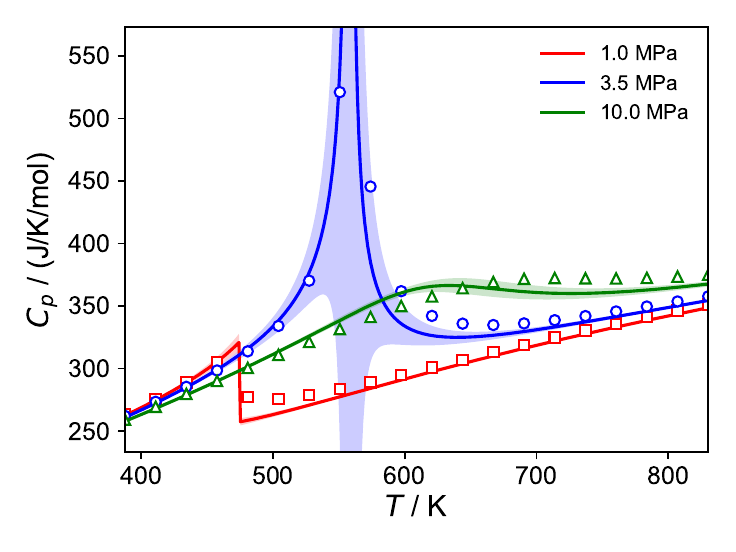}
      \caption{Isobaric heat capacity}
    \end{subfigure}
    \begin{subfigure}[b]{0.49\textwidth}
      \includegraphics[width=1\textwidth]{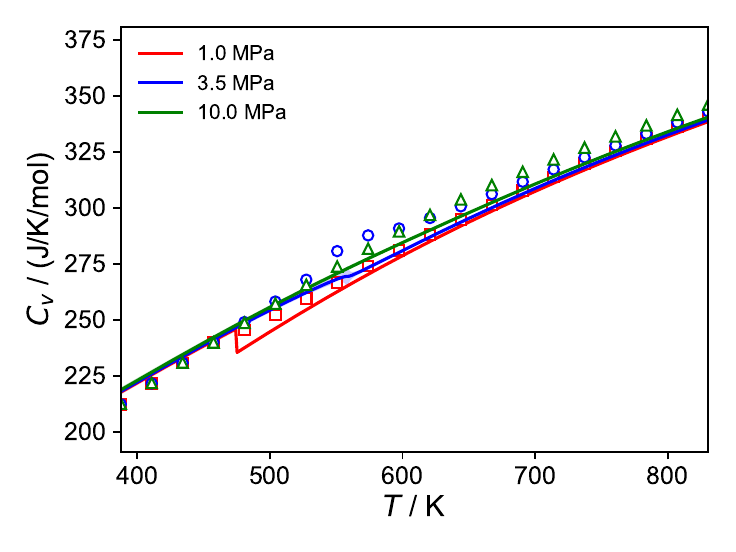}
      \caption{Isochoric heat capacity}
    \end{subfigure}
    \begin{subfigure}[b]{0.49\textwidth}
      \includegraphics[width=1\textwidth]{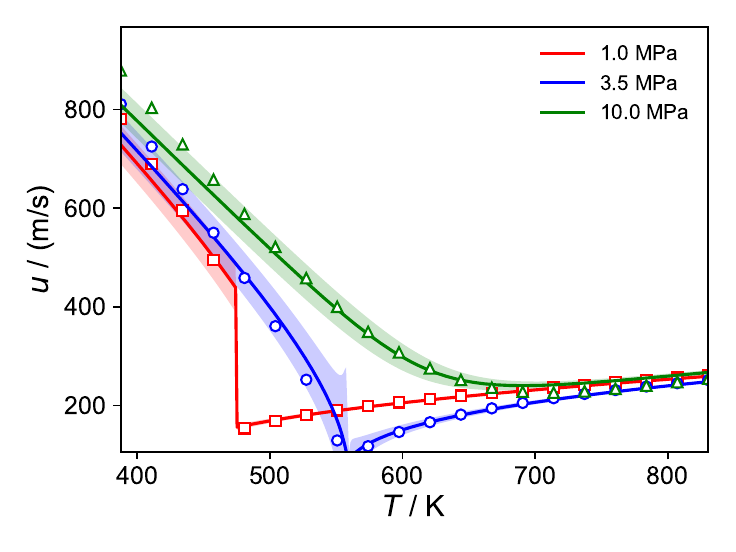}
      \caption{Speed of sound}
    \end{subfigure}
    \begin{subfigure}[b]{0.49\textwidth}
      \includegraphics[width=1\textwidth]{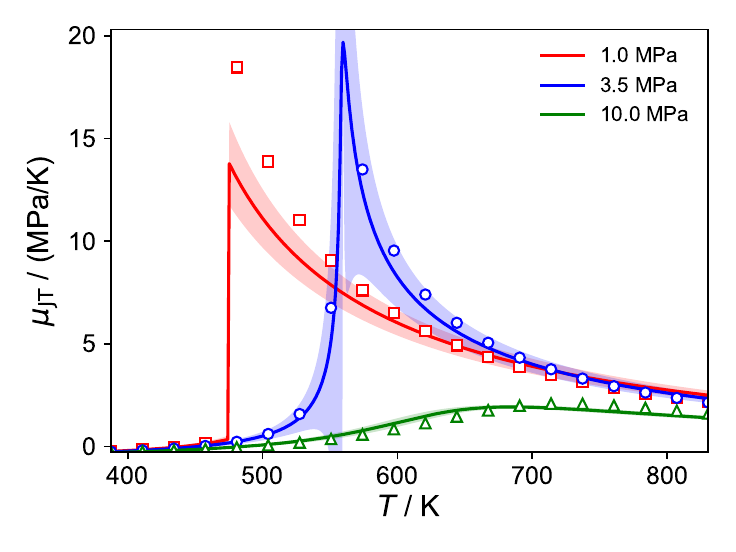}
      \caption{Joule--Thomson coefficient}
    \end{subfigure}
    \caption{Predicted values for the isobaric heat capacity, isochoric heat capacity, speed of sound and Joule--Thomson coefficient for $n$-heptane using PC-SAFT at different pressures. Shaded regions correspond to the uncertainty interval for the predicted properties using the parameters and confidence intervals.}
  \end{figure}
\newpage

\subsection{$n$-octane}
\begin{figure}[H]
  \centering
      \includegraphics[width=0.5\textwidth]{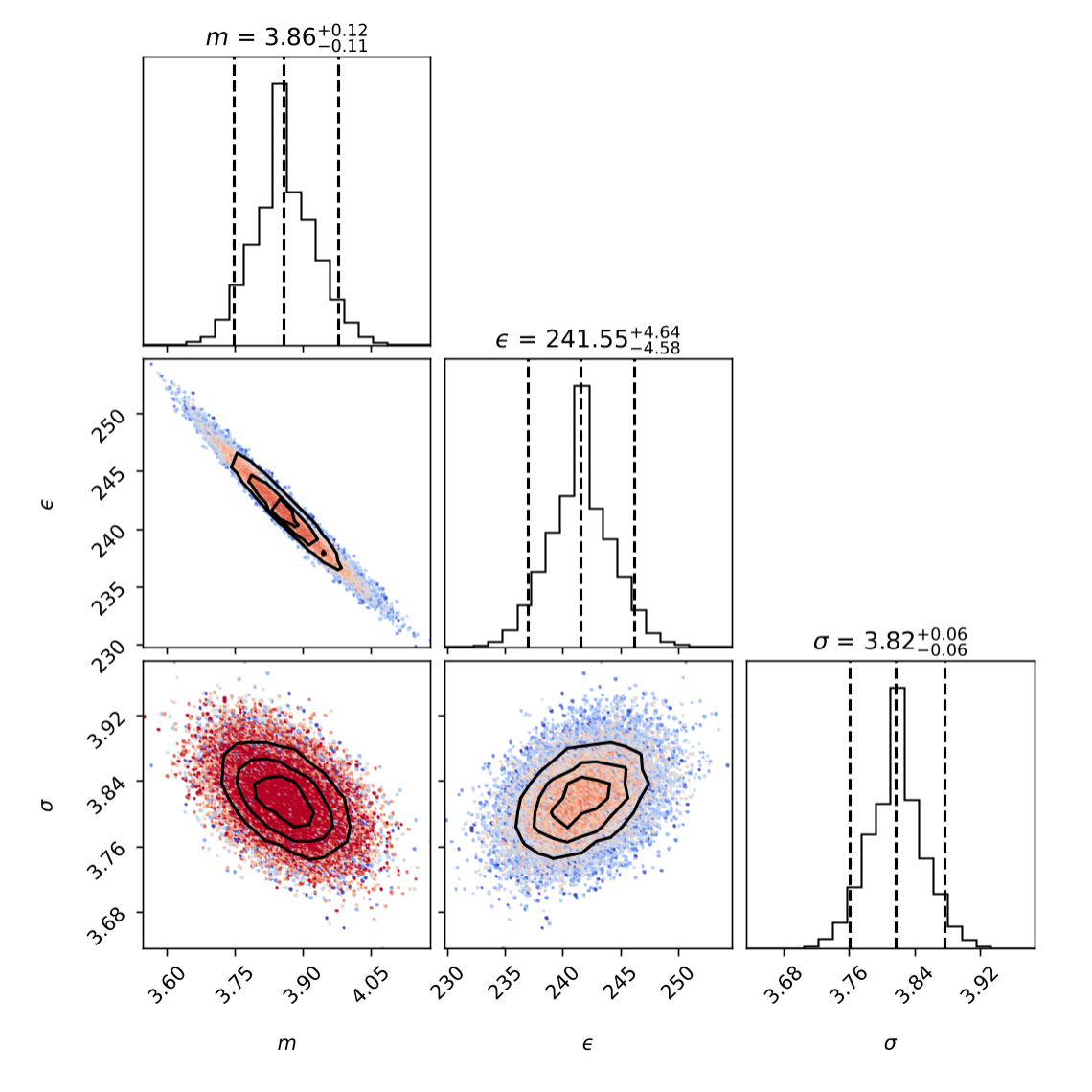}
      \caption{Confidence intervals obtained for the pure component parameters of $n$-octane in PC-SAFT. Colors and styles are identical to figure 2.}
  \end{figure}

\begin{figure}[H]
  \centering
    \begin{subfigure}[b]{0.49\textwidth}
      \includegraphics[width=1\textwidth]{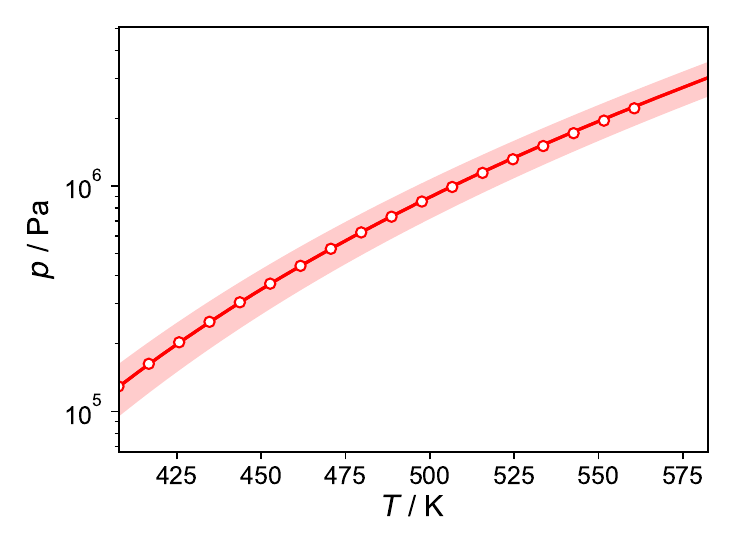}
      \caption{Saturation Curve}
    \end{subfigure}
    \begin{subfigure}[b]{0.49\textwidth}
      \includegraphics[width=1\textwidth]{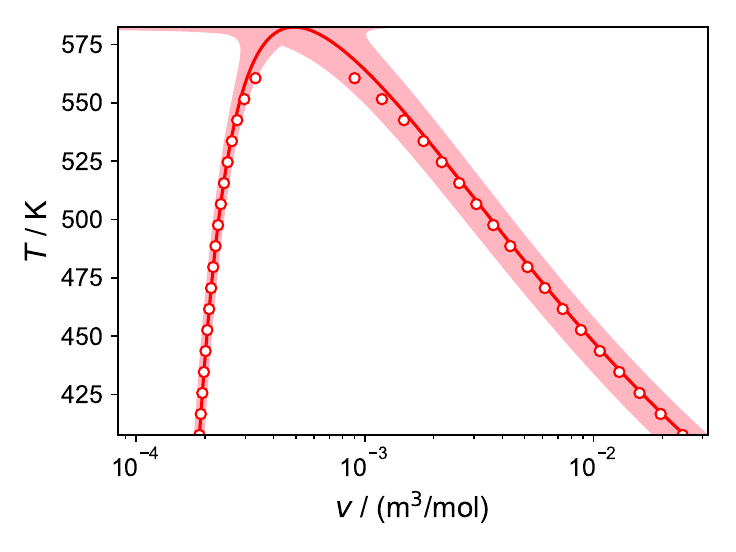}
      \caption{Saturation Envelope}

    \end{subfigure}
    \caption{Predicted values for the saturated volumes and saturation pressure for $n$-octane using PC-SAFT. Shaded regions correspond to the uncertainty interval for the predicted properties using the parameters and confidence intervals.}
  \end{figure}

\begin{figure}[H]
  \centering
    \begin{subfigure}[b]{0.49\textwidth}
      \includegraphics[width=1\textwidth]{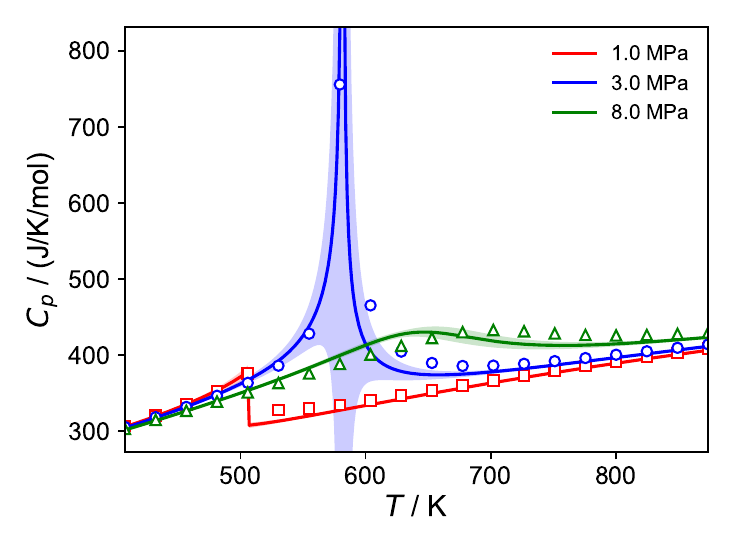}
      \caption{Isobaric heat capacity}
    \end{subfigure}
    \begin{subfigure}[b]{0.49\textwidth}
      \includegraphics[width=1\textwidth]{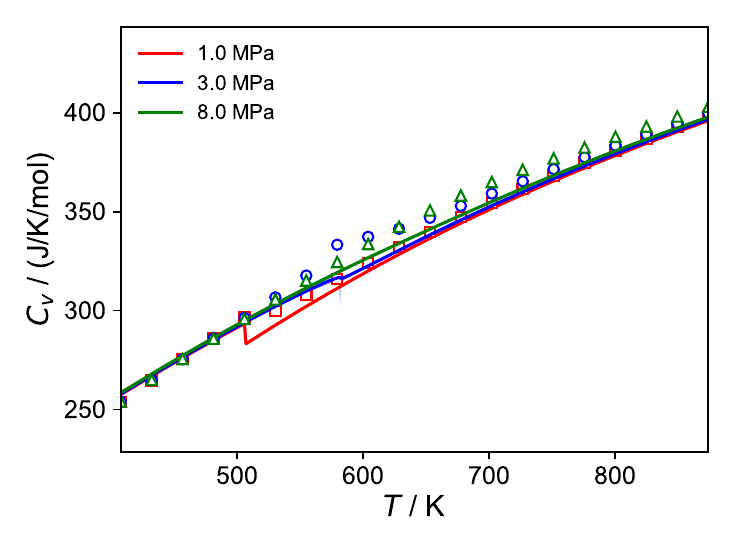}
      \caption{Isochoric heat capacity}
    \end{subfigure}
    \begin{subfigure}[b]{0.49\textwidth}
      \includegraphics[width=1\textwidth]{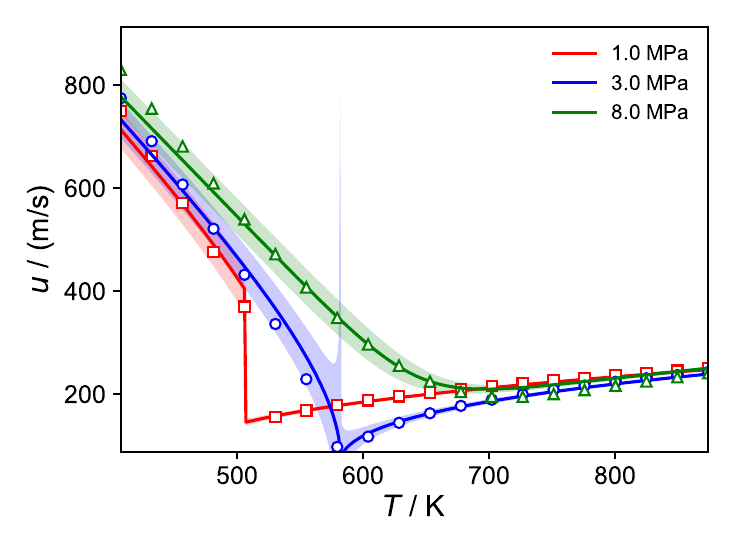}
      \caption{Speed of sound}
    \end{subfigure}
    \begin{subfigure}[b]{0.49\textwidth}
      \includegraphics[width=1\textwidth]{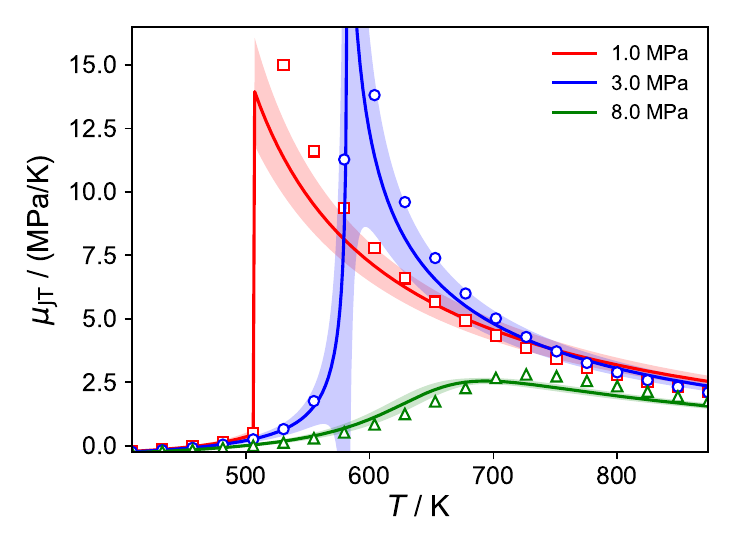}
      \caption{Joule--Thomson coefficient}
    \end{subfigure}
    \caption{Predicted values for the isobaric heat capacity, isochoric heat capacity, speed of sound and Joule--Thomson coefficient for $n$-octane using PC-SAFT at different pressures. Shaded regions correspond to the uncertainty interval for the predicted properties using the parameters and confidence intervals.}
  \end{figure}
\newpage

\subsection{$n$-nonane}
\begin{figure}[H]
  \centering
      \includegraphics[width=0.5\textwidth]{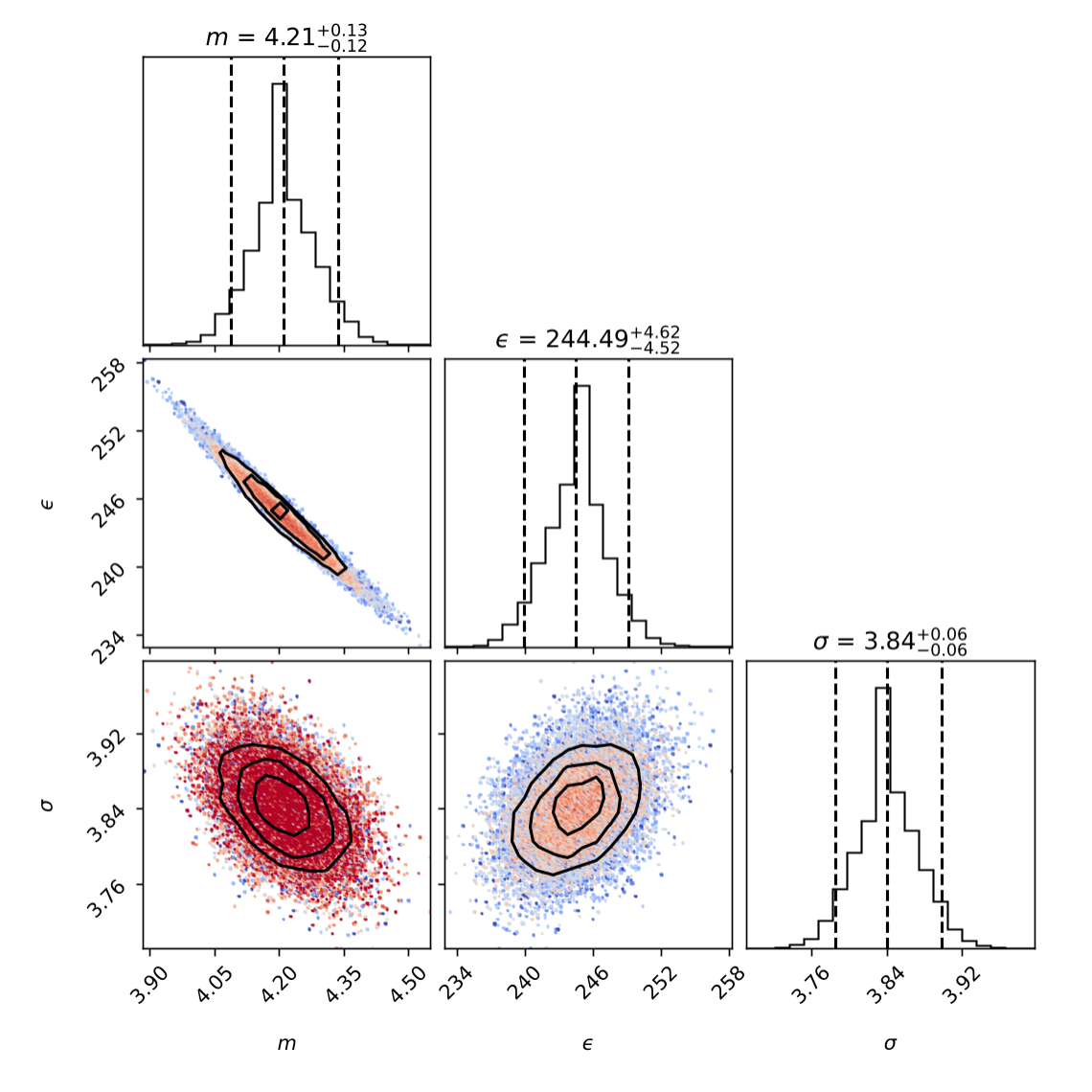}
      \caption{Confidence intervals obtained for the pure component parameters of $n$-nonane in PC-SAFT. Colors and styles are identical to figure 2.}
  \end{figure}

\begin{figure}[H]
  \centering
    \begin{subfigure}[b]{0.49\textwidth}
      \includegraphics[width=1\textwidth]{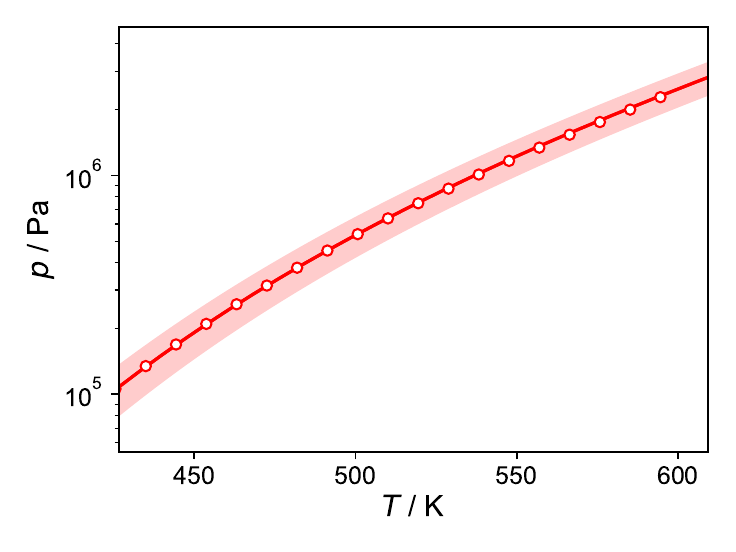}
      \caption{Saturation Curve}
    \end{subfigure}
    \begin{subfigure}[b]{0.49\textwidth}
      \includegraphics[width=1\textwidth]{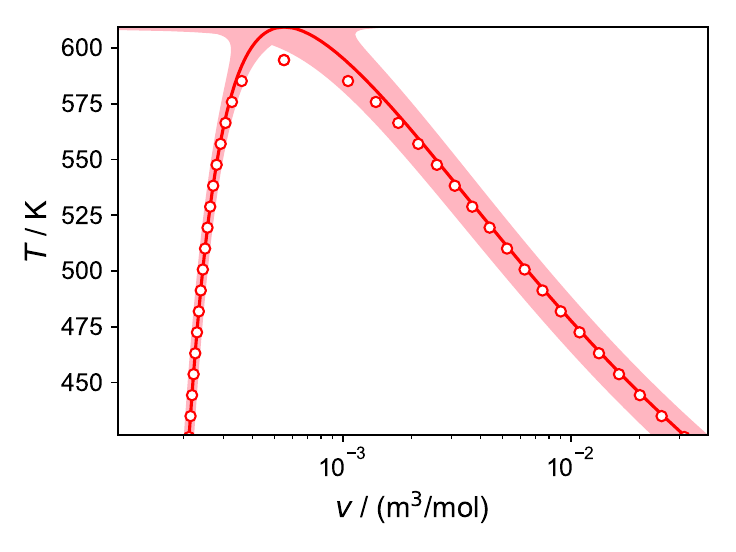}
      \caption{Saturation Envelope}

    \end{subfigure}
    \caption{Predicted values for the saturated volumes and saturation pressure for $n$-nonane using PC-SAFT. Shaded regions correspond to the uncertainty interval for the predicted properties using the parameters and confidence intervals.}
  \end{figure}

\begin{figure}[H]
  \centering
    \begin{subfigure}[b]{0.49\textwidth}
      \includegraphics[width=1\textwidth]{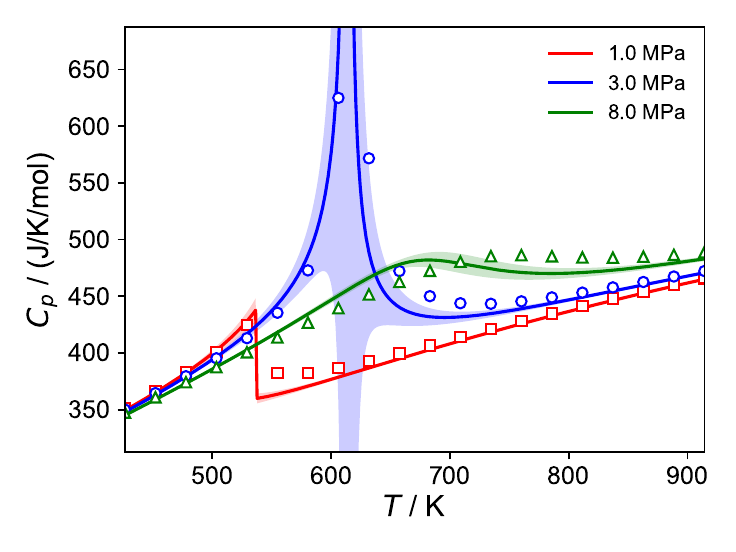}
      \caption{Isobaric heat capacity}
    \end{subfigure}
    \begin{subfigure}[b]{0.49\textwidth}
      \includegraphics[width=1\textwidth]{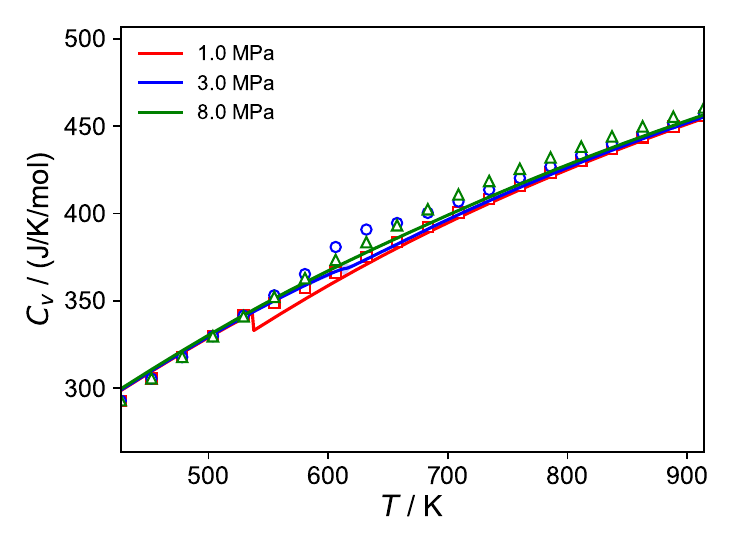}
      \caption{Isochoric heat capacity}
    \end{subfigure}
    \begin{subfigure}[b]{0.49\textwidth}
      \includegraphics[width=1\textwidth]{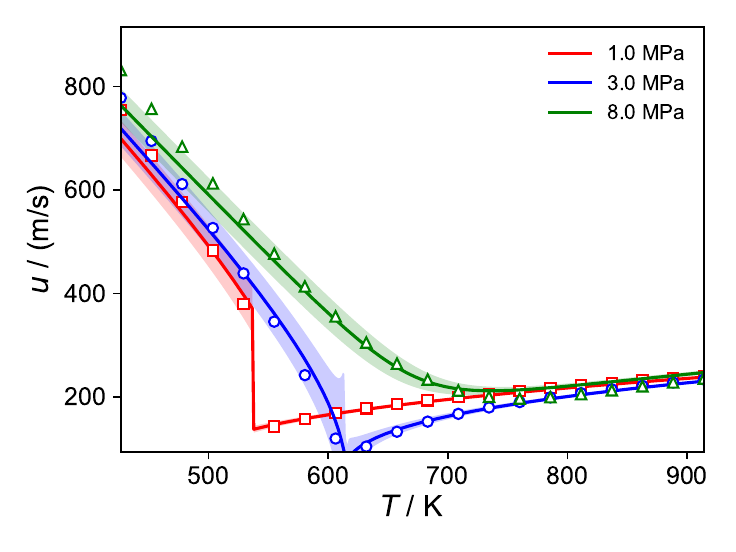}
      \caption{Speed of sound}
    \end{subfigure}
    \begin{subfigure}[b]{0.49\textwidth}
      \includegraphics[width=1\textwidth]{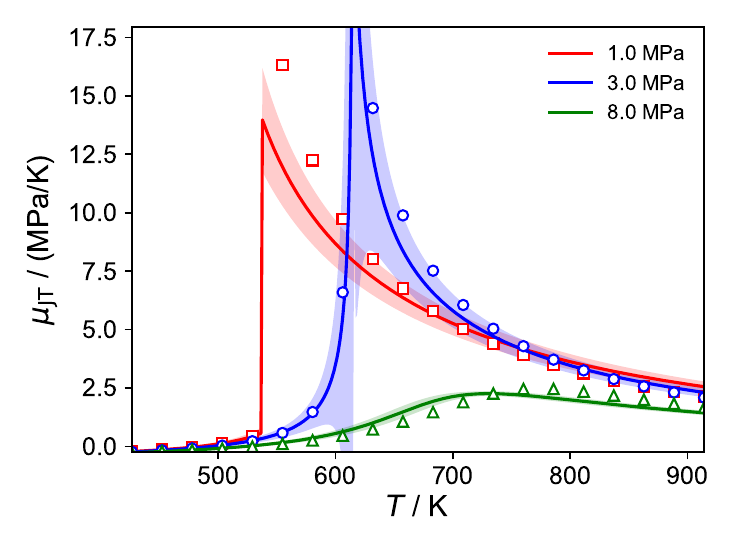}
      \caption{Joule--Thomson coefficient}
    \end{subfigure}
    \caption{Predicted values for the isobaric heat capacity, isochoric heat capacity, speed of sound and Joule--Thomson coefficient for $n$-nonane using PC-SAFT at different pressures. Shaded regions correspond to the uncertainty interval for the predicted properties using the parameters and confidence intervals.}
  \end{figure}
\newpage

\subsection{$n$-decane}
\begin{figure}[H]
  \centering
      \includegraphics[width=0.5\textwidth]{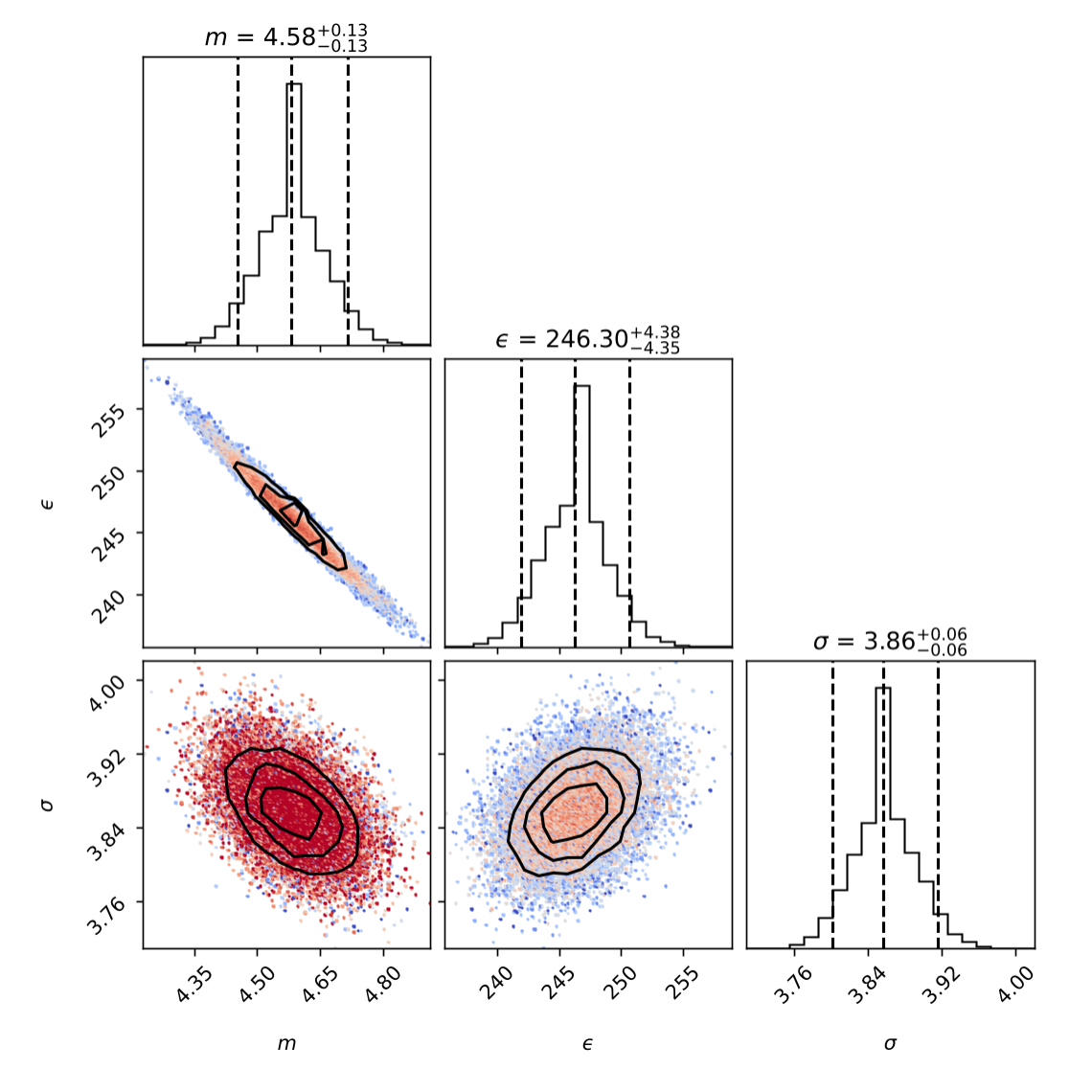}
      \caption{Confidence intervals obtained for the pure component parameters of $n$-decane in PC-SAFT. Colors and styles are identical to figure 2.}
  \end{figure}

\begin{figure}[H]
  \centering
    \begin{subfigure}[b]{0.49\textwidth}
      \includegraphics[width=1\textwidth]{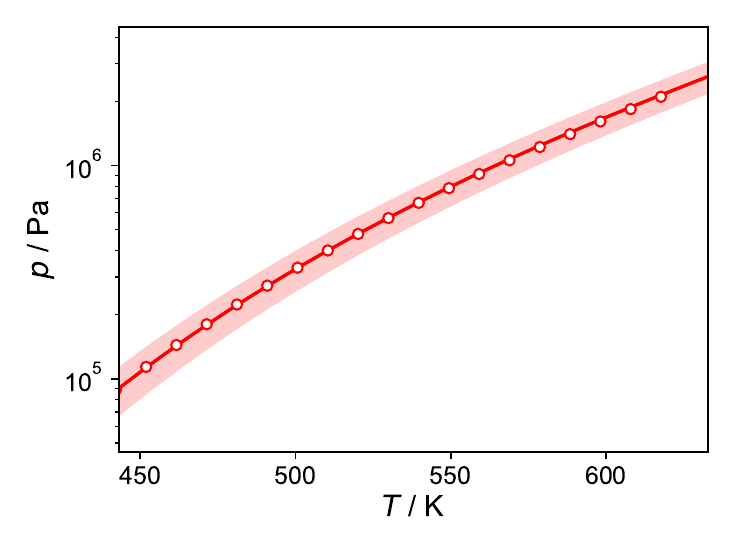}
      \caption{Saturation Curve}
    \end{subfigure}
    \begin{subfigure}[b]{0.49\textwidth}
      \includegraphics[width=1\textwidth]{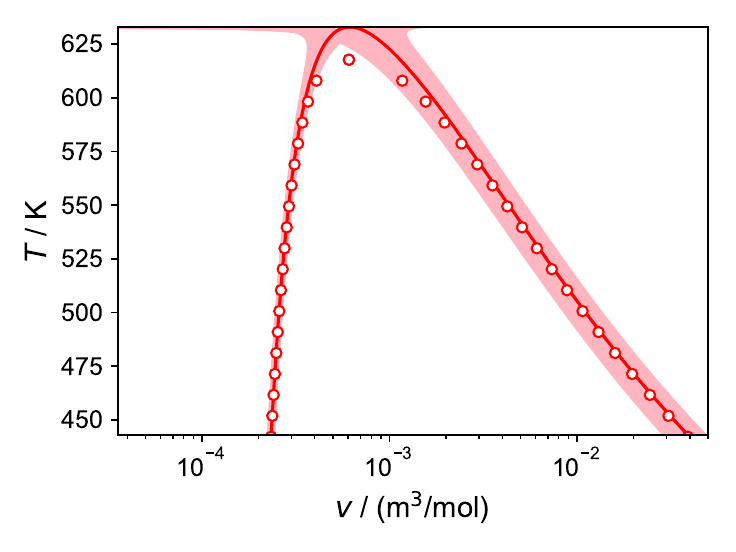}
      \caption{Saturation Envelope}

    \end{subfigure}
    \caption{Predicted values for the saturated volumes and saturation pressure for $n$-decane using PC-SAFT. Shaded regions correspond to the uncertainty interval for the predicted properties using the parameters and confidence intervals.}
  \end{figure}

\begin{figure}[H]
  \centering
    \begin{subfigure}[b]{0.49\textwidth}
      \includegraphics[width=1\textwidth]{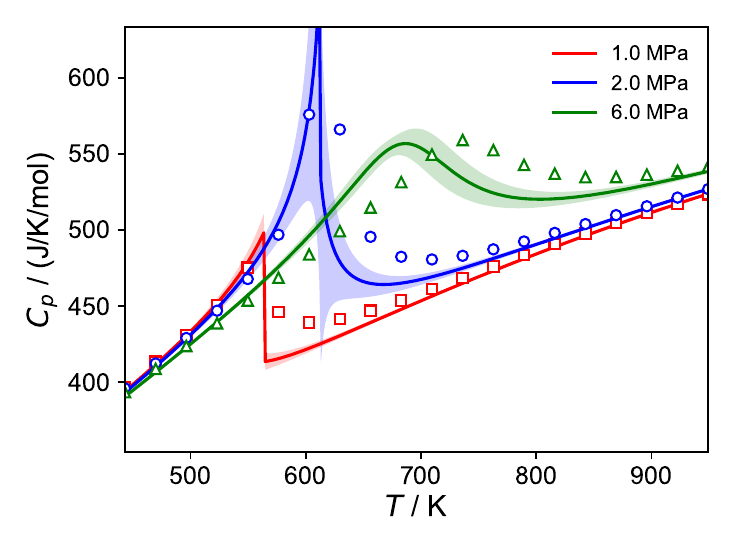}
      \caption{Isobaric heat capacity}
    \end{subfigure}
    \begin{subfigure}[b]{0.49\textwidth}
      \includegraphics[width=1\textwidth]{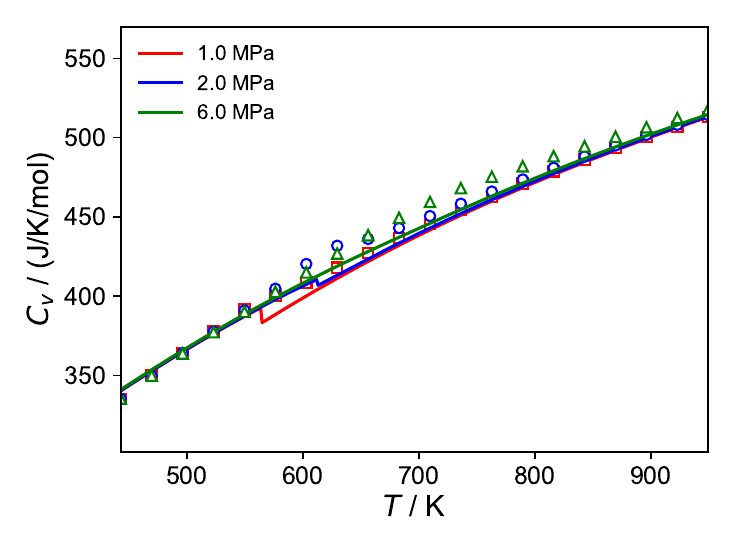}
      \caption{Isochoric heat capacity}
    \end{subfigure}
    \begin{subfigure}[b]{0.49\textwidth}
      \includegraphics[width=1\textwidth]{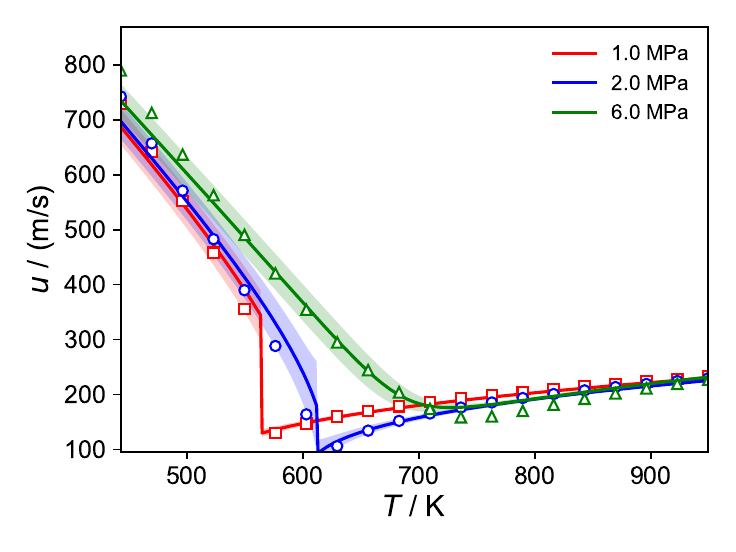}
      \caption{Speed of sound}
    \end{subfigure}
    \begin{subfigure}[b]{0.49\textwidth}
      \includegraphics[width=1\textwidth]{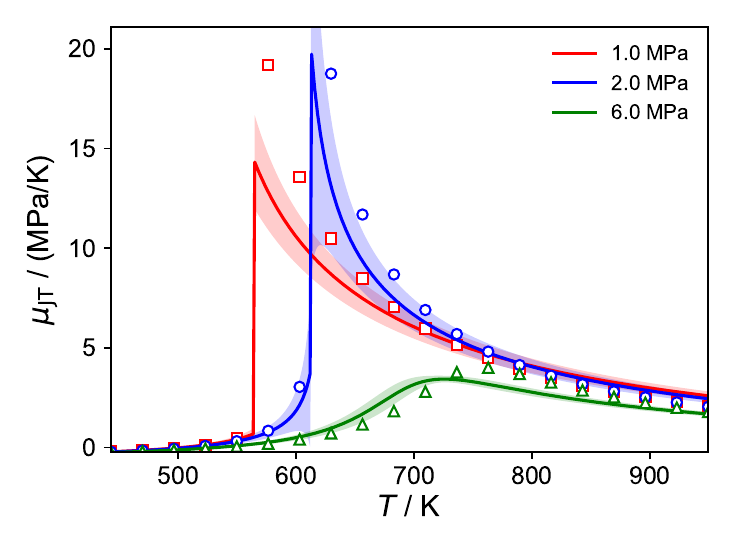}
      \caption{Joule--Thomson coefficient}
    \end{subfigure}
    \caption{Predicted values for the isobaric heat capacity, isochoric heat capacity, speed of sound and Joule--Thomson coefficient for $n$-decane using PC-SAFT at different pressures. Shaded regions correspond to the uncertainty interval for the predicted properties using the parameters and confidence intervals.}
  \end{figure}
\newpage

\subsection{Carbon dioxide}
\begin{figure}[H]
  \centering
      \includegraphics[width=0.5\textwidth]{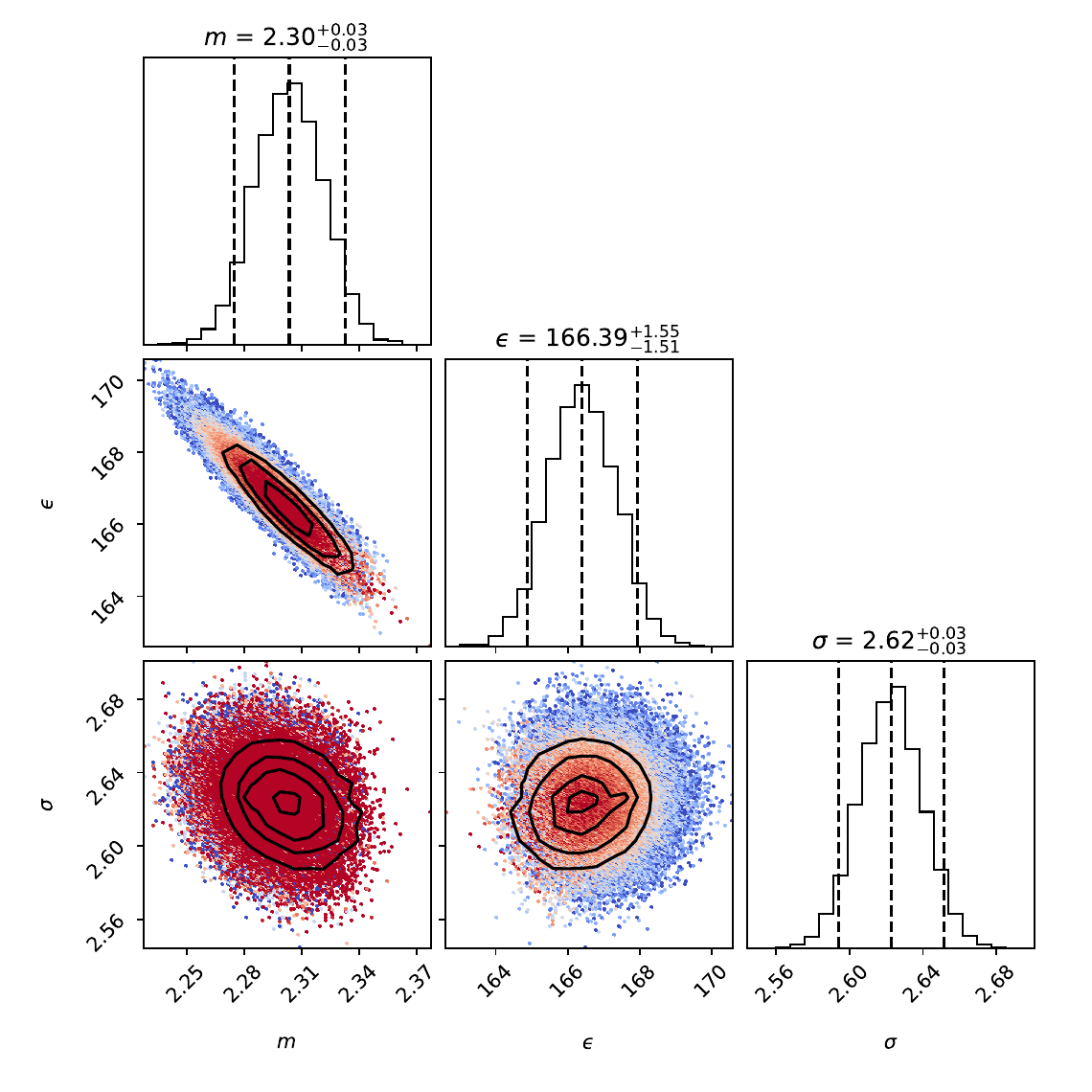}
      \caption{Confidence intervals obtained for the pure component parameters of carbon dioxide in PC-SAFT. Colors and styles are identical to figure 2.}
  \end{figure}

\begin{figure}[H]
  \centering
    \begin{subfigure}[b]{0.49\textwidth}
      \includegraphics[width=1\textwidth]{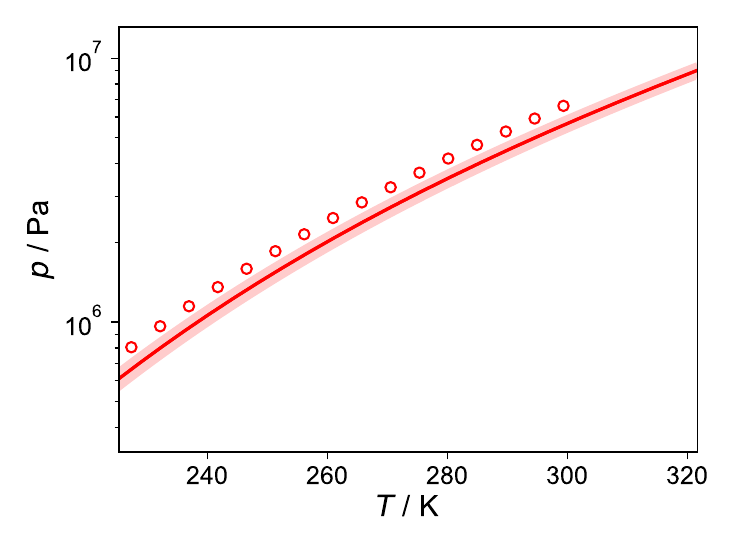}
      \caption{Saturation Curve}
    \end{subfigure}
    \begin{subfigure}[b]{0.49\textwidth}
      \includegraphics[width=1\textwidth]{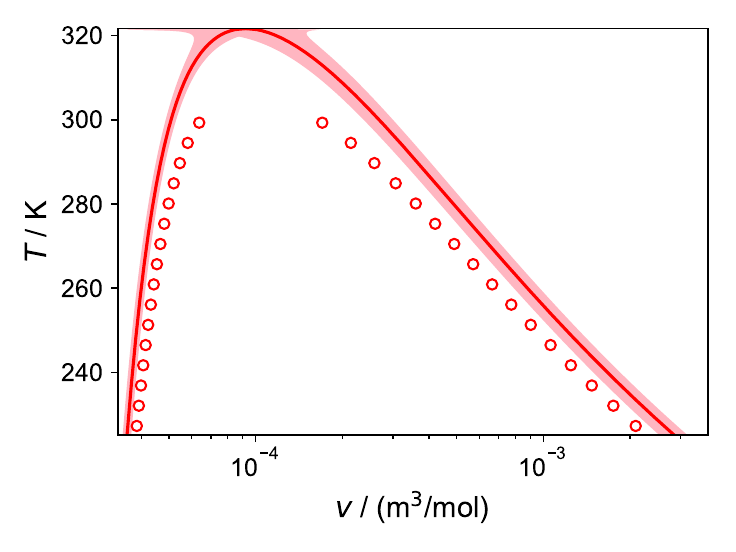}
      \caption{Saturation Envelope}

    \end{subfigure}
    \caption{Predicted values for the saturated volumes and saturation pressure for carbon dioxide using PC-SAFT. Shaded regions correspond to the uncertainty interval for the predicted properties using the parameters and confidence intervals.}
  \end{figure}

\begin{figure}[H]
  \centering
    \begin{subfigure}[b]{0.49\textwidth}
      \includegraphics[width=1\textwidth]{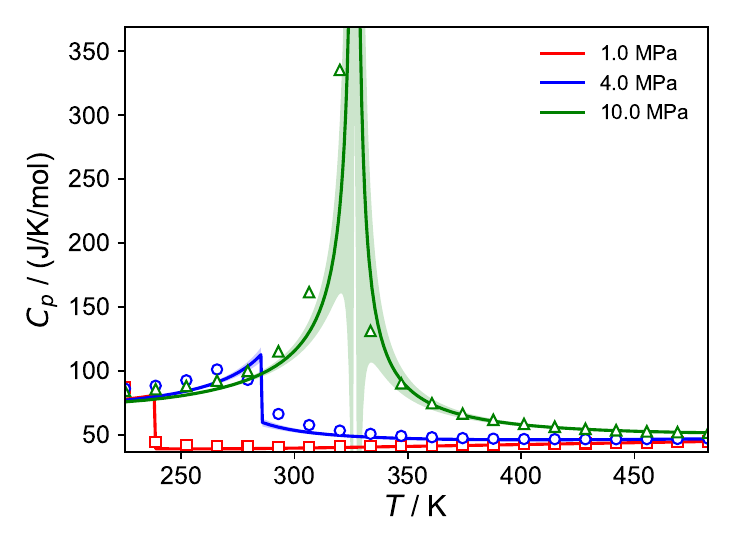}
      \caption{Isobaric heat capacity}
    \end{subfigure}
    \begin{subfigure}[b]{0.49\textwidth}
      \includegraphics[width=1\textwidth]{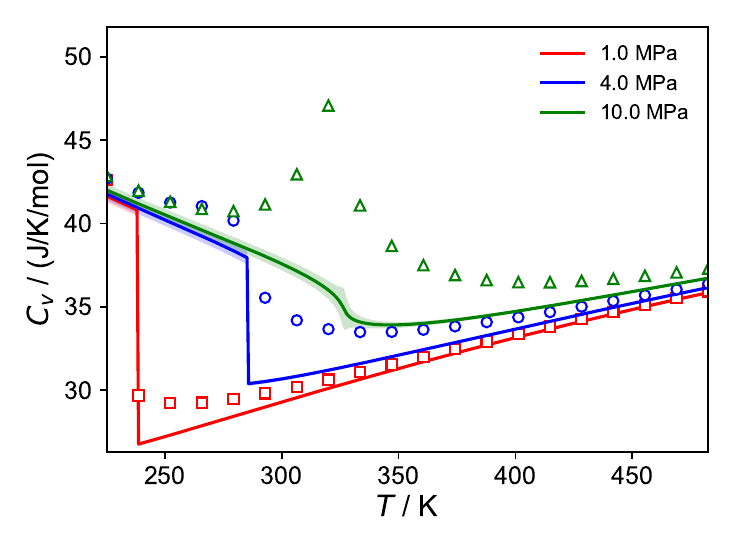}
      \caption{Isochoric heat capacity}
    \end{subfigure}
    \begin{subfigure}[b]{0.49\textwidth}
      \includegraphics[width=1\textwidth]{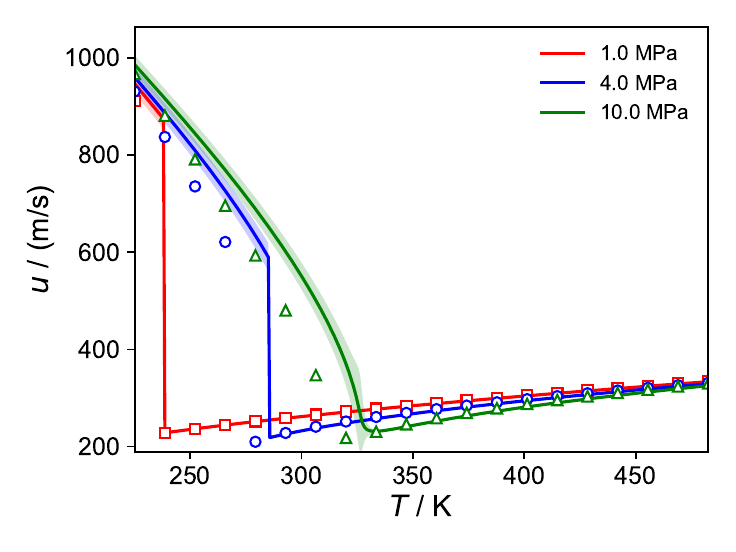}
      \caption{Speed of sound}
    \end{subfigure}
    \begin{subfigure}[b]{0.49\textwidth}
      \includegraphics[width=1\textwidth]{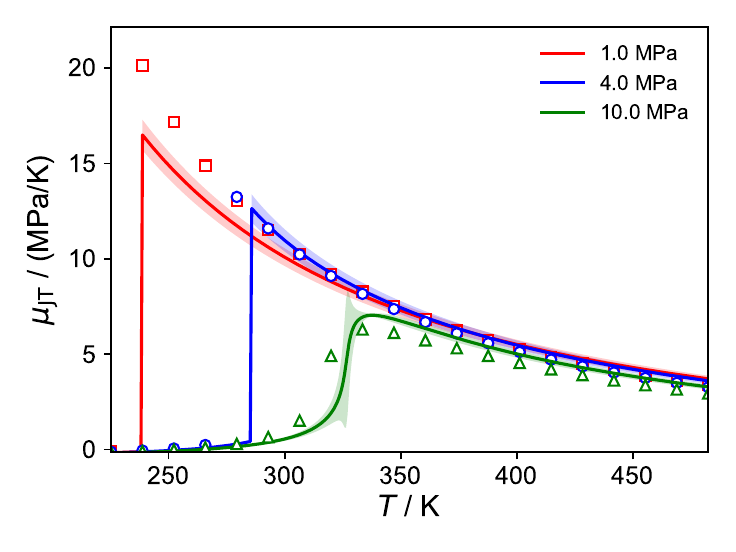}
      \caption{Joule--Thomson coefficient}
    \end{subfigure}
    \caption{Predicted values for the isobaric heat capacity, isochoric heat capacity, speed of sound and Joule--Thomson coefficient for carbon dioxide using PC-SAFT at different pressures. Shaded regions correspond to the uncertainty interval for the predicted properties using the parameters and confidence intervals.}
  \end{figure}
\newpage

\subsection{Argon}
\begin{figure}[H]
  \centering
      \includegraphics[width=0.5\textwidth]{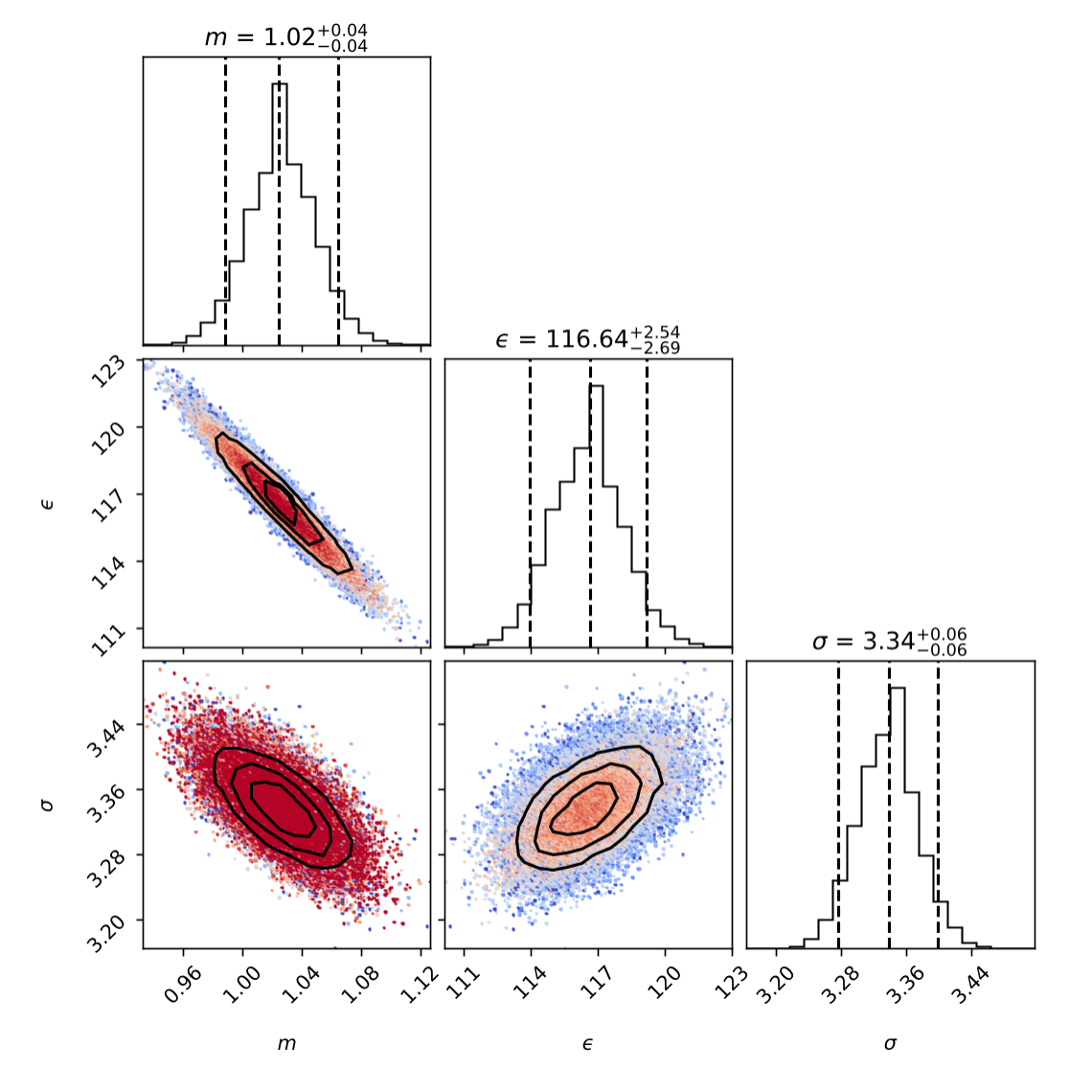}
      \caption{Confidence intervals obtained for the pure component parameters of argon in PC-SAFT. Colors and styles are identical to figure 2.}
  \end{figure}

\begin{figure}[H]
  \centering
    \begin{subfigure}[b]{0.49\textwidth}
      \includegraphics[width=1\textwidth]{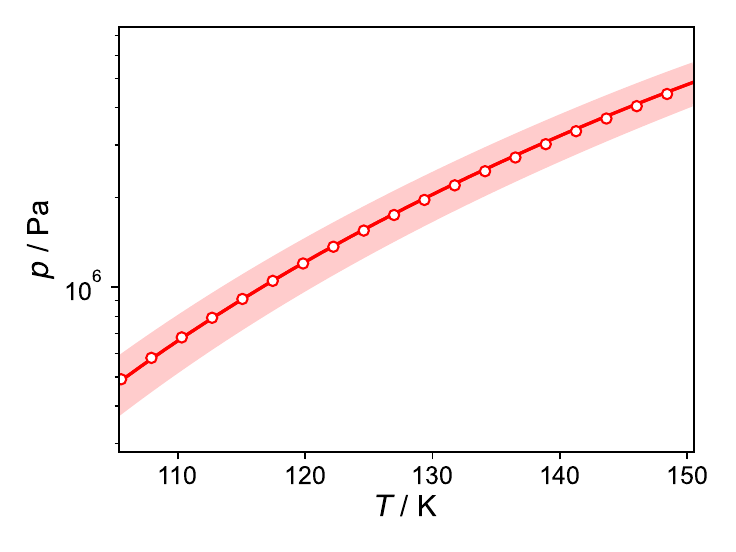}
      \caption{Saturation Curve}
    \end{subfigure}
    \begin{subfigure}[b]{0.49\textwidth}
      \includegraphics[width=1\textwidth]{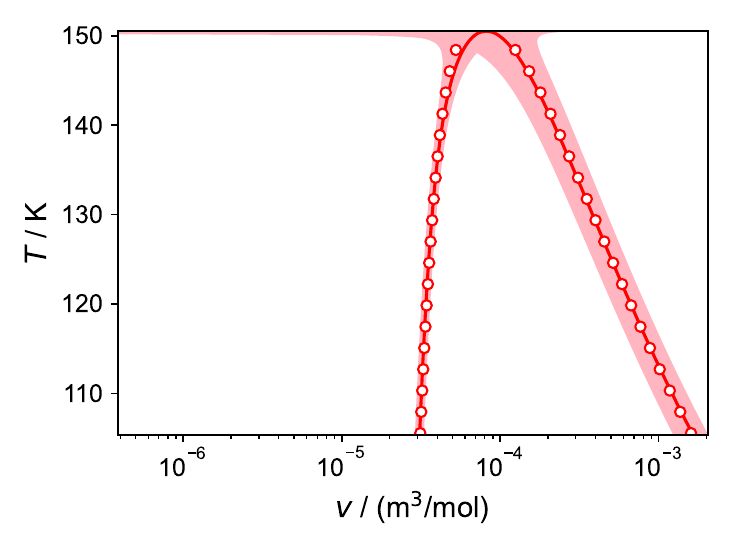}
      \caption{Saturation Envelope}

    \end{subfigure}
    \caption{Predicted values for the saturated volumes and saturation pressure for argon using PC-SAFT. Shaded regions correspond to the uncertainty interval for the predicted properties using the parameters and confidence intervals.}
  \end{figure}

\begin{figure}[H]
  \centering
    \begin{subfigure}[b]{0.49\textwidth}
      \includegraphics[width=1\textwidth]{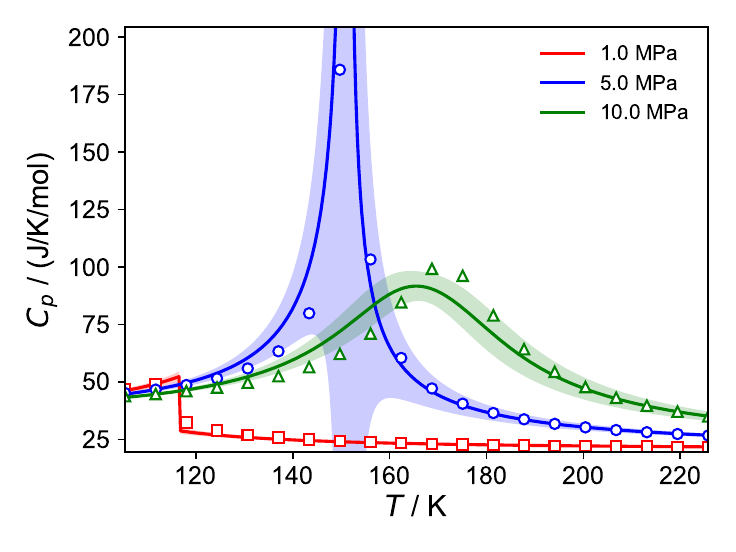}
      \caption{Isobaric heat capacity}
    \end{subfigure}
    \begin{subfigure}[b]{0.49\textwidth}
      \includegraphics[width=1\textwidth]{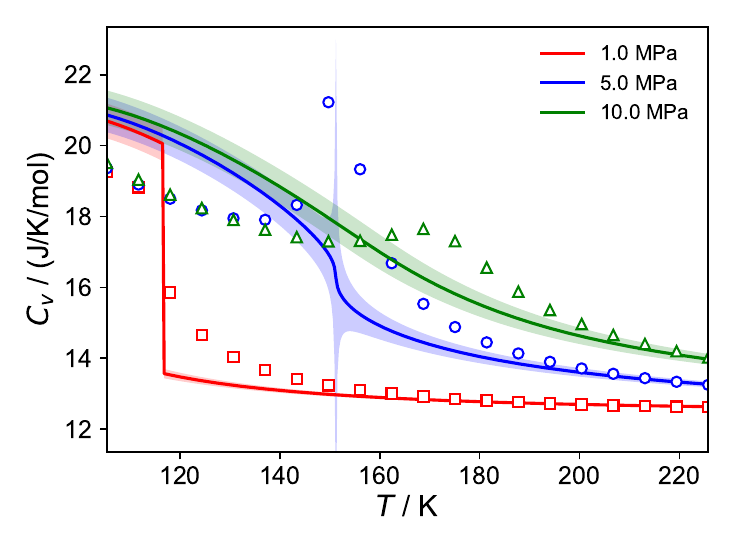}
      \caption{Isochoric heat capacity}
    \end{subfigure}
    \begin{subfigure}[b]{0.49\textwidth}
      \includegraphics[width=1\textwidth]{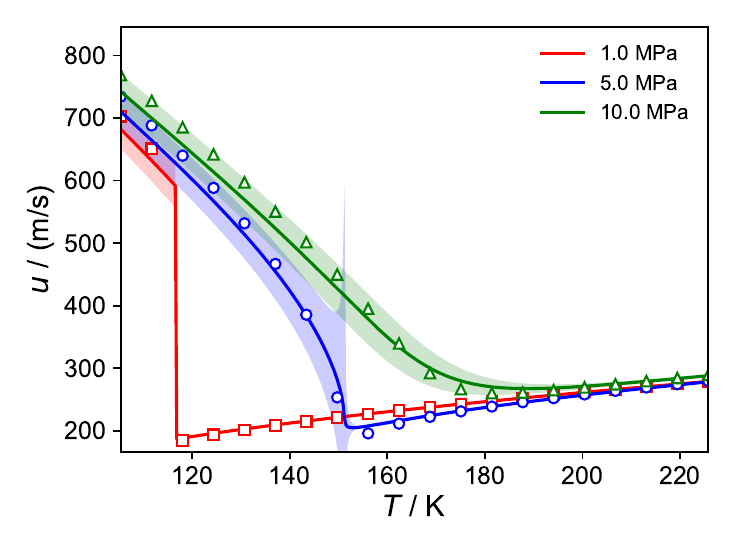}
      \caption{Speed of sound}
    \end{subfigure}
    \begin{subfigure}[b]{0.49\textwidth}
      \includegraphics[width=1\textwidth]{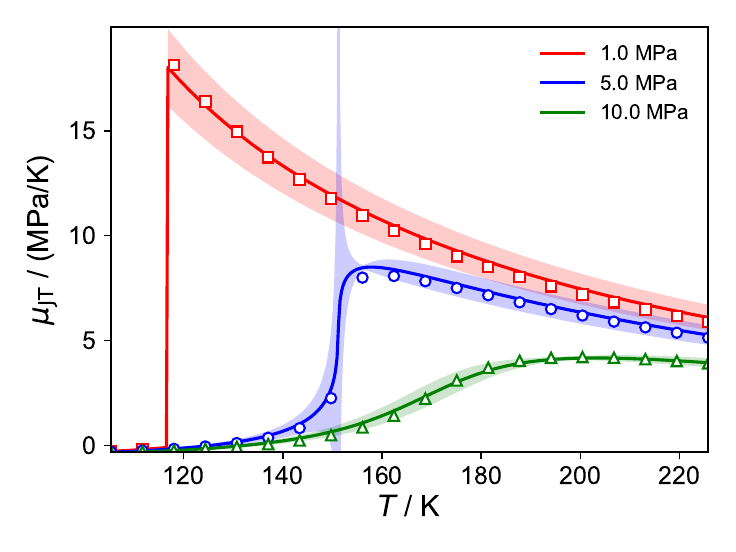}
      \caption{Joule--Thomson coefficient}
    \end{subfigure}
    \caption{Predicted values for the isobaric heat capacity, isochoric heat capacity, speed of sound and Joule--Thomson coefficient for argon using PC-SAFT at different pressures. Shaded regions correspond to the uncertainty interval for the predicted properties using the parameters and confidence intervals.}
  \end{figure}
\newpage

\subsection{Water}
\begin{figure}[H]
  \centering
      \includegraphics[width=0.5\textwidth]{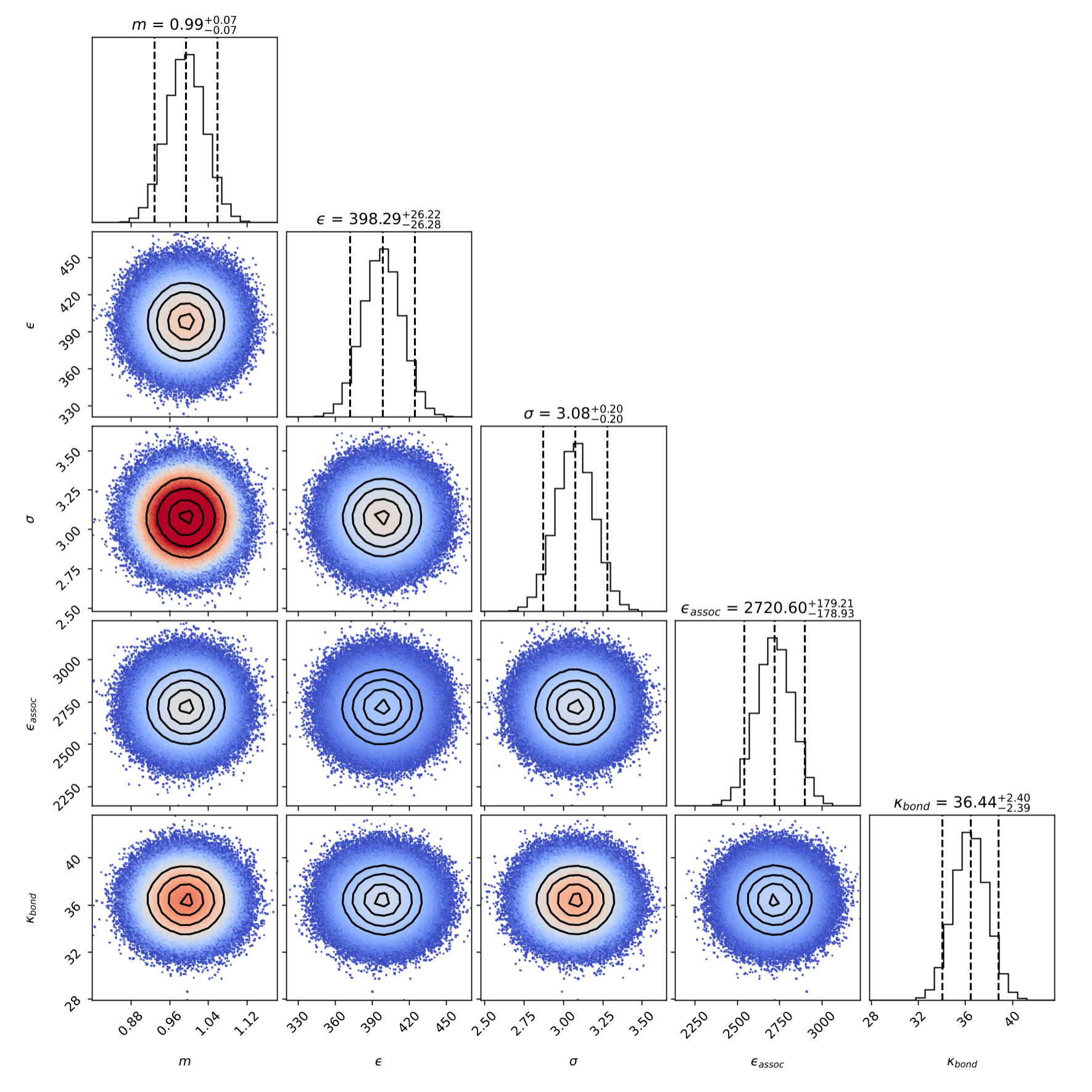}
      \caption{Confidence intervals obtained for the pure component parameters of water in PC-SAFT. Colors and styles are identical to figure 2.}
  \end{figure}

\begin{figure}[H]
  \centering
    \begin{subfigure}[b]{0.49\textwidth}
      \includegraphics[width=1\textwidth]{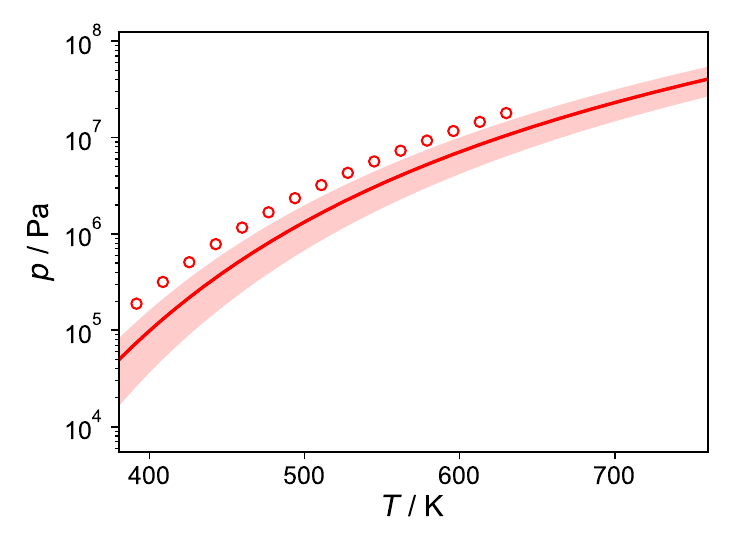}
      \caption{Saturation Curve}
    \end{subfigure}
    \begin{subfigure}[b]{0.49\textwidth}
      \includegraphics[width=1\textwidth]{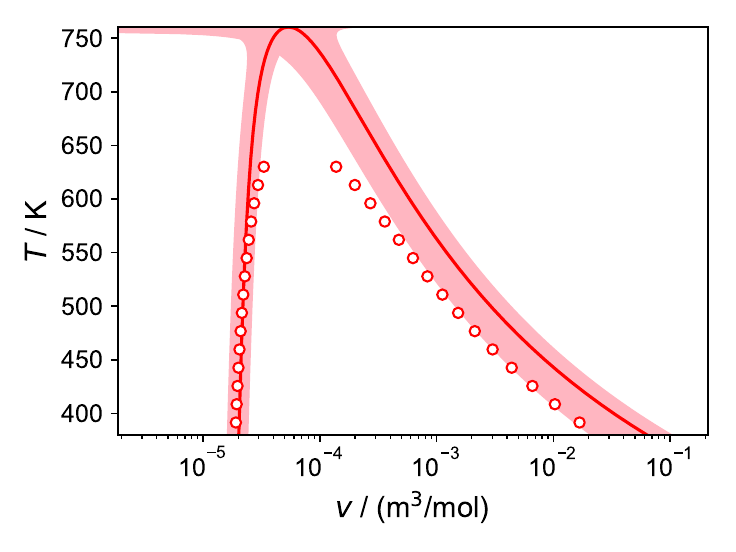}
      \caption{Saturation Envelope}

    \end{subfigure}
    \caption{Predicted values for the saturated volumes and saturation pressure for water using PC-SAFT. Shaded regions correspond to the uncertainty interval for the predicted properties using the parameters and confidence intervals.}
  \end{figure}

\begin{figure}[H]
  \centering
    \begin{subfigure}[b]{0.49\textwidth}
      \includegraphics[width=1\textwidth]{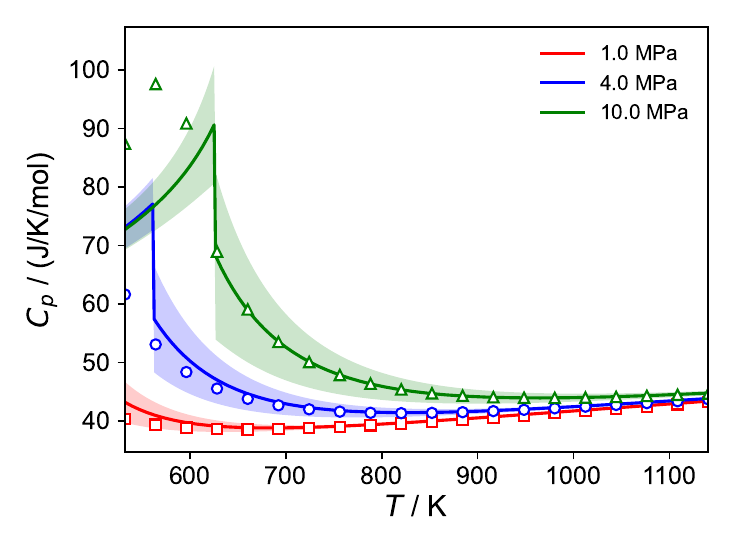}
      \caption{Isobaric heat capacity}
    \end{subfigure}
    \begin{subfigure}[b]{0.49\textwidth}
      \includegraphics[width=1\textwidth]{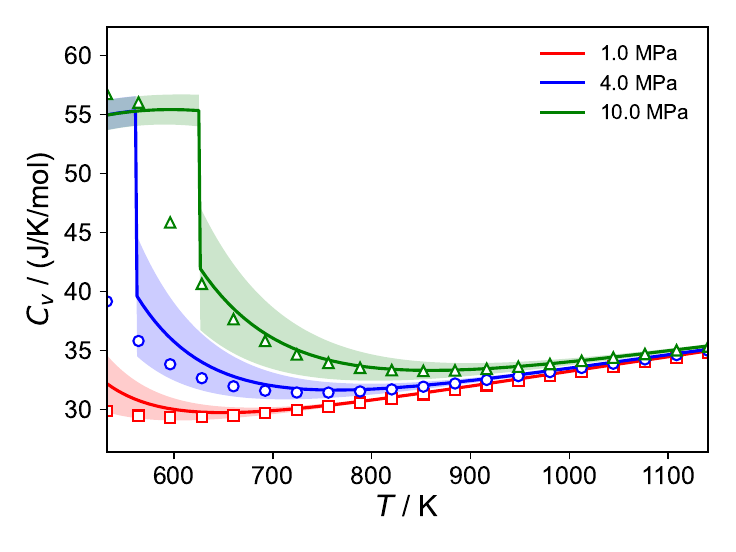}
      \caption{Isochoric heat capacity}
    \end{subfigure}
    \begin{subfigure}[b]{0.49\textwidth}
      \includegraphics[width=1\textwidth]{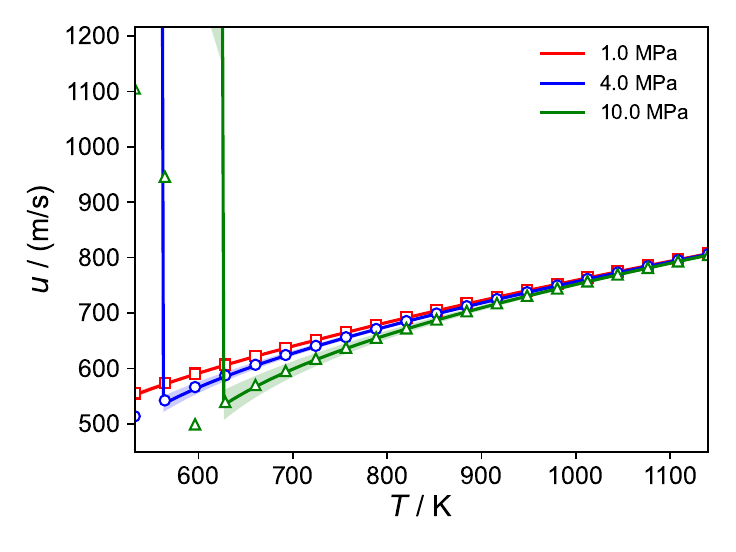}
      \caption{Speed of sound}
    \end{subfigure}
    \begin{subfigure}[b]{0.49\textwidth}
      \includegraphics[width=1\textwidth]{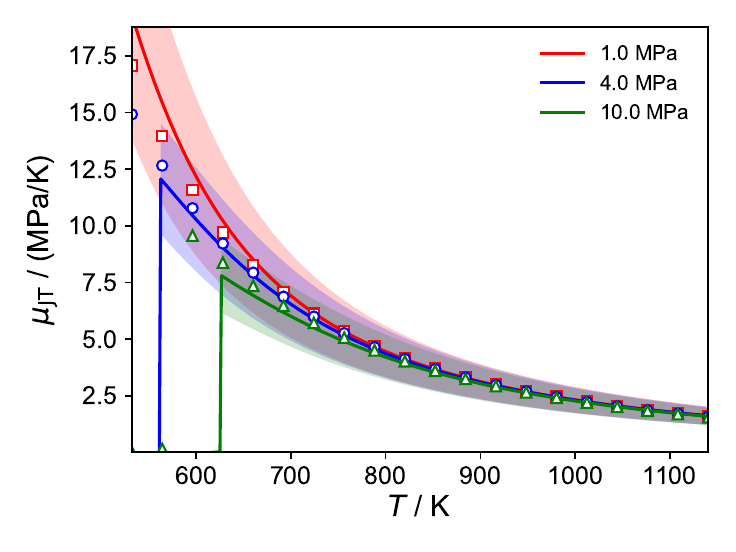}
      \caption{Joule--Thomson coefficient}
    \end{subfigure}
    \caption{Predicted values for the isobaric heat capacity, isochoric heat capacity, speed of sound and Joule--Thomson coefficient for water using PC-SAFT at different pressures. Shaded regions correspond to the uncertainty interval for the predicted properties using the parameters and confidence intervals.}
  \end{figure}
\newpage

\subsection{Methanol}
\begin{figure}[H]
  \centering
      \includegraphics[width=0.5\textwidth]{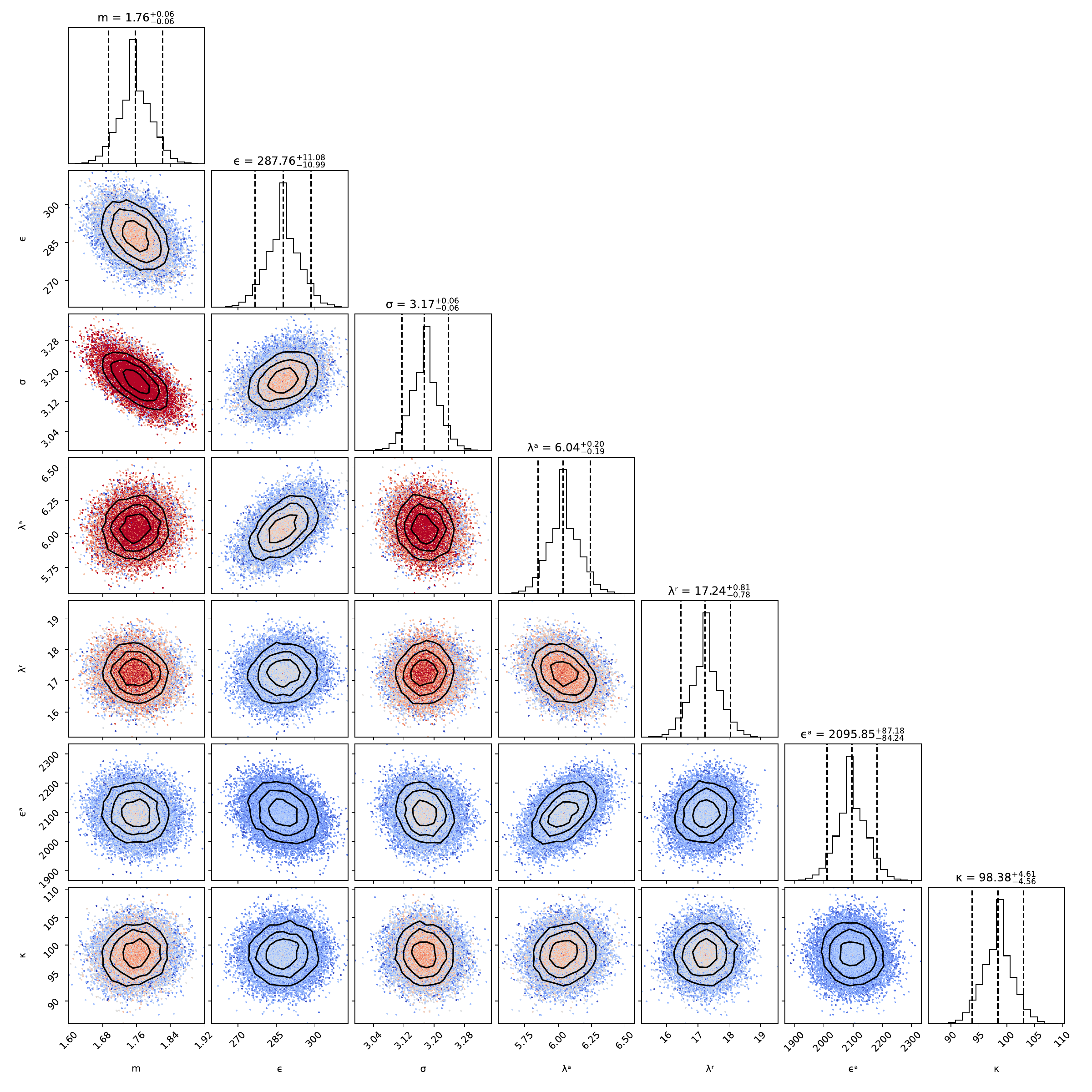}
      \caption{Confidence intervals obtained for the pure component parameters of methanol in PC-SAFT. Colors and styles are identical to figure 2.}
  \end{figure}
\section{SAFT-VR Mie}
\subsection{Methane}

\begin{figure}[H]
  \centering
      \includegraphics[width=0.5\textwidth]{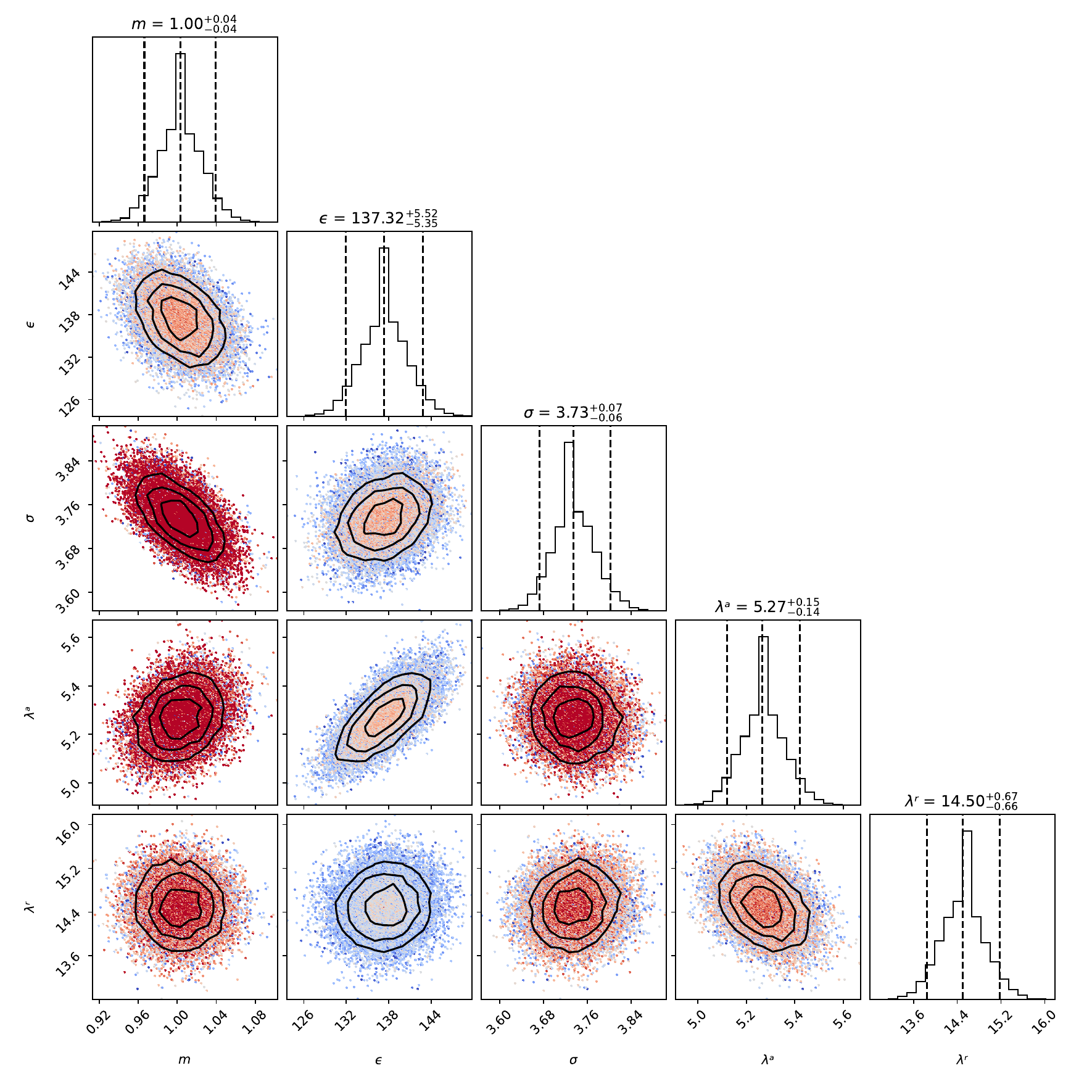}
      \caption{Confidence intervals obtained for the pure component parameters of methane in SAFT-VR Mie. Colors and styles are identical to figure 2.}
  \end{figure}

\begin{figure}[H]
  \centering
    \begin{subfigure}[b]{0.49\textwidth}
      \includegraphics[width=1\textwidth]{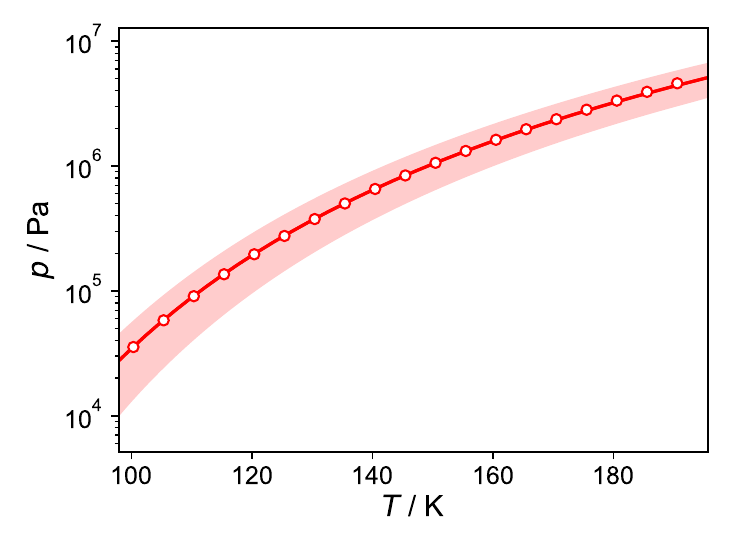}
      \caption{Saturation Curve}
    \end{subfigure}
    \begin{subfigure}[b]{0.49\textwidth}
      \includegraphics[width=1\textwidth]{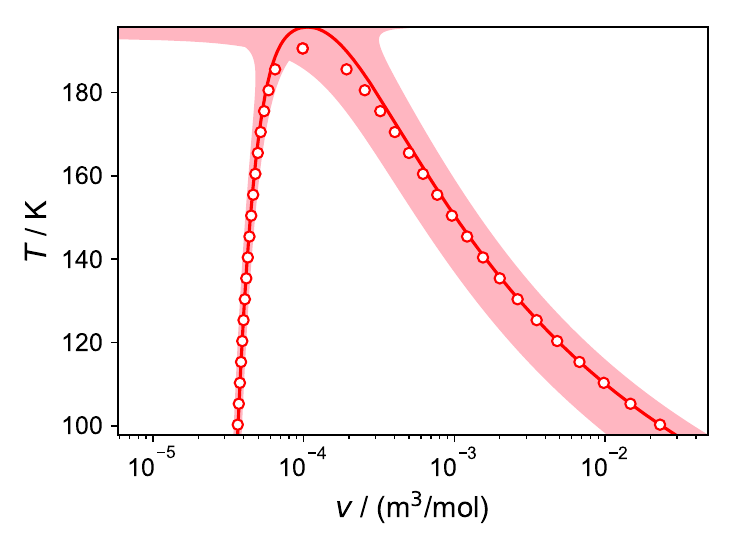}
      \caption{Saturation Envelope}

    \end{subfigure}
    \caption{Predicted values for the saturated volumes and saturation pressure for methane using SAFT-VR Mie. Shaded regions correspond to the uncertainty interval for the predicted properties using the parameters and confidence intervals.}
  \end{figure}

\begin{figure}[H]
  \centering
    \begin{subfigure}[b]{0.49\textwidth}
      \includegraphics[width=1\textwidth]{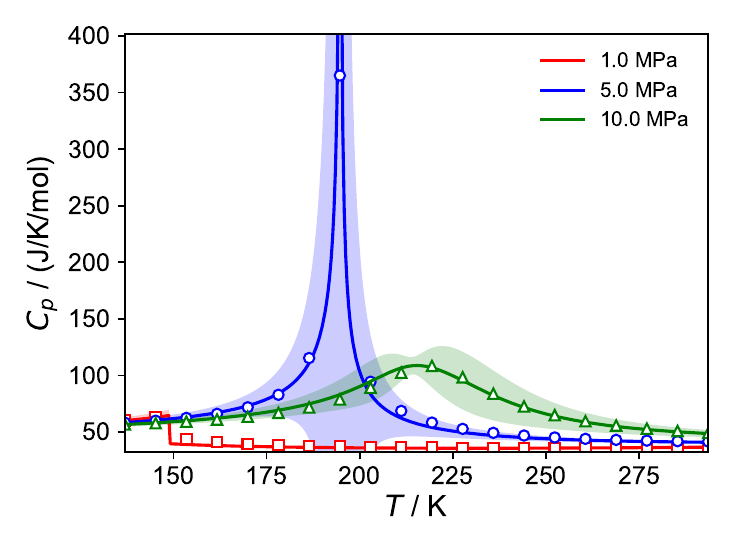}
      \caption{Isobaric heat capacity}
    \end{subfigure}
    \begin{subfigure}[b]{0.49\textwidth}
      \includegraphics[width=1\textwidth]{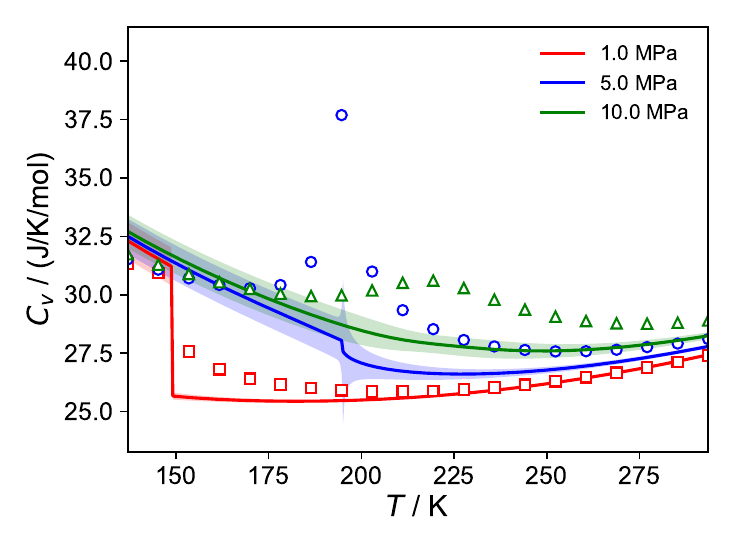}
      \caption{Isochoric heat capacity}
    \end{subfigure}
    \begin{subfigure}[b]{0.49\textwidth}
      \includegraphics[width=1\textwidth]{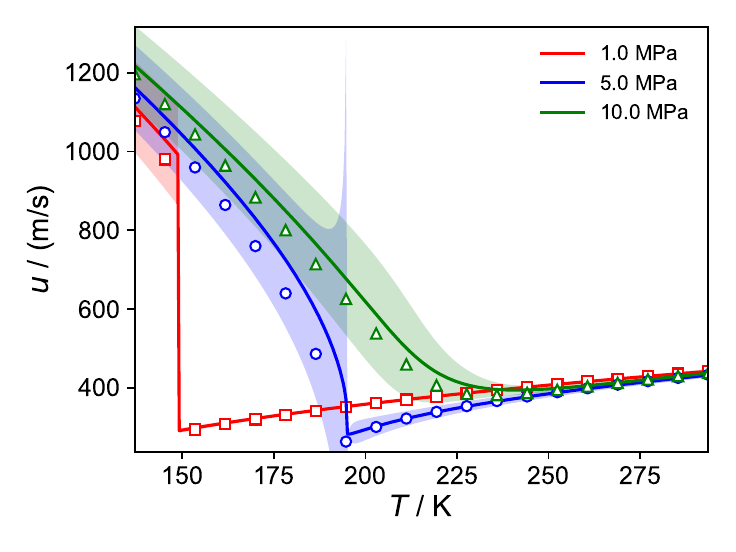}
      \caption{Speed of sound}
    \end{subfigure}
    \begin{subfigure}[b]{0.49\textwidth}
      \includegraphics[width=1\textwidth]{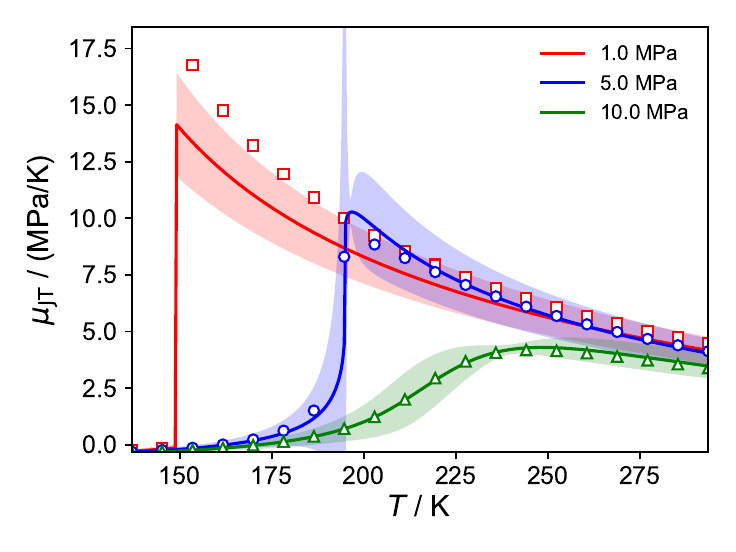}
      \caption{Joule--Thomson coefficient}
    \end{subfigure}
    \caption{Predicted values for the isobaric heat capacity, isochoric heat capacity, speed of sound and Joule--Thomson coefficient for methane using SAFT-VR Mie at different pressures. Shaded regions correspond to the uncertainty interval for the predicted properties using the parameters and confidence intervals.}
  \end{figure}

\newpage
\subsection{Ethane}
\begin{figure}[H]
  \centering
      \includegraphics[width=0.5\textwidth]{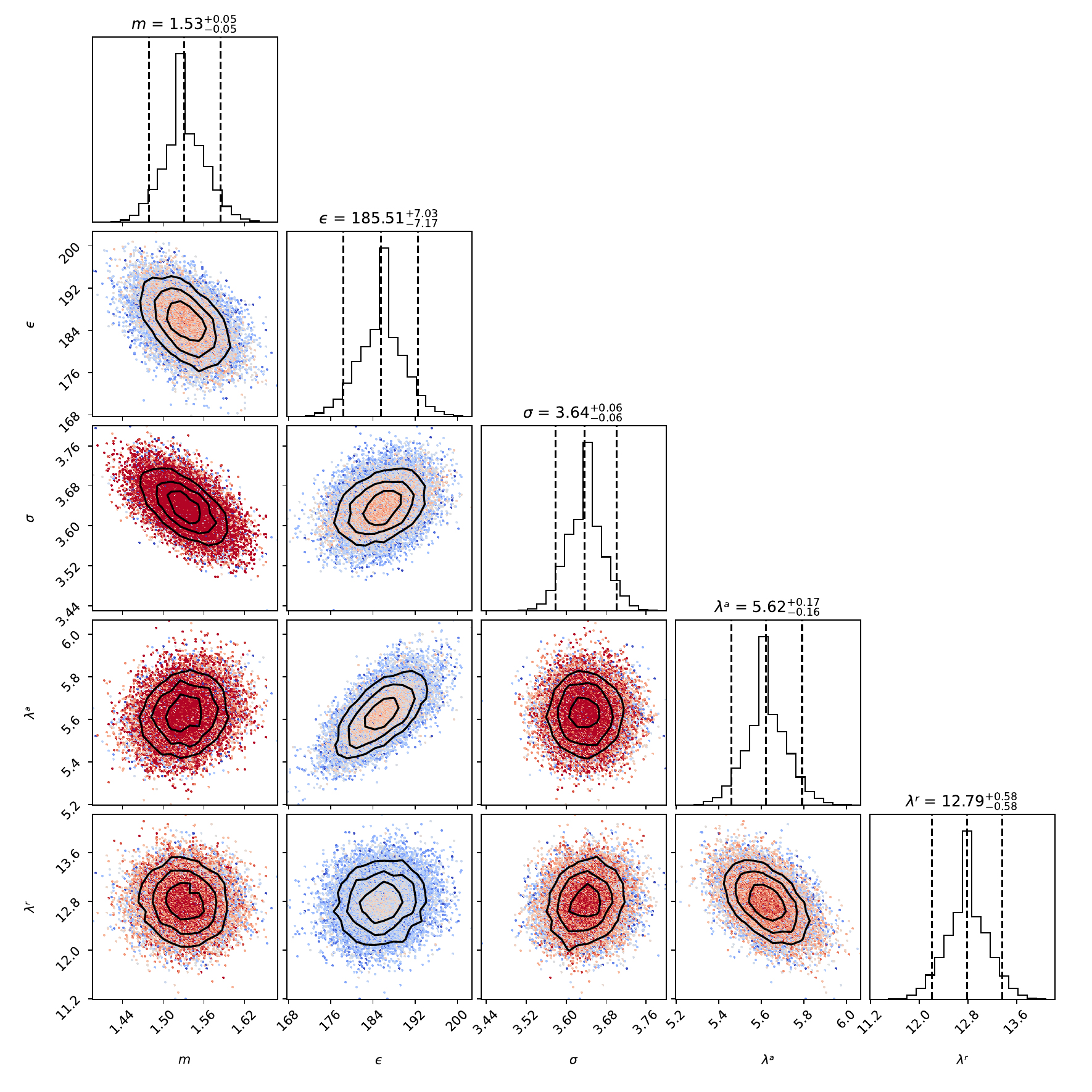}
      \caption{Confidence intervals obtained for the pure component parameters of ethane in SAFT-VR Mie. Colors and styles are identical to figure 2.}
  \end{figure}

\begin{figure}[H]
  \centering
    \begin{subfigure}[b]{0.49\textwidth}
      \includegraphics[width=1\textwidth]{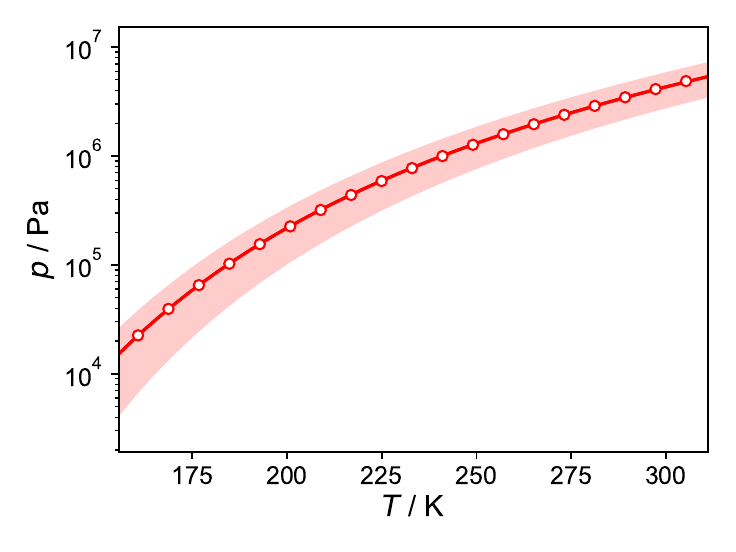}
      \caption{Saturation Curve}
    \end{subfigure}
    \begin{subfigure}[b]{0.49\textwidth}
      \includegraphics[width=1\textwidth]{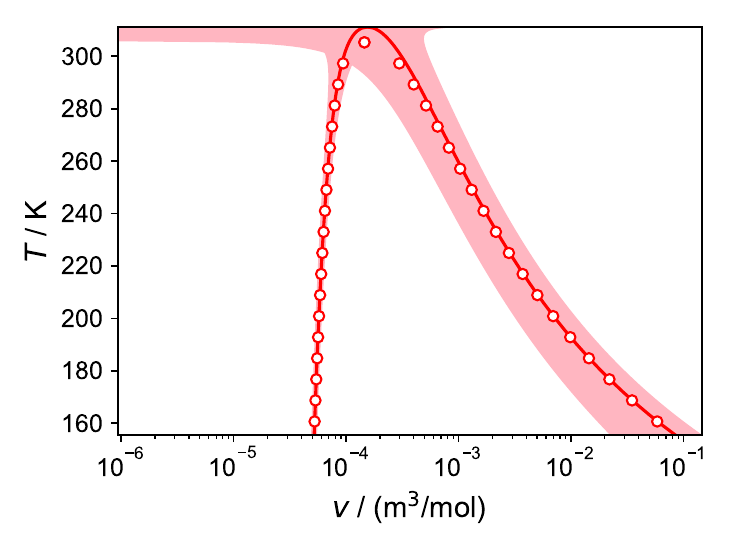}
      \caption{Saturation Envelope}

    \end{subfigure}
    \caption{Predicted values for the saturated volumes and saturation pressure for ethane using SAFT-VR Mie. Shaded regions correspond to the uncertainty interval for the predicted properties using the parameters and confidence intervals.}
  \end{figure}

\begin{figure}[H]
  \centering
    \begin{subfigure}[b]{0.49\textwidth}
      \includegraphics[width=1\textwidth]{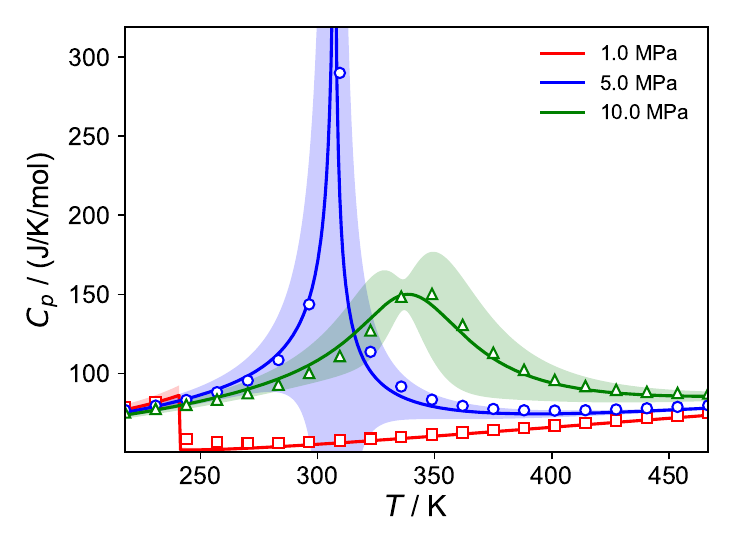}
      \caption{Isobaric heat capacity}
    \end{subfigure}
    \begin{subfigure}[b]{0.49\textwidth}
      \includegraphics[width=1\textwidth]{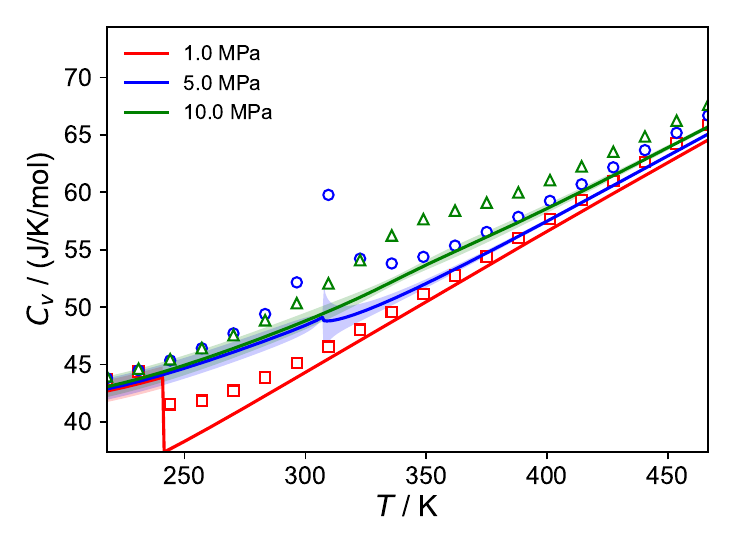}
      \caption{Isochoric heat capacity}
    \end{subfigure}
    \begin{subfigure}[b]{0.49\textwidth}
      \includegraphics[width=1\textwidth]{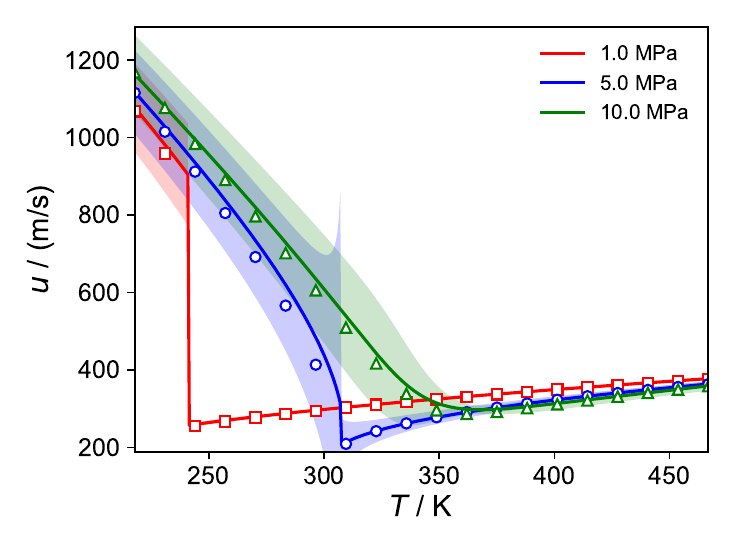}
      \caption{Speed of sound}
    \end{subfigure}
    \begin{subfigure}[b]{0.49\textwidth}
      \includegraphics[width=1\textwidth]{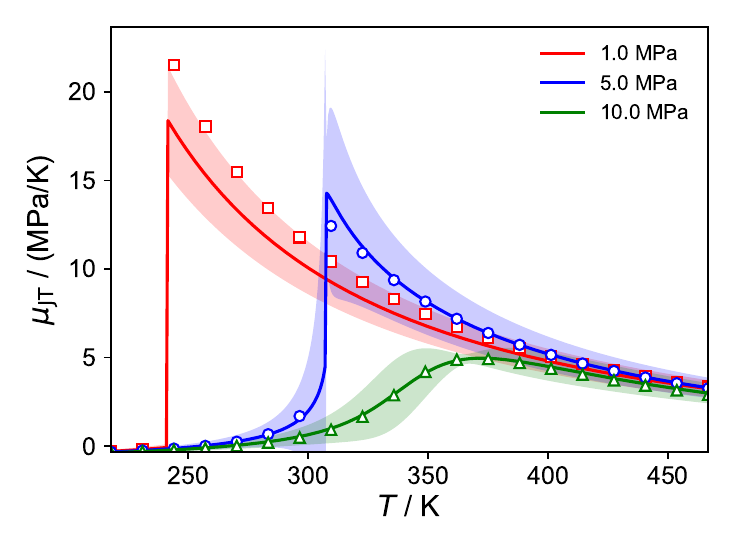}
      \caption{Joule--Thomson coefficient}
    \end{subfigure}
    \caption{Predicted values for the isobaric heat capacity, isochoric heat capacity, speed of sound and Joule--Thomson coefficient for ethane using SAFT-VR Mie at different pressures. Shaded regions correspond to the uncertainty interval for the predicted properties using the parameters and confidence intervals.}
  \end{figure}
\newpage
\subsection{Propane}
\begin{figure}[H]
  \centering
      \includegraphics[width=0.5\textwidth]{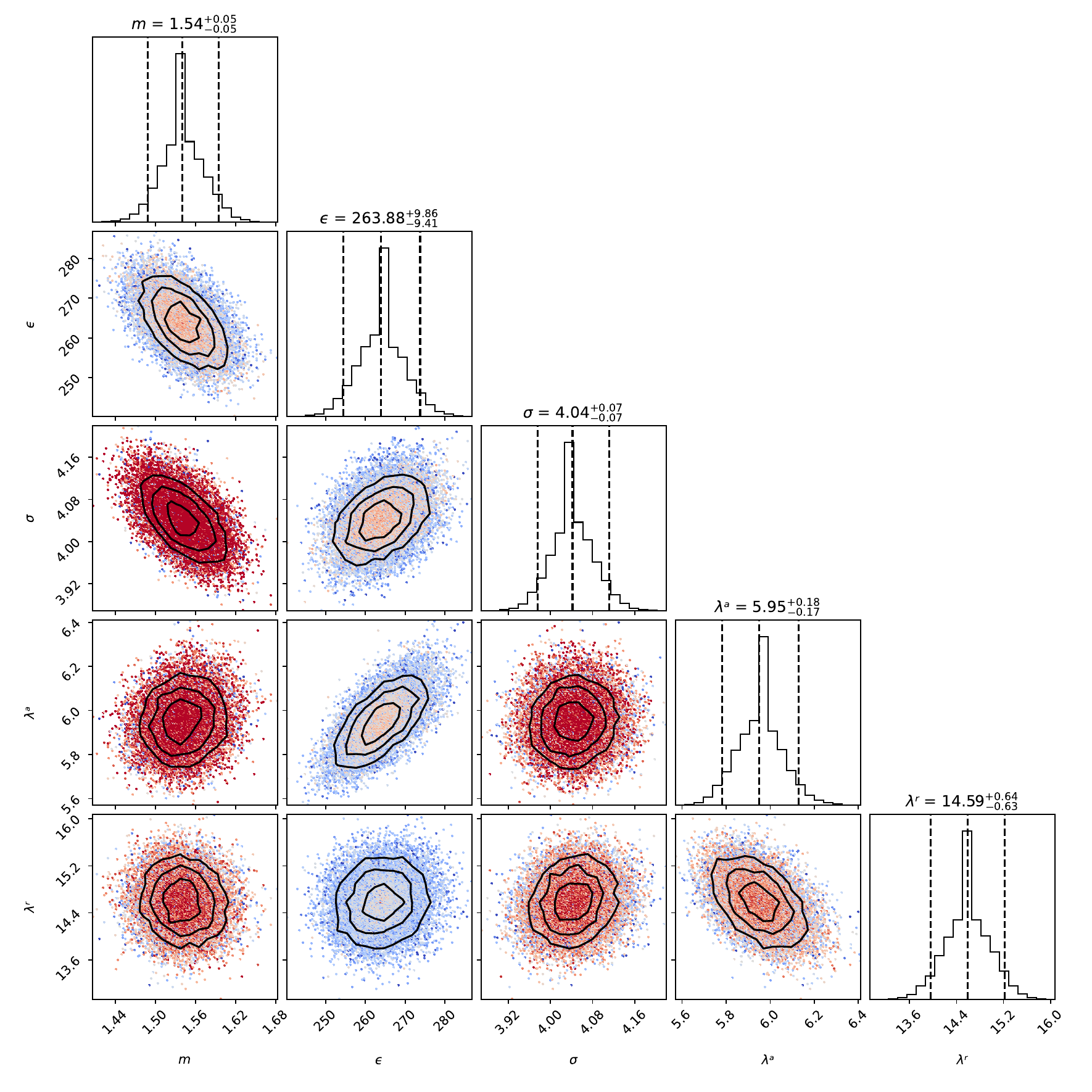}
      \caption{Confidence intervals obtained for the pure component parameters of propane in SAFT-VR Mie. Colors and styles are identical to figure 2.}
  \end{figure}

\begin{figure}[H]
  \centering
    \begin{subfigure}[b]{0.49\textwidth}
      \includegraphics[width=1\textwidth]{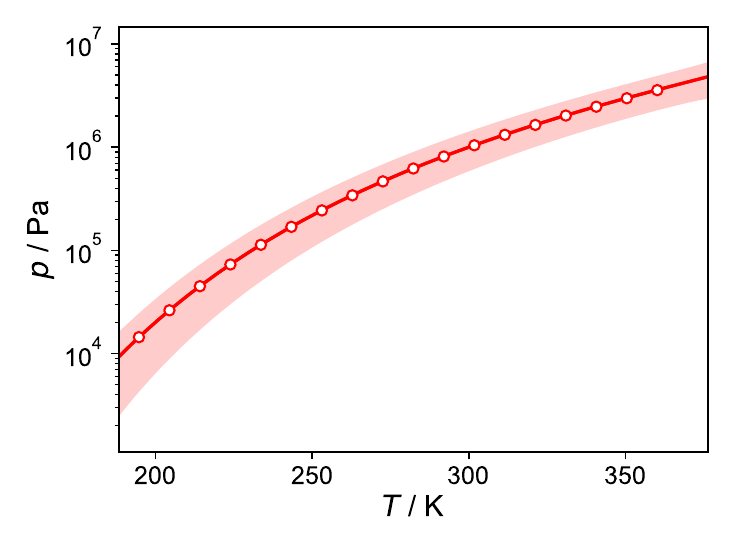}
      \caption{Saturation Curve}
    \end{subfigure}
    \begin{subfigure}[b]{0.49\textwidth}
      \includegraphics[width=1\textwidth]{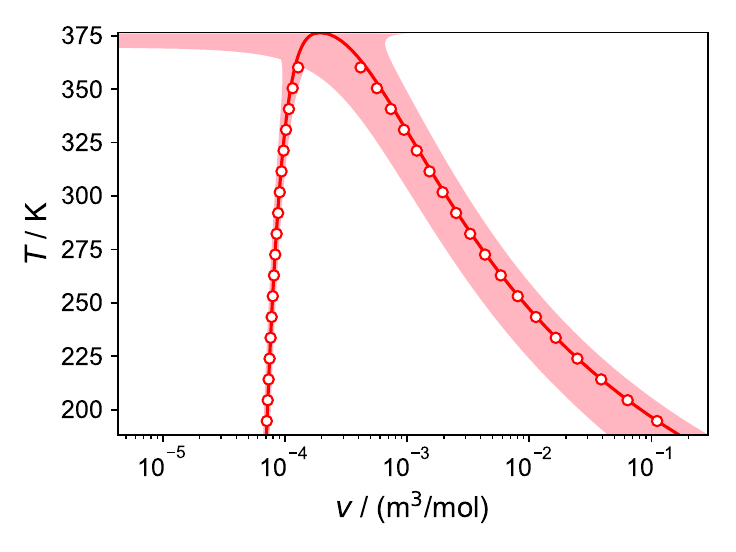}
      \caption{Saturation Envelope}

    \end{subfigure}
    \caption{Predicted values for the saturated volumes and saturation pressure for propane using SAFT-VR Mie. Shaded regions correspond to the uncertainty interval for the predicted properties using the parameters and confidence intervals.}
  \end{figure}

\begin{figure}[H]
  \centering
    \begin{subfigure}[b]{0.49\textwidth}
      \includegraphics[width=1\textwidth]{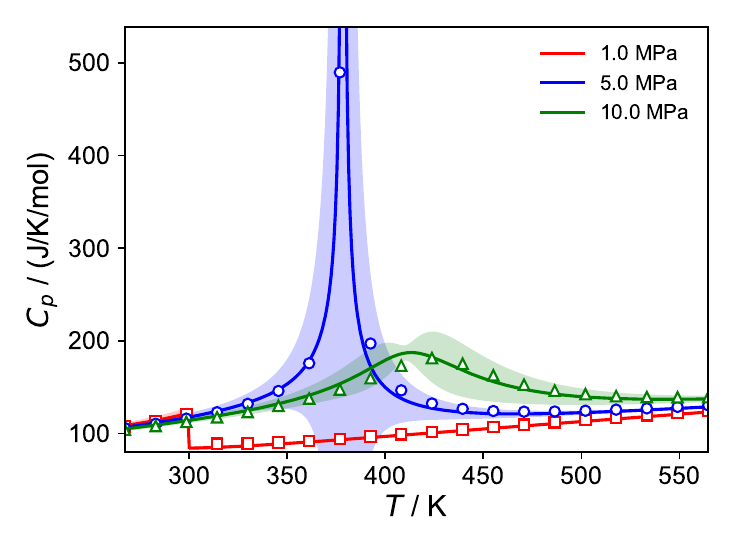}
      \caption{Isobaric heat capacity}
    \end{subfigure}
    \begin{subfigure}[b]{0.49\textwidth}
      \includegraphics[width=1\textwidth]{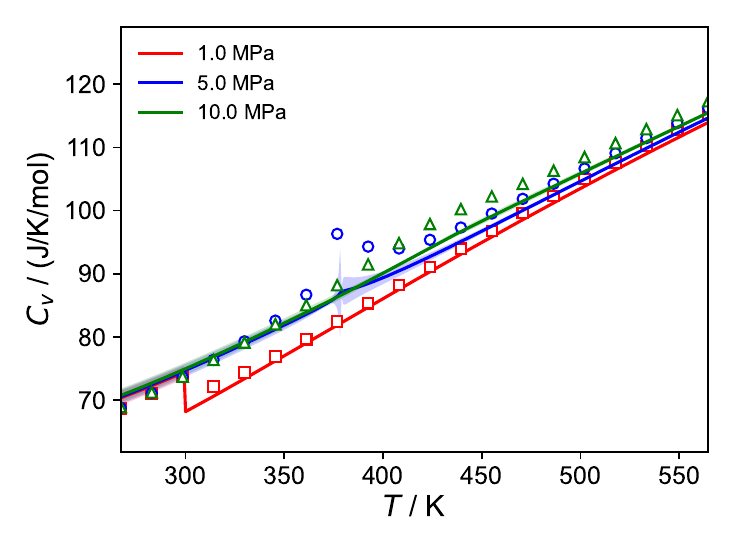}
      \caption{Isochoric heat capacity}
    \end{subfigure}
    \begin{subfigure}[b]{0.49\textwidth}
      \includegraphics[width=1\textwidth]{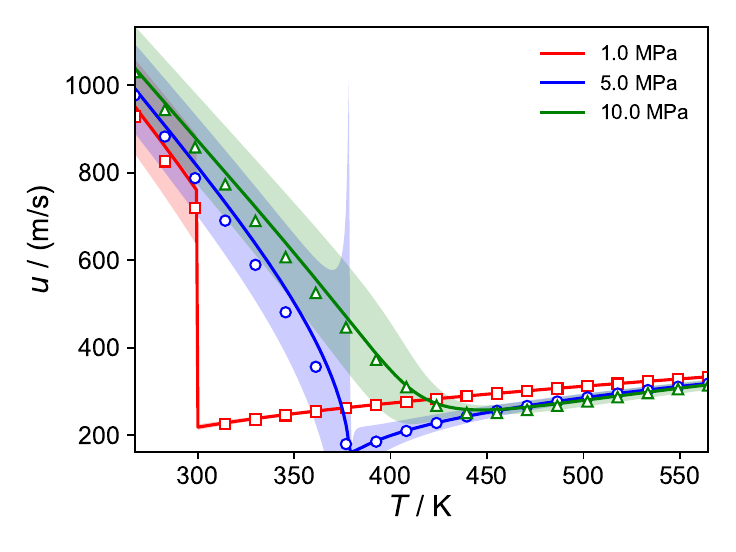}
      \caption{Speed of sound}
    \end{subfigure}
    \begin{subfigure}[b]{0.49\textwidth}
      \includegraphics[width=1\textwidth]{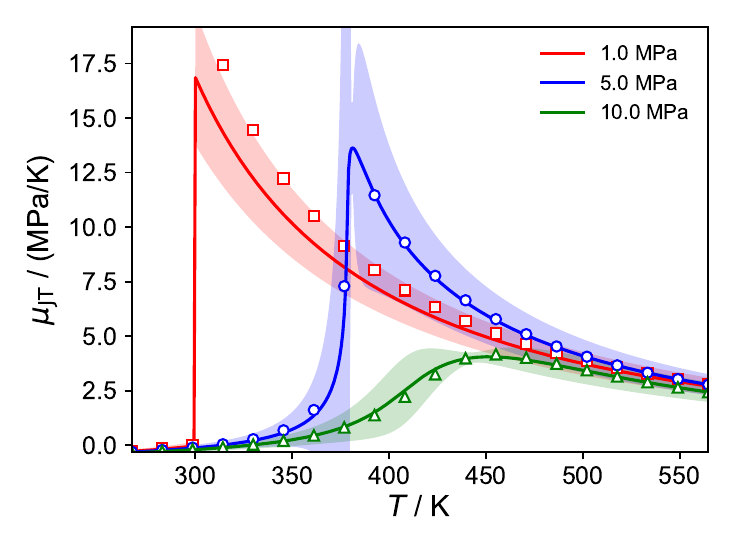}
      \caption{Joule--Thomson coefficient}
    \end{subfigure}
    \caption{Predicted values for the isobaric heat capacity, isochoric heat capacity, speed of sound and Joule--Thomson coefficient for propane using SAFT-VR Mie at different pressures. Shaded regions correspond to the uncertainty interval for the predicted properties using the parameters and confidence intervals.}
  \end{figure}
\newpage

\subsection{$n$-butane}
\begin{figure}[H]
  \centering
      \includegraphics[width=0.5\textwidth]{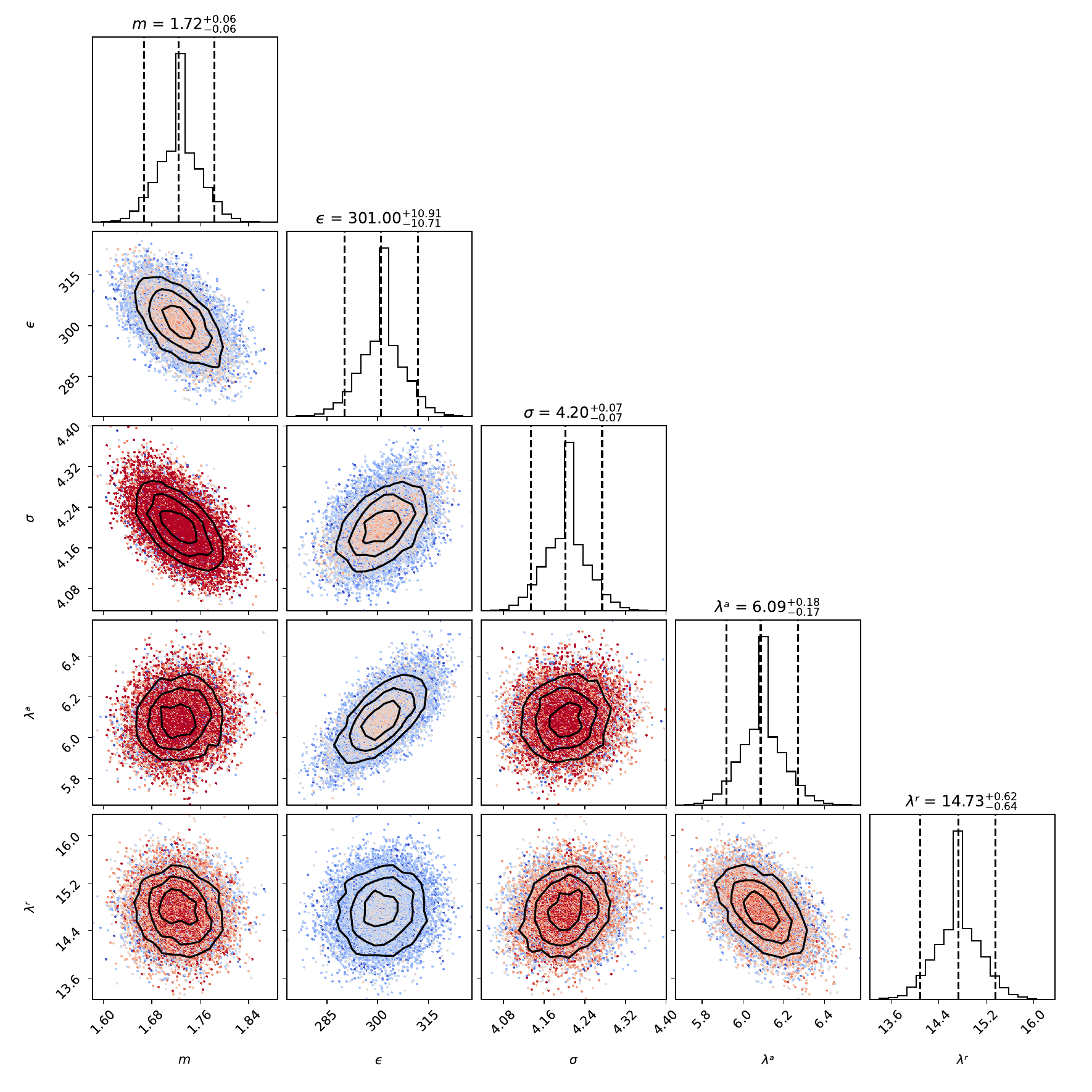}
      \caption{Confidence intervals obtained for the pure component parameters of $n$-butane in SAFT-VR Mie. Colors and styles are identical to figure 2.}
  \end{figure}

\begin{figure}[H]
  \centering
    \begin{subfigure}[b]{0.49\textwidth}
      \includegraphics[width=1\textwidth]{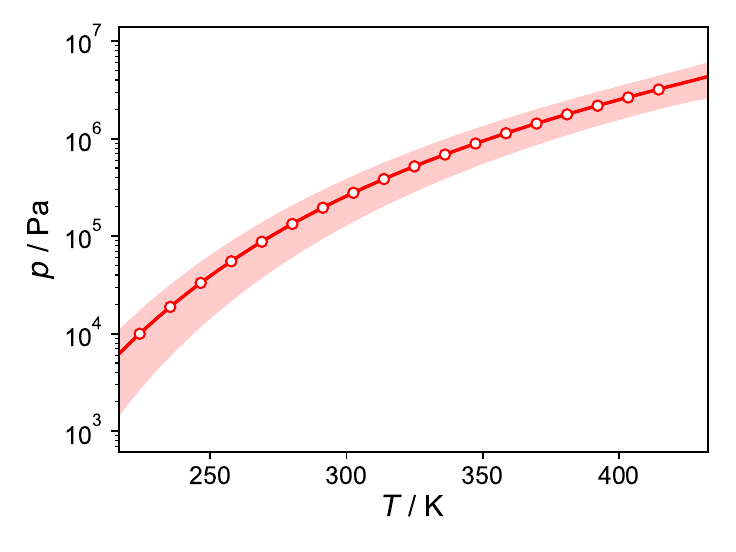}
      \caption{Saturation Curve}
    \end{subfigure}
    \begin{subfigure}[b]{0.49\textwidth}
      \includegraphics[width=1\textwidth]{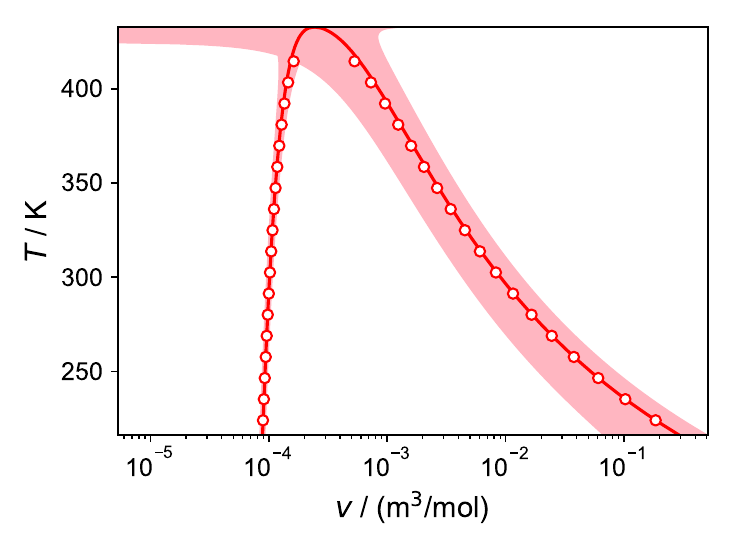}
      \caption{Saturation Envelope}

    \end{subfigure}
    \caption{Predicted values for the saturated volumes and saturation pressure for $n$-butane using SAFT-VR Mie. Shaded regions correspond to the uncertainty interval for the predicted properties using the parameters and confidence intervals.}
  \end{figure}

\begin{figure}[H]
  \centering
    \begin{subfigure}[b]{0.49\textwidth}
      \includegraphics[width=1\textwidth]{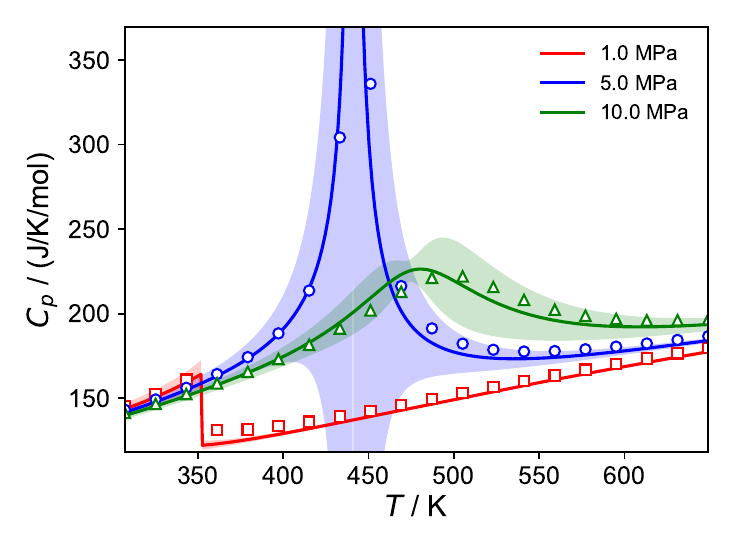}
      \caption{Isobaric heat capacity}
    \end{subfigure}
    \begin{subfigure}[b]{0.49\textwidth}
      \includegraphics[width=1\textwidth]{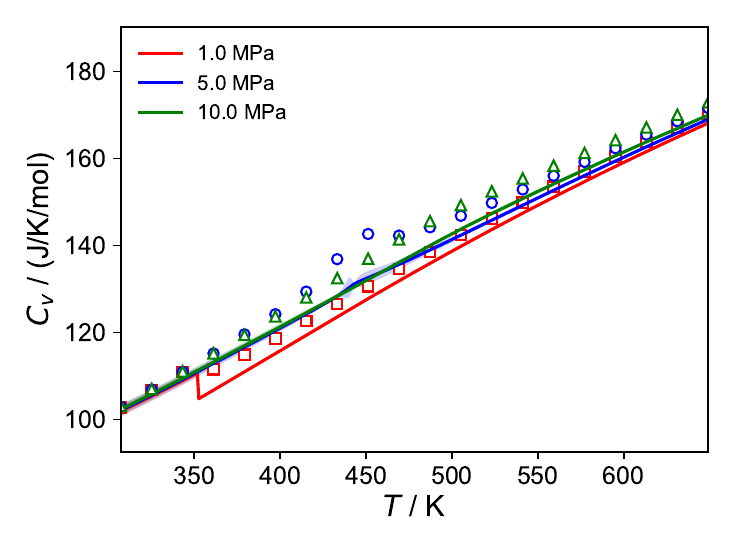}
      \caption{Isochoric heat capacity}
    \end{subfigure}
    \begin{subfigure}[b]{0.49\textwidth}
      \includegraphics[width=1\textwidth]{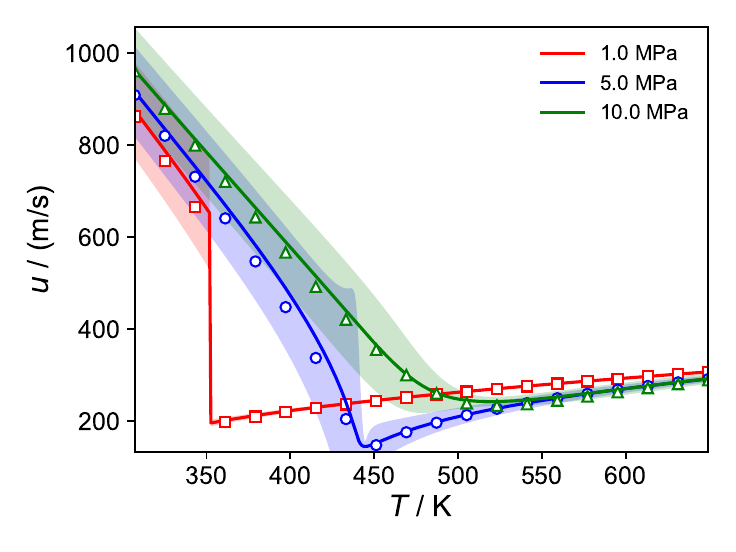}
      \caption{Speed of sound}
    \end{subfigure}
    \begin{subfigure}[b]{0.49\textwidth}
      \includegraphics[width=1\textwidth]{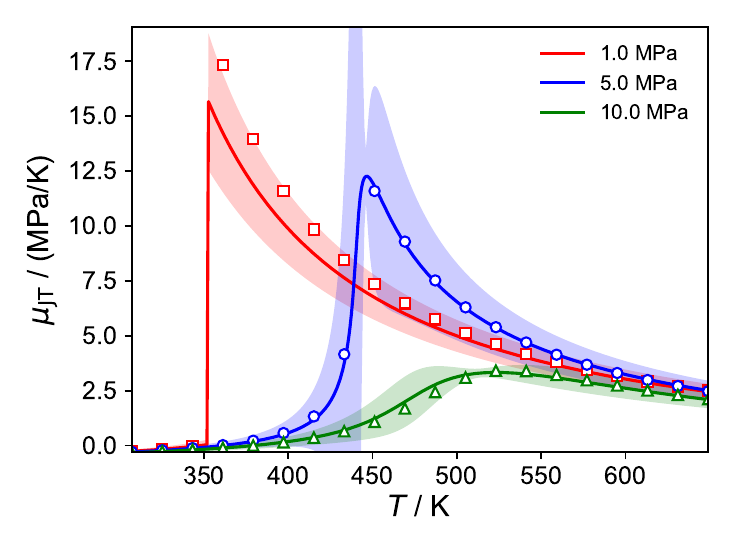}
      \caption{Joule--Thomson coefficient}
    \end{subfigure}
    \caption{Predicted values for the isobaric heat capacity, isochoric heat capacity, speed of sound and Joule--Thomson coefficient for $n$-butane using SAFT-VR Mie at different pressures. Shaded regions correspond to the uncertainty interval for the predicted properties using the parameters and confidence intervals.}
  \end{figure}
\newpage

\subsection{$n$-pentane}
\begin{figure}[H]
  \centering
      \includegraphics[width=0.5\textwidth]{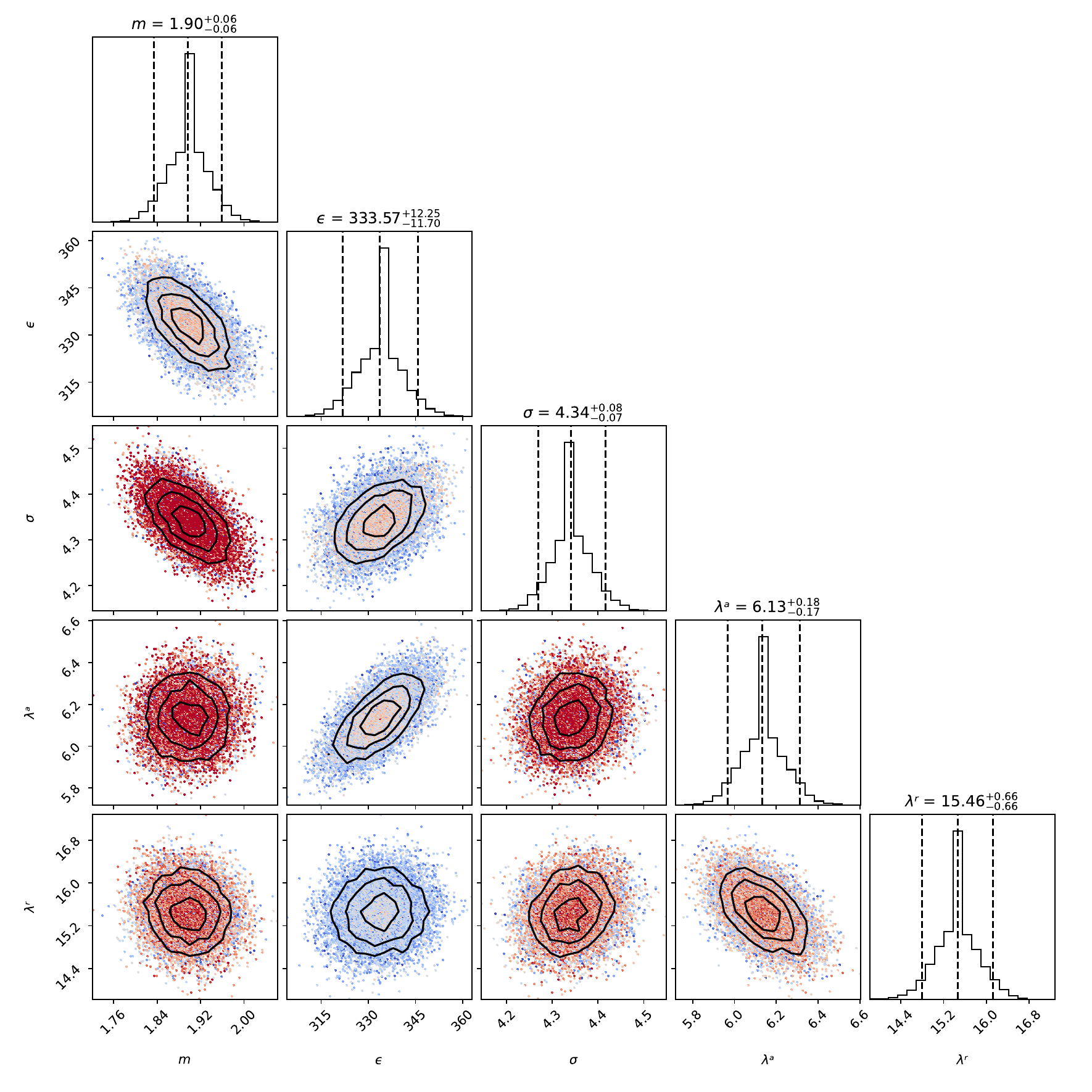}
      \caption{Confidence intervals obtained for the pure component parameters of $n$-pentane in SAFT-VR Mie. Colors and styles are identical to figure 2.}
  \end{figure}

\begin{figure}[H]
  \centering
    \begin{subfigure}[b]{0.49\textwidth}
      \includegraphics[width=1\textwidth]{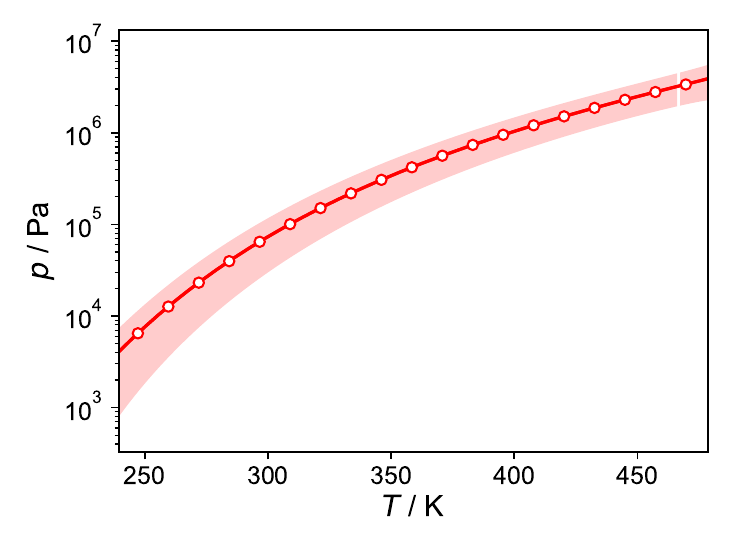}
      \caption{Saturation Curve}
    \end{subfigure}
    \begin{subfigure}[b]{0.49\textwidth}
      \includegraphics[width=1\textwidth]{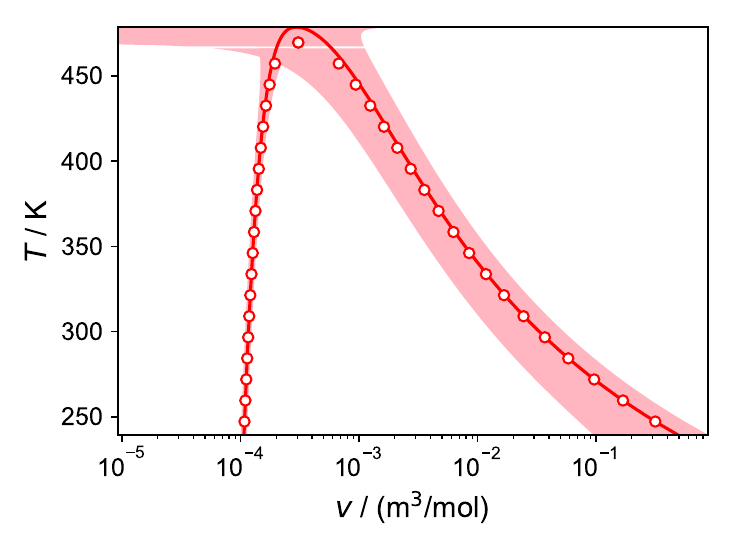}
      \caption{Saturation Envelope}

    \end{subfigure}
    \caption{Predicted values for the saturated volumes and saturation pressure for $n$-pentane using SAFT-VR Mie. Shaded regions correspond to the uncertainty interval for the predicted properties using the parameters and confidence intervals.}
  \end{figure}

\begin{figure}[H]
  \centering
    \begin{subfigure}[b]{0.49\textwidth}
      \includegraphics[width=1\textwidth]{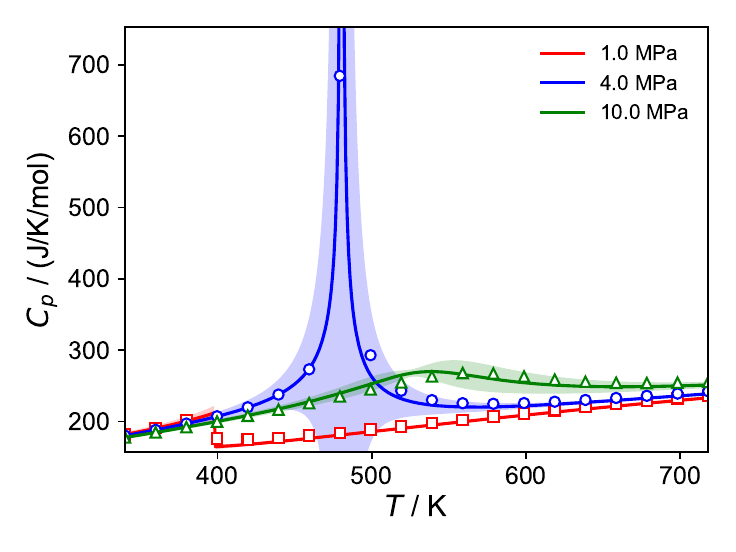}
      \caption{Isobaric heat capacity}
    \end{subfigure}
    \begin{subfigure}[b]{0.49\textwidth}
      \includegraphics[width=1\textwidth]{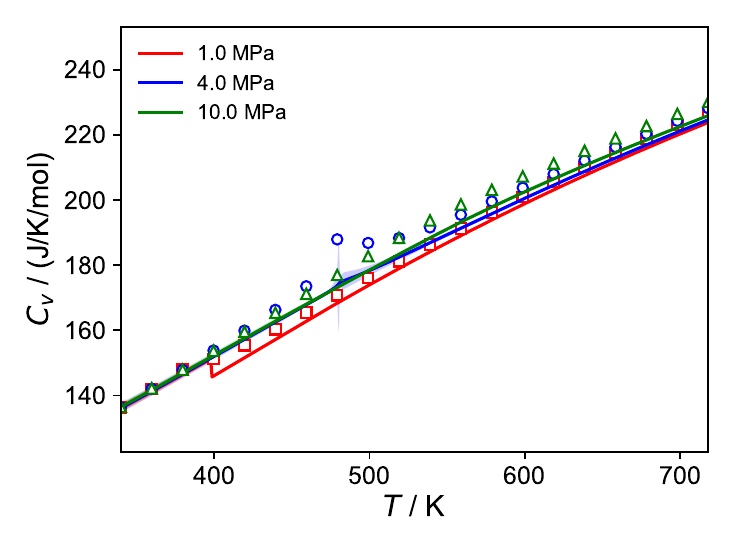}
      \caption{Isochoric heat capacity}
    \end{subfigure}
    \begin{subfigure}[b]{0.49\textwidth}
      \includegraphics[width=1\textwidth]{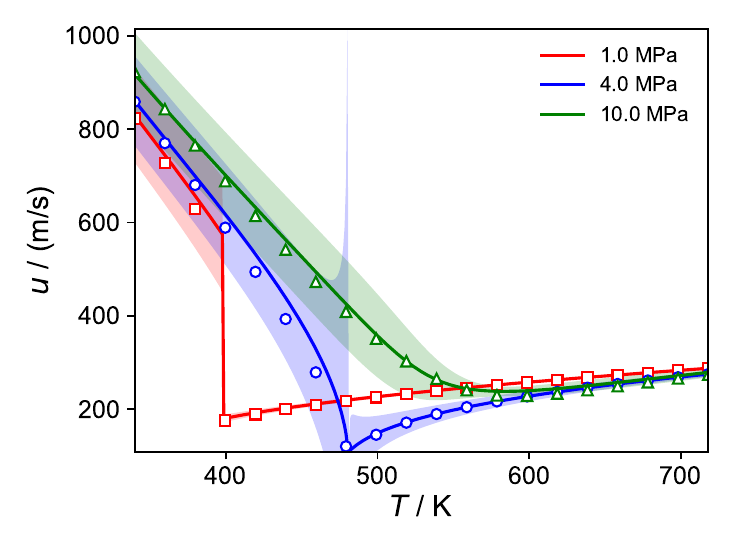}
      \caption{Speed of sound}
    \end{subfigure}
    \begin{subfigure}[b]{0.49\textwidth}
      \includegraphics[width=1\textwidth]{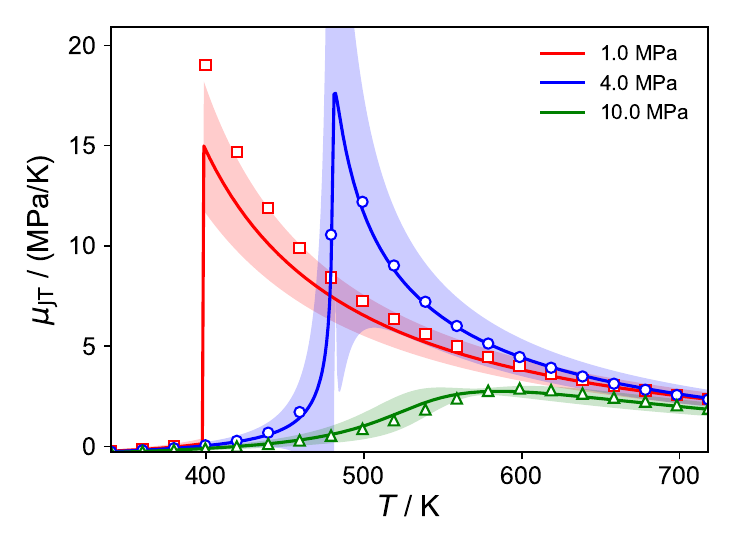}
      \caption{Joule--Thomson coefficient}
    \end{subfigure}
    \caption{Predicted values for the isobaric heat capacity, isochoric heat capacity, speed of sound and Joule--Thomson coefficient for $n$-pentane using SAFT-VR Mie at different pressures. Shaded regions correspond to the uncertainty interval for the predicted properties using the parameters and confidence intervals.}
  \end{figure}
\newpage

\subsection{$n$-hexane}
\begin{figure}[H]
  \centering
      \includegraphics[width=0.5\textwidth]{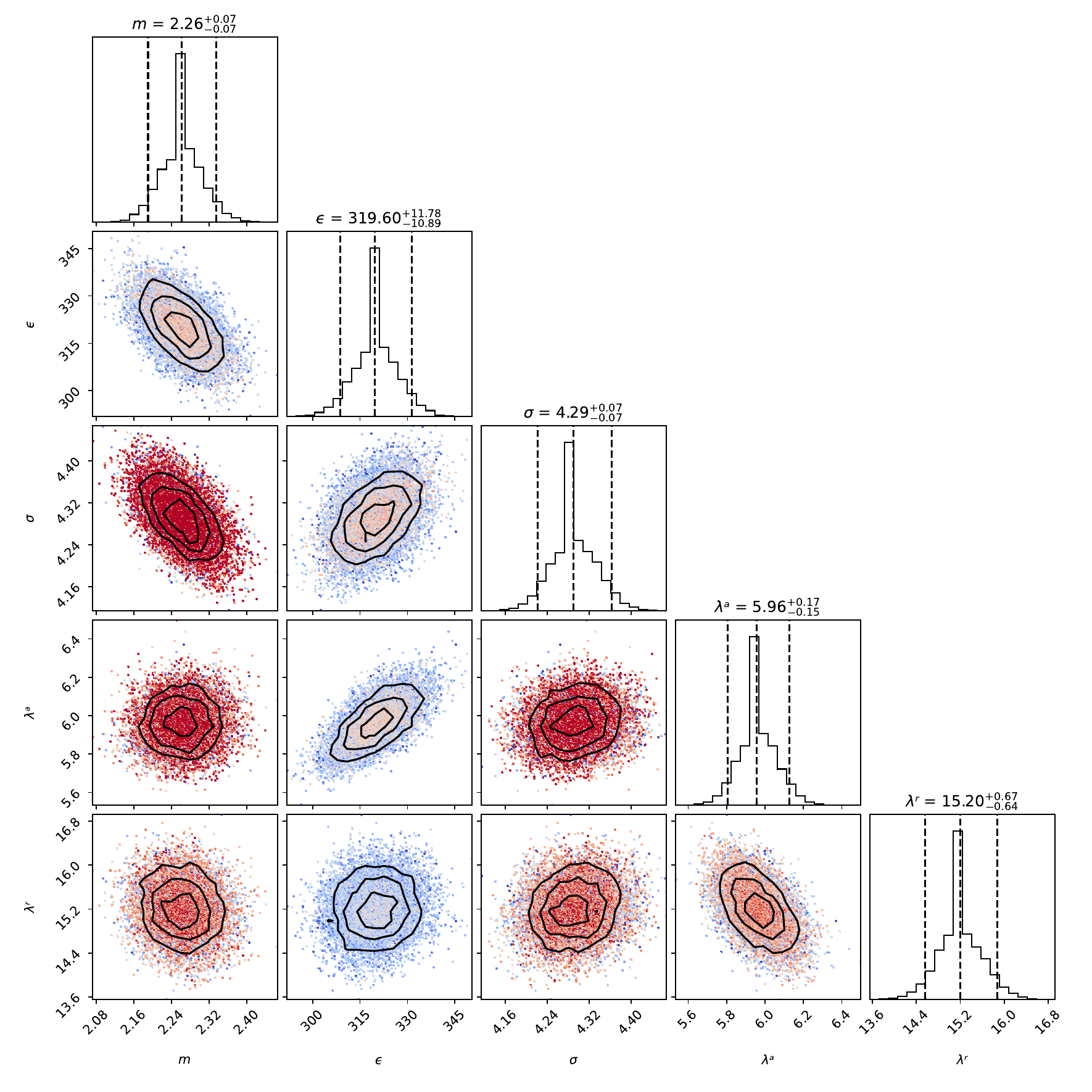}
      \caption{Confidence intervals obtained for the pure component parameters of $n$-hexane in SAFT-VR Mie. Colors and styles are identical to figure 2.}
  \end{figure}

\begin{figure}[H]
  \centering
    \begin{subfigure}[b]{0.49\textwidth}
      \includegraphics[width=1\textwidth]{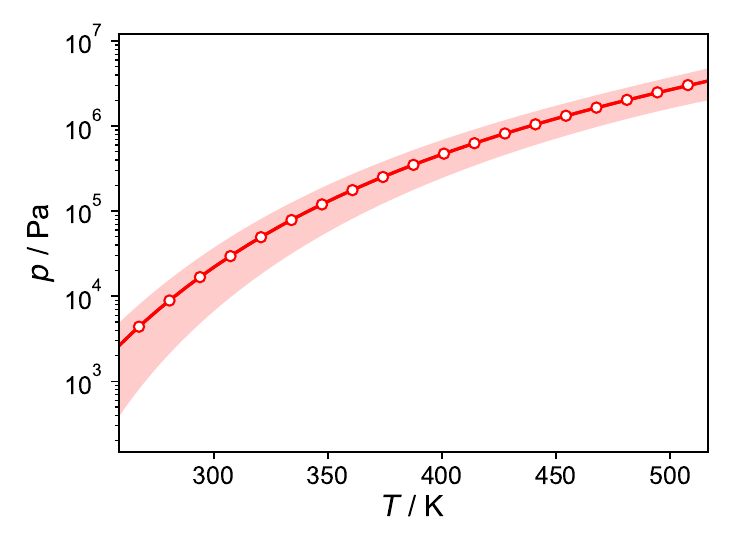}
      \caption{Saturation Curve}
    \end{subfigure}
    \begin{subfigure}[b]{0.49\textwidth}
      \includegraphics[width=1\textwidth]{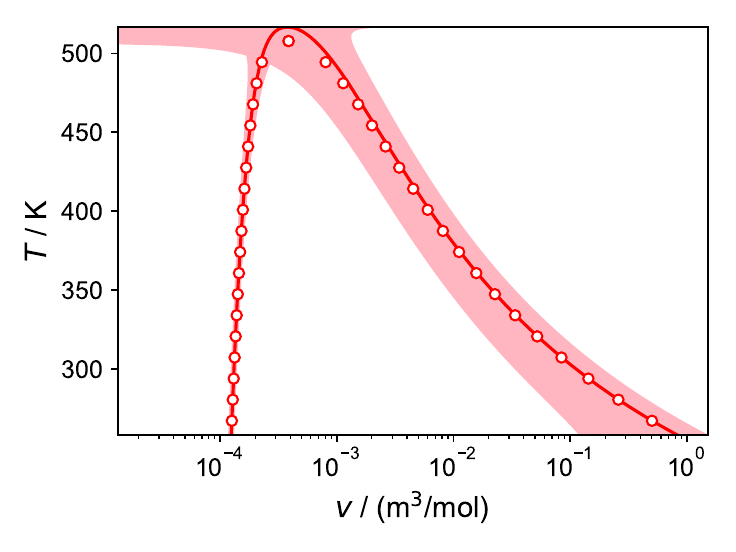}
      \caption{Saturation Envelope}

    \end{subfigure}
    \caption{Predicted values for the saturated volumes and saturation pressure for $n$-hexane using SAFT-VR Mie. Shaded regions correspond to the uncertainty interval for the predicted properties using the parameters and confidence intervals.}
  \end{figure}

\begin{figure}[H]
  \centering
    \begin{subfigure}[b]{0.49\textwidth}
      \includegraphics[width=1\textwidth]{Figures/SAFTVRMie/Error/hexane_cp_error.pdf}
      \caption{Isobaric heat capacity}
    \end{subfigure}
    \begin{subfigure}[b]{0.49\textwidth}
      \includegraphics[width=1\textwidth]{Figures/SAFTVRMie/Error/hexane_cv_error.pdf}
      \caption{Isochoric heat capacity}
    \end{subfigure}
    \begin{subfigure}[b]{0.49\textwidth}
      \includegraphics[width=1\textwidth]{Figures/SAFTVRMie/Error/hexane_sos_error.pdf}
      \caption{Speed of sound}
    \end{subfigure}
    \begin{subfigure}[b]{0.49\textwidth}
      \includegraphics[width=1\textwidth]{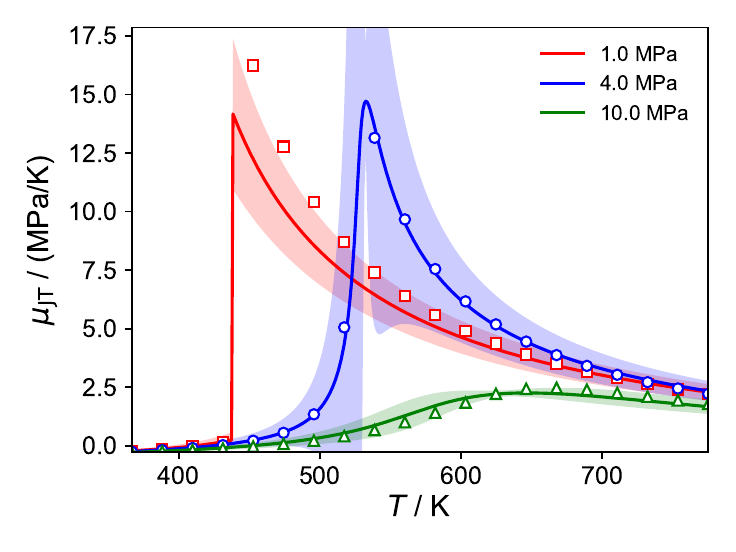}
      \caption{Joule--Thomson coefficient}
    \end{subfigure}
    \caption{Predicted values for the isobaric heat capacity, isochoric heat capacity, speed of sound and Joule--Thomson coefficient for $n$-hexane using SAFT-VR Mie at different pressures. Shaded regions correspond to the uncertainty interval for the predicted properties using the parameters and confidence intervals.}
  \end{figure}
\newpage

\subsection{$n$-heptane}
\begin{figure}[H]
  \centering
      \includegraphics[width=0.5\textwidth]{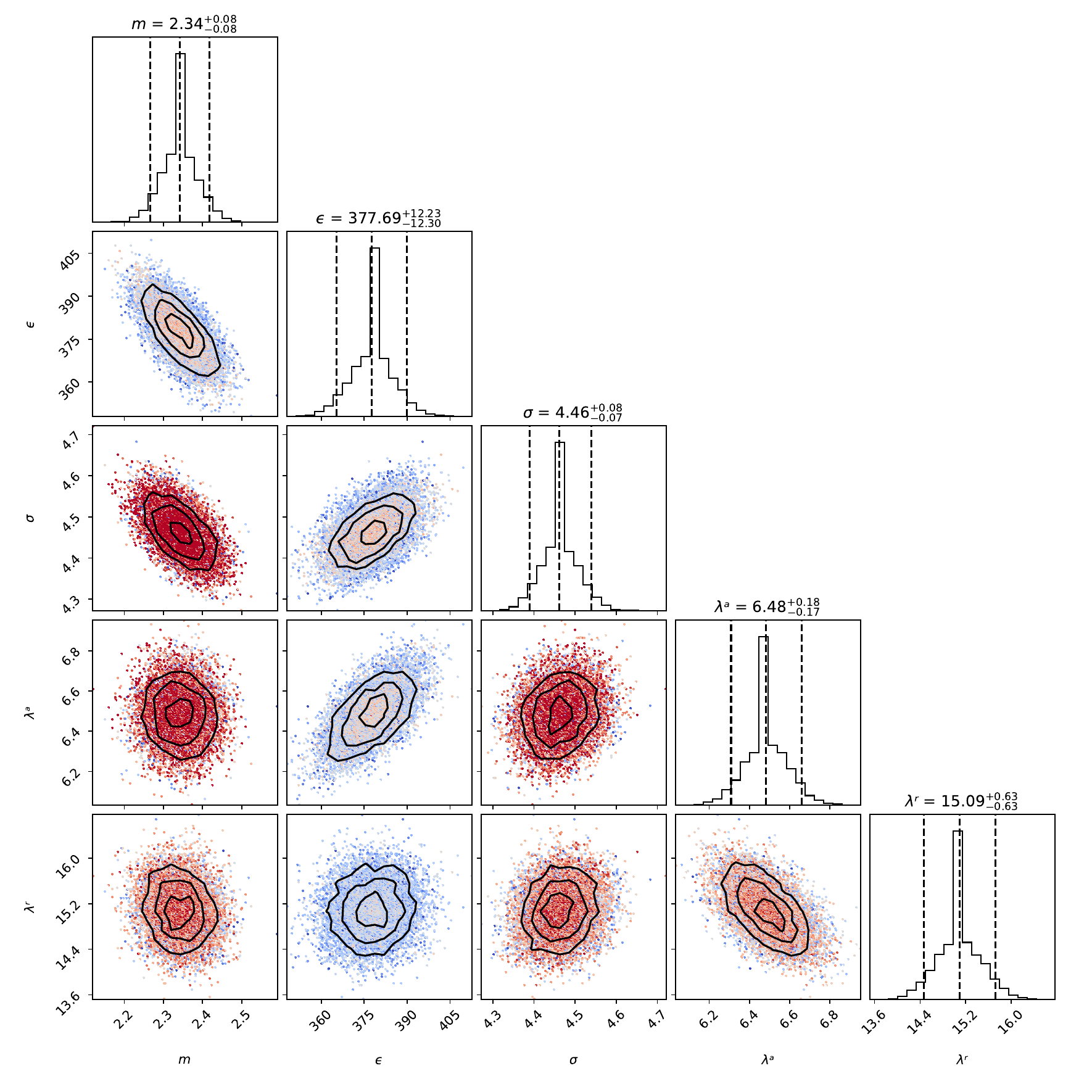}
      \caption{Confidence intervals obtained for the pure component parameters of $n$-heptane in SAFT-VR Mie. Colors and styles are identical to figure 2.}
  \end{figure}

\begin{figure}[H]
  \centering
    \begin{subfigure}[b]{0.49\textwidth}
      \includegraphics[width=1\textwidth]{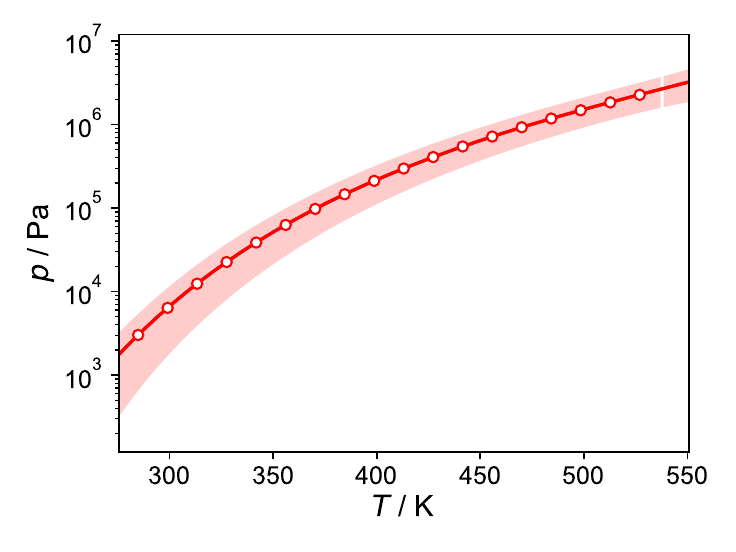}
      \caption{Saturation Curve}
    \end{subfigure}
    \begin{subfigure}[b]{0.49\textwidth}
      \includegraphics[width=1\textwidth]{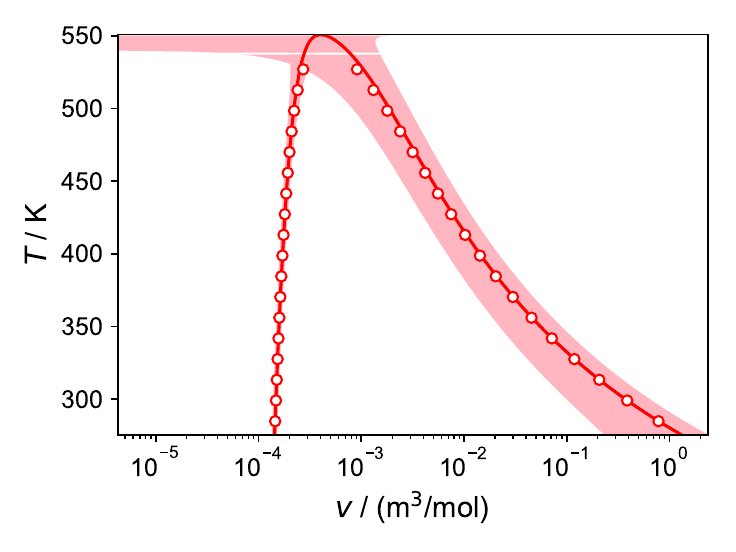}
      \caption{Saturation Envelope}

    \end{subfigure}
    \caption{Predicted values for the saturated volumes and saturation pressure for $n$-heptane using SAFT-VR Mie. Shaded regions correspond to the uncertainty interval for the predicted properties using the parameters and confidence intervals.}
  \end{figure}

\begin{figure}[H]
  \centering
    \begin{subfigure}[b]{0.49\textwidth}
      \includegraphics[width=1\textwidth]{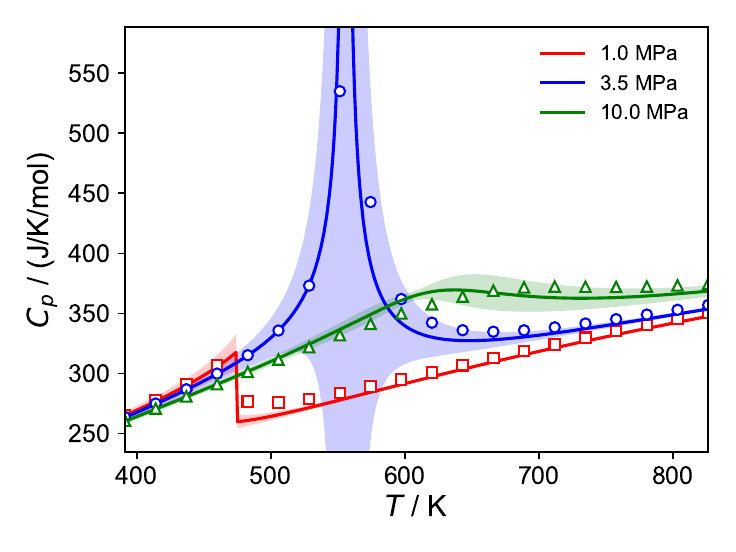}
      \caption{Isobaric heat capacity}
    \end{subfigure}
    \begin{subfigure}[b]{0.49\textwidth}
      \includegraphics[width=1\textwidth]{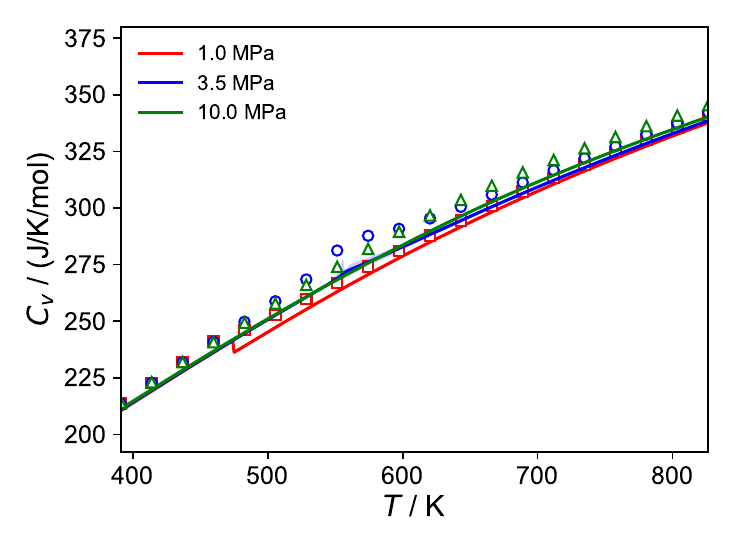}
      \caption{Isochoric heat capacity}
    \end{subfigure}
    \begin{subfigure}[b]{0.49\textwidth}
      \includegraphics[width=1\textwidth]{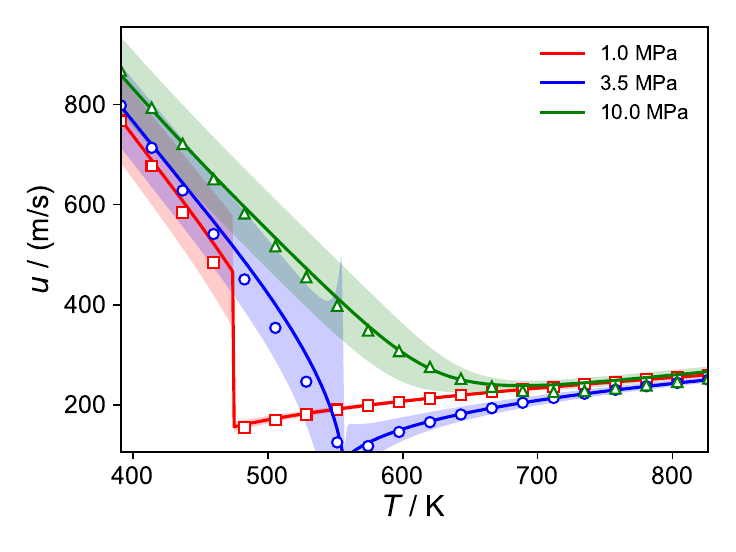}
      \caption{Speed of sound}
    \end{subfigure}
    \begin{subfigure}[b]{0.49\textwidth}
      \includegraphics[width=1\textwidth]{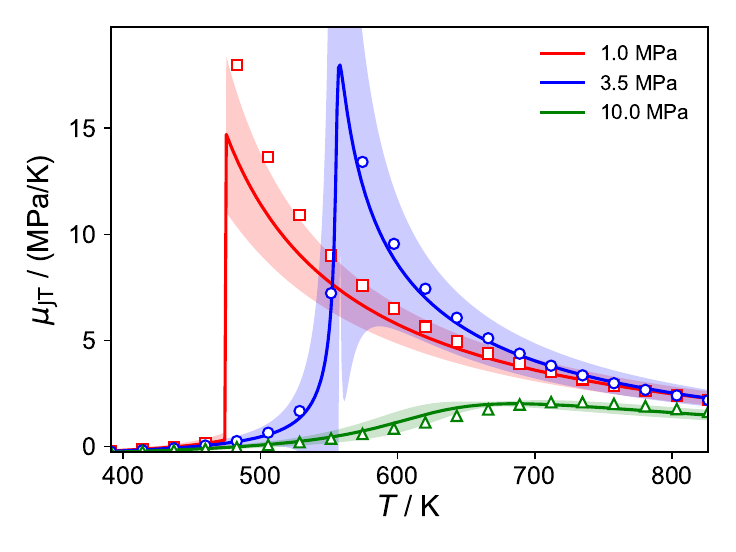}
      \caption{Joule--Thomson coefficient}
    \end{subfigure}
    \caption{Predicted values for the isobaric heat capacity, isochoric heat capacity, speed of sound and Joule--Thomson coefficient for $n$-heptane using SAFT-VR Mie at different pressures. Shaded regions correspond to the uncertainty interval for the predicted properties using the parameters and confidence intervals.}
  \end{figure}
\newpage

\subsection{$n$-octane}
\begin{figure}[H]
  \centering
      \includegraphics[width=0.5\textwidth]{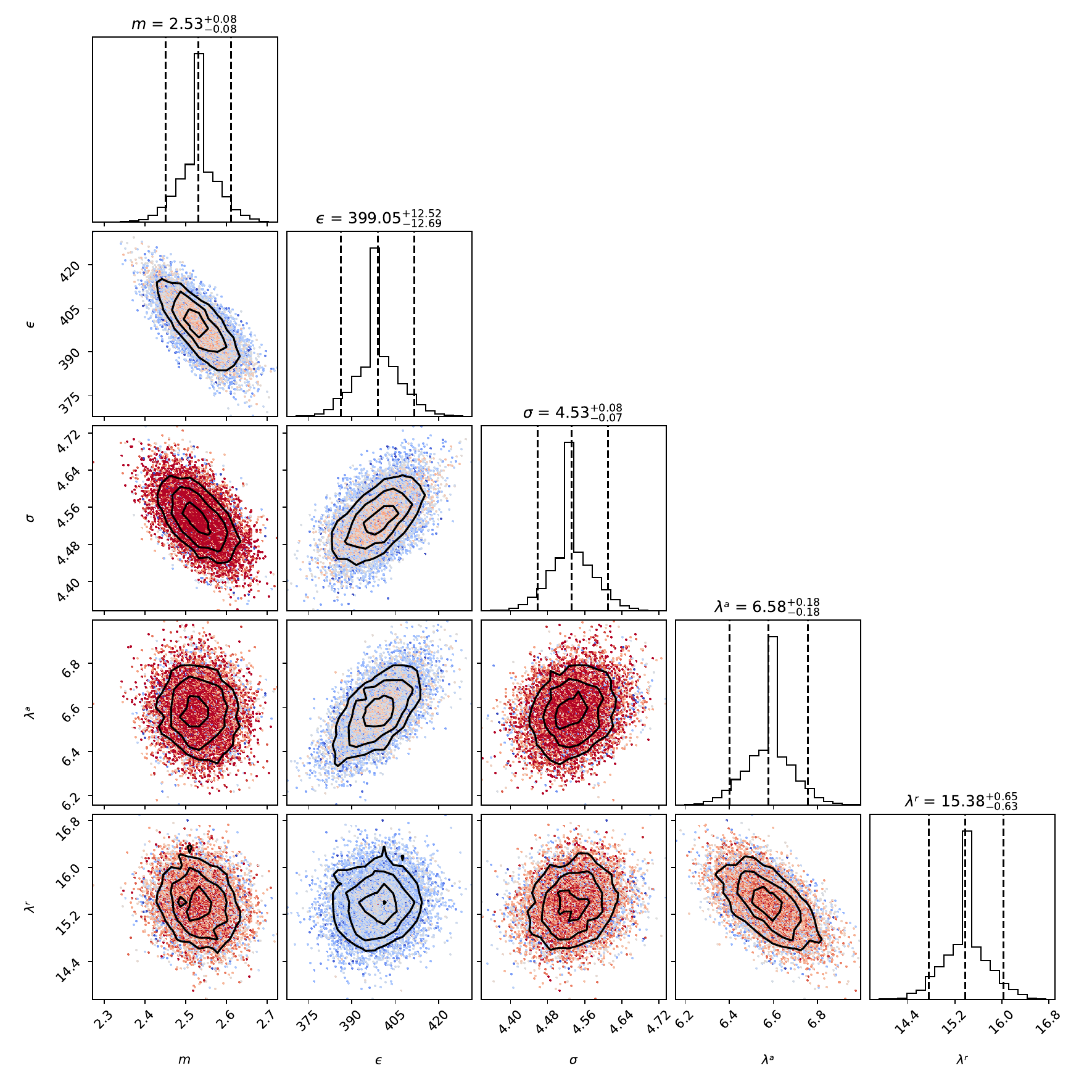}
      \caption{Confidence intervals obtained for the pure component parameters of $n$-octane in SAFT-VR Mie. Colors and styles are identical to figure 2.}
  \end{figure}

\begin{figure}[H]
  \centering
    \begin{subfigure}[b]{0.49\textwidth}
      \includegraphics[width=1\textwidth]{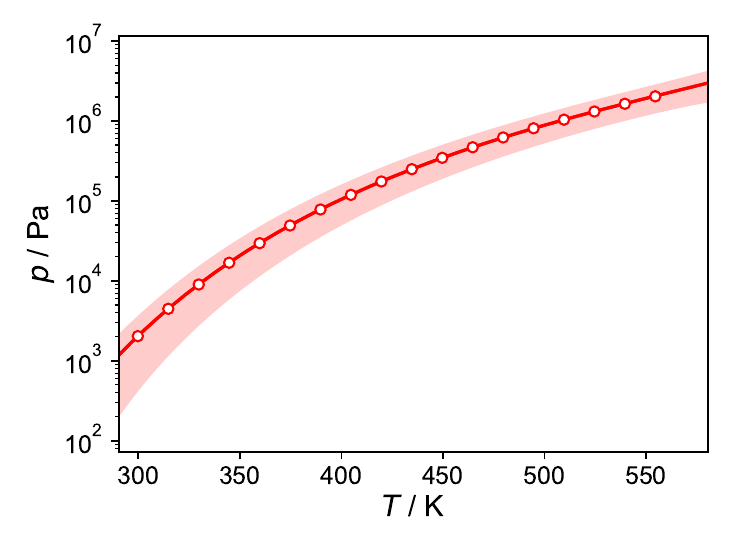}
      \caption{Saturation Curve}
    \end{subfigure}
    \begin{subfigure}[b]{0.49\textwidth}
      \includegraphics[width=1\textwidth]{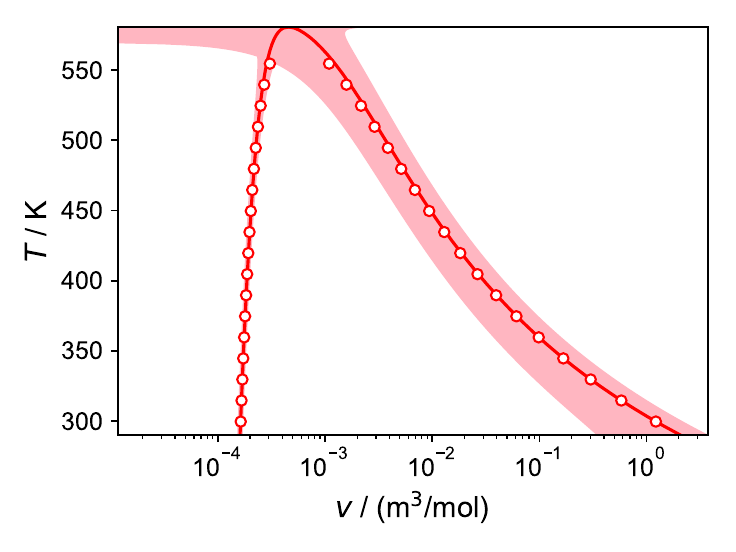}
      \caption{Saturation Envelope}

    \end{subfigure}
    \caption{Predicted values for the saturated volumes and saturation pressure for $n$-octane using SAFT-VR Mie. Shaded regions correspond to the uncertainty interval for the predicted properties using the parameters and confidence intervals.}
  \end{figure}

\begin{figure}[H]
  \centering
    \begin{subfigure}[b]{0.49\textwidth}
      \includegraphics[width=1\textwidth]{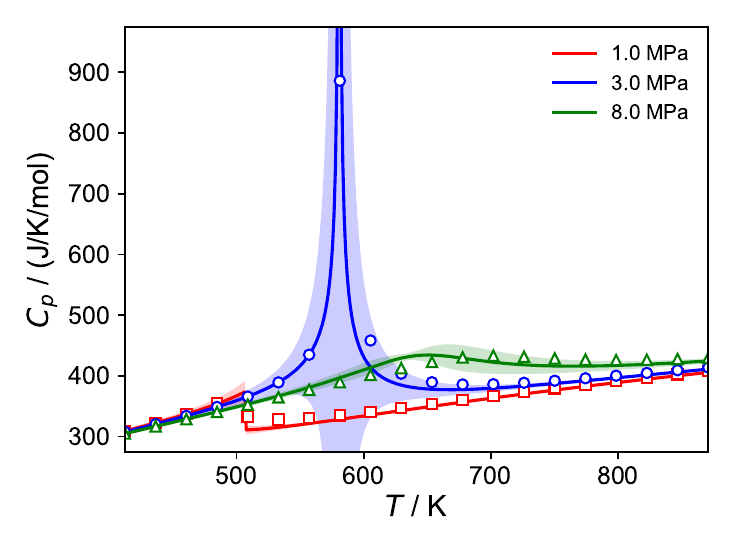}
      \caption{Isobaric heat capacity}
    \end{subfigure}
    \begin{subfigure}[b]{0.49\textwidth}
      \includegraphics[width=1\textwidth]{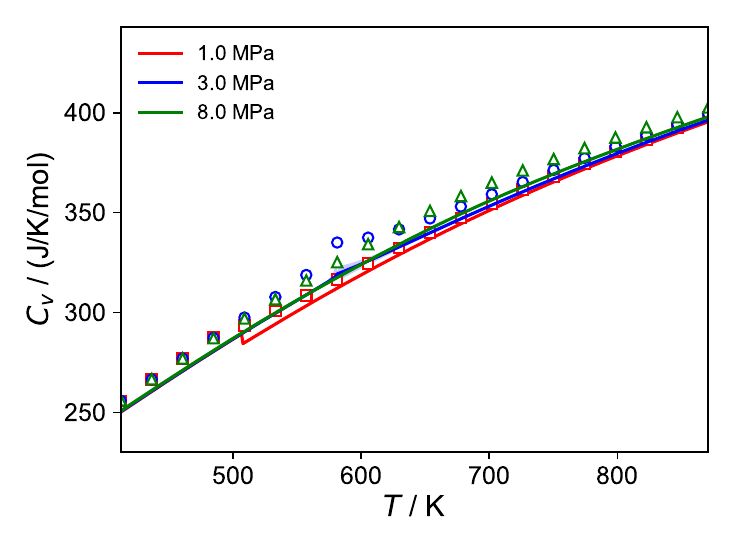}
      \caption{Isochoric heat capacity}
    \end{subfigure}
    \begin{subfigure}[b]{0.49\textwidth}
      \includegraphics[width=1\textwidth]{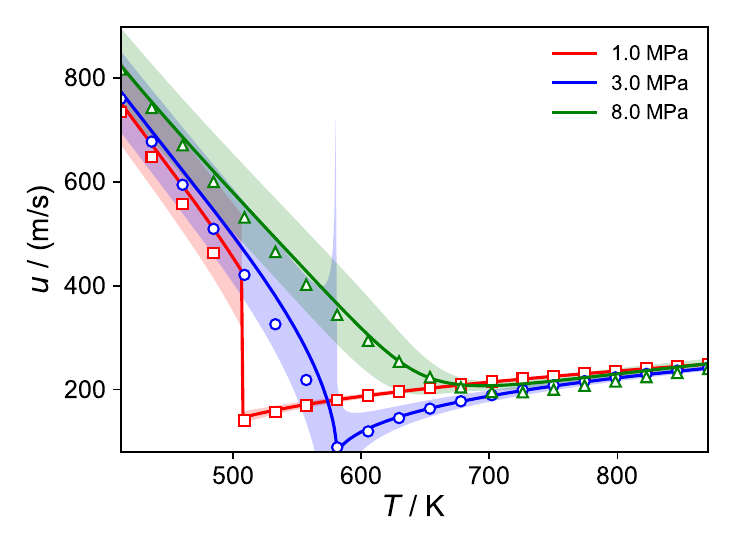}
      \caption{Speed of sound}
    \end{subfigure}
    \begin{subfigure}[b]{0.49\textwidth}
      \includegraphics[width=1\textwidth]{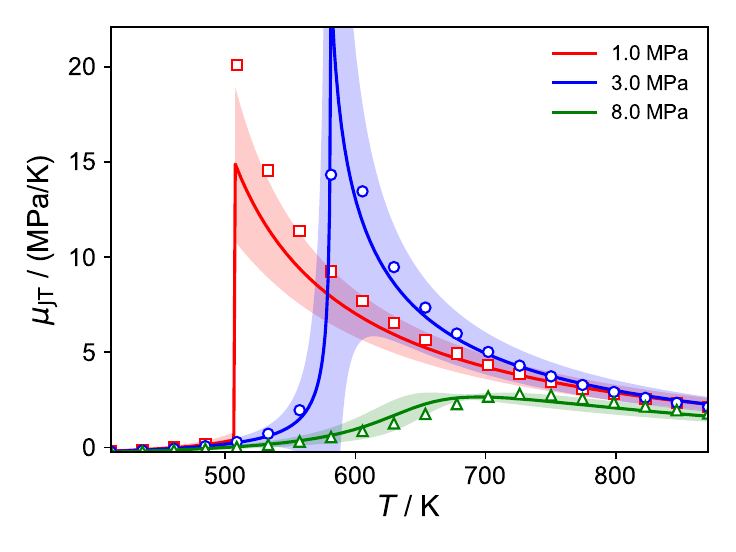}
      \caption{Joule--Thomson coefficient}
    \end{subfigure}
    \caption{Predicted values for the isobaric heat capacity, isochoric heat capacity, speed of sound and Joule--Thomson coefficient for $n$-octane using SAFT-VR Mie at different pressures. Shaded regions correspond to the uncertainty interval for the predicted properties using the parameters and confidence intervals.}
  \end{figure}
\newpage

\subsection{$n$-nonane}
\begin{figure}[H]
  \centering
      \includegraphics[width=0.5\textwidth]{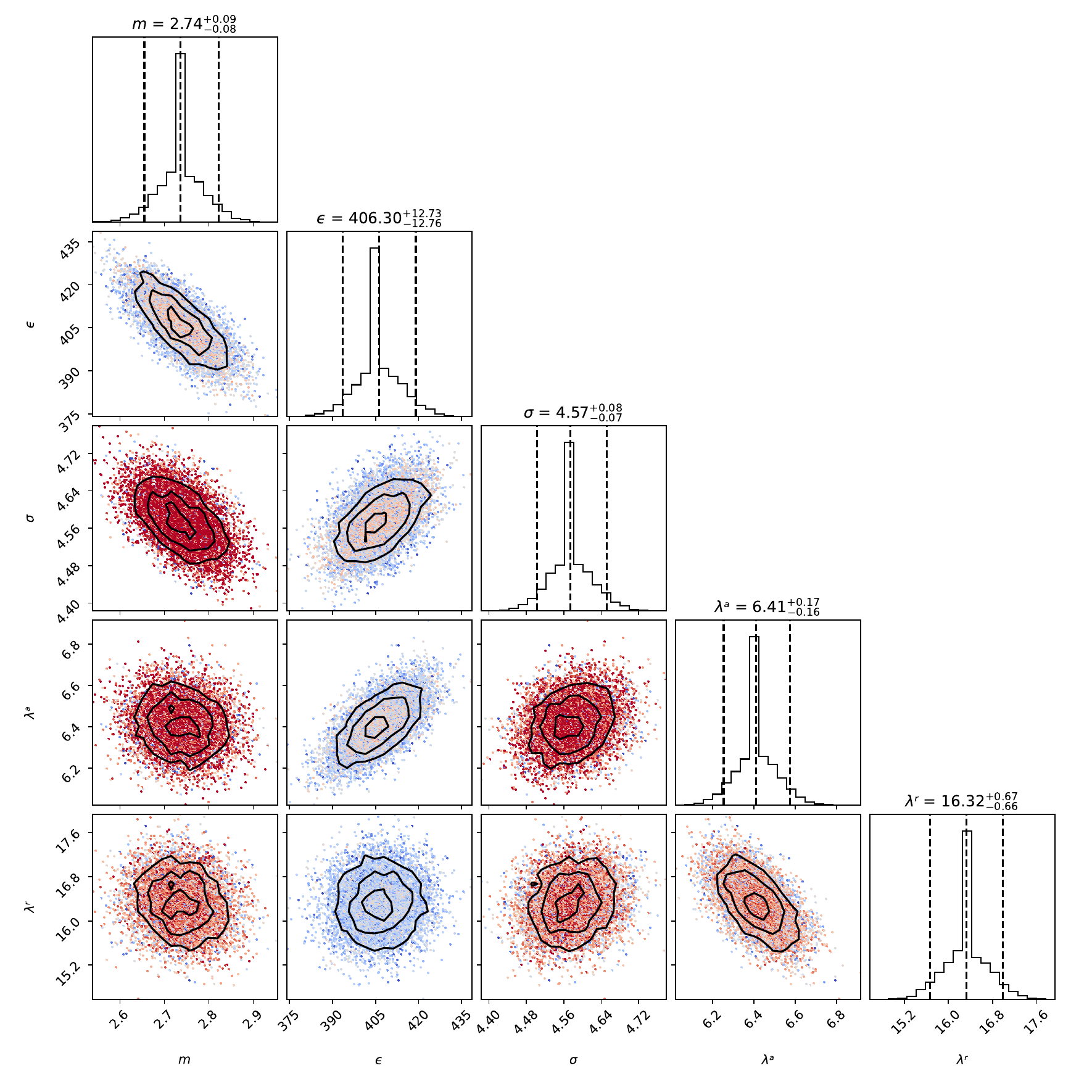}
      \caption{Confidence intervals obtained for the pure component parameters of $n$-nonane in SAFT-VR Mie. Colors and styles are identical to figure 2.}
  \end{figure}

\begin{figure}[H]
  \centering
    \begin{subfigure}[b]{0.49\textwidth}
      \includegraphics[width=1\textwidth]{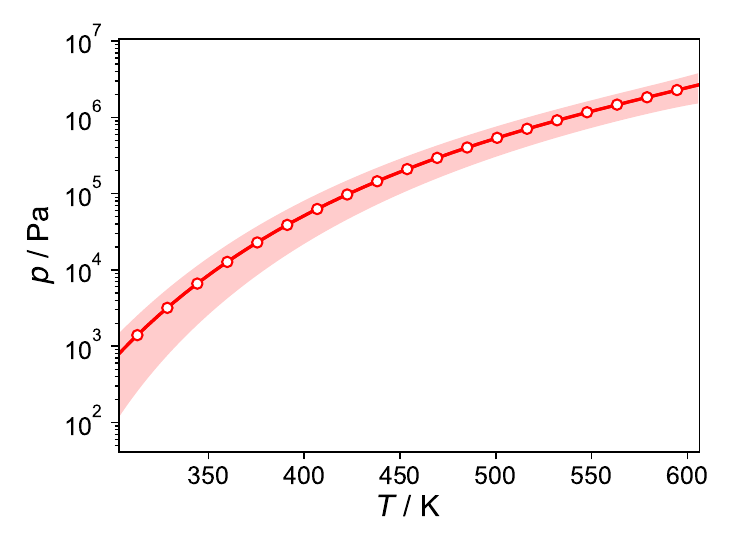}
      \caption{Saturation Curve}
    \end{subfigure}
    \begin{subfigure}[b]{0.49\textwidth}
      \includegraphics[width=1\textwidth]{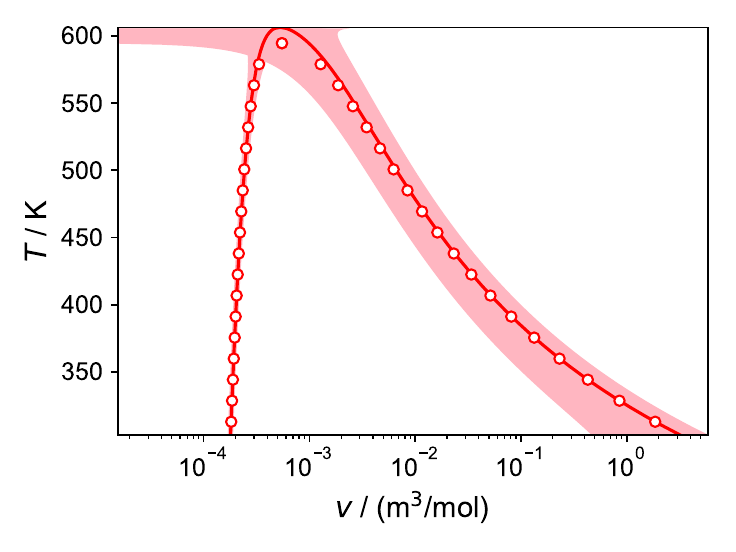}
      \caption{Saturation Envelope}

    \end{subfigure}
    \caption{Predicted values for the saturated volumes and saturation pressure for $n$-nonane using SAFT-VR Mie. Shaded regions correspond to the uncertainty interval for the predicted properties using the parameters and confidence intervals.}
  \end{figure}

\begin{figure}[H]
  \centering
    \begin{subfigure}[b]{0.49\textwidth}
      \includegraphics[width=1\textwidth]{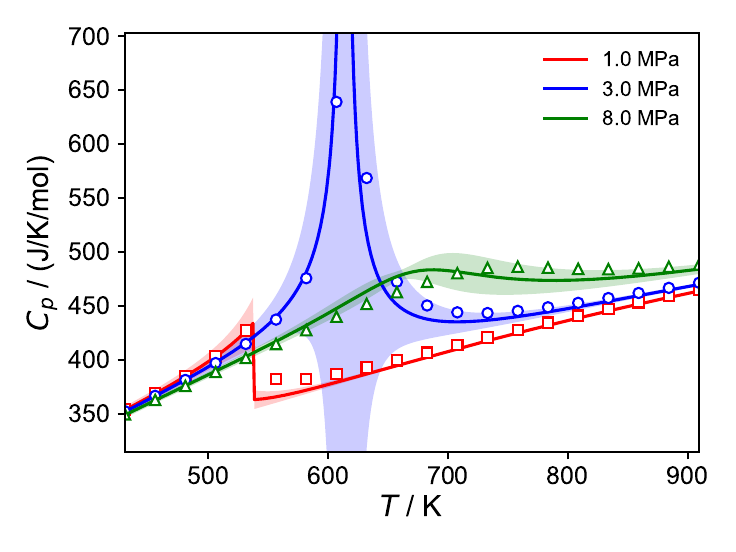}
      \caption{Isobaric heat capacity}
    \end{subfigure}
    \begin{subfigure}[b]{0.49\textwidth}
      \includegraphics[width=1\textwidth]{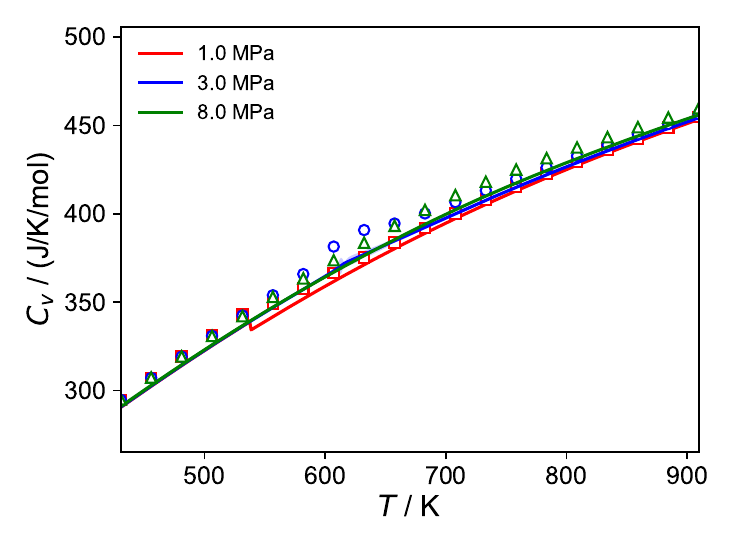}
      \caption{Isochoric heat capacity}
    \end{subfigure}
    \begin{subfigure}[b]{0.49\textwidth}
      \includegraphics[width=1\textwidth]{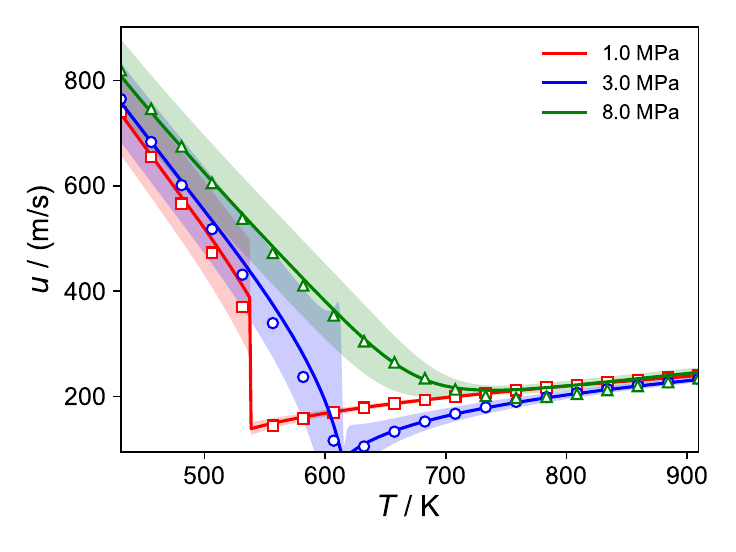}
      \caption{Speed of sound}
    \end{subfigure}
    \begin{subfigure}[b]{0.49\textwidth}
      \includegraphics[width=1\textwidth]{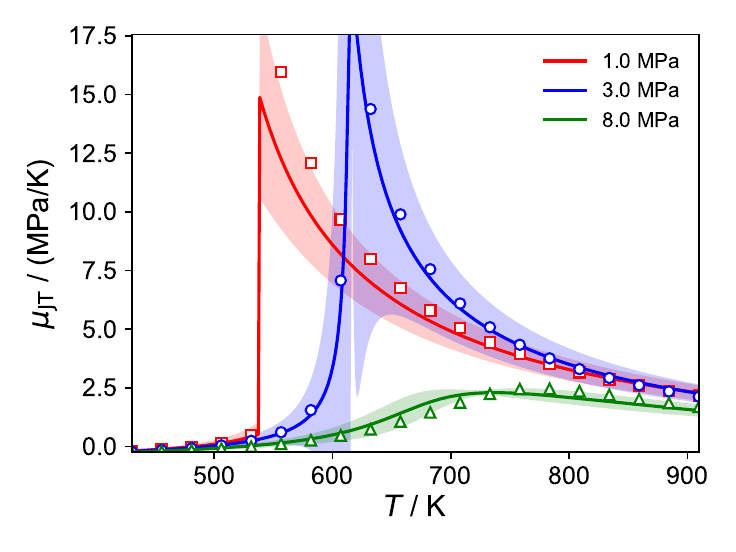}
      \caption{Joule--Thomson coefficient}
    \end{subfigure}
    \caption{Predicted values for the isobaric heat capacity, isochoric heat capacity, speed of sound and Joule--Thomson coefficient for $n$-nonane using SAFT-VR Mie at different pressures. Shaded regions correspond to the uncertainty interval for the predicted properties using the parameters and confidence intervals.}
  \end{figure}
\newpage

\subsection{$n$-decane}
\begin{figure}[H]
  \centering
      \includegraphics[width=0.5\textwidth]{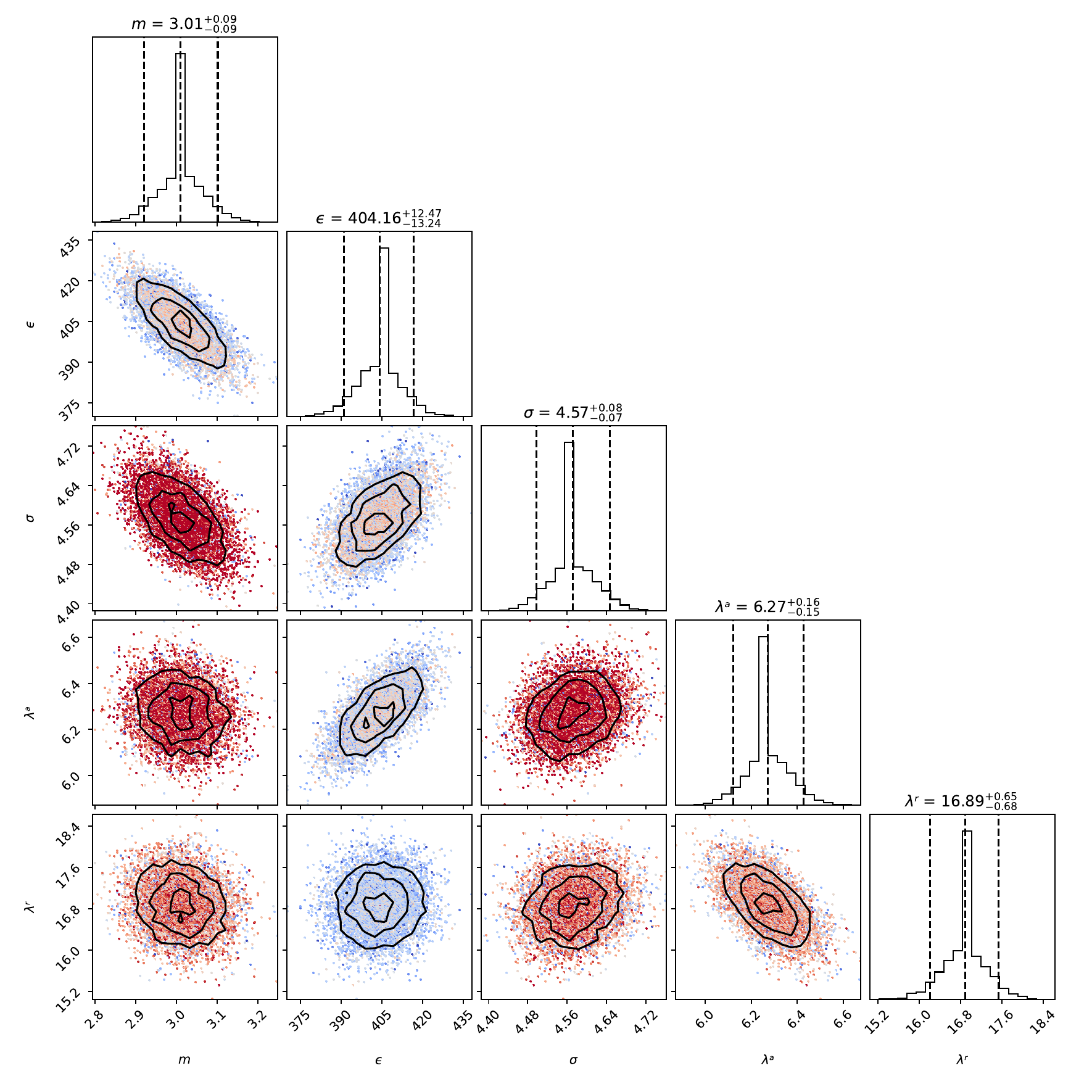}
      \caption{Confidence intervals obtained for the pure component parameters of $n$-decane in SAFT-VR Mie. Colors and styles are identical to figure 2.}
  \end{figure}

\begin{figure}[H]
  \centering
    \begin{subfigure}[b]{0.49\textwidth}
      \includegraphics[width=1\textwidth]{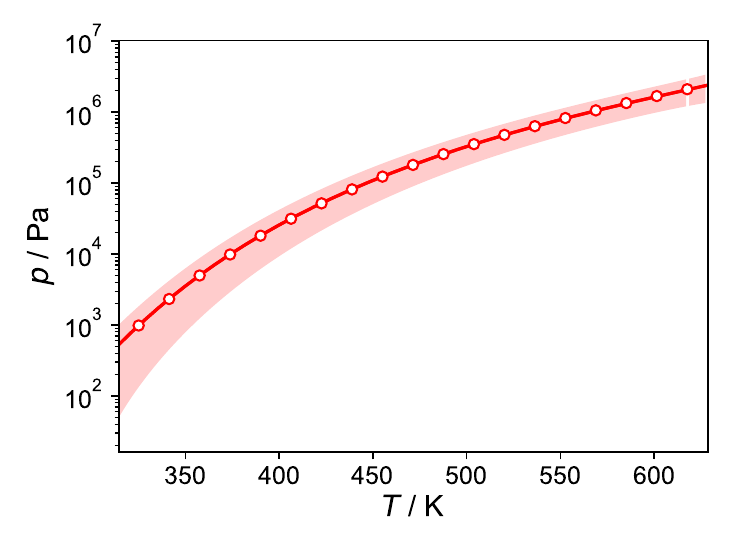}
      \caption{Saturation Curve}
    \end{subfigure}
    \begin{subfigure}[b]{0.49\textwidth}
      \includegraphics[width=1\textwidth]{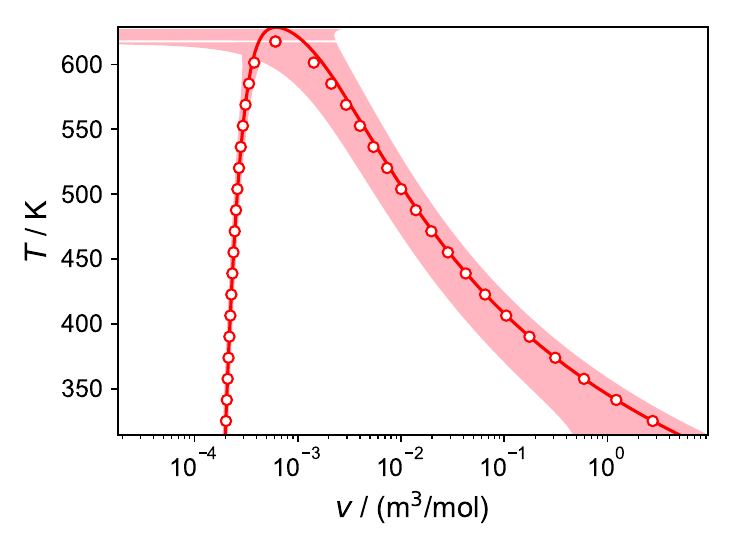}
      \caption{Saturation Envelope}

    \end{subfigure}
    \caption{Predicted values for the saturated volumes and saturation pressure for $n$-decane using SAFT-VR Mie. Shaded regions correspond to the uncertainty interval for the predicted properties using the parameters and confidence intervals.}
  \end{figure}

\begin{figure}[H]
  \centering
    \begin{subfigure}[b]{0.49\textwidth}
      \includegraphics[width=1\textwidth]{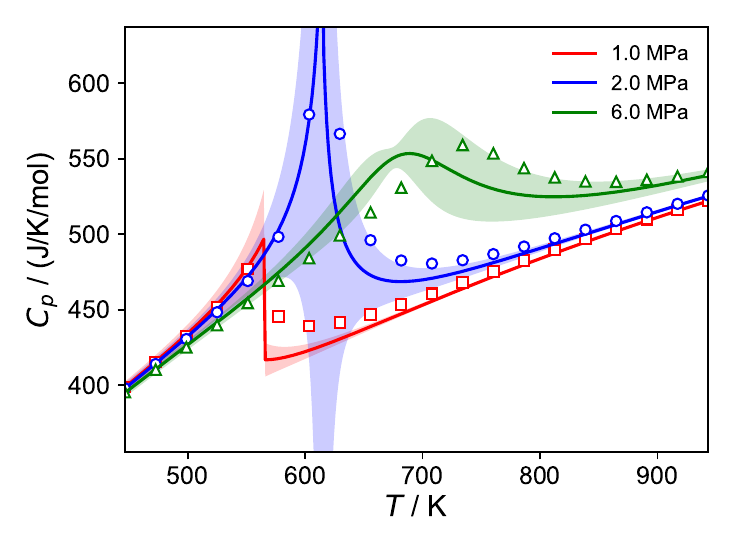}
      \caption{Isobaric heat capacity}
    \end{subfigure}
    \begin{subfigure}[b]{0.49\textwidth}
      \includegraphics[width=1\textwidth]{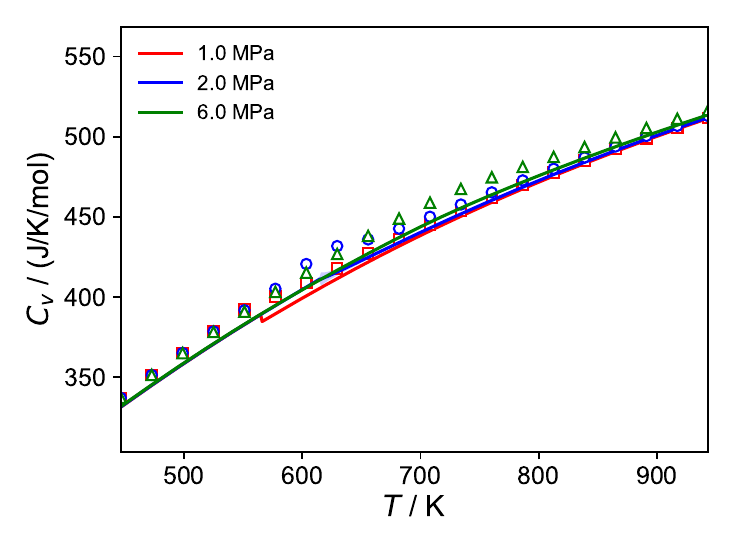}
      \caption{Isochoric heat capacity}
    \end{subfigure}
    \begin{subfigure}[b]{0.49\textwidth}
      \includegraphics[width=1\textwidth]{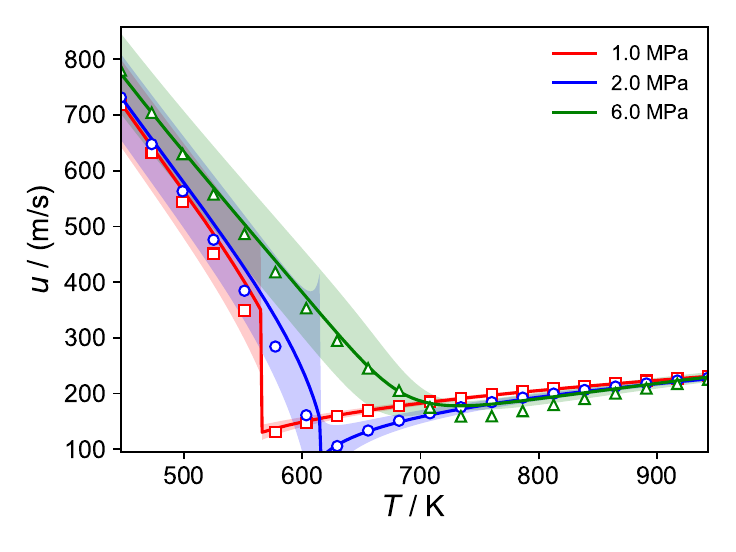}
      \caption{Speed of sound}
    \end{subfigure}
    \begin{subfigure}[b]{0.49\textwidth}
      \includegraphics[width=1\textwidth]{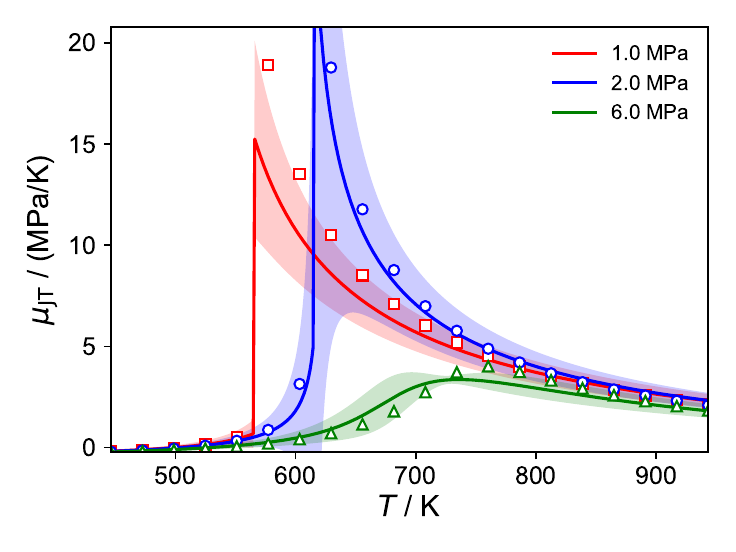}
      \caption{Joule--Thomson coefficient}
    \end{subfigure}
    \caption{Predicted values for the isobaric heat capacity, isochoric heat capacity, speed of sound and Joule--Thomson coefficient for $n$-decane using SAFT-VR Mie at different pressures. Shaded regions correspond to the uncertainty interval for the predicted properties using the parameters and confidence intervals.}
  \end{figure}
\newpage

\subsection{Water}
\begin{figure}[H]
  \centering
      \includegraphics[width=0.5\textwidth]{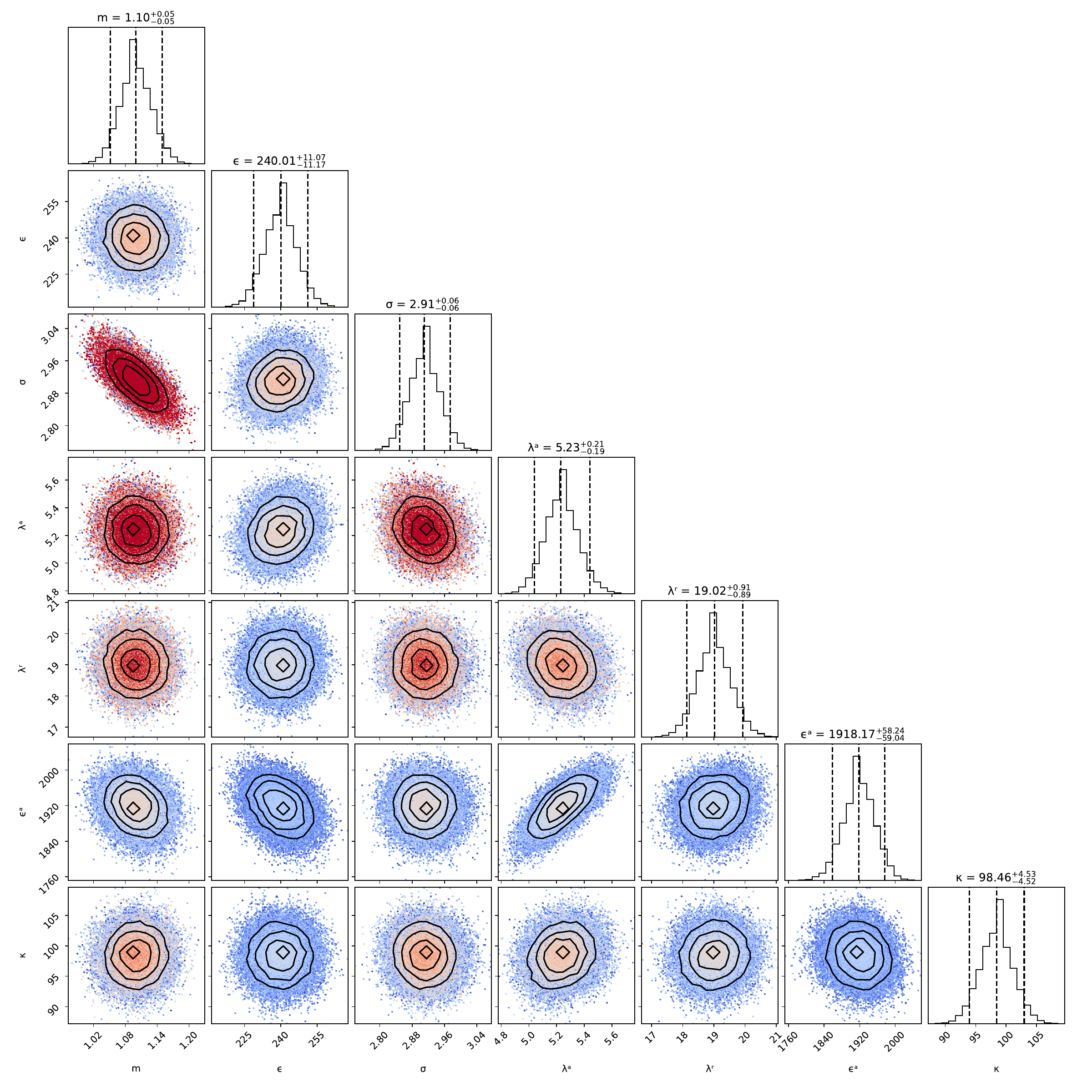}
      \caption{Confidence intervals obtained for the pure component parameters of water in SAFT-VR Mie. Colors and styles are identical to figure 2.}
  \end{figure}

\begin{figure}[H]
  \centering
    \begin{subfigure}[b]{0.49\textwidth}
      \includegraphics[width=1\textwidth]{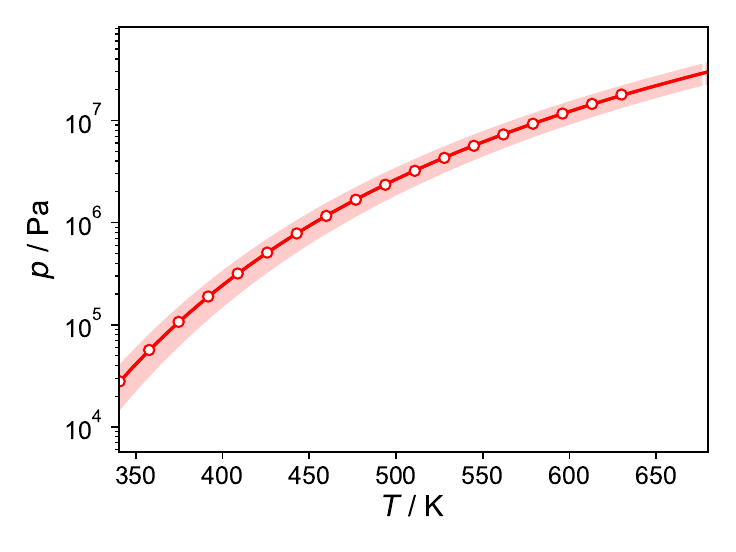}
      \caption{Saturation Curve}
    \end{subfigure}
    \begin{subfigure}[b]{0.49\textwidth}
      \includegraphics[width=1\textwidth]{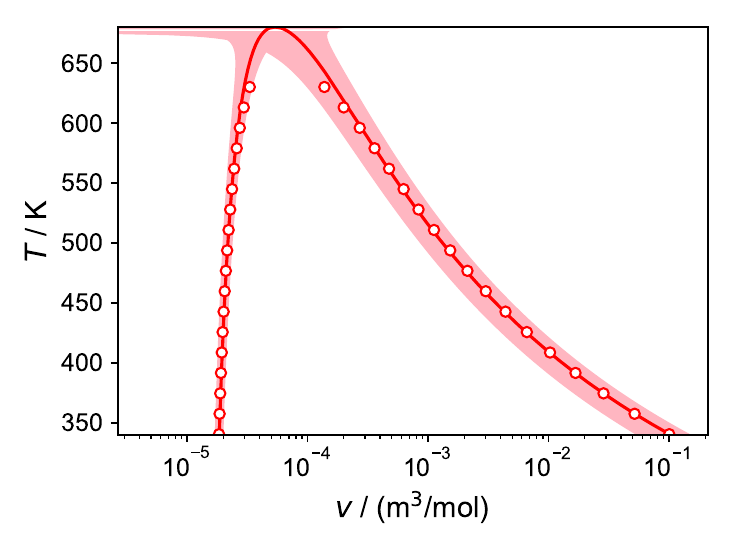}
      \caption{Saturation Envelope}

    \end{subfigure}
    \caption{Predicted values for the saturated volumes and saturation pressure for water using SAFT-VR Mie. Shaded regions correspond to the uncertainty interval for the predicted properties using the parameters and confidence intervals.}
  \end{figure}

\begin{figure}[H]
  \centering
    \begin{subfigure}[b]{0.49\textwidth}
      \includegraphics[width=1\textwidth]{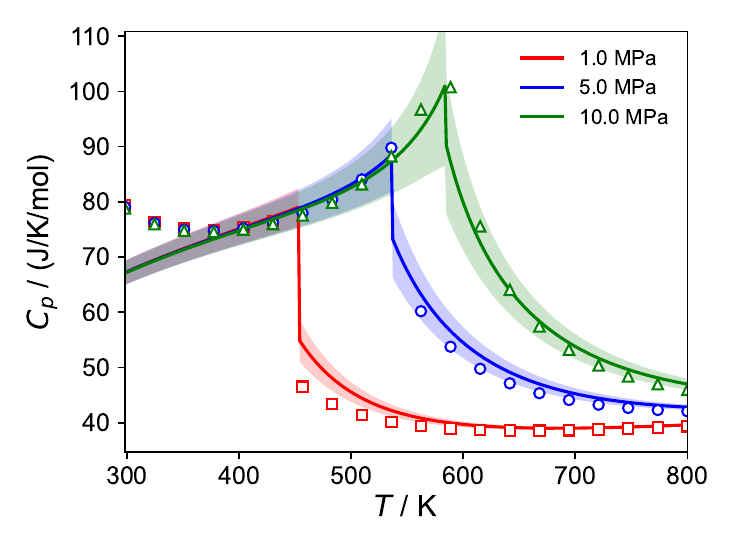}
      \caption{Isobaric heat capacity}
    \end{subfigure}
    \begin{subfigure}[b]{0.49\textwidth}
      \includegraphics[width=1\textwidth]{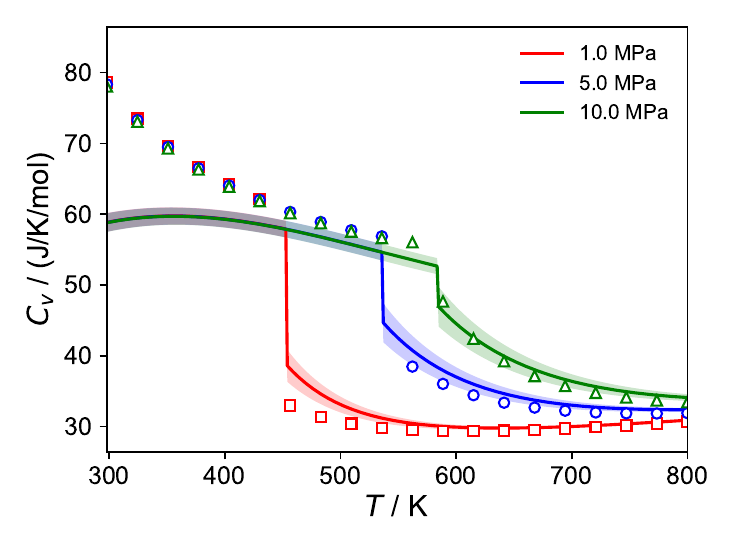}
      \caption{Isochoric heat capacity}
    \end{subfigure}
    \begin{subfigure}[b]{0.49\textwidth}
      \includegraphics[width=1\textwidth]{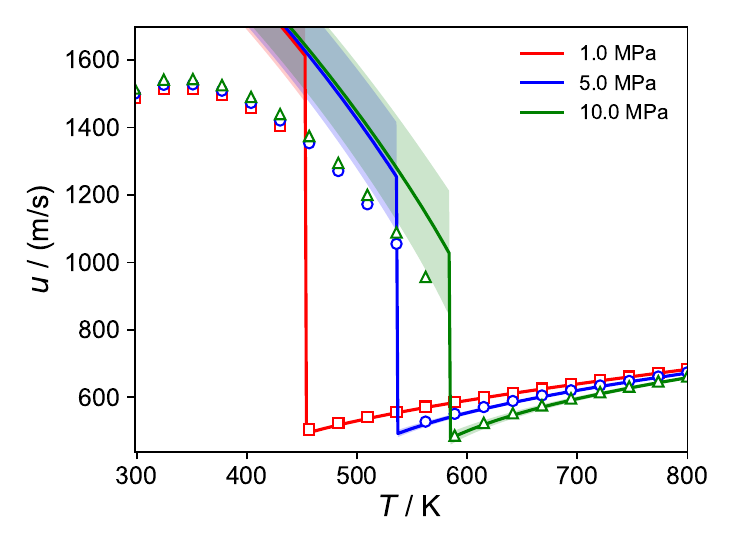}
      \caption{Speed of sound}
    \end{subfigure}
    \begin{subfigure}[b]{0.49\textwidth}
      \includegraphics[width=1\textwidth]{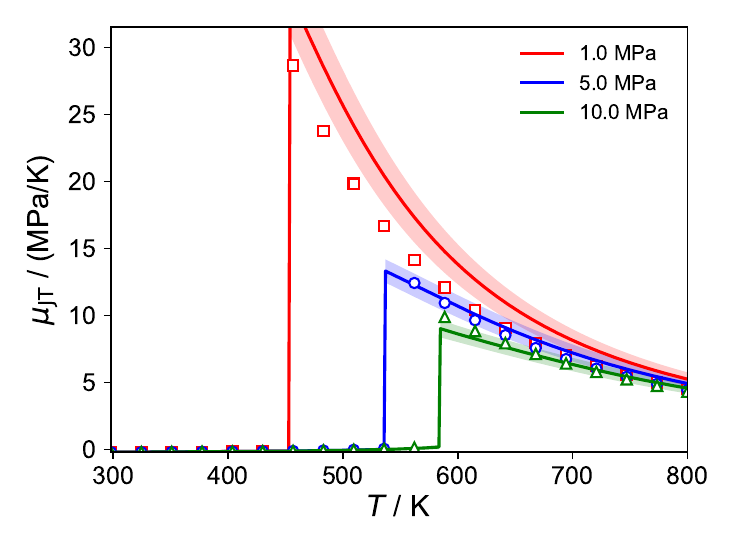}
      \caption{Joule--Thomson coefficient}
    \end{subfigure}
    \caption{Predicted values for the isobaric heat capacity, isochoric heat capacity, speed of sound and Joule--Thomson coefficient for water using SAFT-VR Mie at different pressures. Shaded regions correspond to the uncertainty interval for the predicted properties using the parameters and confidence intervals.}
  \end{figure}
\newpage
\subsection{Methanol}
\begin{figure}[H]
  \centering
      \includegraphics[width=0.5\textwidth]{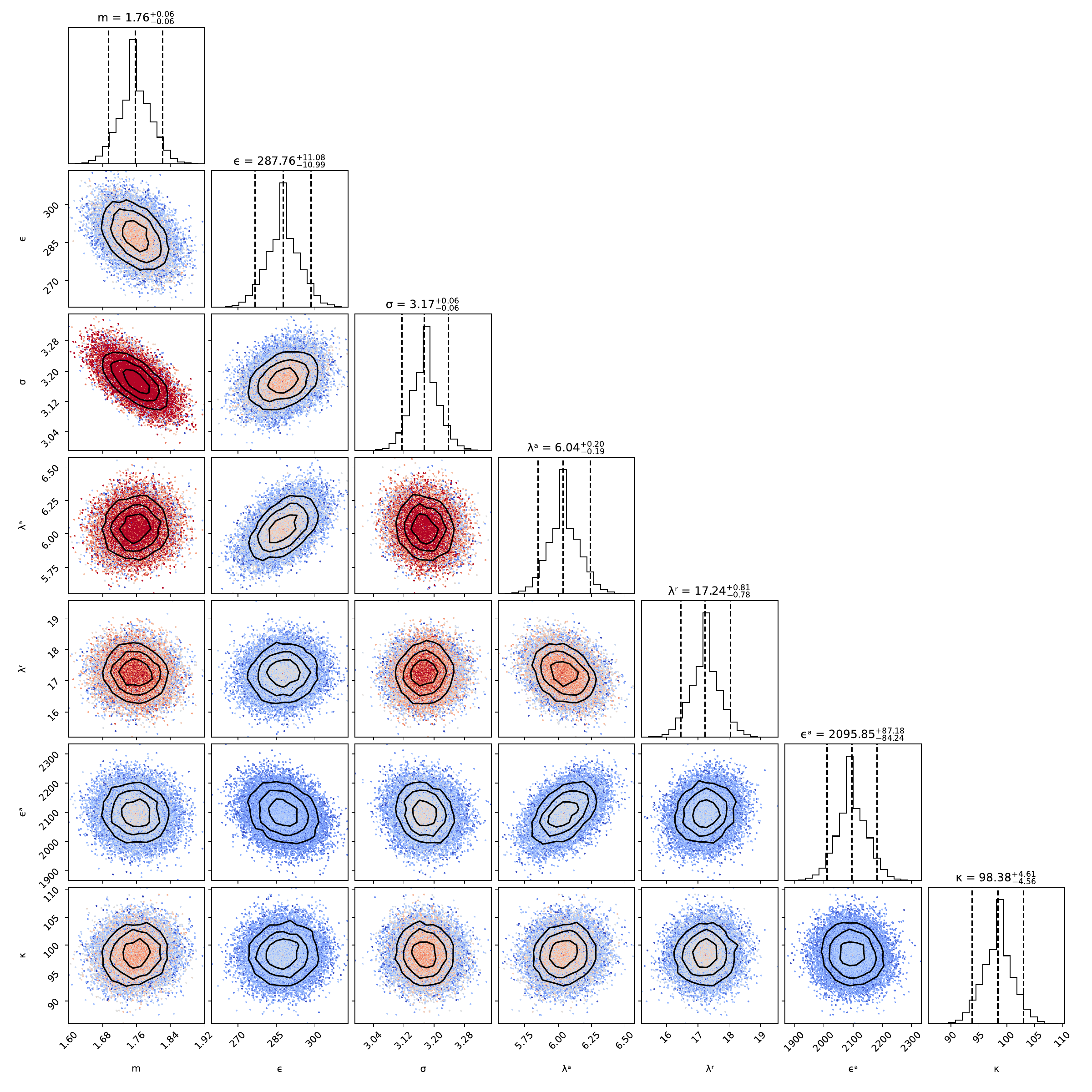}
      \caption{Confidence intervals obtained for the pure component parameters of methanol in SAFT-VR Mie. Colors and styles are identical to figure 2.}
  \end{figure}